\documentclass[twocolumn]{aastex63}

\usepackage{color}
\usepackage{graphicx}
\usepackage{amsmath}
\usepackage{amssymb}
\usepackage{appendix}
\usepackage{longtable}
\usepackage{CJK}
\hypersetup{pdfauthor={Name}}

\newcommand{\totpos}{107}

\newcommand{\totobsconstraint}{17}

\newcommand{\totsamp}{90}

\newcommand{\totbrightphot}{245}

\newcommand{\totbrightphotnum}{49}

\newcommand{\totlitphotnum}{45}

\newcommand{\totarchivephotnum}{11}

\newcommand{\totfits}{335}

\newcommand{\totassoc}{84}

\newcommand{\totnewspec}{25}

\newcommand{\totspec}{40}

\newcommand{\totoffset}{83}

\def\ra#1#2#3{#1$^{\rm h}$#2$^{\rm m}$#3$^{\rm s}$}
\def\dec#1#2#3{$#1^\circ#2'#3''$}
\def\nod{\nodata}
\def\swift{{\it Swift}}

\graphicspath{{./}{figures/}}

\shorttitle{Short GRB Hosts I: Catalog}
\shortauthors{Fong et al.}

\begin{document}
\begin{CJK*}{UTF8}{gbsn}

\title{Short GRB Host Galaxies I: Photometric and Spectroscopic Catalogs, Host Associations, and Galactocentric Offsets}

\correspondingauthor{W. Fong}
\email{wfong@northwestern.edu}

\newcommand{\NU}{\affiliation{Center for Interdisciplinary Exploration and Research in Astrophysics (CIERA) and Department of Physics and Astronomy, Northwestern University, Evanston, IL 60208, USA}}

\newcommand{\CfA}{\affiliation{Center for Astrophysics\:$|$\:Harvard \& Smithsonian, 60 Garden St. Cambridge, MA 02138, USA}}

\newcommand{\PSU}{\affiliation{Department of Astronomy \& Astrophysics, The Pennsylvania State University, University Park, PA 16802, USA}}

\newcommand{\CM}{\affiliation{College of Marin, Kentfield, CA 94904, USA}}

\newcommand{\ICDS}{\affiliation{Institute for Computational \& Data Sciences, The Pennsylvania State University, University Park, PA, USA}}

\newcommand{\LANL}{\affiliation{Center for Theoretical Astrophysics, Los Alamos National Laboratory, Los Alamos, NM, 87545, USA}}

\newcommand{\UCB}{\affiliation{Department of Astronomy, University of California, Berkeley, CA 94720-3411, USA}}

\newcommand{\SU}{\affiliation{The Oskar Klein Centre, Department of Astronomy, Stockholm University, AlbaNova, SE-106 91 Stockholm, Sweden}}

\newcommand{\UA}{\affiliation{University of Arizona, Steward Observatory, 933~N.~Cherry~Ave., Tucson, AZ 85721, USA}}

\newcommand{\bham}{\affiliation{Birmingham Institute for Gravitational Wave Astronomy and School of Physics and Astronomy, University of Birmingham, Birmingham B15 2TT, UK}}

\newcommand{\mpia}{\affiliation{Max-Planck-Institut f\"{u}r Astronomie (MPIA), K\"{o}nigstuhl 17, 69117 Heidelberg, Germany}}

\newcommand{\GWU}{\affiliation{Department of Physics, The George Washington University, Washington, DC 20052, USA}}

\newcommand{\LJMU}{\affiliation{Astrophysics Research Institute, Liverpool John Moores University, IC2, Liverpool Science Park, 146 Brownlow Hill, Liverpool L3 5RF, UK}}

\newcommand{\GNL}{\affiliation{Gemini Observatory/NSF's NOIRLab, 670 N. A'ohoku Place, Hilo, HI 96720, USA}}

\newcommand{\Radboud}{\affiliation{Department of Astrophysics/IMAPP, Radboud University, 6525 AJ Nijmegen, The Netherlands}}

\newcommand{\Leicester}{\affiliation{School of Physics and Astronomy, University of Leicester, University Road, Leicester, LE1 7RH, UK}}

\newcommand{\LCO}{\affiliation{Las Cumbres Observatory, 6740 Cortona Drive, Suite 102, Goleta, CA 93117-5575, USA}}

\newcommand{\UCSB}{\affiliation{Department of Physics, University of California, Santa Barbara, CA 93106-9530, USA}}

\author[0000-0002-7374-935X]{Wen-fai Fong}
\NU

\author[0000-0002-2028-9329]{Anya E. Nugent}
\NU

\author[0000-0002-9363-8606]{Yuxin Dong}
\NU

\author[0000-0002-9392-9681]{Edo Berger}
\CfA

\author[0000-0001-8340-3486]{Kerry Paterson}\mpia

\author[0000-0002-7706-5668]{Ryan Chornock}
\UCB

\author[0000-0001-7821-9369]{Andrew Levan}\Radboud

\author[0000-0003-0526-2248]{Peter Blanchard}
\NU

\author[0000-0002-8297-2473]{Kate D. Alexander}
\NU

\author[0000-0003-0123-0062]{Jennifer Andrews}
\GNL

\author[0000-0002-9118-9448]{Bethany E. Cobb}
\GWU

\author[0000-0001-6455-5660]{Antonino Cucchiara}
\CM

\author[0000-0002-3714-672X]{Derek Fox}
\PSU

\author[0000-0003-2624-0056]{Chris L. Fryer}
\LANL

\author[0000-0002-5025-4645]{Alexa C. Gordon}
\NU

\author[0000-0002-5740-7747]{Charles D. Kilpatrick}\NU

\author[0000-0001-9454-4639]{Ragnhild Lunnan} 
\SU

\author[0000-0003-4768-7586]{Raffaella Margutti}
\UCB

\author[0000-0001-9515-478X]{Adam Miller}
\NU

\author[0000-0002-0370-157X]{Peter Milne}\UA

\author[0000-0002-2555-3192]{Matt Nicholl}\bham

\author[0000-0001-8472-1996]{Daniel Perley}\LJMU

\author[0000-0002-9267-6213]{Jillian Rastinejad}
\NU

\author[0000-0003-3937-0618]{Alicia Rouco Escorial}
\NU

\author[0000-0001-9915-8147]{Genevieve Schroeder}\NU

\author[0000-0001-5510-2424]{Nathan Smith}\UA

\author[0000-0003-3274-6336]{Nial Tanvir}\Leicester

\author[0000-0003-0794-5982]{Giacomo Terreran}
\LCO\UCSB

\begin{abstract}
We present a comprehensive optical and near-infrared census of the fields of \totsamp\ short gamma-ray bursts (GRBs) discovered in 2005-2021, constituting all short GRBs for which host galaxy associations are feasible ($\approx 60\%$ of the total {\it Swift} short GRB population). We contribute \totbrightphot\ new multi-band imaging observations across \totbrightphotnum\ distinct GRBs and \totnewspec\ spectra of their host galaxies. Supplemented by literature and archival survey data, the catalog contains \totfits\ photometric and \totspec\ spectroscopic data sets. The photometric catalog reaches $3\sigma$ depths of $\gtrsim 24-27$~mag and $\gtrsim 23-26$~mag for the optical and near-infrared bands, respectively. We identify host galaxies for \totassoc\ bursts, in which the most robust associations make up 54\% (49/\totsamp) of events, while only a small fraction, 6.7\%, have inconclusive host associations. Based on new spectroscopy, we determine 17 host spectroscopic redshifts with a range of $z\approx 0.15-1.6$ and find that $\approx 25-44\%$ of {\it Swift} short GRBs originate from $z>1$. We also present the galactocentric offset catalog for \totoffset\ short GRBs. Taking into account the large range of individual measurement uncertainties, we find a median of projected offset of $\approx 7.9$~kpc, for which the bursts with the most robust associations have a smaller median of $\approx 4.9$~kpc. Our catalog captures more high-redshift and low-luminosity hosts, and more highly-offset bursts than previously found, thereby diversifying the population of known short GRB hosts and properties. In terms of locations and host luminosities, the populations of short GRBs with and without detectable extended emission are statistically indistinguishable. This suggests that they arise from the same progenitors, or from multiple progenitors which form and evolve in similar environments. All of the data products are available on the {\it Broadband Repository for Investigating Gamma-ray burst Host Traits} (BRIGHT) website.
\end{abstract}


\section{Introduction}
\label{sec:intro}

The host galaxies of astrophysical transients provide crucial insight on the nature of their progenitors. For instance, core-collapse supernovae (CCSNe), long-duration $\gamma$-ray bursts (GRBs), and superluminous SNe are almost exclusively found to occur in star-forming galaxies (e.g., \citealt{sgl09,slt+10,lcl+11,lcb+14,pqy+16,sys+21}), helping to establish their progenitors as massive stars. The rate of these events thus traces recent star formation (e.g., \citealt{pks+16}). In contrast, Type Ia SNe originate in a mix of star-forming and quiescent galaxies \citep{vlf05}, consistent with an evolved progenitor and an event rate that traces both stellar mass and star formation \citep{slp+06}.

Short-duration GRBs are relativistic explosions with prompt gamma-ray emission durations of $\lesssim\!2$~seconds \citep{kmf+93}, beaming-corrected total energy scales of $\approx10^{50}$~erg \citep{fbm+15}, and synchroton afterglow radiation across the electromagnetic spectrum. Launched in 2004, NASA's Neil Gehrels {\it Swift} Observatory \citep{ggg+04} has served as the primary workhorse for short GRB discovery and precise localization. The detection of X-ray afterglows allows {\it Swift} to localize $\approx 70\%$ of short GRBs which it discovers to within just a few arcseconds, resulting in $\sim 8$ such short GRBs per year \citep{lsb+16}. Critically, these localizations enable robust associations between GRBs and their host galaxies. With host galaxies, one can discern fundamental properties such as redshifts and burst energy scales, as well as properties of the environment on sub-galactic to kiloparsec scales.

The locations of transients with respect to their host galaxies also provide crucial diagnostics into their origins. While transients originating from massive stars (long GRBs, CCSNe, SLSNe) are typically located in or proximal to regions of active star formation \citep{fls+06,kkp08,lcb+15,bbf16,llt+17,ama+20}, short GRBs often occur several kiloparsecs from their host galaxies, with locations only weakly correlated with the stellar light distributions of their hosts \citep{ber10,fb13,tlt+14,zkn+20,otd+22}. Moreover, high angular resolution observations have revealed weak correlations between short GRB locations and the distributions of their host stellar mass or star formation \citep{fb13}. These studies provide some of the most compelling indirect evidence to date that their progenitors migrate from their birth sites to explosion sites, matching the hallmark prediction of binary neutron star (BNS) and neutron star-black hole (NSBH) mergers \citep{fwh99}.

With the 2017 joint detection of the BNS merger GW170817 in conjunction with a short GRB, we now have direct evidence that BNS mergers are the progenitors of at least some short GRBs \citep{gw170817grb,gvb+17,sfk+17,mc21}. With the ground-breaking discoveries of the first definitive BNS mergers provided by gravitational wave (GW) facilities \citep{gw170817,gw190425}, and the promise of more to come in the very near future \citep{GWProspects20}, it is especially timely to perform a uniform and careful study of {\it Swift} short GRB environments, which serve as a cosmological comparison dataset out to a redshift of $z\approx 2.5$ \citep{ber14,skm+18,pfn+20,otd+22}. At present, there is heightened interest in short GRBs, their inference on heavy element nucleosynthesis, and the crucial role they play in understanding the evolution of mergers over cosmic time.

The existing sample of cosmological short GRB ($z \approx 0.1$-$2.5$) is much larger than the two confirmed BNS merger detections from GWs to date \citep{gw170817,gw190425}, and it will be many years before GW-detected mergers yield a comparable sample of well-localized events based on expected rates \citep{gwtc-1}. Aside from GW events, nearly all of our observational constraints on BNS systems originate from the $\sim$19 known Galactic BNS systems \citep{wkh04,wwk10,tkf+17,vns+18}, a population that suffers from various selection biases. For instance, it is challenging to identify tight (short delay-time) Galactic BNS systems due to signal-smearing, as well as systems with large orbital separations (long delay-times) due to the small relative changes in proper motions \citep{tkf+17}. On the other hand, short GRBs are detected to cosmological redshifts via $\gamma$-ray emission, and represent a large and complementary data set of merging systems. Thus, to fully understand how BNS/NSBH binaries form and merge across cosmic history, and to provide a legacy comparison data set for future GW events, it is critical to identify and characterize the host galaxies of as many short GRBs as possible.

The first decade of short GRB host galaxy studies primarily concentrated on those bursts with sub-arcsecond localizations via the detection of optical afterglows \citep{vlr+05,ffp+05,bfp+07,dmc+09,fbc+13,ber14,pas19}. However, the selection based on optical afterglows may bias the host galaxy sample and interpretation for their progenitors, as has been shown for long GRB hosts (e.g., \citealt{hmj+12}). Now, we are well into the second decade of afterglow discoveries, and are equipped with over 100 short GRBs localized primarily by X-ray afterglows.

Here, we present a comprehensive census of the locations and environments of the {\it Swift} short GRB population, representing a decade-long observational campaign to identify and characterize as many short GRB host galaxies as possible, irrespective of the detection of an optical afterglow. This work represents the first of a series of two papers. Paper I focuses on the photometric and spectroscopic catalogs, host galaxy associations, spectroscopic redshifts, and galactocentric offsets. Paper II, \citet{BRIGHT-II}, focuses on spectral energy distribution (SED) modeling of these data, their inferred stellar population properties and implications for the progenitors. We house all of the data and modeling products in these works on the Broadband Repository for Investigating Gamma-ray burst Host Traits (BRIGHT) website\footnote{\url{http://bright.ciera.northwestern.edu}}. 

In Section~\ref{sec:sample}, we describe our sample of \totsamp\ events. In Section~\ref{sec:obs}, we introduce \totbrightphot\ photometric observations across \totbrightphotnum\ distinct short GRBs taken with 4-m to 10-m ground-based telescopes and the {\it Hubble Space Telescope} ({\it HST}). In Section~\ref{sec:assoc}, we describe the process for host association and report associations to \totassoc\ events with varying degrees of robustness. In Section~\ref{sec:specobs}, we present \totnewspec\ spectroscopic observations of short GRB hosts and redshifts for 17 events. In Section~\ref{sec:offsets}, we report the galactocentric offsets for \totoffset\ events (angular, physical and host-normalized when available), and compare the distributions to long GRBs and NS merger models. We discuss the implications of the results, selection effects, and assessment of contamination to our sample in Section~\ref{sec:disc}. We conclude in Section~\ref{sec:conc}.

Unless otherwise stated, all observations are reported in the AB magnitude system and have been corrected for Galactic extinction in the direction of the GRB \citep{sf11} and employed the \citet{ccm89} extinction
law. We employ a standard cosmology of $H_{0}$ = 69.6~km~s$^{-1}$~Mpc$^{-1}$, $\Omega_{M}$ = 0.286, $\Omega_{vac}$ = 0.714 \citep{blw+14}.

\section{Sample Description}
\label{sec:sample}

\begin{figure*}[t]
\centering
\includegraphics[width=0.45\textwidth]{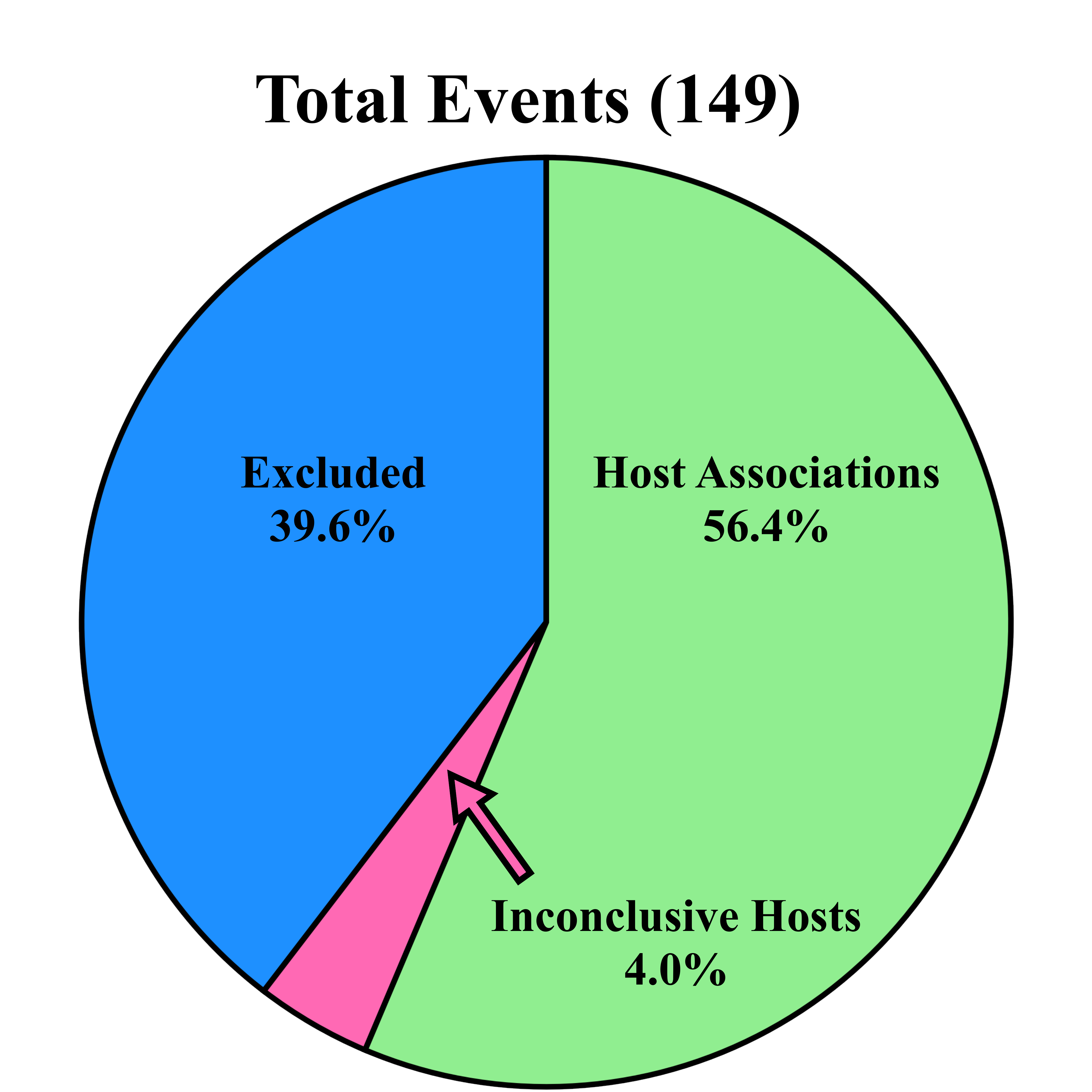}
\includegraphics[width=0.45\textwidth]{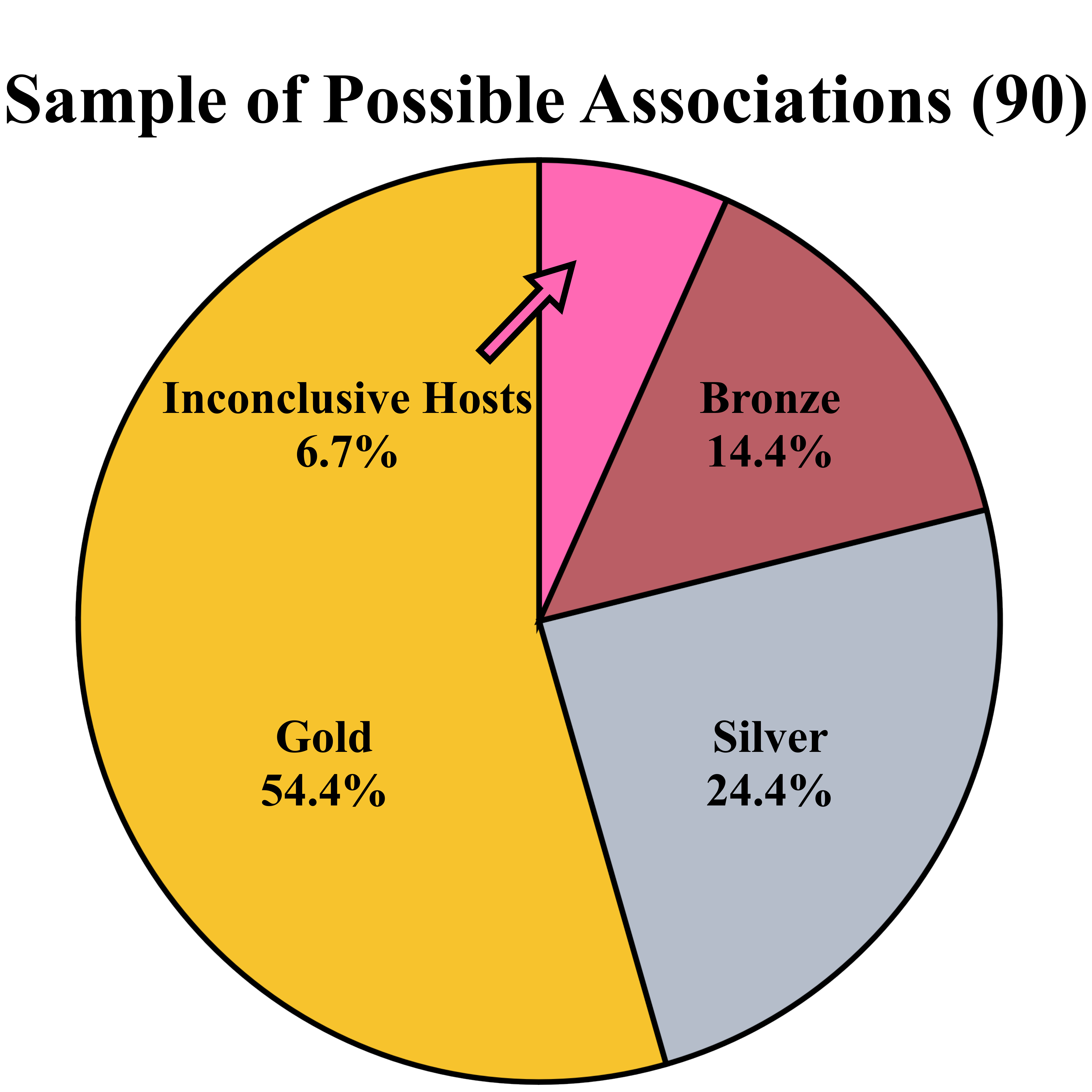}
\caption{{\it Left:} Classification of all short GRBs in the starting sample of short GRBs primarily discovered by {\it Swift}. Of the 149 total detected {\it Swift} short GRBs, host associations exist for $\approx 56\%$ of the population, while only $\approx 4\%$ have inconclusive hosts. We exclude $\approx 40\%$ of the total population because they have no reported afterglows, afterglow localizations too large to enable meaningful host galaxy searches, or are events which are subject to optical observing constraints. {\it Right:} Classification of the \totsamp\ short GRBs in our sample for which meaningful host searches and follow-up is feasible. We report host associations for $\approx 93.3\%$ of events (\totassoc); the most robust associations ($P_{\rm cc,min}\lesssim 0.02$) comprise over half of events (Gold sample). We cannot make conclusive host associations ($P_{\rm cc,min}\gtrsim 0.25$) for $\approx 6.7\%$ of events. 
\label{fig:pie}}
\end{figure*}

We begin with all short GRBs discovered by NASA's \swift\ Observatory \citep{ggg+04} since its launch in 2004, as well as two short GRBs discovered by the High Energy Transient Explorer (HETE-2), but with $\lesssim$~1 arcsec localizations via the detection of afterglows (GRBs\,050709 and 060121). We include events which meet both of the following criteria: (i) GRBs with {\it Swift} $\gamma$-ray durations of $T_{90} \leq 2$~s (15-350~keV) or those which are classified as short or possibly-short with extended emission (short GRB-EE; according to \citealt{nb06,lsb+16}), and (ii) bursts with detected afterglows with $\lesssim 5''$-radius precision, which typically enable associations to  host galaxies. Such afterglow discoveries primarily come from the {\it Swift} X-ray Telescope (XRT), which routinely provides localizations of $\approx 2-5\arcsec$ in radius (90\% confidence; \citealt{ebp+09}), as well as ground-based optical facilities and the {\it Chandra X-ray Observatory} ({\it CXO}). We note that by using observable, as opposed to rest-frame quantities, we are able to be inclusive of the sizable fraction of bursts with unknown redshift. We note that we explore possible selection effects and contamination in Section~\ref{sec:disc}.

We additionally include three events which nominally have long durations but likely do not originate from massive star progenitors: GRBs\,060614, 160303A, and 211211A. GRB\,060614 has $T_{90}=108.7$~sec and is classified as a possible short GRB-EE \citep{lsb+16}, with a spectral lag and $\gamma$-ray luminosity completely consistent with the short-hard GRB population \citep{gnb+06,lsb+16}. Additionally, with the lack of associated SN to deep limits, there is a general consensus that this event did not originate from a massive star collapse \citep{dcp+06,gfp+06,fwt+06}.  Moreover, there is tentative evidence for a photometric excess interpreted as a kilonova \citep{jlc+15}. Thus, we include this burst in our sample. The {\it Swift}/BAT light curve of GRB\,160303A exhibits a $\sim0.4$~s spike followed by a low-significance tail to $\sim 5$~s, with a spectral lag of (measured in the 100-350 keV to 25-50 keV bands) of $24 \pm 24$~ms, yielding inconclusive results as to its classification from the $\gamma$-ray properties alone \citep{gcn19148}. However, the lack of clear emission lines from the host galaxy in its afterglow spectrum \citep{gcn19154}, coupled with its large offset (Section~\ref{sec:offset_distribution}) indicate a GRB with an older stellar progenitor, and we thus include this burst in our sample. Finally, GRB\,211211A has $T_{90} \approx 51.4$~s and is in the long GRB class based on its $\gamma$-ray hardness and duration. However, this event was followed by near-infrared (NIR) transient emission interpreted as an $r$-process kilonova and also has deep limits on an associated supernova. Taken together, this event likely originated from a NS merger \citep{rgl+22}. Including these three bursts, there are \totpos\ events discovered over 2004-2021 which meet the criteria for our starting sample.

From this sample, we exclude \totobsconstraint\ short GRBs with sight-lines that are subject to observing constraints which prevent meaningful host galaxy follow-up. Such constraints include significant contamination of the afterglow position by a foreground star, high-extinction sight-lines from the Galaxy ($A_V \gtrsim 2$~mag; \citealt{sf11}), or crowded fields. Thus, our starting sample of \totsamp\ short GRBs with positions and sight-lines that enable host galaxy searches and follow-up corresponds to $\approx 60\%$ of the total {\it Swift} population (Figure~\ref{fig:pie}).

\section{Photometric Catalog \& Observations}
\label{sec:obs}

The first goal of our study is to build a multi-band photometric catalog for the locations of the \totsamp\ short GRBs in the sample in order to identify new host galaxies or confirm previously reported ones. Once we establish a host galaxy, we obtain imaging in multiple filters and/or spectroscopy (Section~\ref{sec:specobs}) to characterize their spectral energy distributions (SEDs). The photometric part of the catalog is comprised of ground-based and {\it HST} observations, supplemented by published literature and archival survey data. Here we describe the imaging data in the catalog, data reduction, and photometric methods.

In total, we newly contribute \totbrightphot\ observations in various bands across \totbrightphotnum\ distinct short GRBs. We supplement this with literature and archival data for a total of \totfits\ photometric data points and imaging products across \totsamp\ events in the photometric catalog. The photometry and host galaxy positions are listed in Appendix Table~\ref{tab:phot}.

\subsection{Ground-based Imaging for Host Galaxy Searches}

\begin{deluxetable*}{lllc}
\linespread{1.2}
\tabletypesize{\scriptsize}
\tablecaption{Telescopes, Instruments, Photometric or Spectroscopic Set-ups \label{tab:telescopes}}
\tablecolumns{4}
\tablewidth{0pt}
\tablehead{
\colhead{Telescope} &
\colhead{Instrument} &
\colhead{Mode} &
\colhead{Set-ups or Filters}
}
\startdata
Gemini-South & Gemini Multi-Object Spectrograph (GMOS) & Imaging & $griz$ \\
& & Spectroscopy & $R400$, $B600$ \\
& FLAMINGOS-2 & Spectroscopy & $JH$ \\
Gemini-North & Gemini Multi-Object Spectrograph (GMOS) & Imaging & $griz$ \\
& & Spectroscopy & $R400$, $B600$ \\
& Near-Infrared Imager & $JHK$ \\
Keck I & Low Resolution Imaging Spectrometer (LRIS) & Imaging & $GRI$, $RG850$ \\
& & Spectroscopy & 400/3400, 400/8500 \\
& Multi-Object Spectrometer For Infra-Red Exploration (MOSFIRE) & Imaging & $YJHK_s$ \\
Keck II & DEep Imaging Multi-Object Spectrograph (DEIMOS) & $VRIZ$ \\
Large Binocular Telescope & Large Binocular Camera (LBC) & $ugriz$ \\
 & Multi-Object Double CCD Spectrographs (MODS1, MODS2) & $ugriz$ \\
Magellan Baade & Inamori Magellan Areal Camera and Spectrograph (IMACS) & f/2 Imaging & $griz$ \\
& & f/2 Spectroscopy & 200/15.0, 300/17.5, 300/26.7 \\
& FourStar & Imaging & $JHK_s$ \\
Magellan Clay & Low Dispersion Survey Spectrograph~3 (LDSS3) & Imaging & $griz$ \\
& & Spectroscopy & VPH-ALL \\
& Persson's Auxiliary Nasmyth Infrared Camera (PANIC) & $JK_S$ \\
MMT & Binospec & Imaging & $griz$ \\
& & Spectroscopy & 270l \\
 & MMTCam & Imaging & $gri$ \\
 & Magellan Infrared Spectrograph (MMIRS) & Imaging &  $YJHK$ \\
United Kingdom InfraRed Telescope & Wide Field Camera (WFCAM) & Imaging & $YJHK$ \\
& UKIRT Fast-Track Imager (UFTI) & Imaging & $JHK$ \\
Hubble Space Telescope (HST) & Wide Field Survey Camera~3 (WFC3) & Imaging & F814W, F110W, F160W \\
& Advanced Camera for Surveys (ACS) & Imaging & F606W
\enddata
\tablecomments{Telescope, instrument suites, imaging filters, and spectroscopic gratings and grisms used in the new host galaxy data presented in this paper (Section~\ref{sec:obs} and Section~\ref{sec:specobs}). Literature or archival data that supplement this sample comprise a larger variety of telescopes and instruments not listed here.}
\end{deluxetable*}

We first present new ground-based data for events in our sample that were discovered since the launch of {\it Swift} in 2004 until December 2021. We attempted a photometric host galaxy search for every short GRB discovered during these years, except for those which already have published identified hosts, are difficult to access with our telescope resources, or have observing constraints that would prevent a meaningful search. In some cases for bursts with published hosts, we obtained imaging in complementary filters to characterize the host SEDs.

To search for host galaxies, we initiated an initial round of deep ground-based imaging for \totbrightphotnum\ bursts. In general, we obtained optical imaging centered on the most precise available afterglow position in the $r$- or $i$-bands. If this imaging did not yield a plausible host galaxy candidate at or proximal to the afterglow position, we obtained near-infrared (NIR) imaging in the $J$- or $K$-bands to search for a reddened host (potentially due to a dusty or higher redshift origin). If either set of optical or NIR initial imaging revealed a plausible host galaxy (see Section~\ref{sec:pcc}), we obtained 1-10 bands of additional multi-band observations in any of the $ugriz$, $UVRI$, $RG850$, $YJHK$ or $K_s$ filters to characterize the putative host galaxy SED. If neither our optical nor NIR imaging yielded a plausible host galaxy, we used these ground-based limits to place constraints on the luminosity and redshift of spatially-coincident hosts. For five of these events, we obtained follow-up {\it HST} observations to perform a more sensitive search for spatially-coincident hosts (Section~\ref{sec:hst}).

We obtained these observations with the twin 6.5-m Magellan/Baade and Clay telescopes (PIs: Berger, Blanchard), 8-m Gemini-North and Gemini-South telescopes (PIs: Fong, Cucchiara), 6.5-m MMT (PIs: Fong, Nugent), twin 10-m Keck I and II telescopes (PIs: Paterson, Fong, Terreran, Miller), the 3.8-m United Kingdom InfraRed Telescope (UKIRT; PI: Fong), and the twin 10.2-m Large Binocular Telescopes (LBT; PI: Fong, Smith). We used 18 distinct instruments across these facilities for imaging. The telescopes, instruments and filters used for our catalog are listed in Table~\ref{tab:telescopes}. This imaging typically reaches $3\sigma$ limits of $m_{\rm AB,opt} \!\gtrsim\!24-26$~mag and $m_{\rm AB, NIR}\!\gtrsim 22-23.5$~mag. 

\begin{figure*}[t]
\centering
\includegraphics[width=0.245\textwidth]{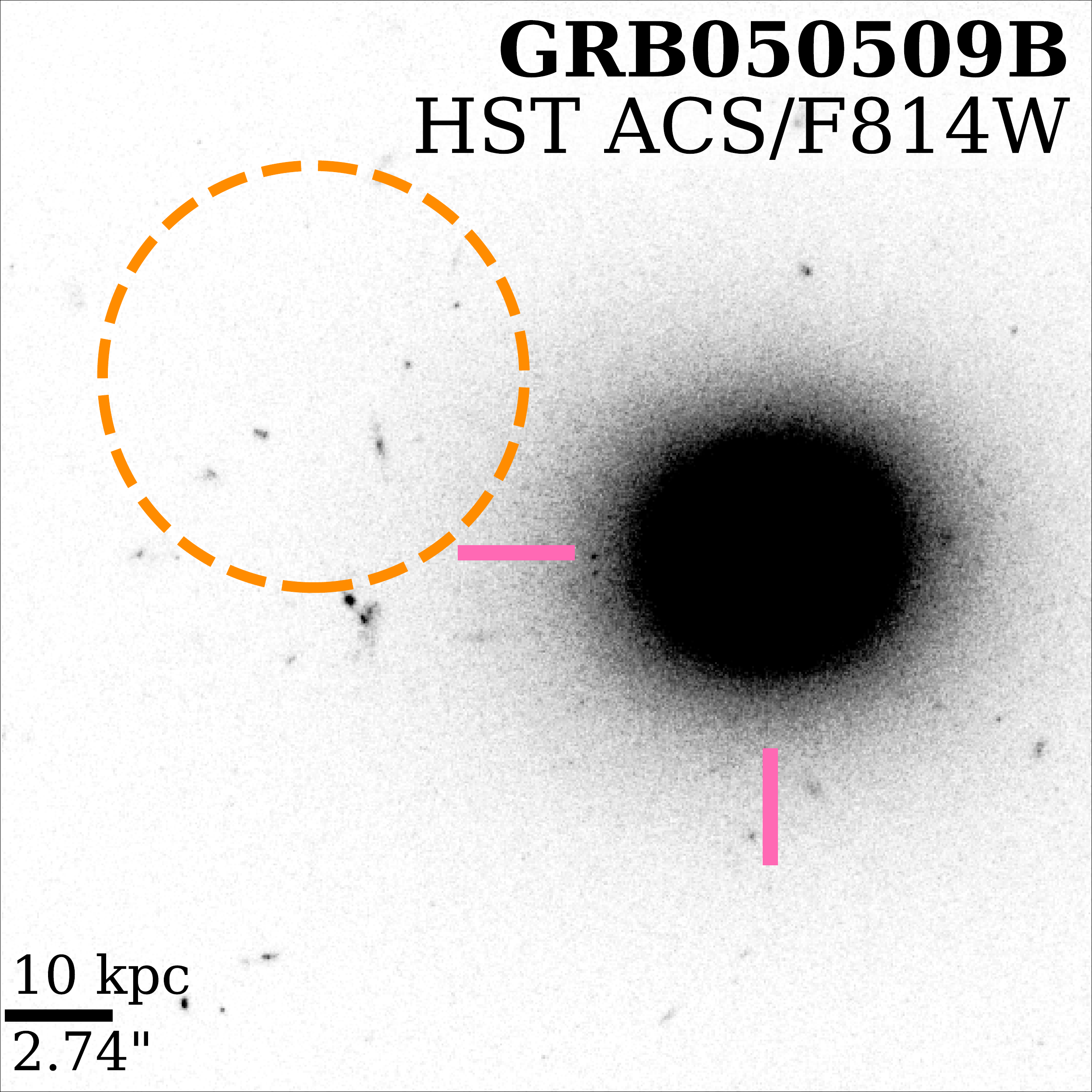}
\includegraphics[width=0.245\textwidth]{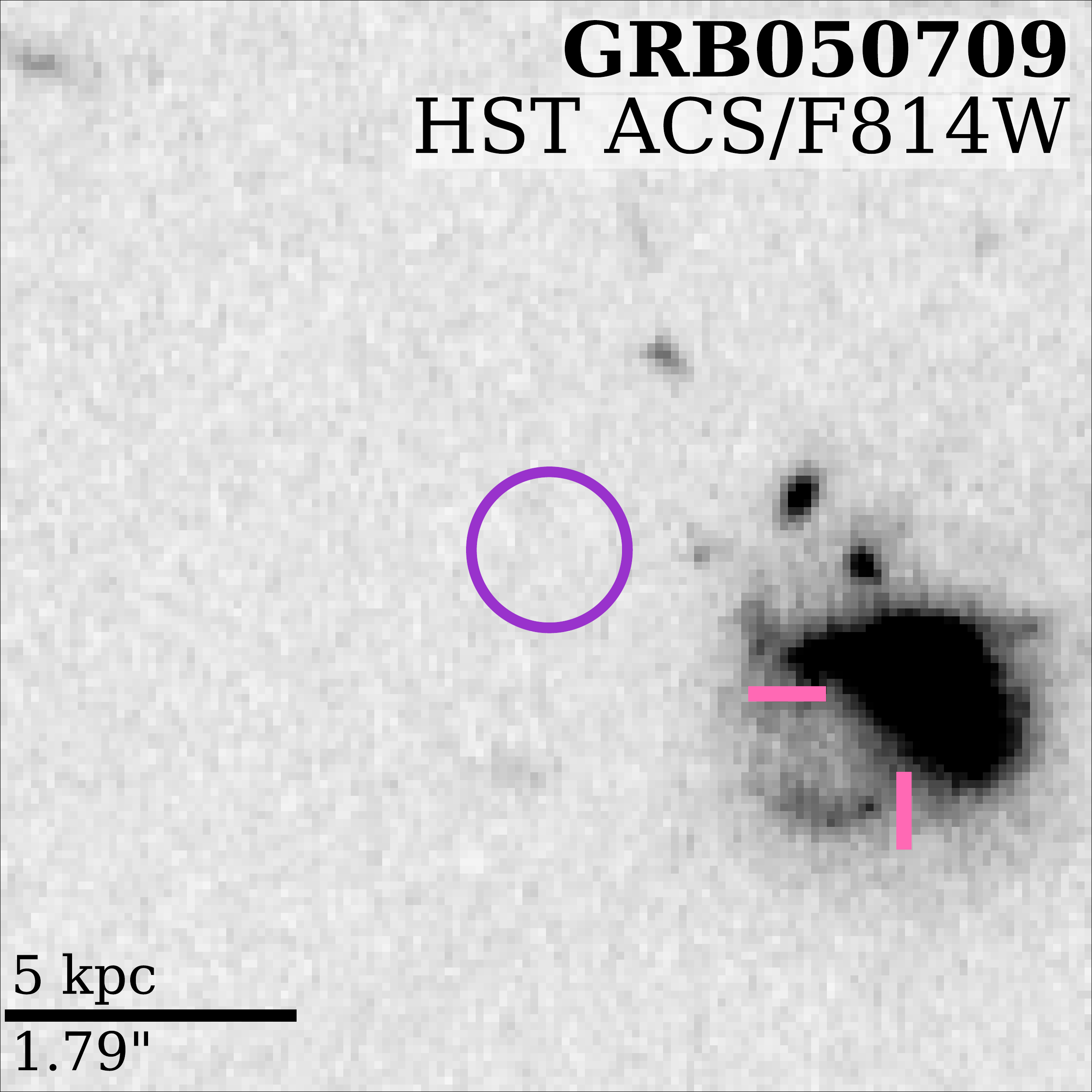}
\includegraphics[width=0.245\textwidth]{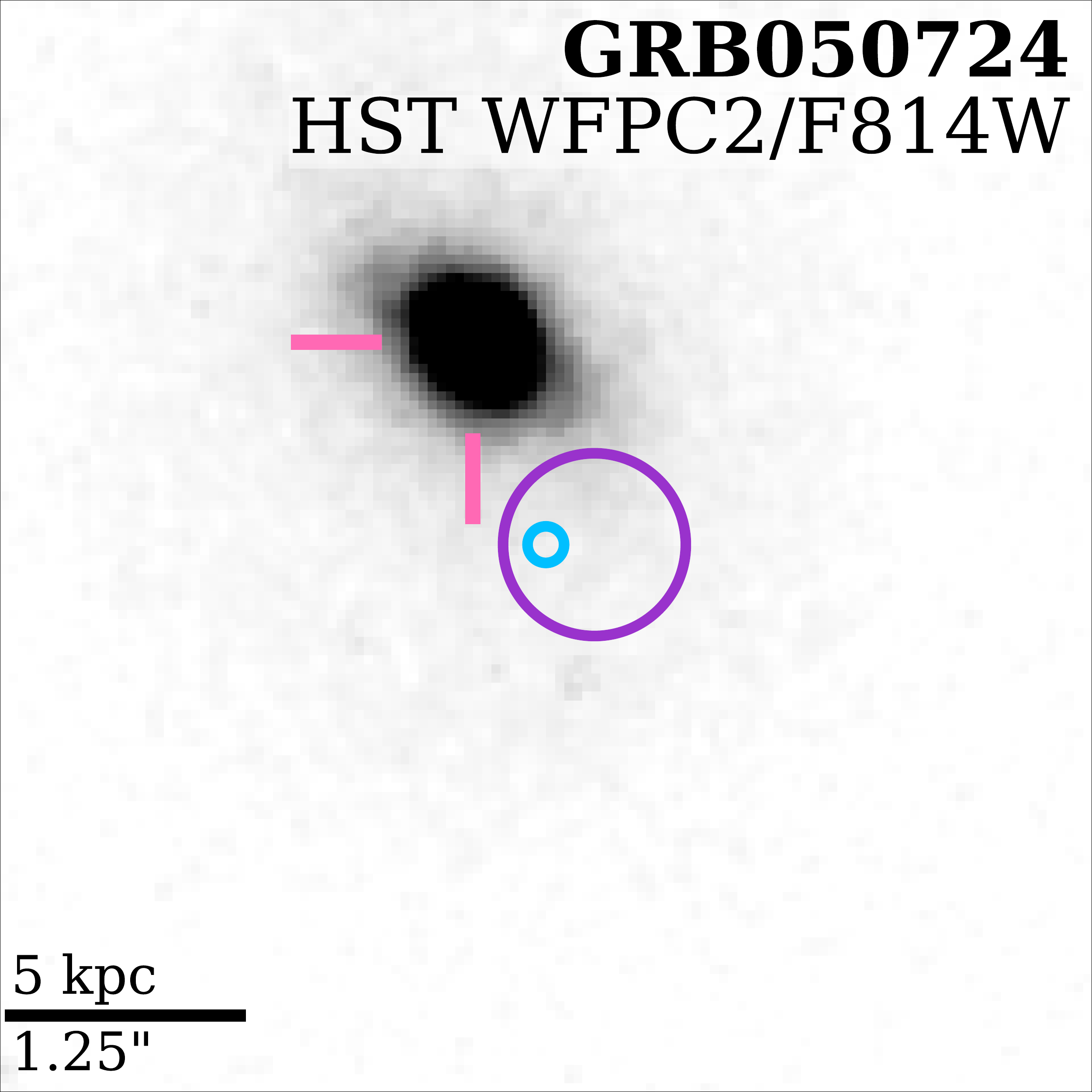}
\includegraphics[width=0.245\textwidth]{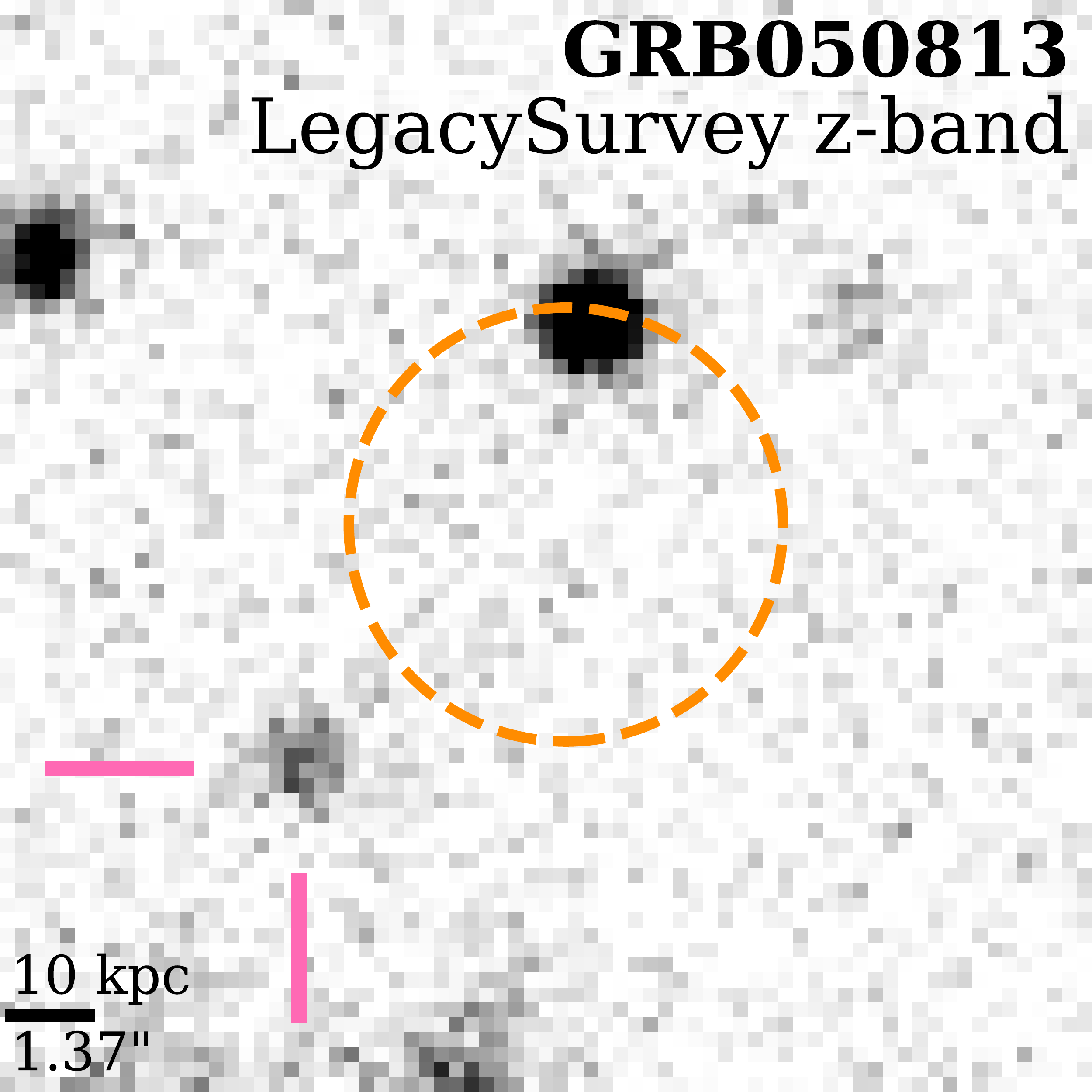}
\includegraphics[width=0.245\textwidth]{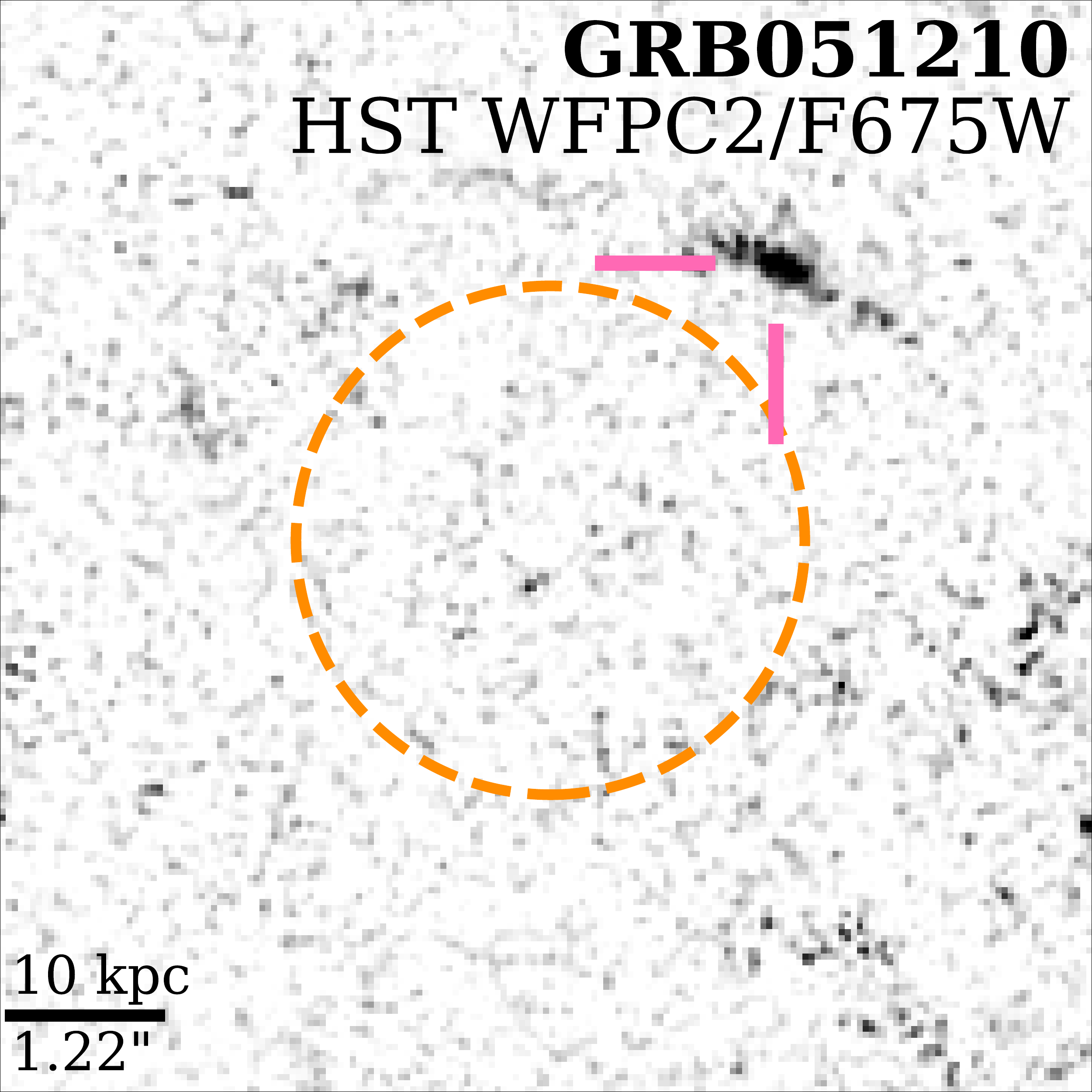}
\includegraphics[width=0.245\textwidth]{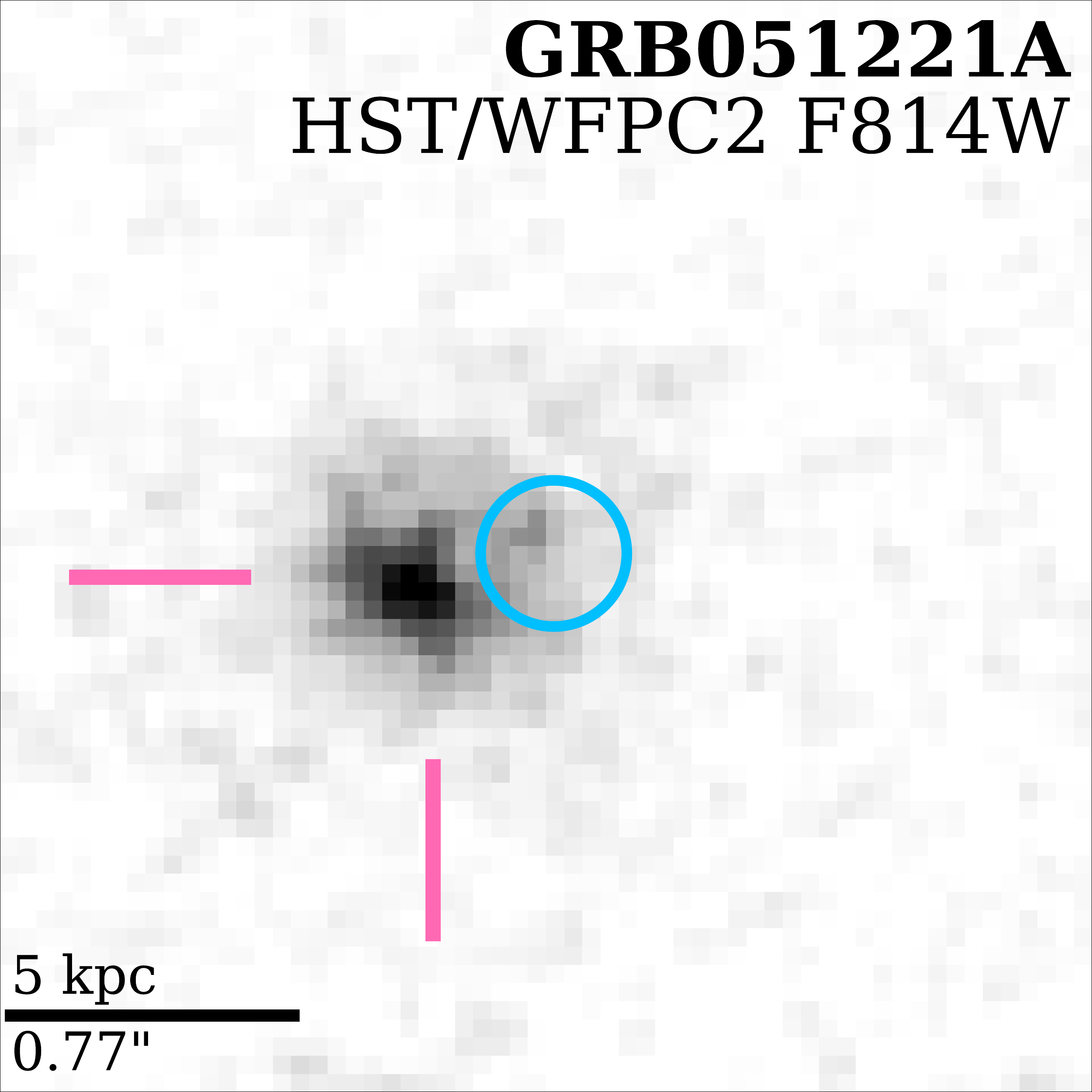}
\includegraphics[width=0.245\textwidth]{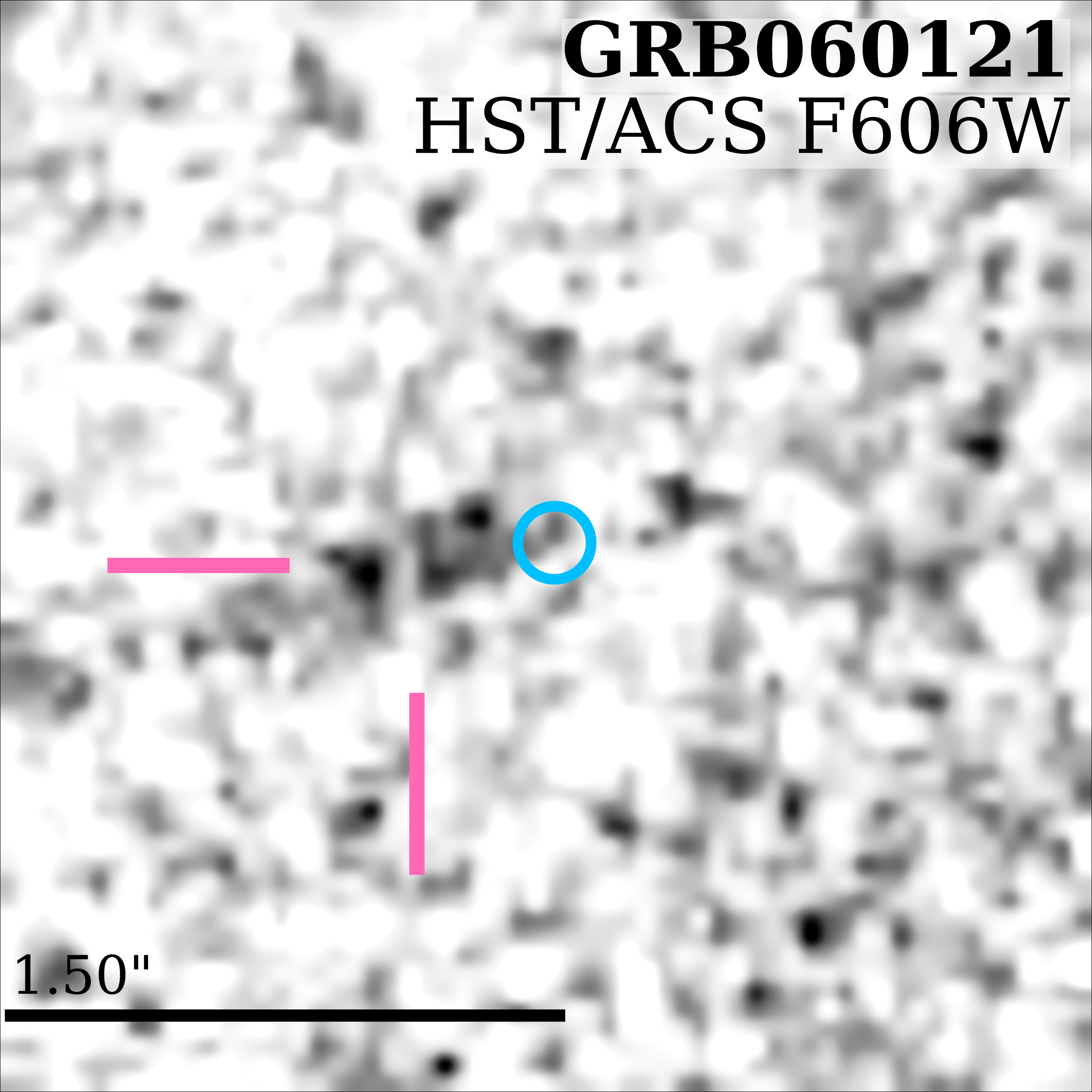}
\includegraphics[width=0.245\textwidth]{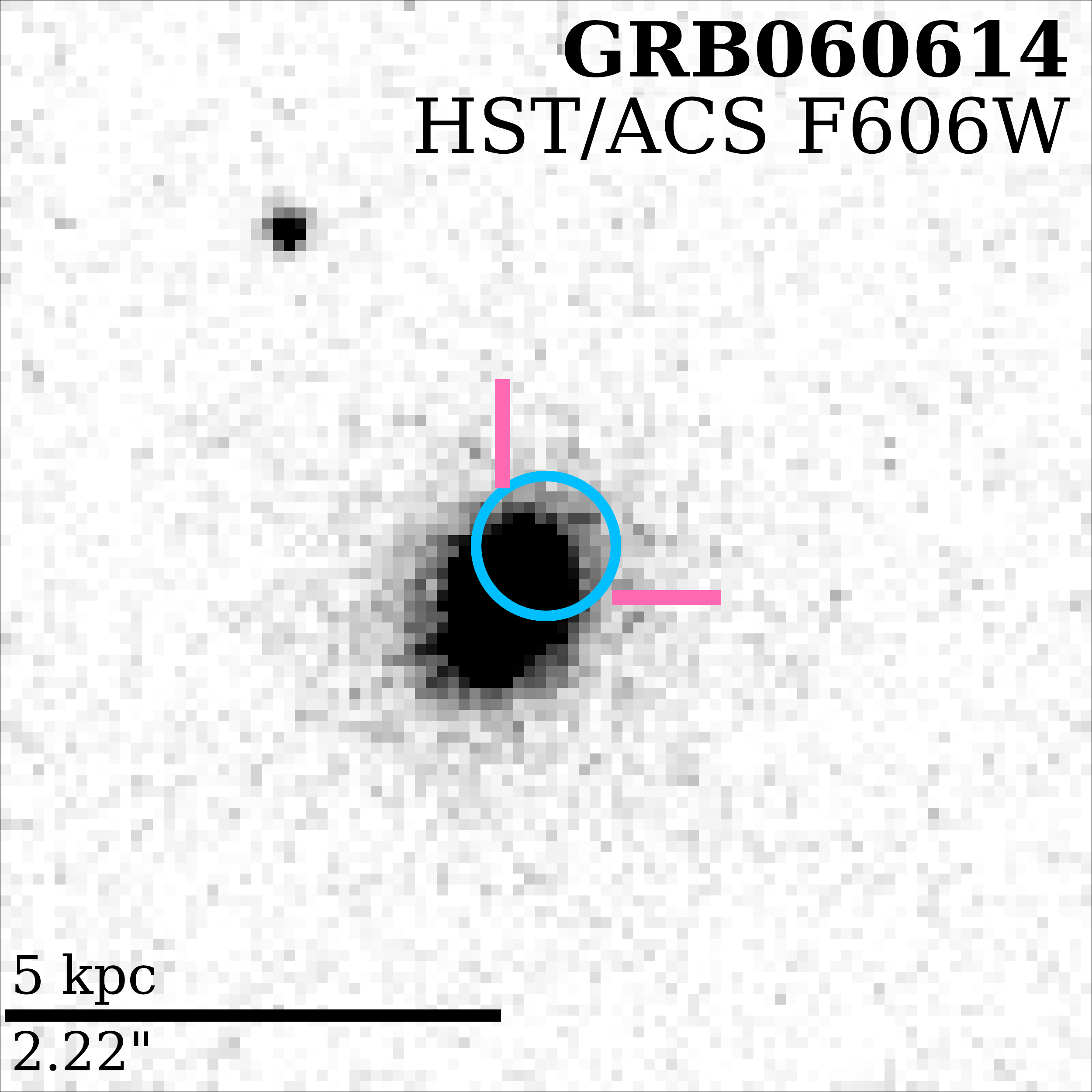}
\includegraphics[width=0.245\textwidth]{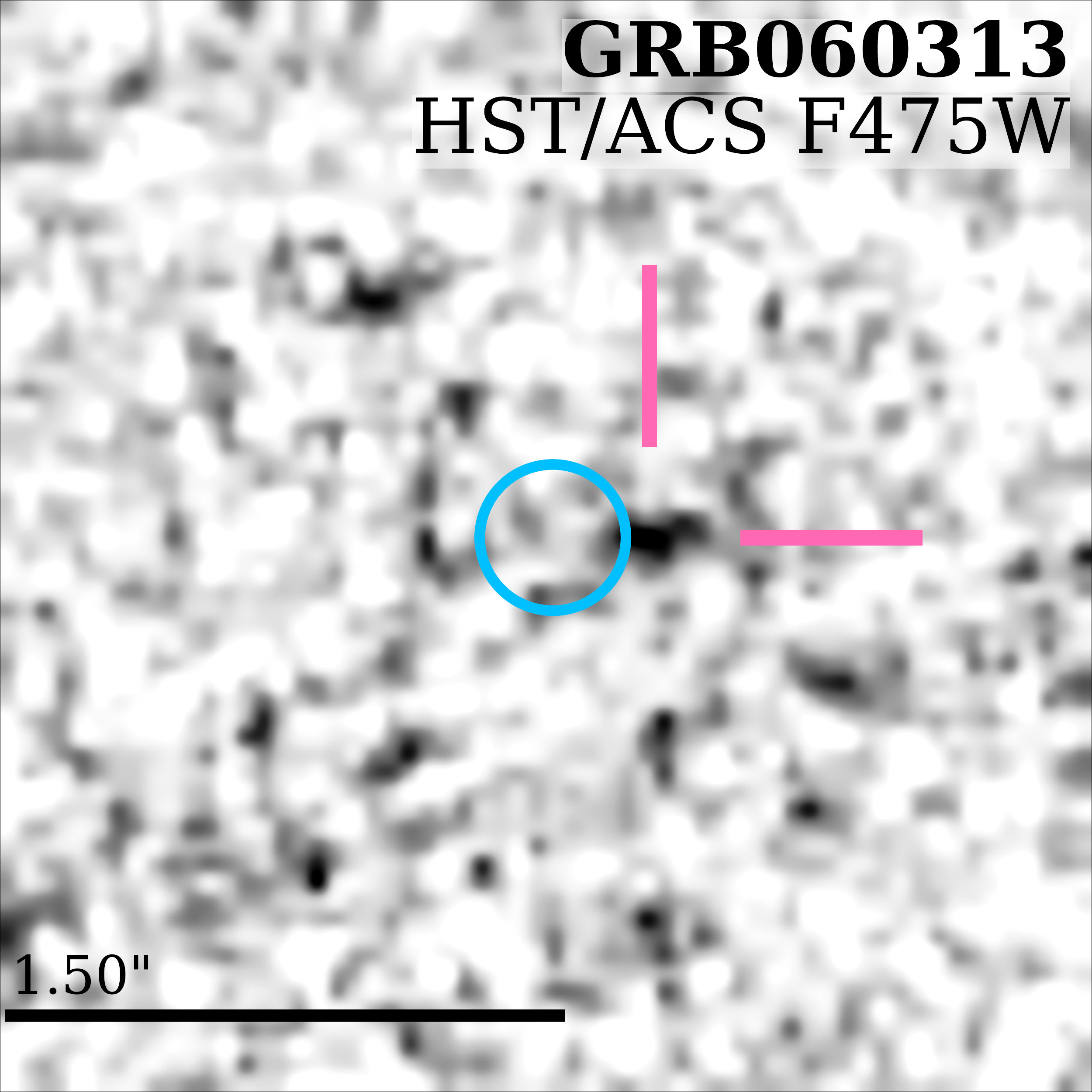}
\includegraphics[width=0.245\textwidth]{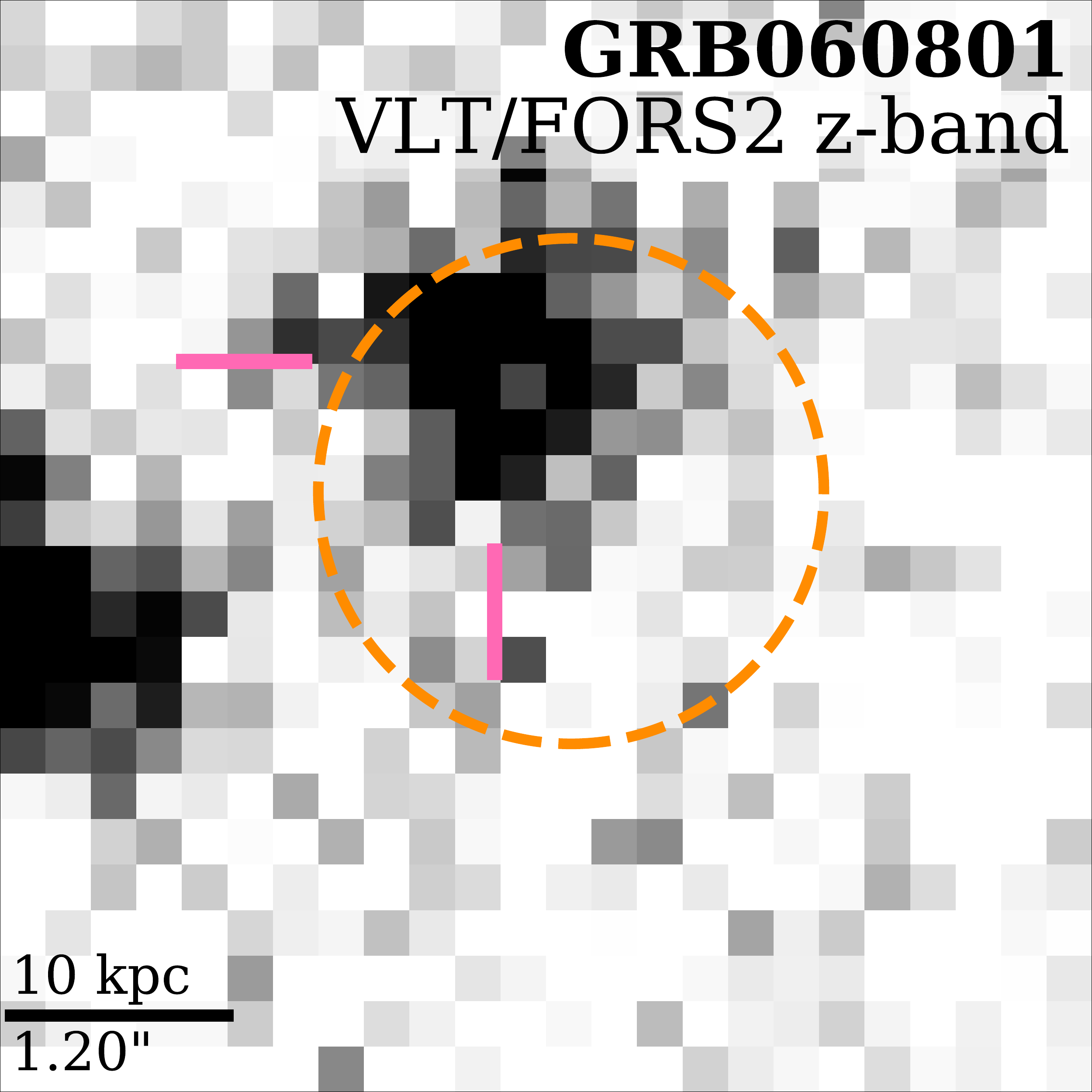}
\includegraphics[width=0.245\textwidth]{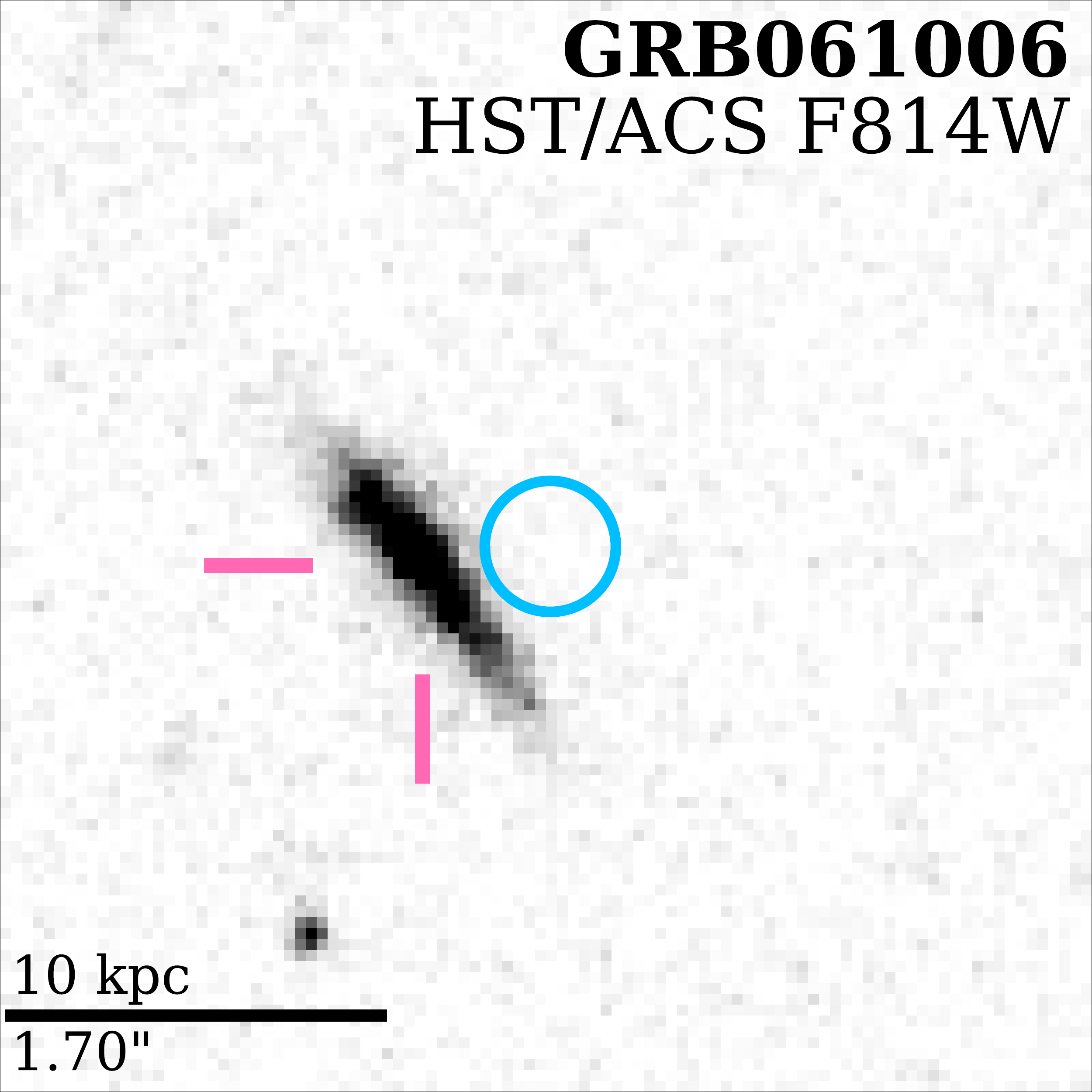}
\includegraphics[width=0.245\textwidth]{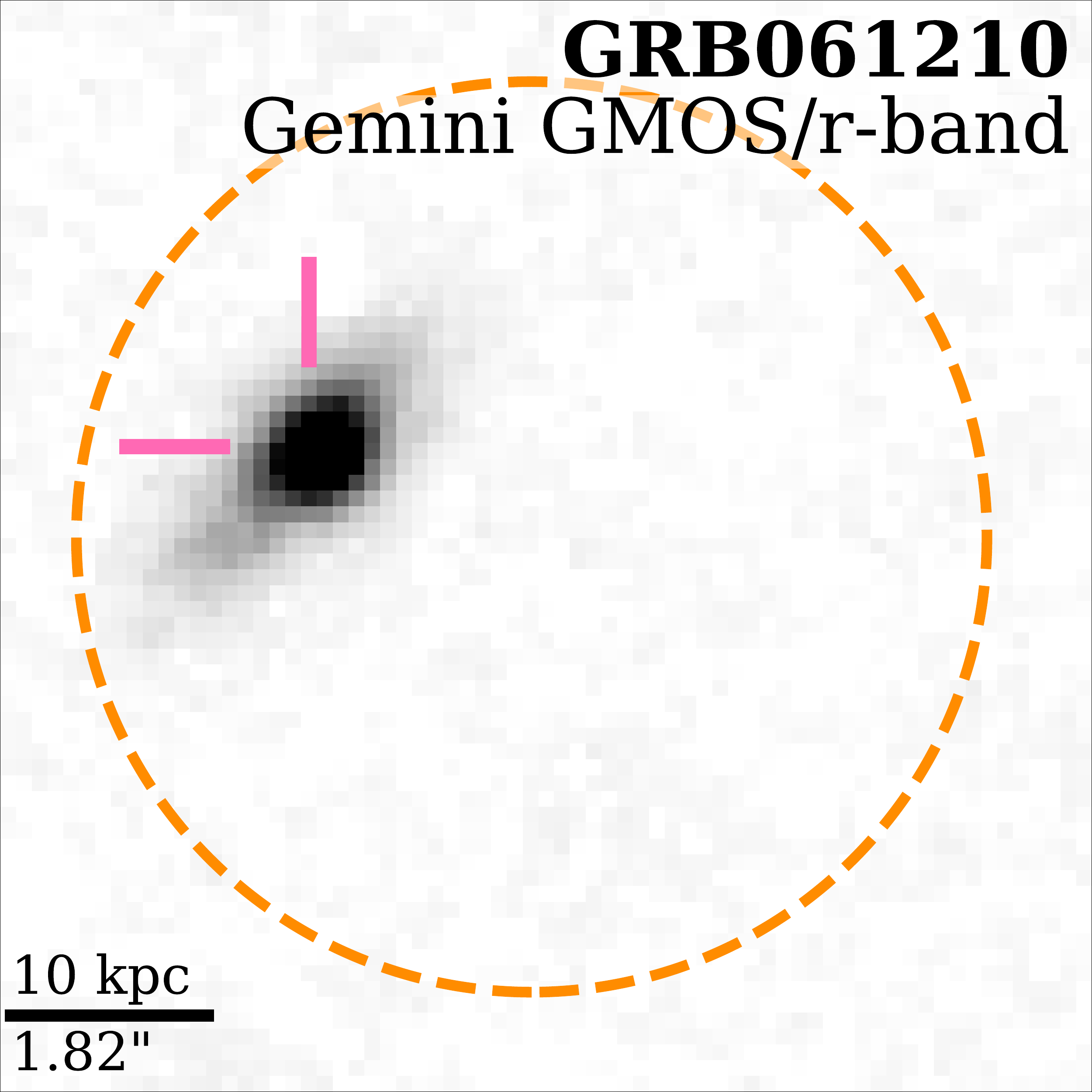}
\includegraphics[width=0.245\textwidth]{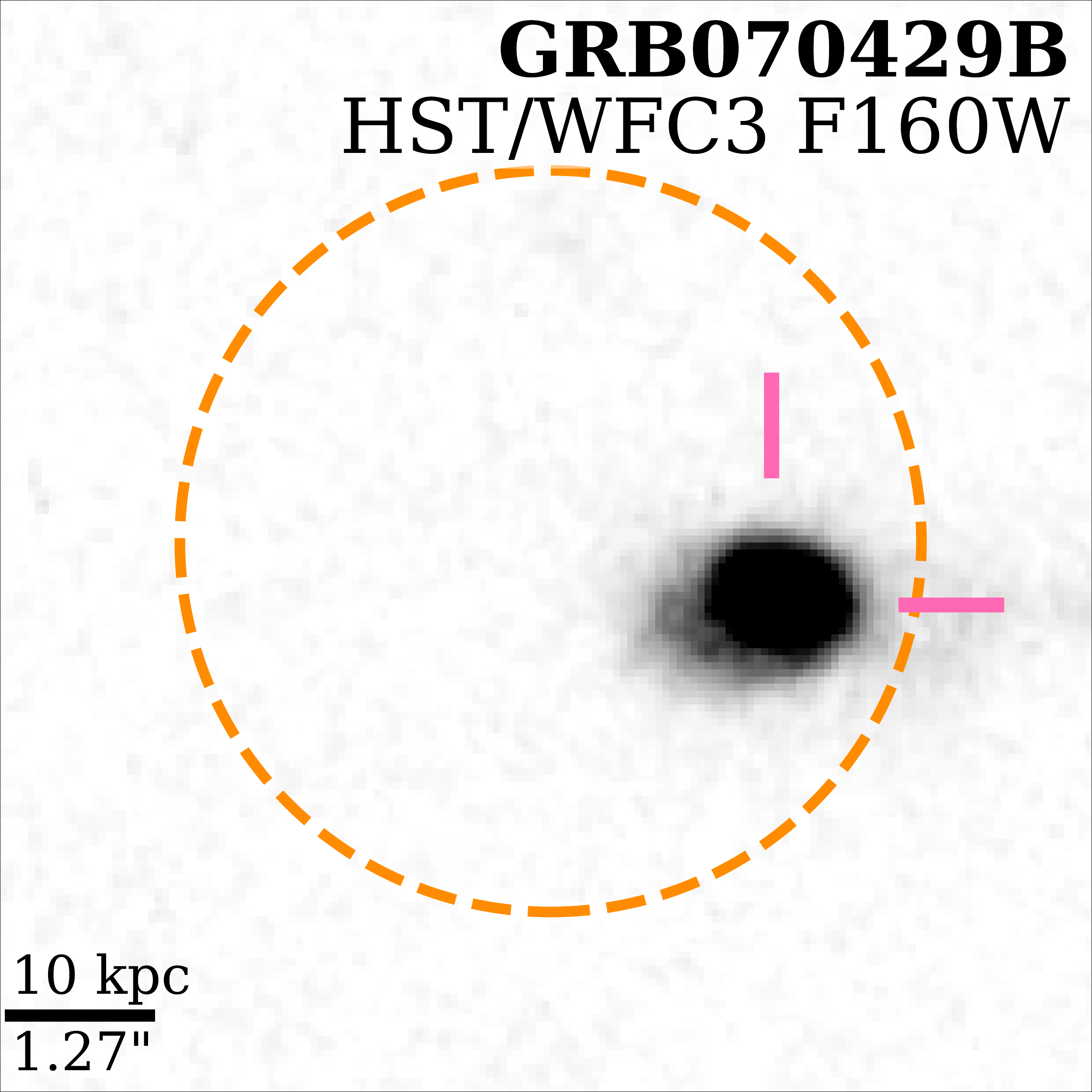}
\includegraphics[width=0.245\textwidth]{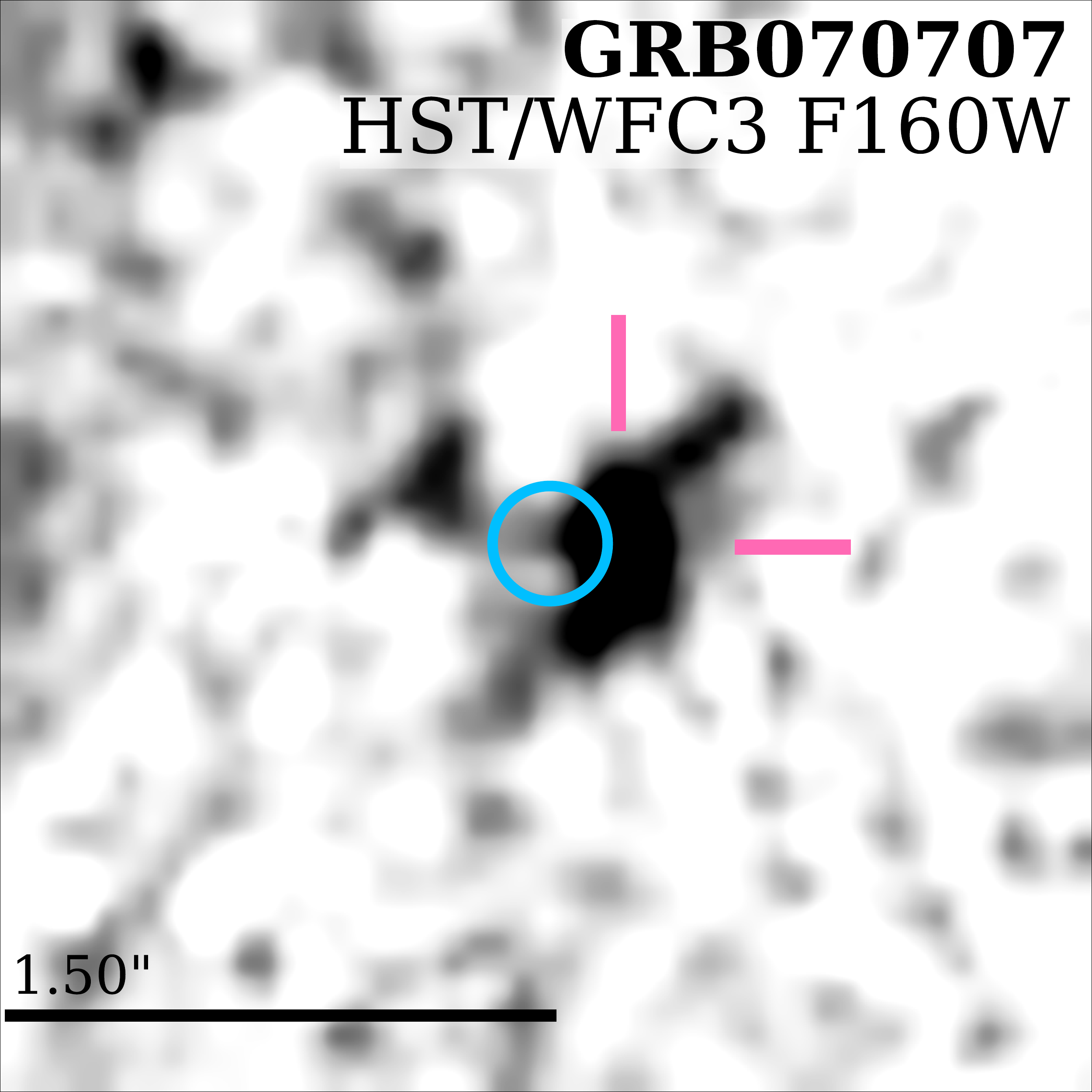}
\includegraphics[width=0.245\textwidth]{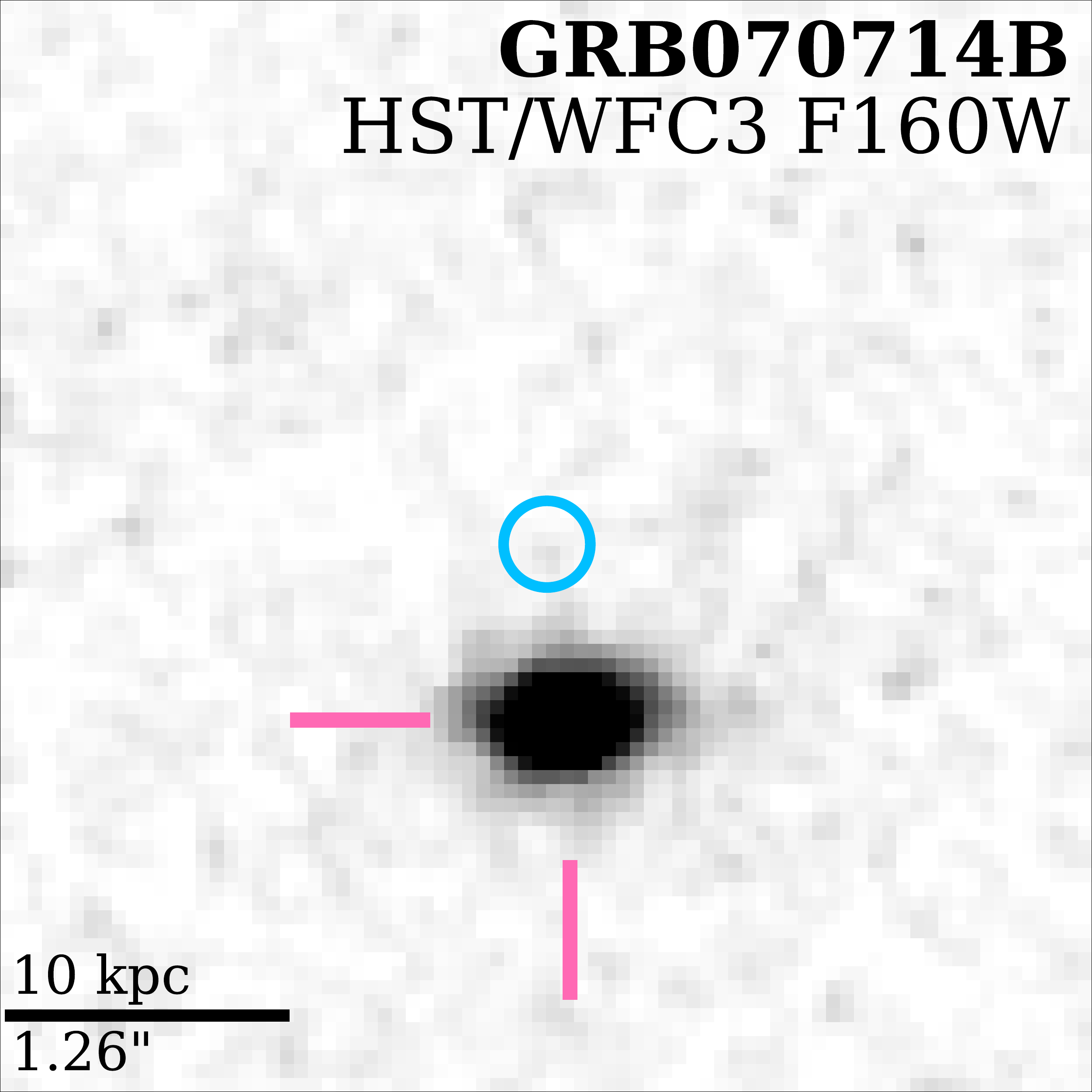}
\includegraphics[width=0.245\textwidth]{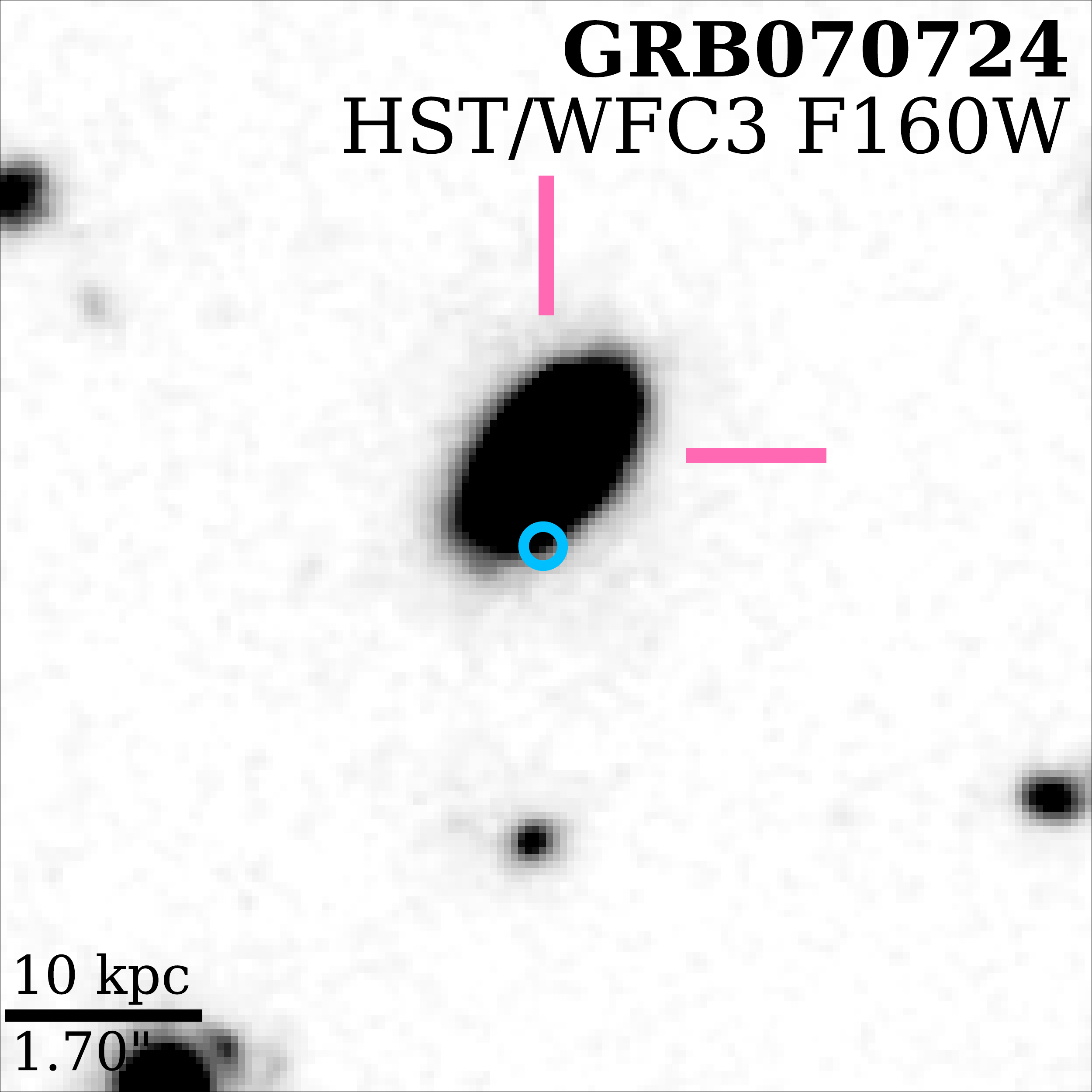}
\includegraphics[width=0.245\textwidth]{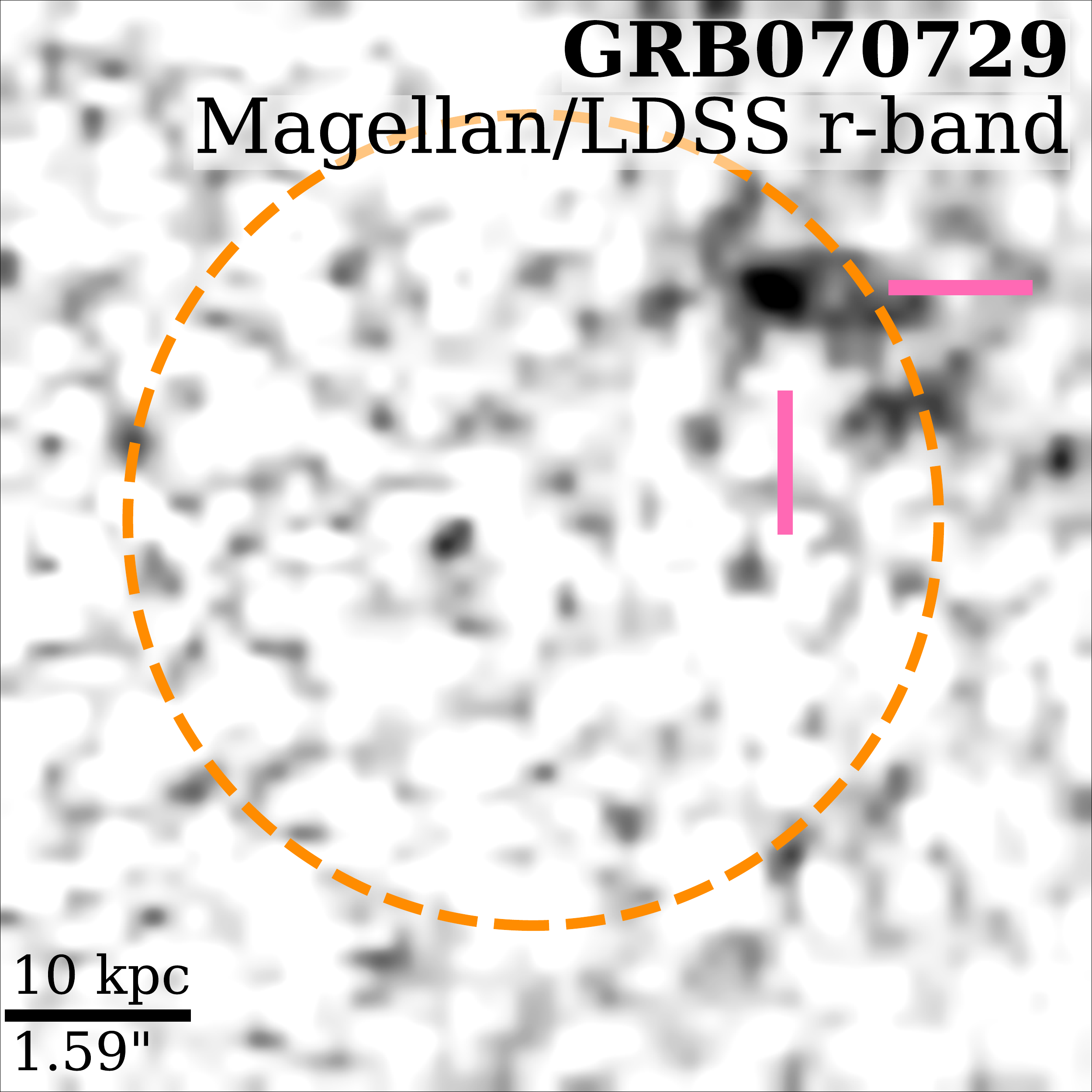}
\includegraphics[width=0.245\textwidth]{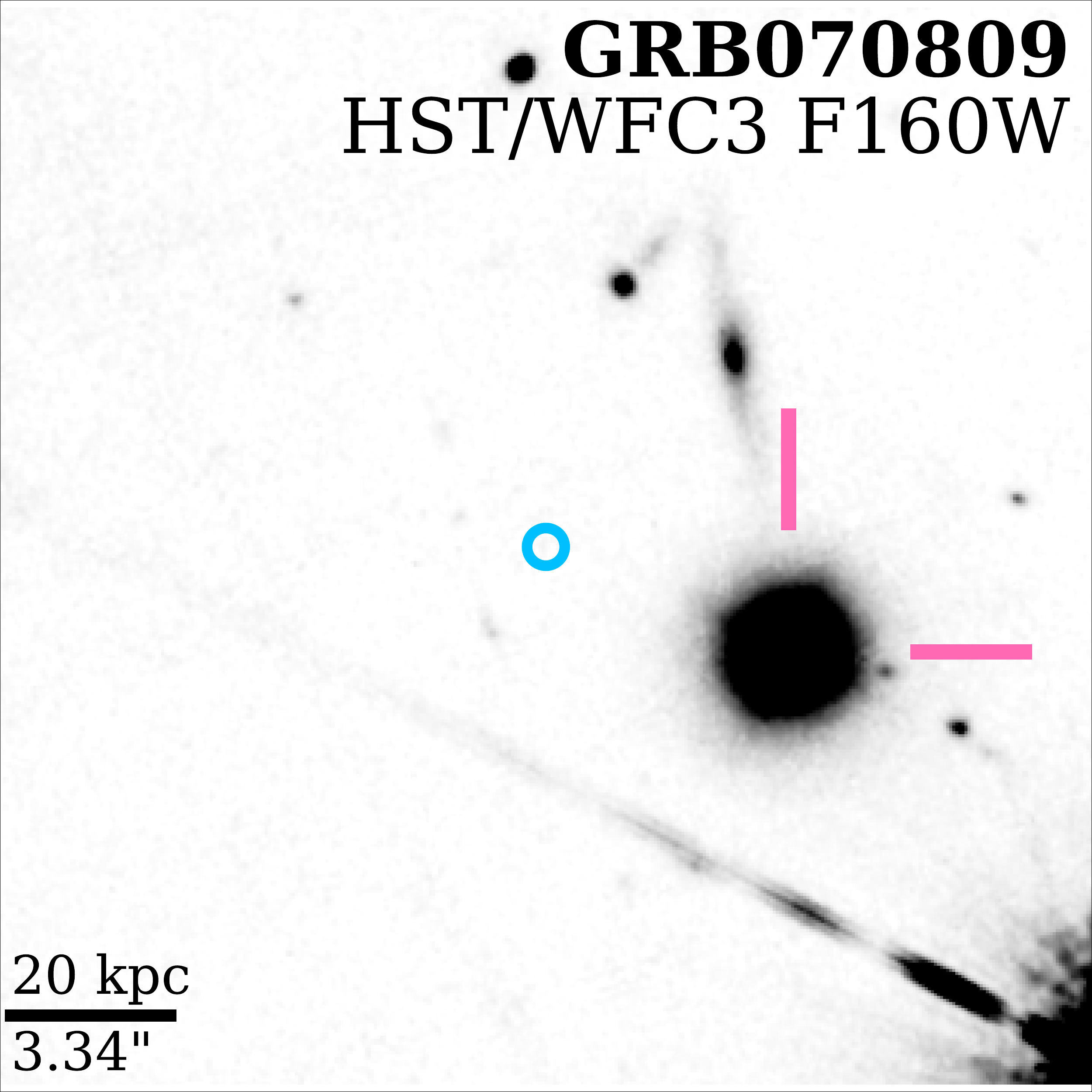}
\includegraphics[width=0.245\textwidth]{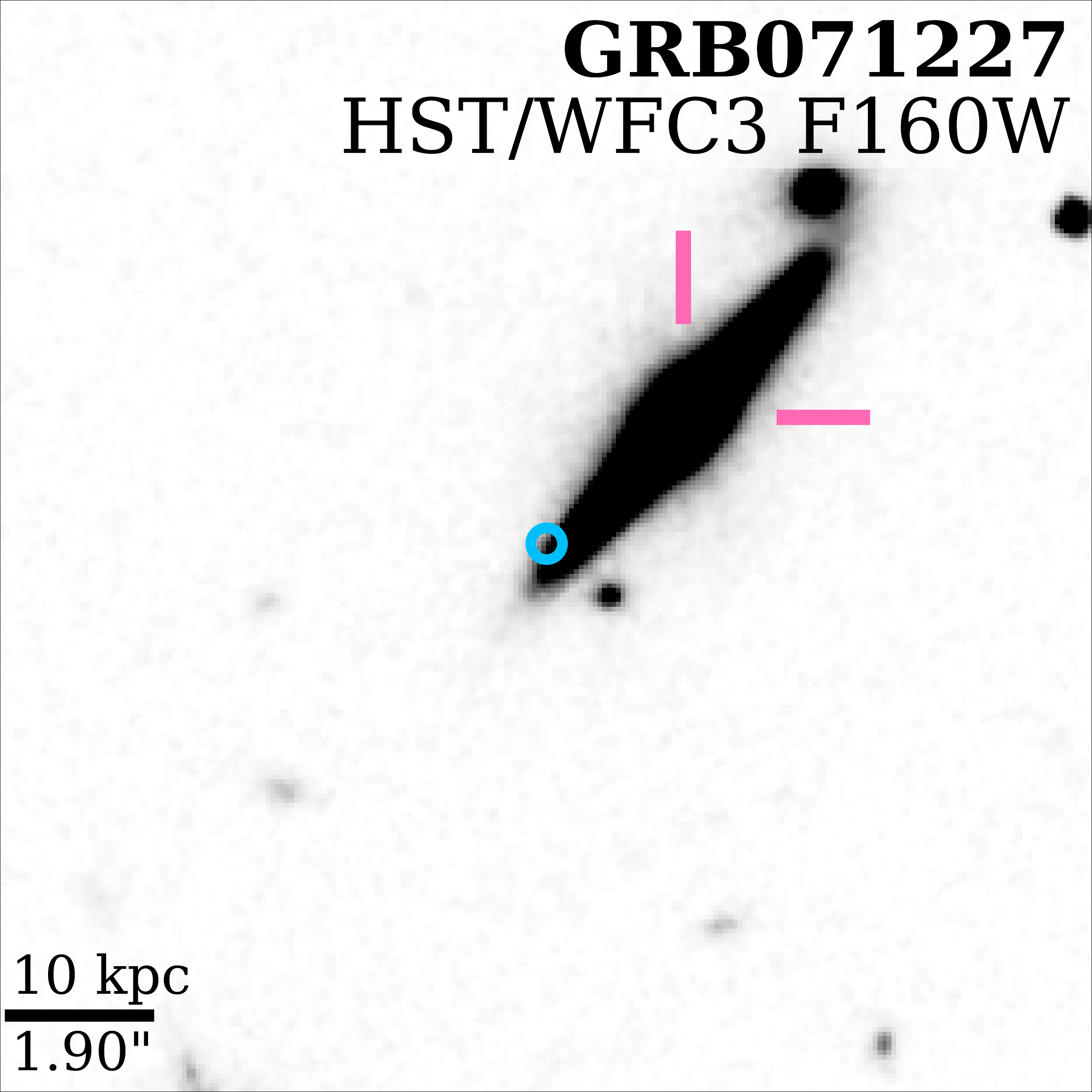}
\includegraphics[width=0.245\textwidth]{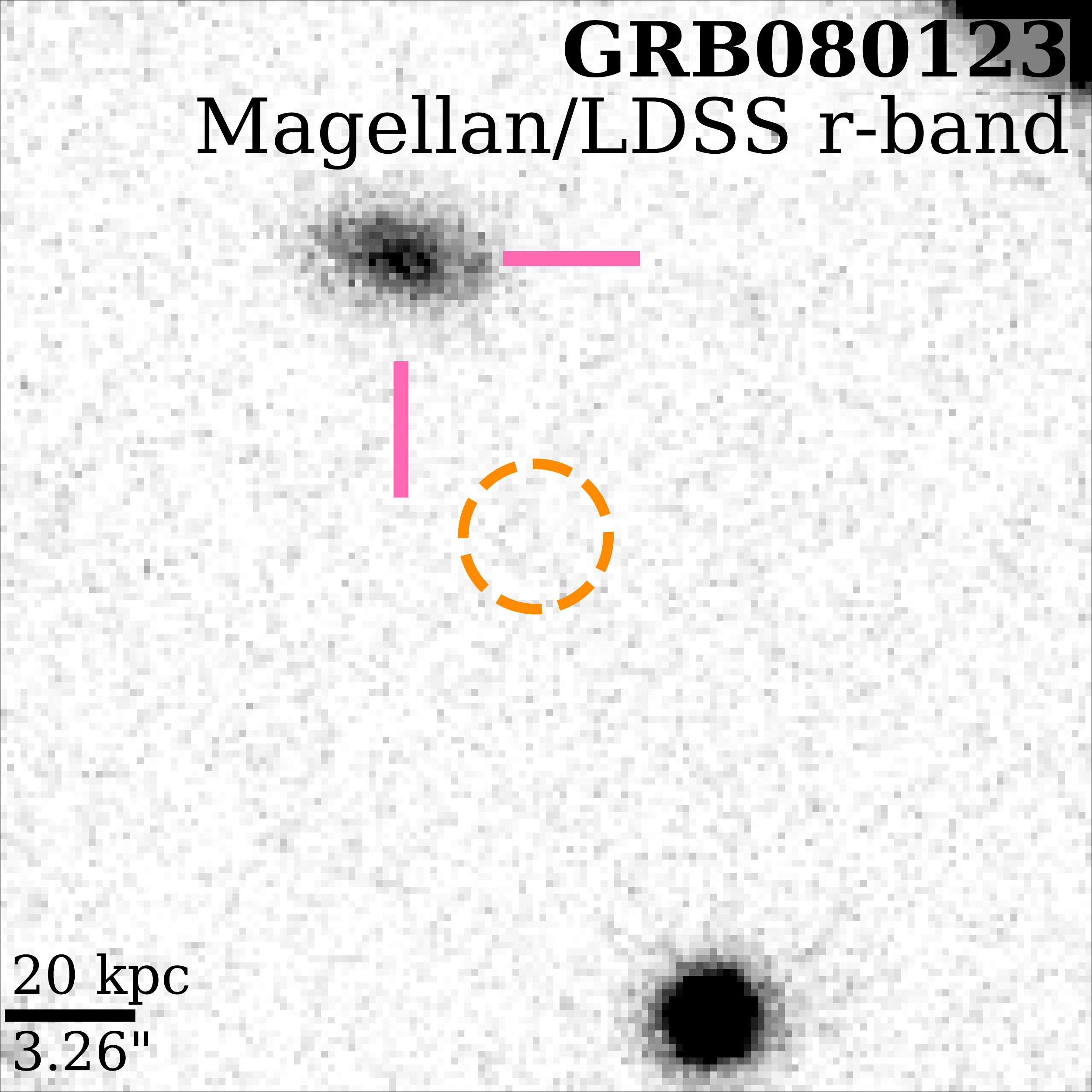}
\caption{Representative images of the host galaxies of the short GRBs in our catalog. In each panel, the most precise afterglow localization(s) for each burst is/are plotted (XRT 90\%: orange dashed, optical $1\sigma$: blue, {\it Chandra} or VLA $1\sigma$: purple). The putative host galaxy is denoted by the pink cross-hairs. All images are oriented North up and East to the left.
\label{fig:imagepanel}}
\end{figure*}

For data reduction and co-addition, we use a combination of standard tasks in the IRAF/{\tt ccdred} package (\citealt{iraf1}; for Magellan, LBT, MMT/MMTCam data), observatory-specific pipelines (for Gemini data), and the {\tt POTPyRI} software\footnote{\url{https://github.com/CIERA-Transients/POTPyRI}} (for Keck, MMT/Binospec and MMIRS data). For optical data, we apply bias corrections, flat-field corrections using either dome or twilight flats, and dark current corrections when relevant. For NIR data, we additionally apply sky subtraction using coeval on-sky frames. For UKIRT/WFCAM data, we obtain pre-processed images from the WFCAM Science Archive \citep{hcc+08} which are already corrected for bias, flat-field, and dark current by the Cambridge Astronomical Survey Unit\footnote{http://casu.ast.cam.ac.uk/}. For each epoch and filter, we co-add the images and perform astrometry relative to 2MASS using a combination of tasks in Starlink\footnote{http://starlink.eao.hawaii.edu/starlink} and IRAF.  The full listing is available in Appendix Table~\ref{tab:phot} and the BRIGHT website\footnote{\url{https://bright.ciera.northwestern.edu/}}. The images are shown in Figures~\ref{fig:imagepanel}-\ref{fig:hostless}.

\renewcommand{\thefigure}{\arabic{figure} (Cont.)}
\addtocounter{figure}{-1}

\begin{figure*}[t]
\centering
\includegraphics[width=0.245\textwidth]{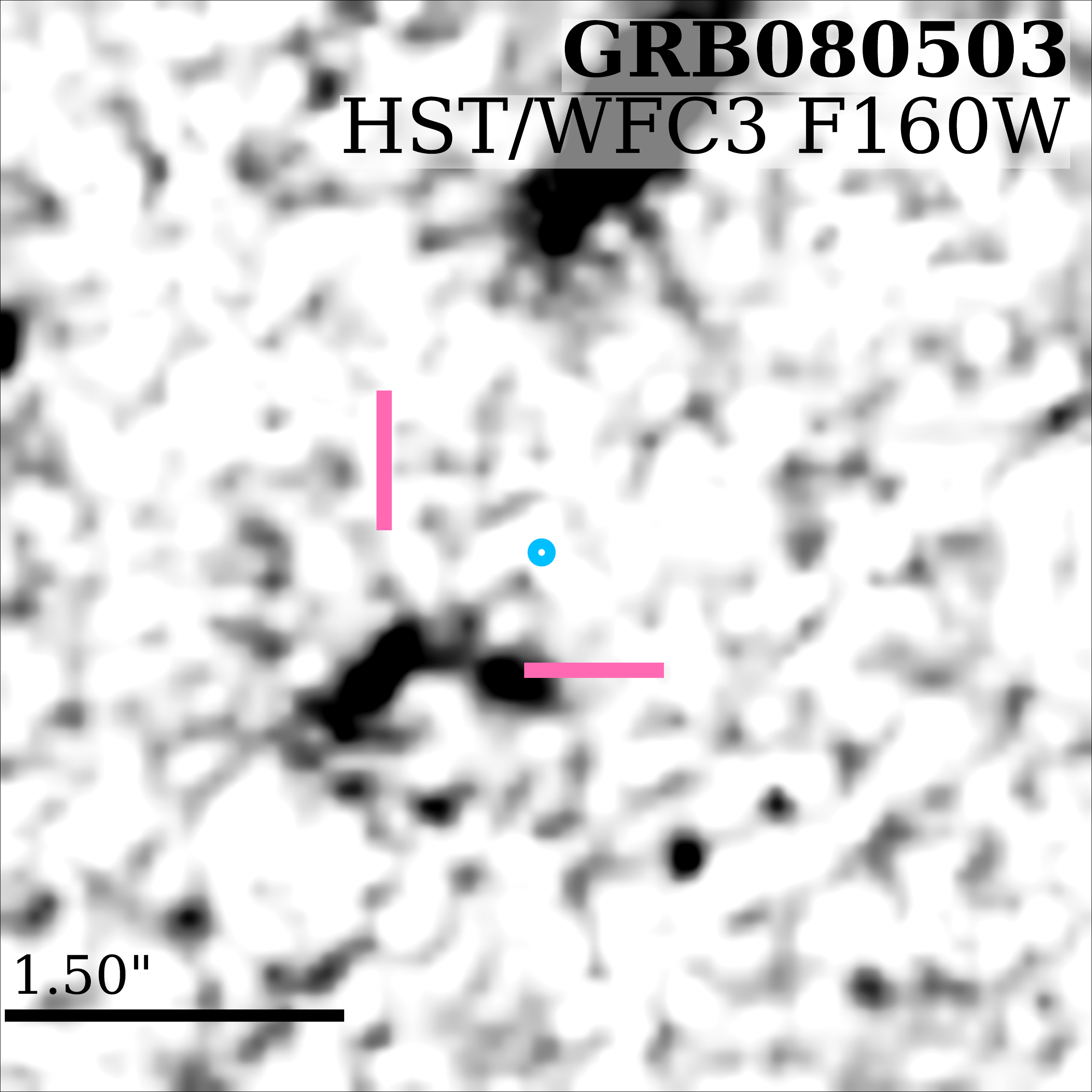}
\includegraphics[width=0.245\textwidth]{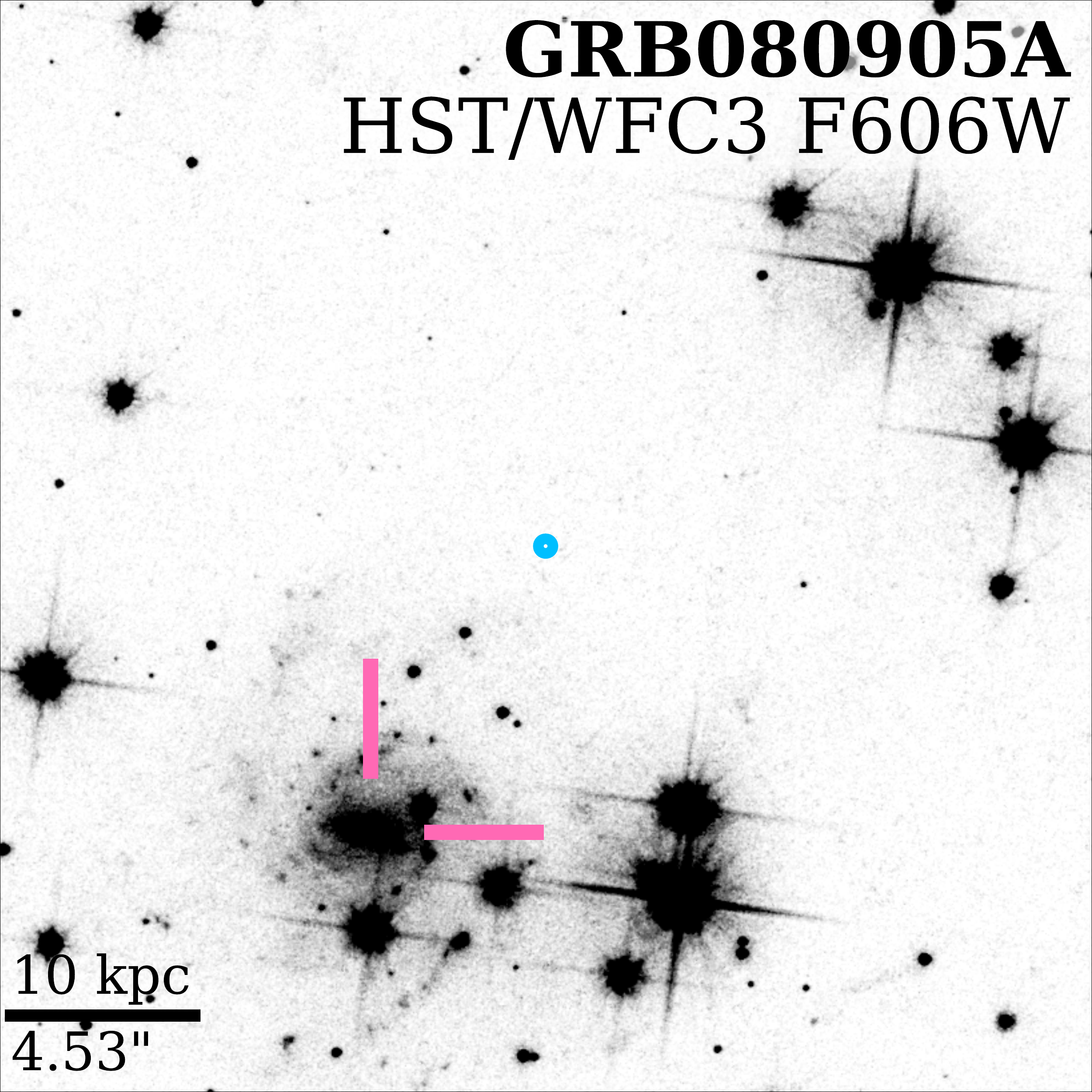}
\includegraphics[width=0.245\textwidth]{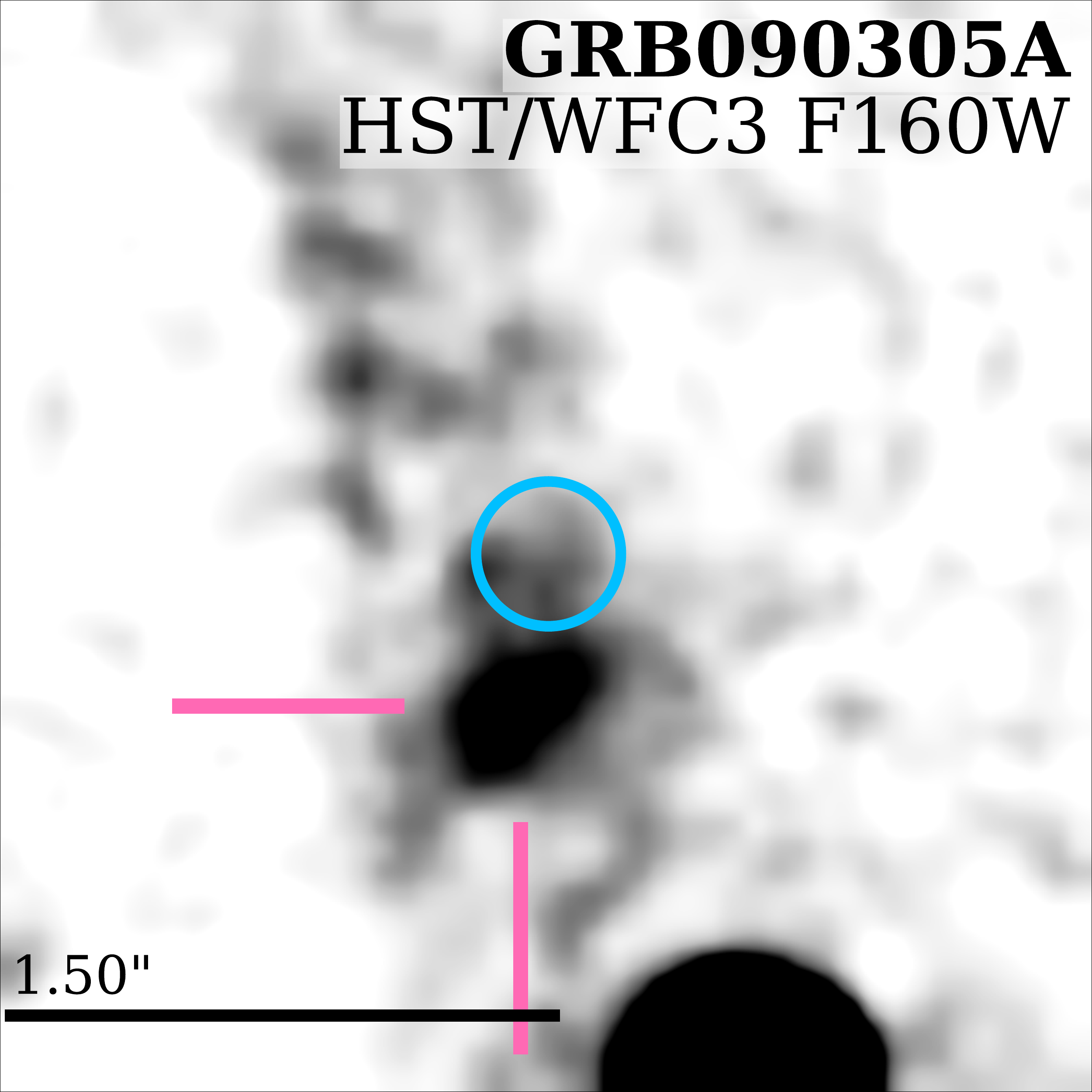}
\includegraphics[width=0.245\textwidth]{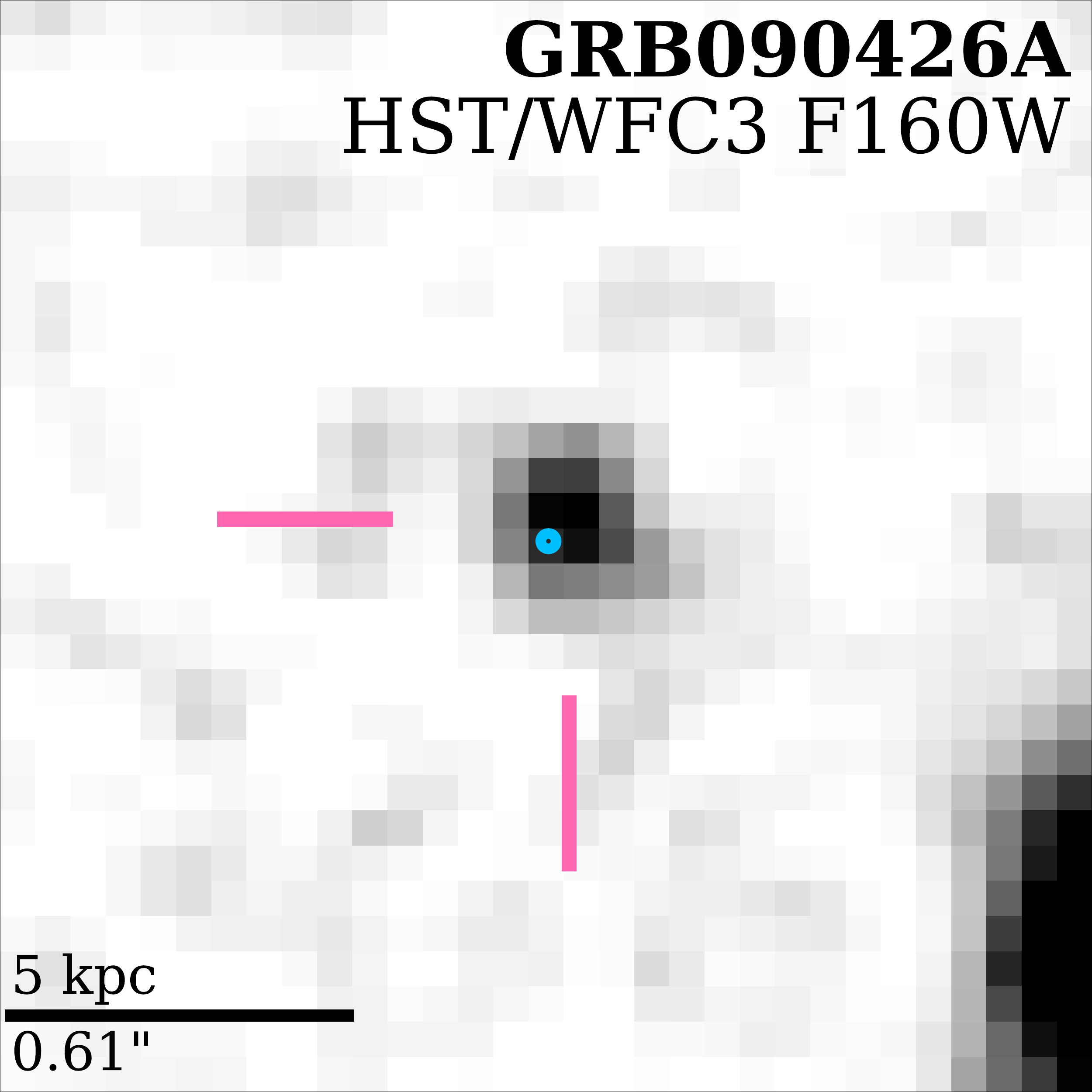}
\includegraphics[width=0.245\textwidth]{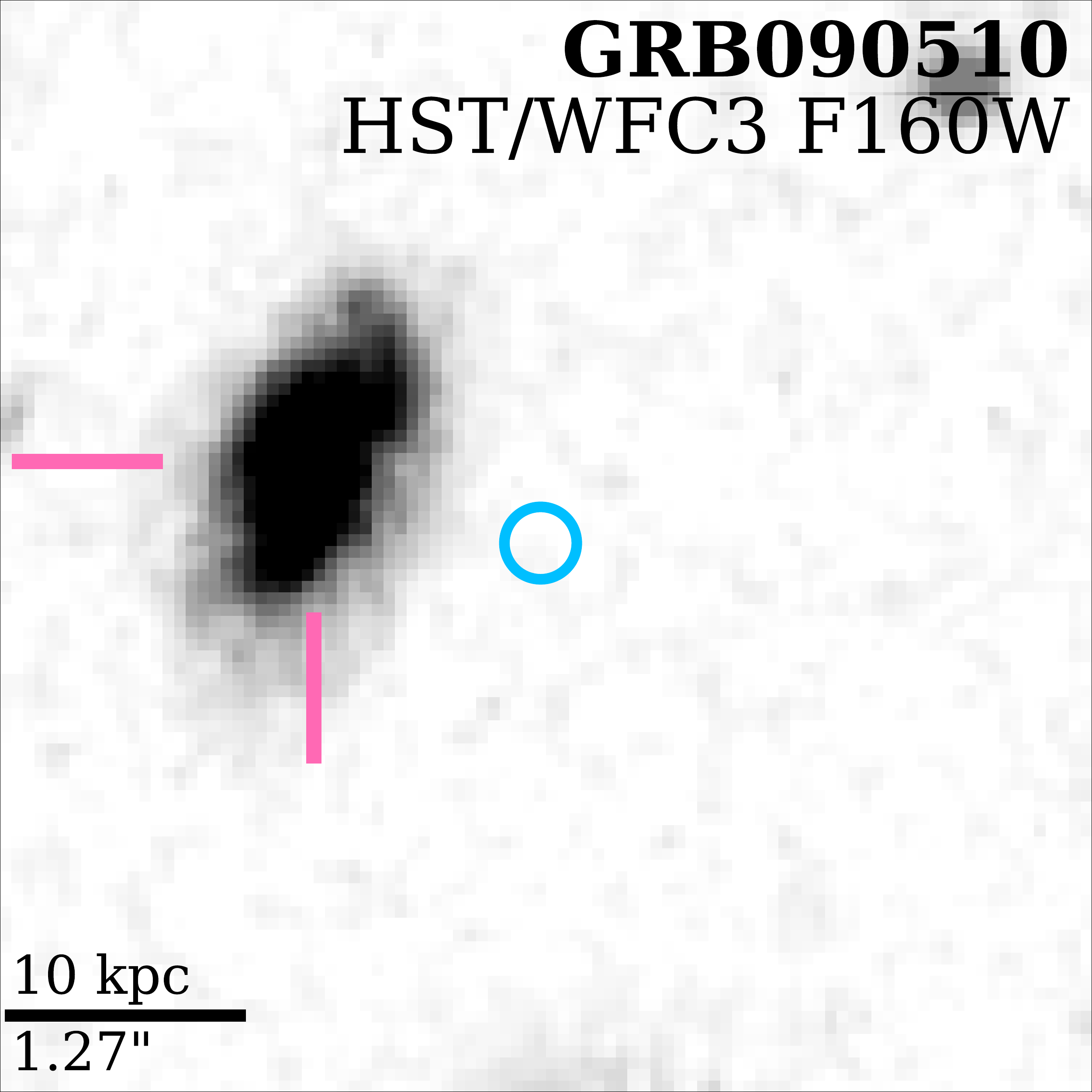}
\includegraphics[width=0.245\textwidth]{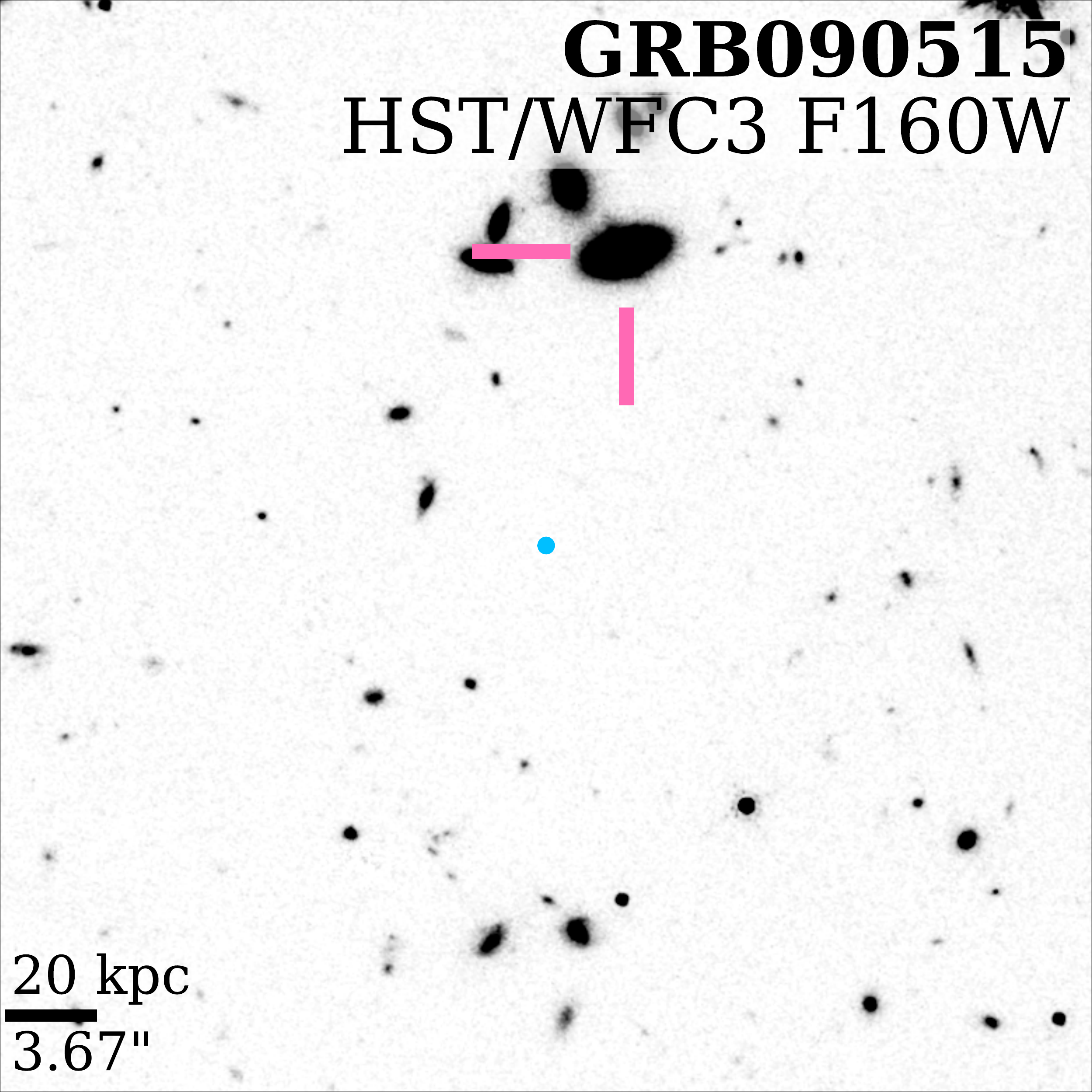}
\includegraphics[width=0.245\textwidth]{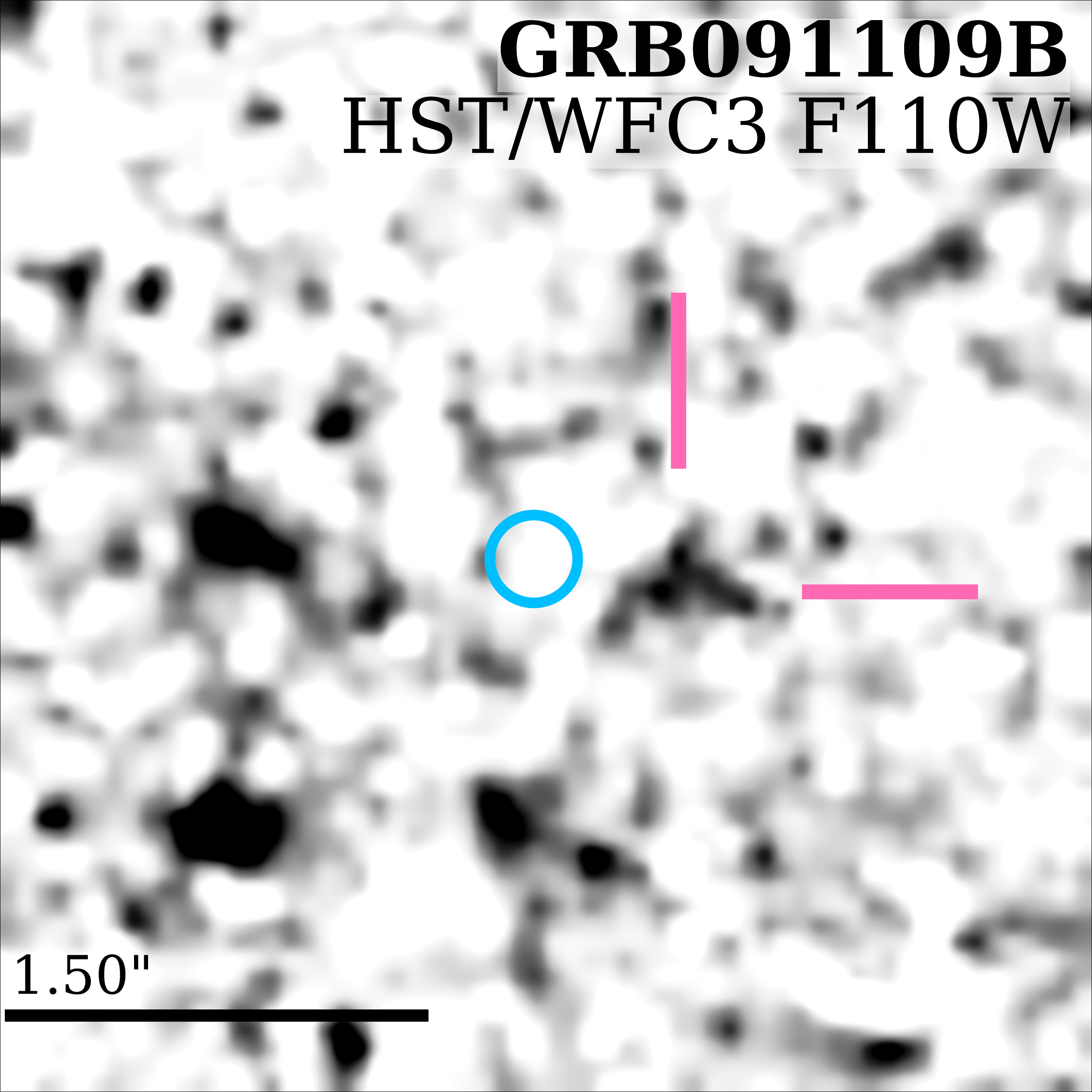}
\includegraphics[width=0.245\textwidth]{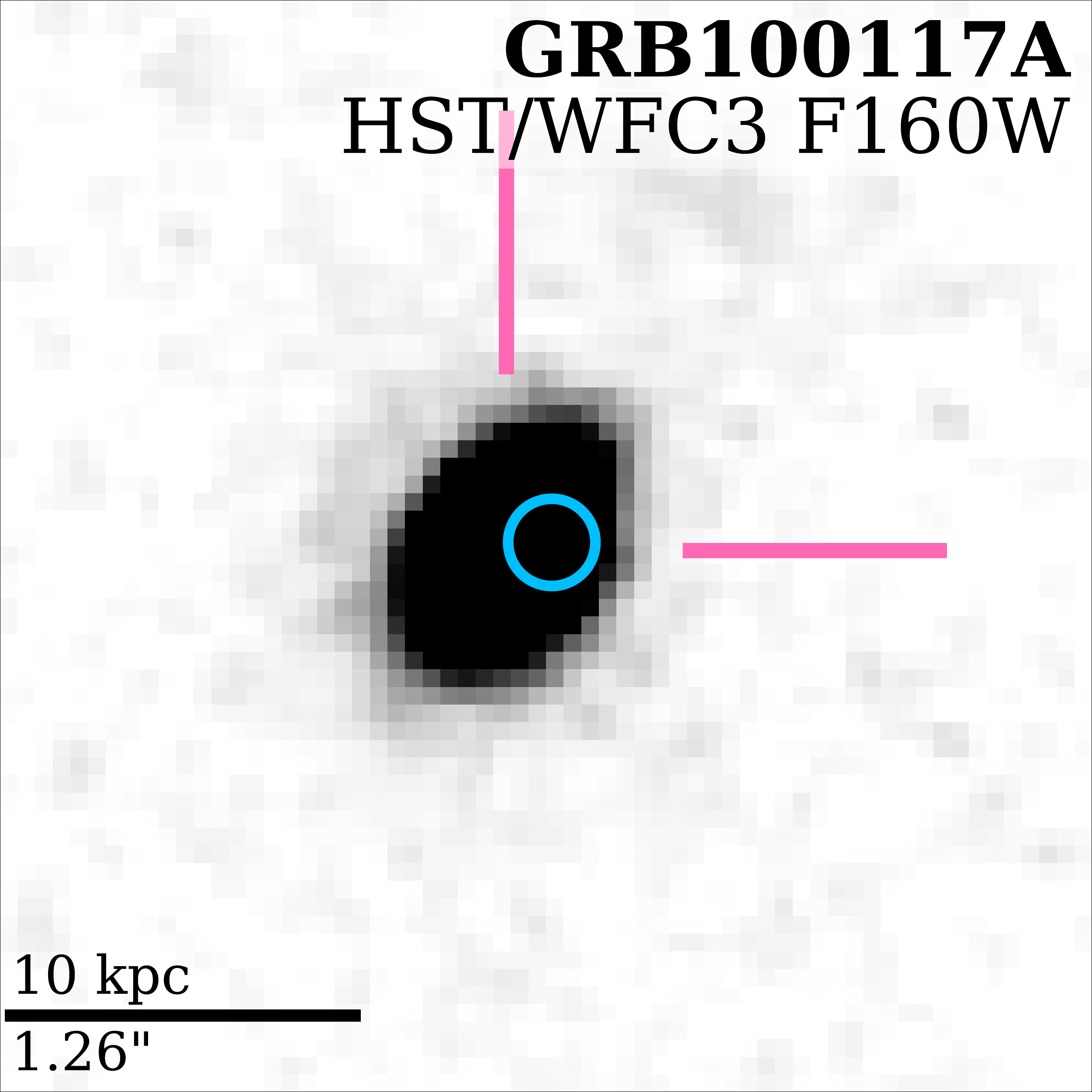}
\includegraphics[width=0.245\textwidth]{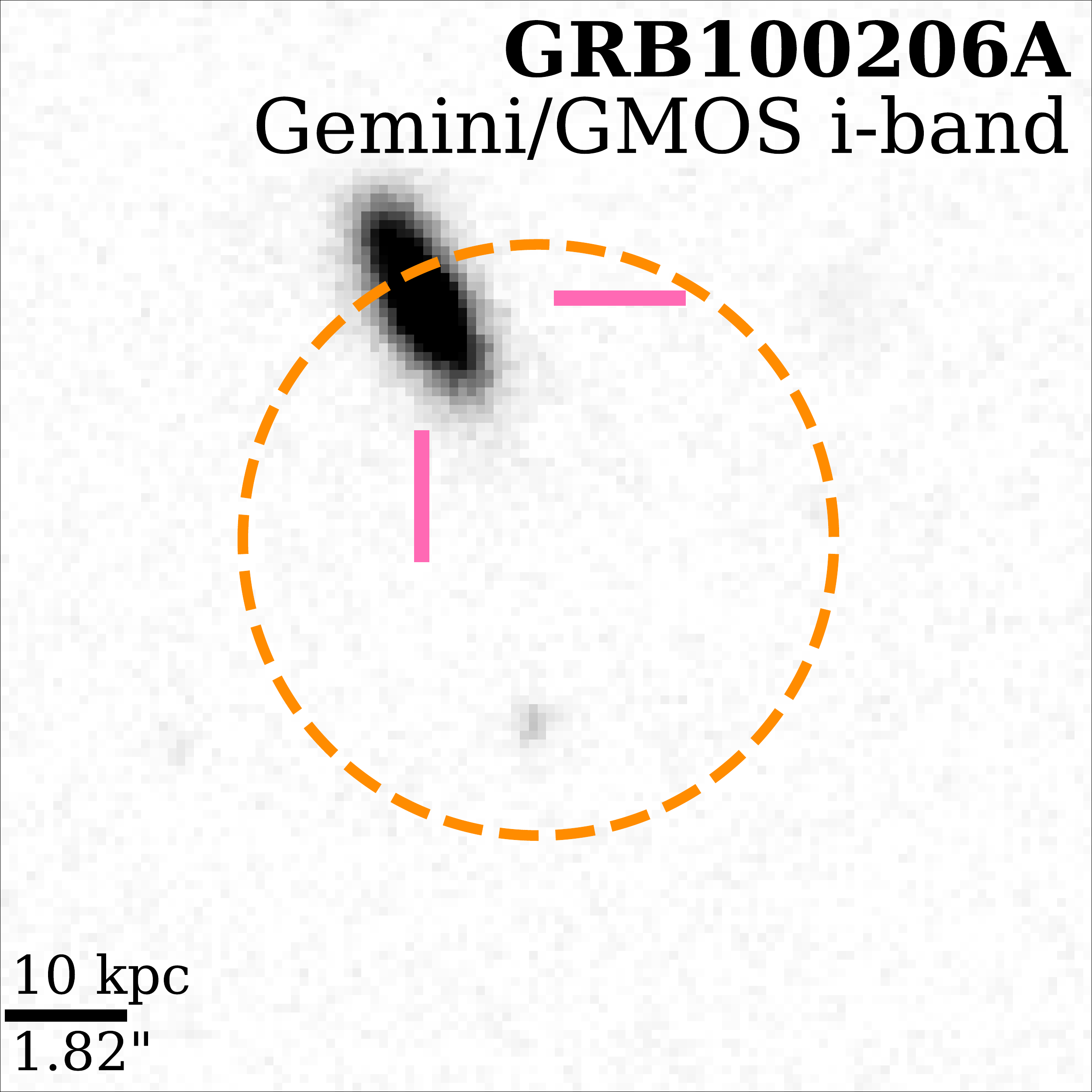}
\includegraphics[width=0.245\textwidth]{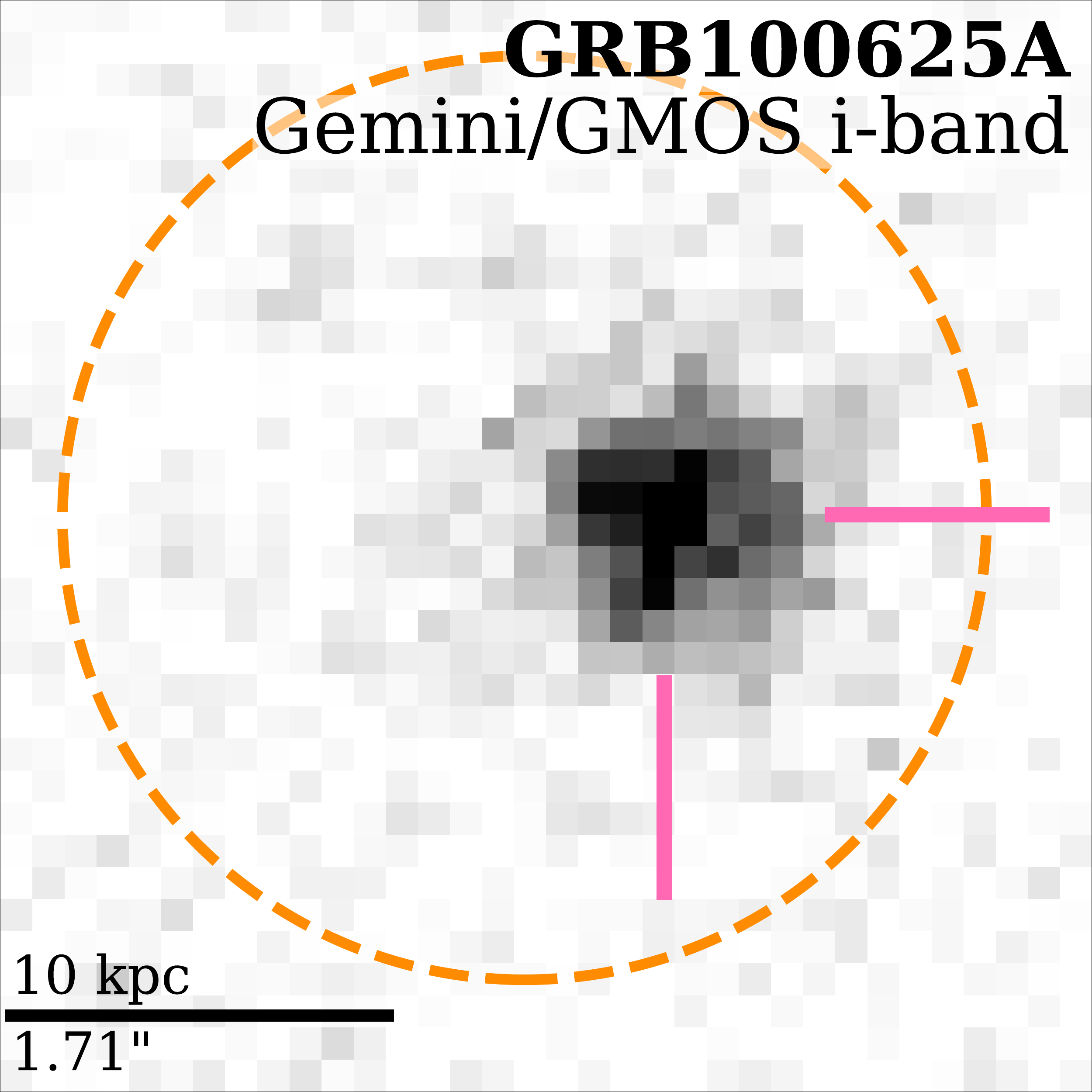}
\includegraphics[width=0.245\textwidth]{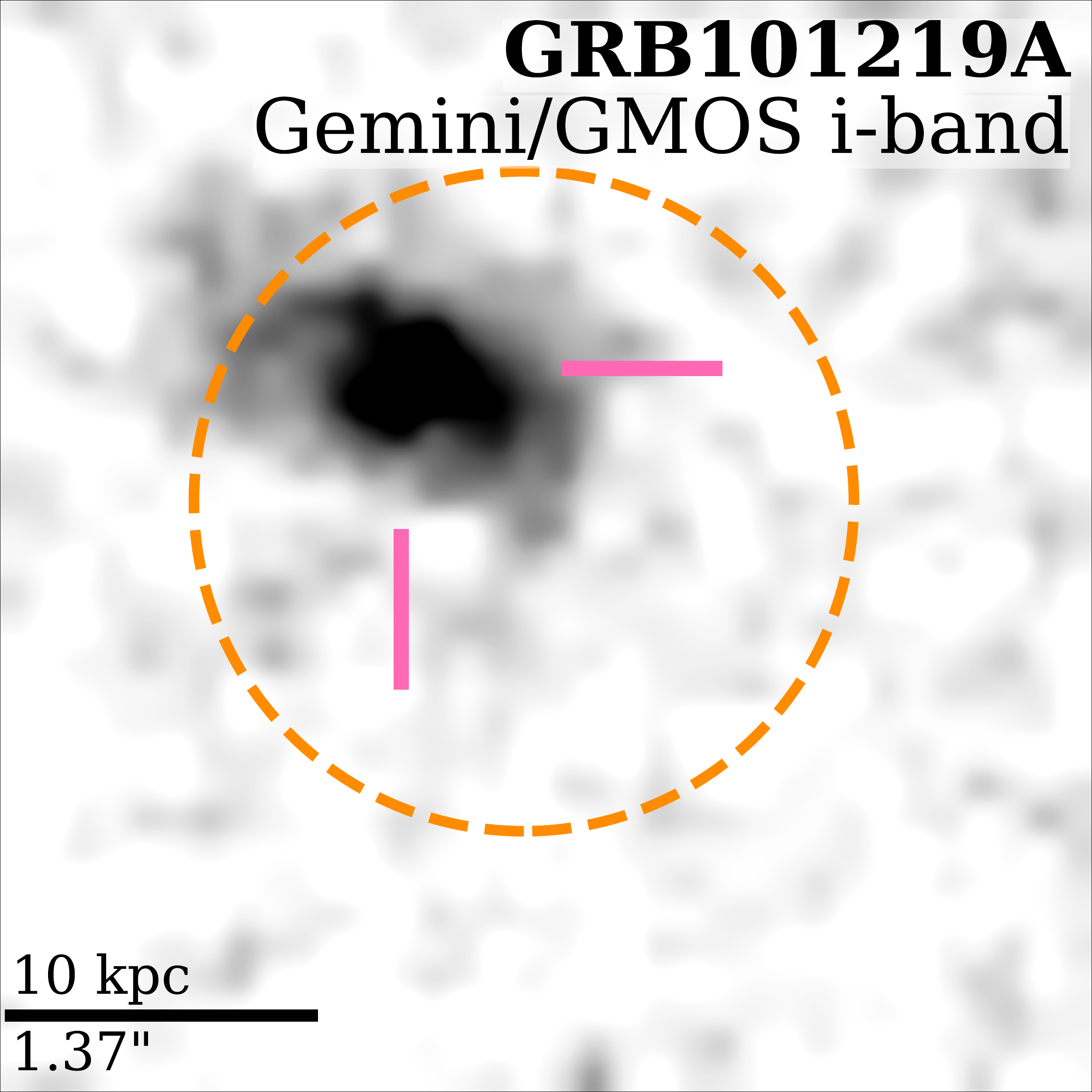}
\includegraphics[width=0.245\textwidth]{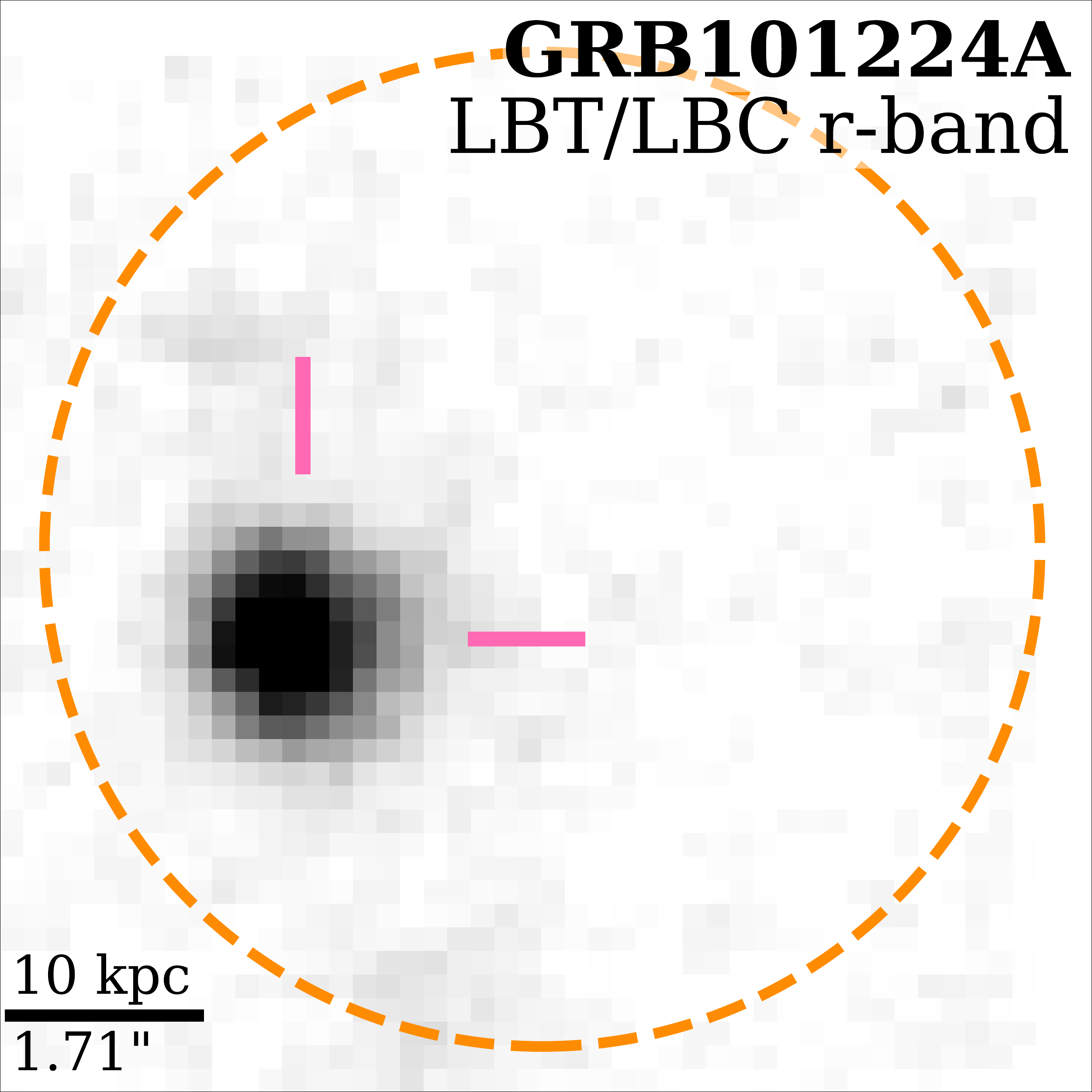}
\includegraphics[width=0.245\textwidth]{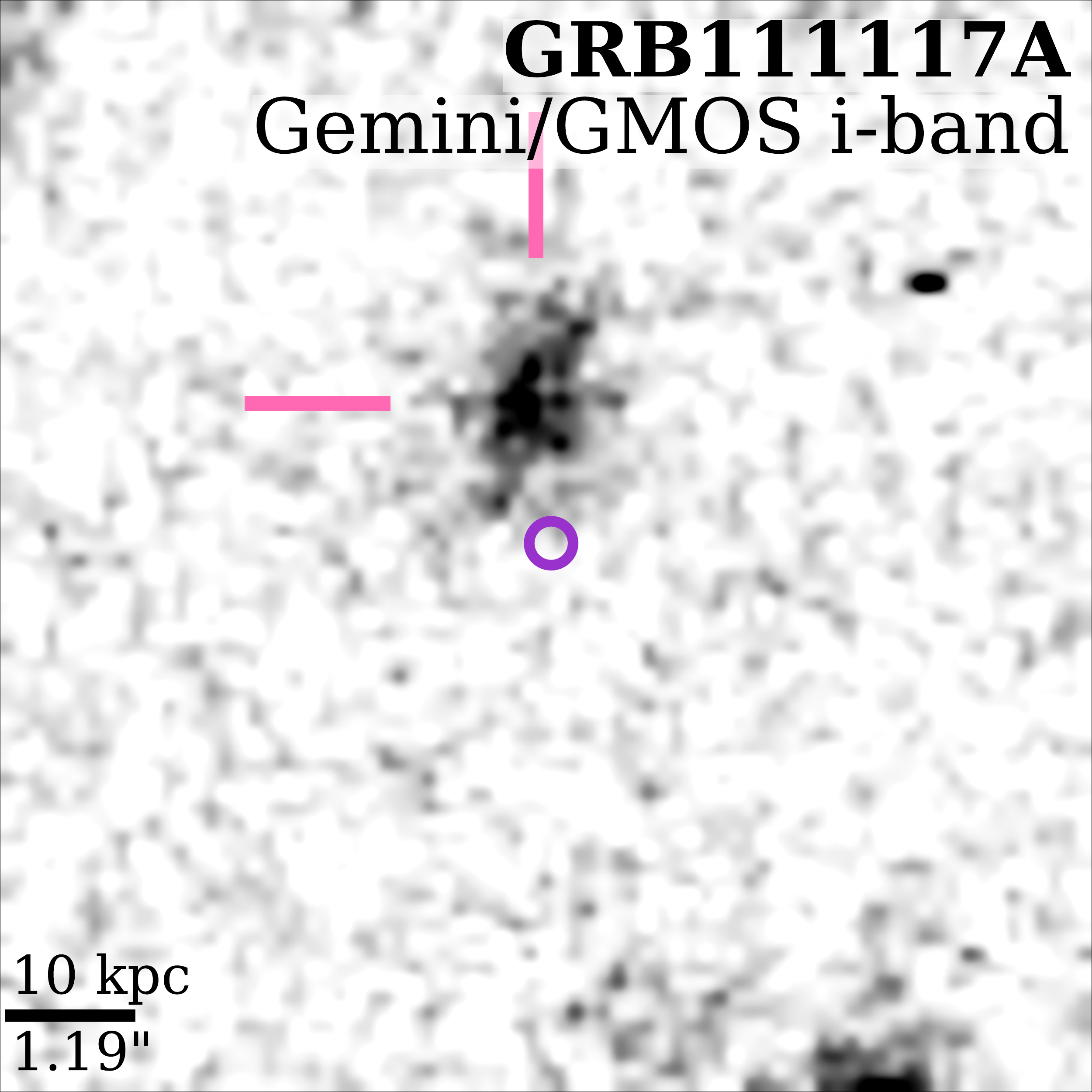}
\includegraphics[width=0.245\textwidth]{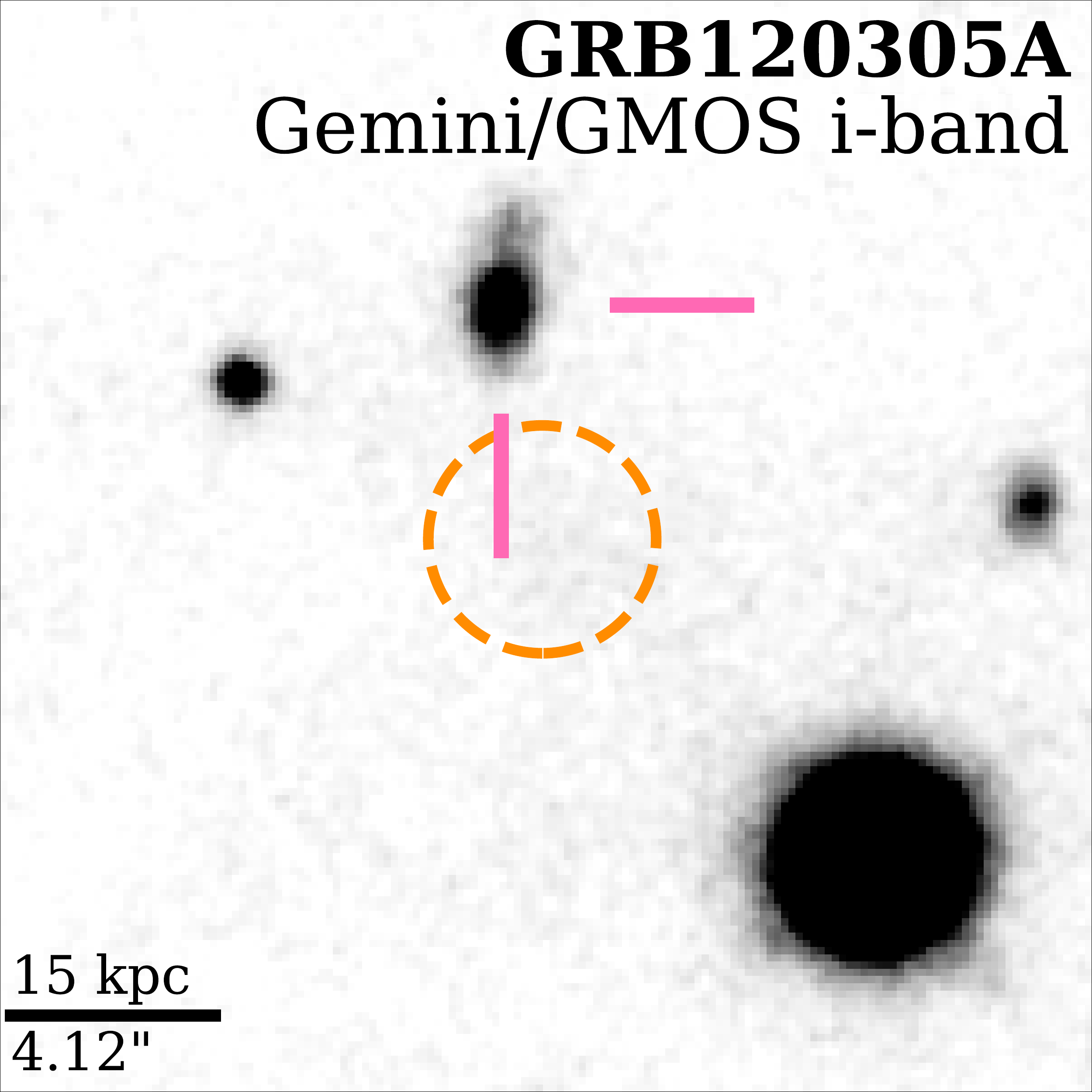}
\includegraphics[width=0.245\textwidth]{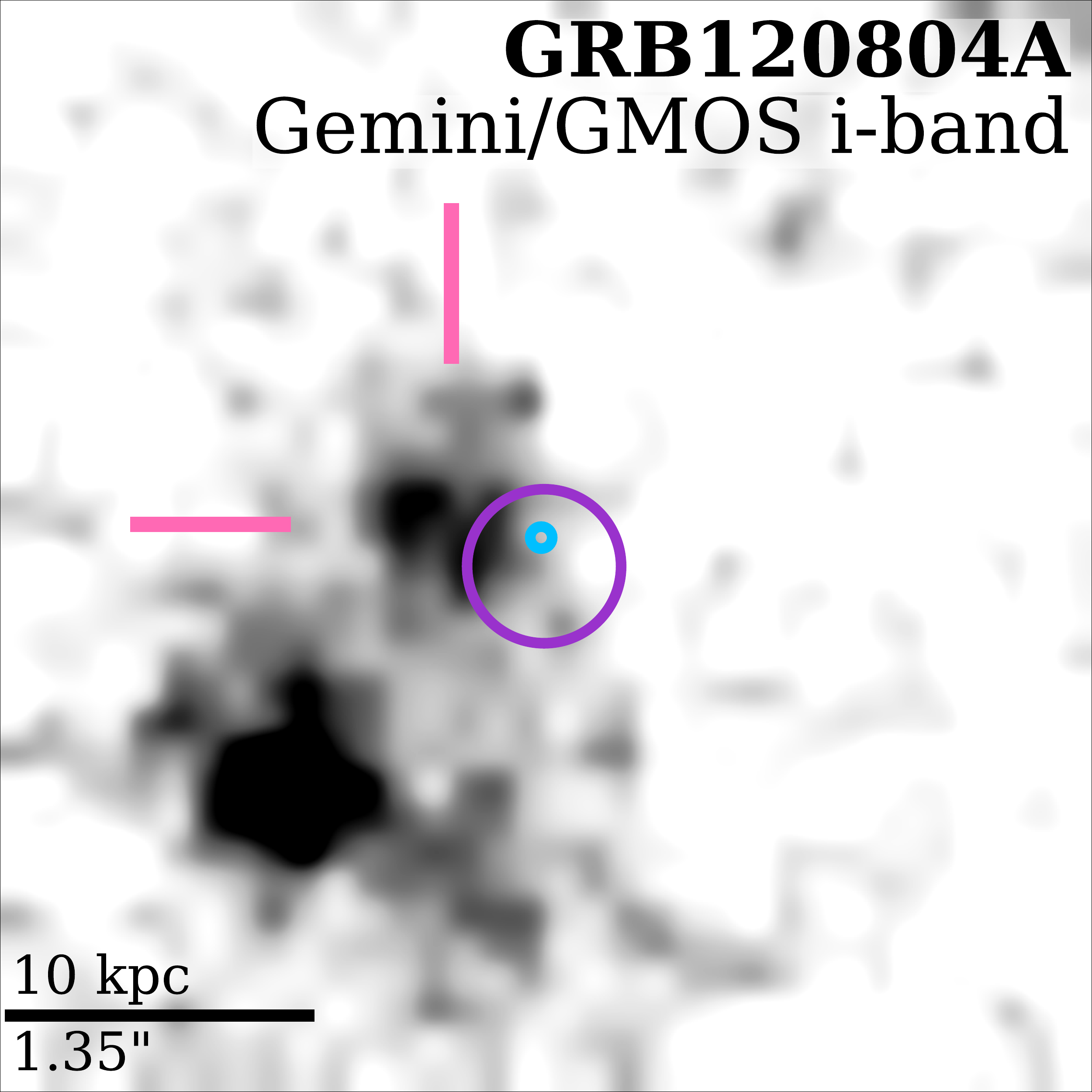}
\includegraphics[width=0.245\textwidth]{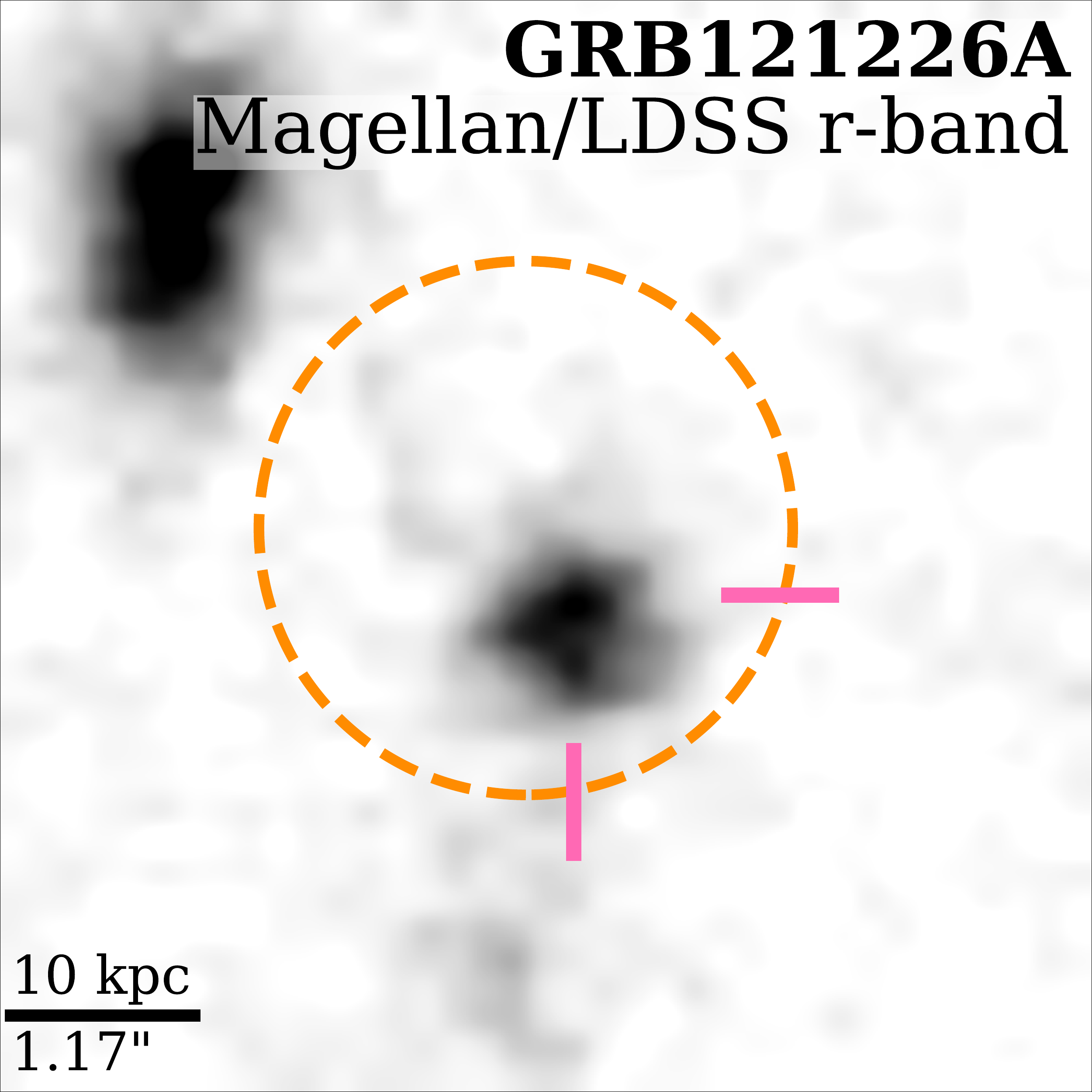}
\includegraphics[width=0.245\textwidth]{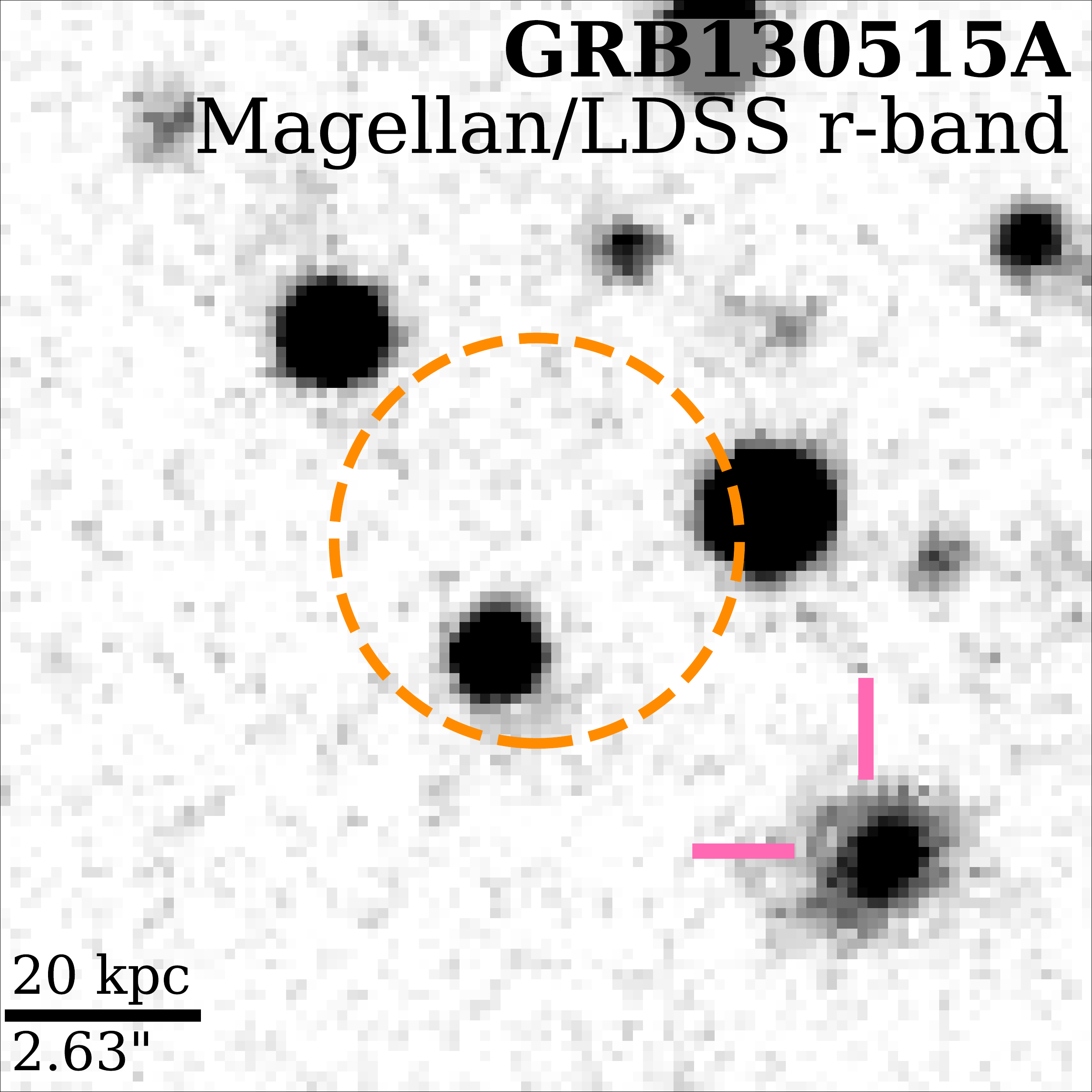}
\includegraphics[width=0.245\textwidth]{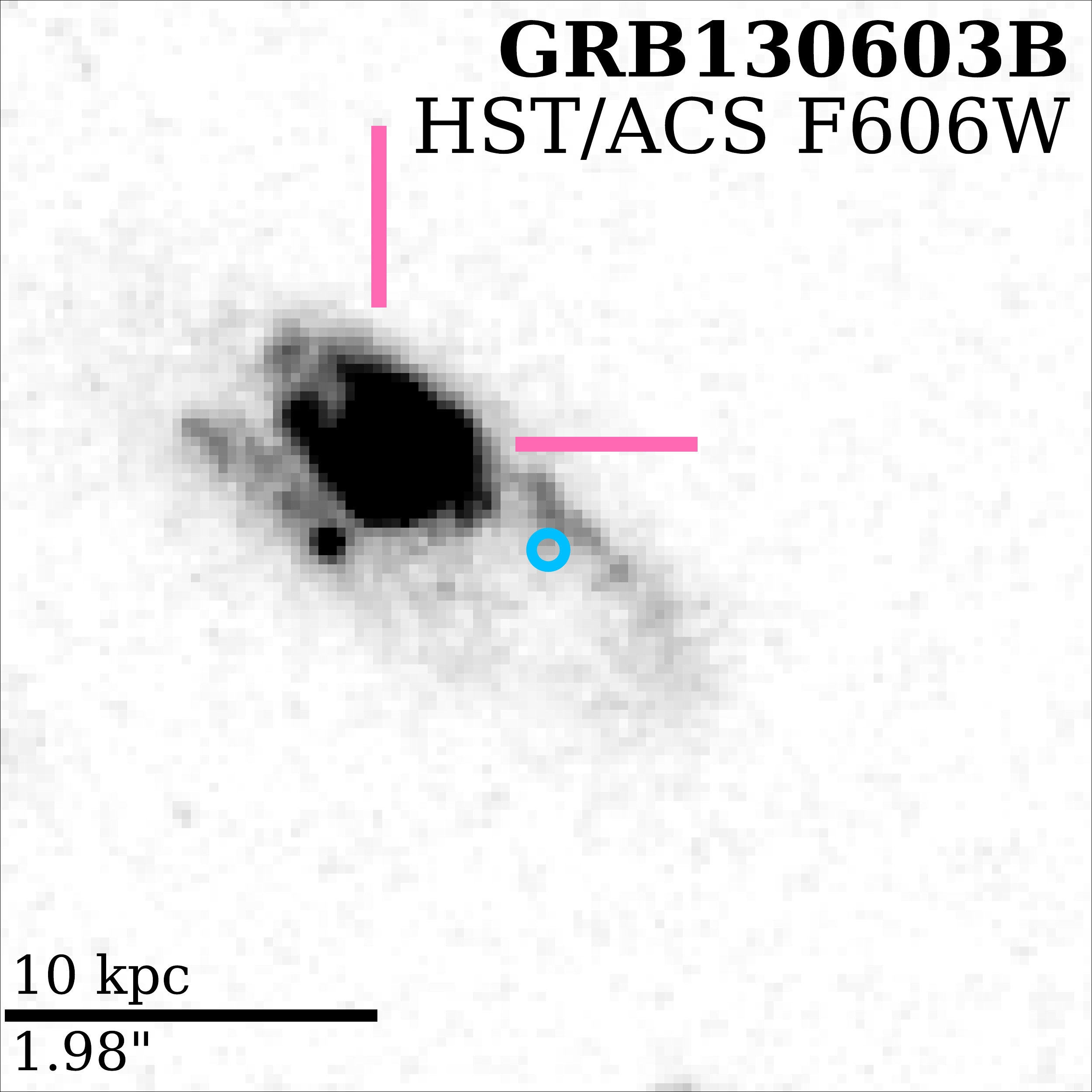}
\includegraphics[width=0.245\textwidth]{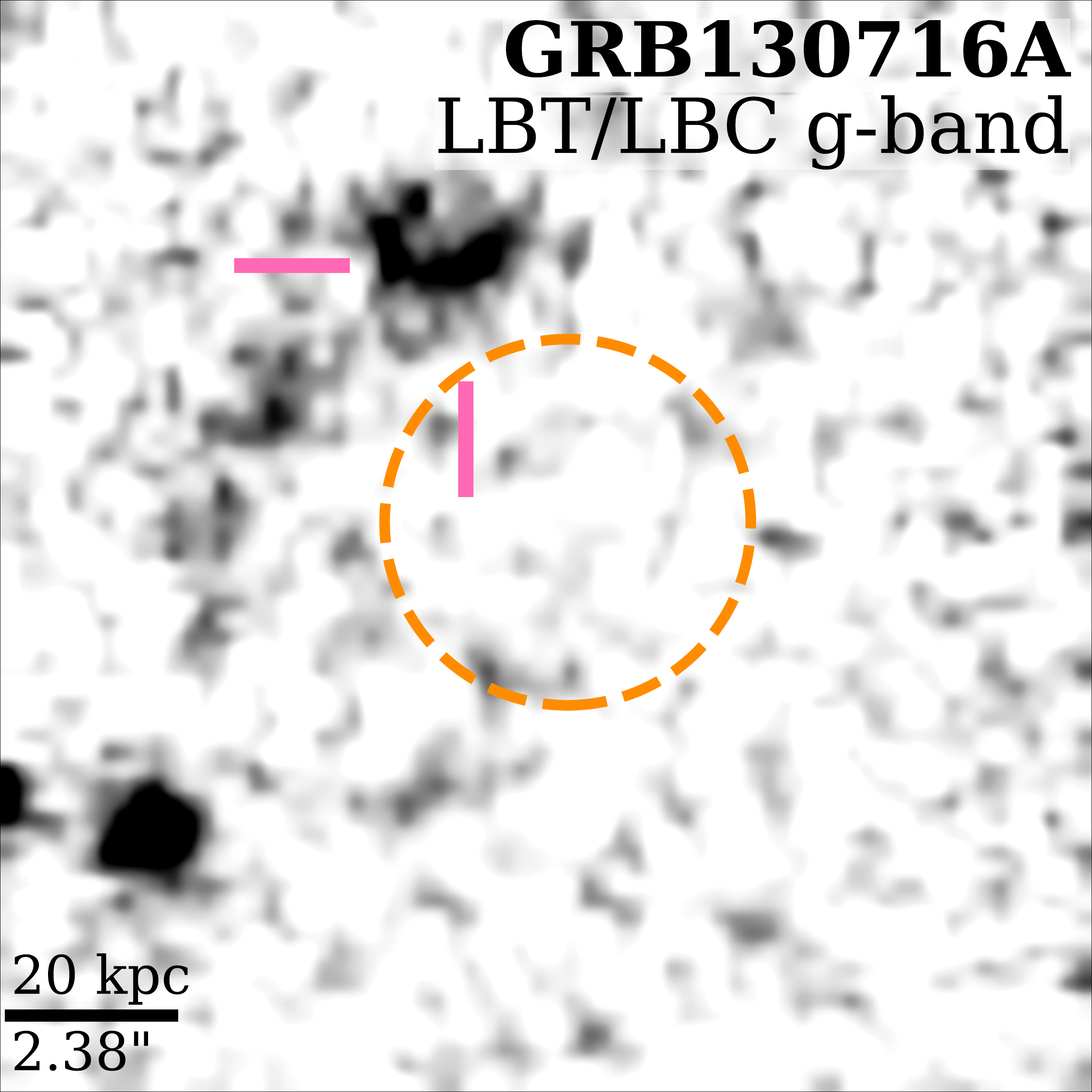}
\includegraphics[width=0.245\textwidth]{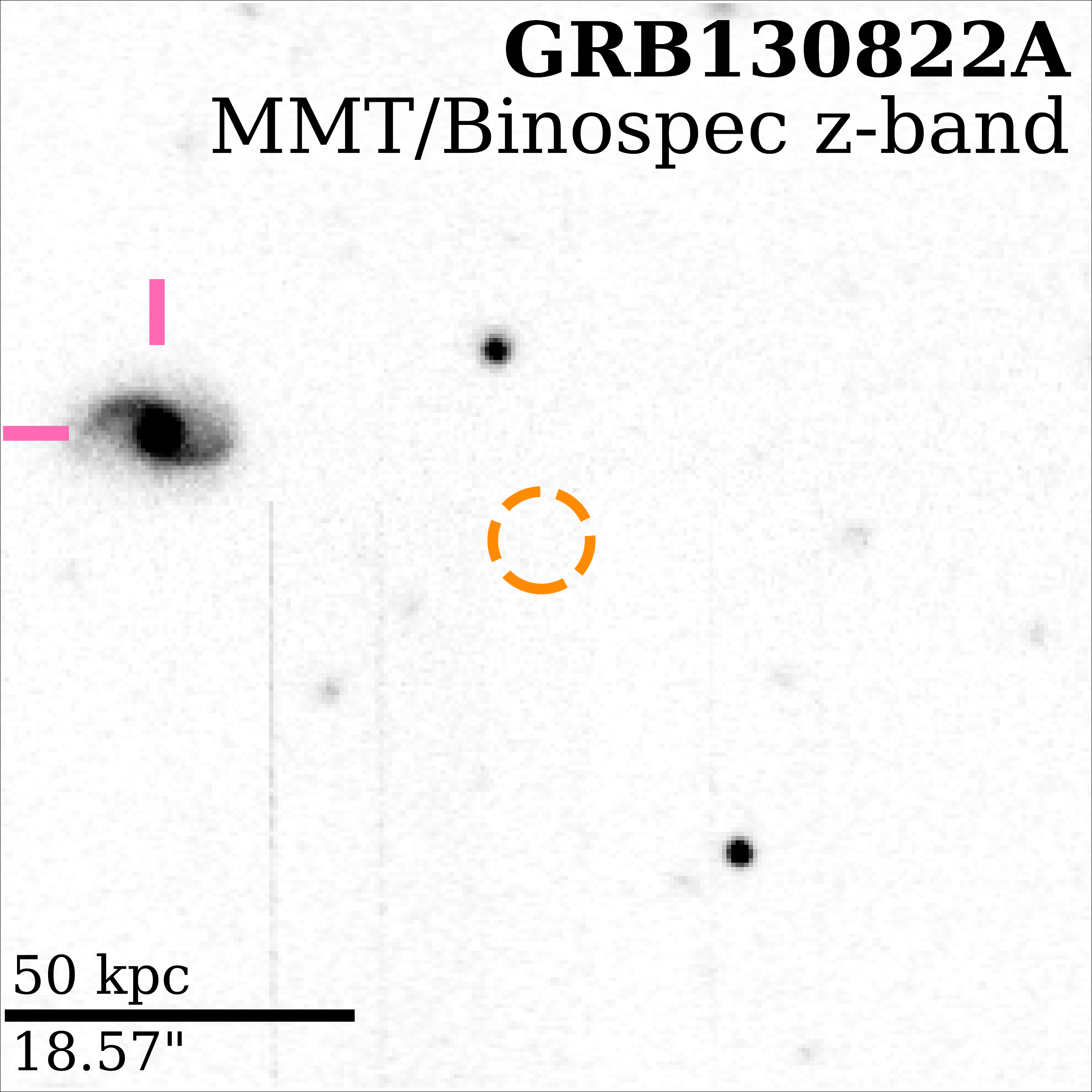}
\caption{Representative images of the host galaxies of the short GRBs in our catalog. In each panel, the most precise afterglow localization(s) for each burst is/are plotted (XRT 90\%: orange dashed, optical $1\sigma$: blue, {\it Chandra} or VLA $1\sigma$: purple). The putative host galaxy is denoted by the pink cross-hairs. All images are oriented North up and East to the left.}
\end{figure*}

\addtocounter{figure}{-1}

\begin{figure*}[t]
\centering
\includegraphics[width=0.245\textwidth]{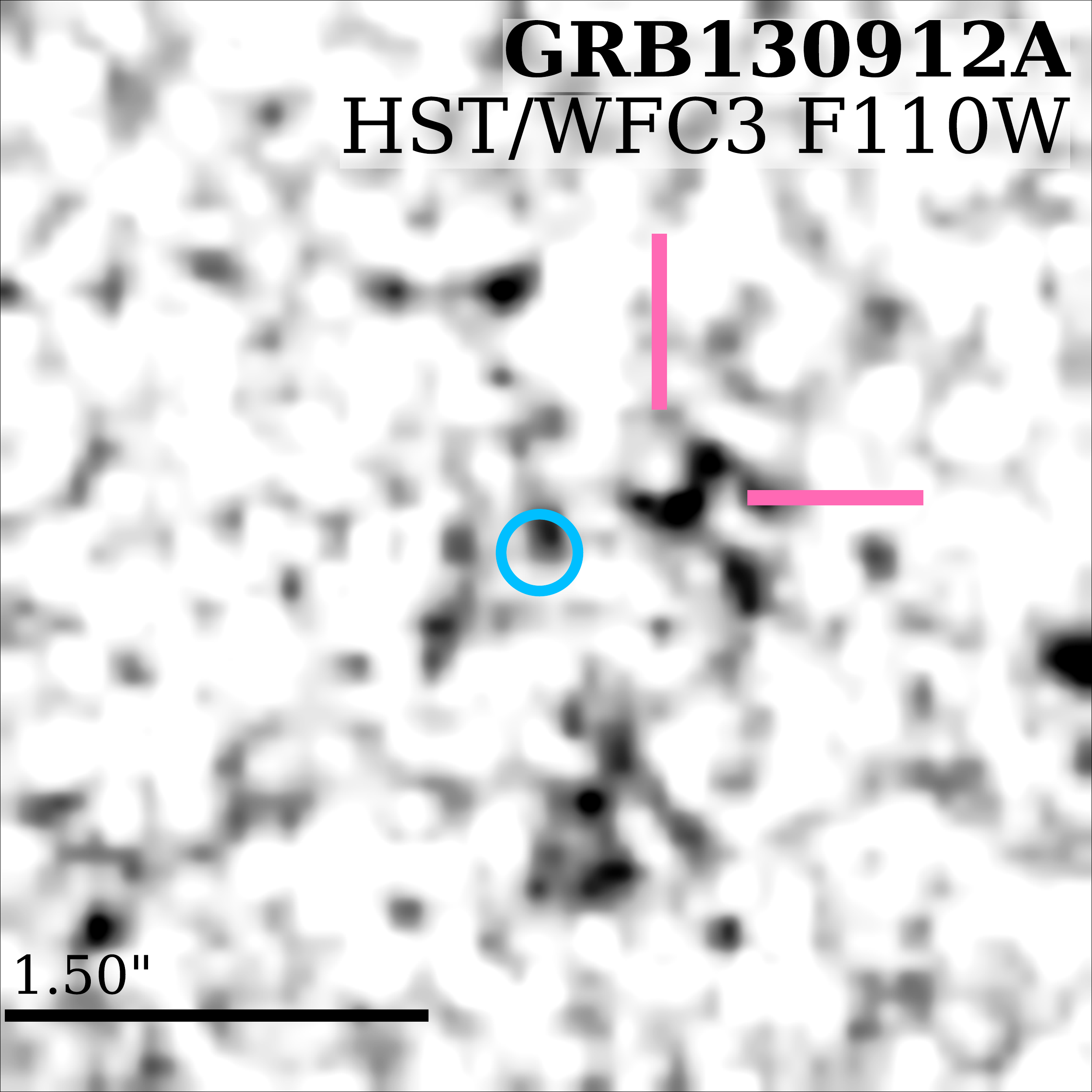}
\includegraphics[width=0.245\textwidth]{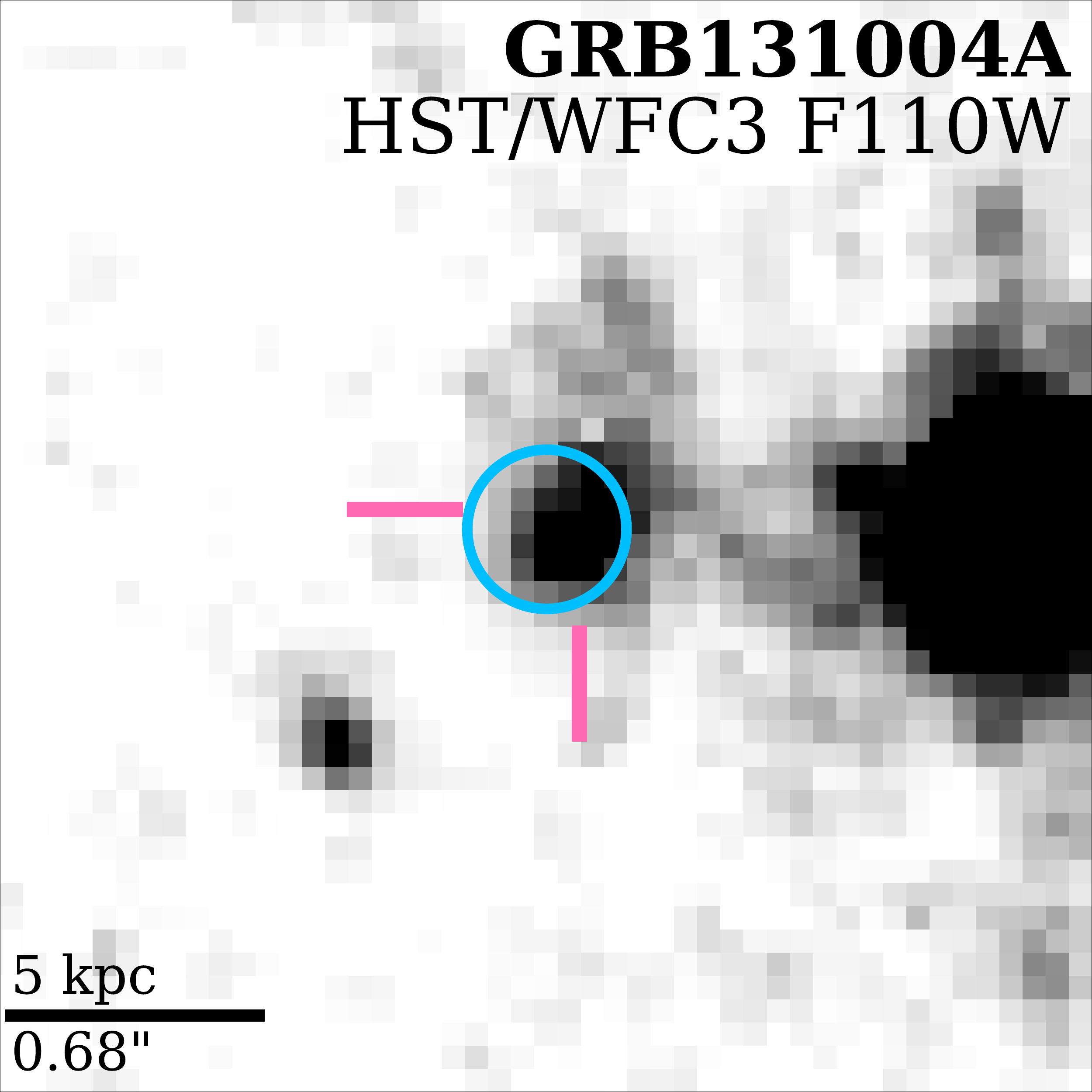}
\includegraphics[width=0.245\textwidth]{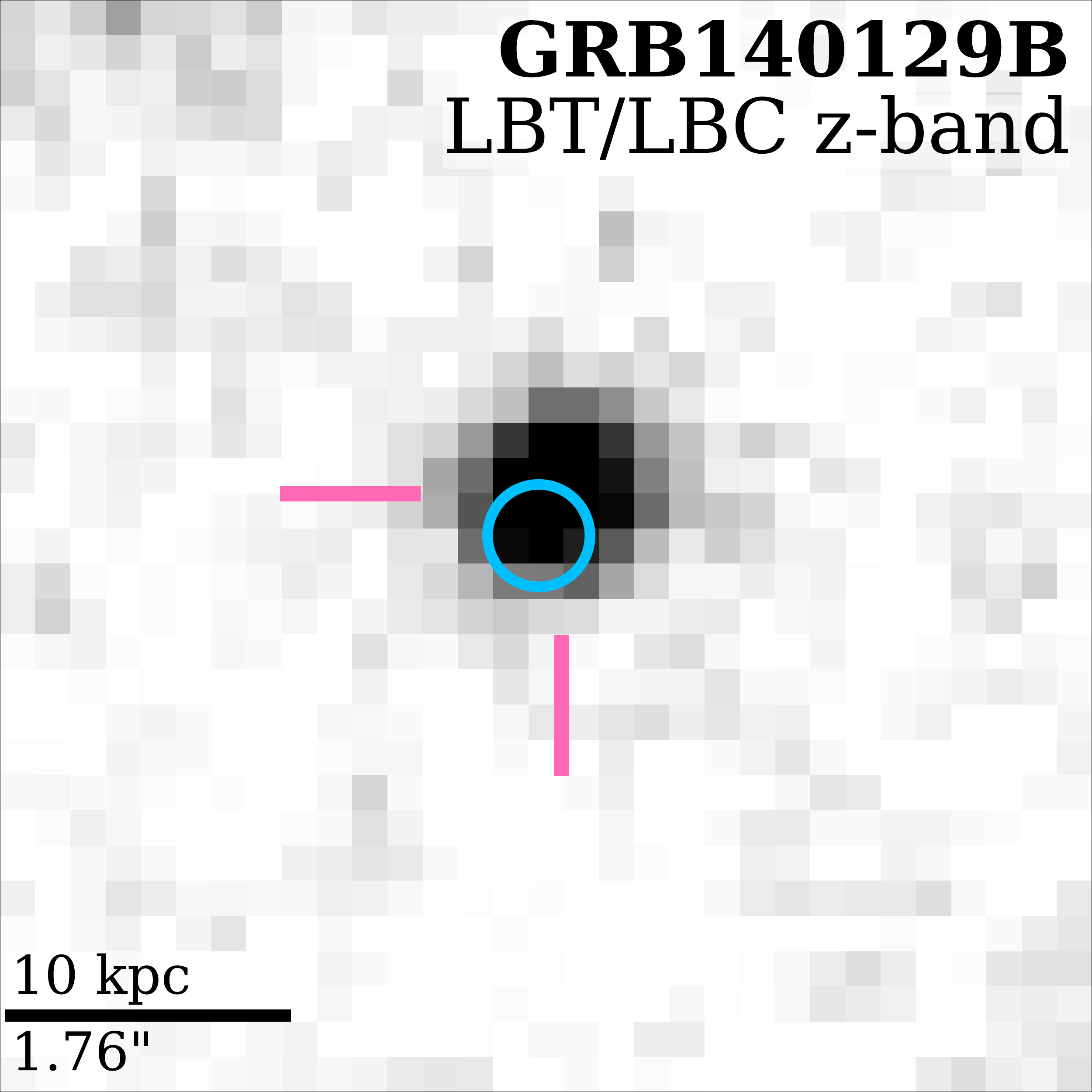}
\includegraphics[width=0.245\textwidth]{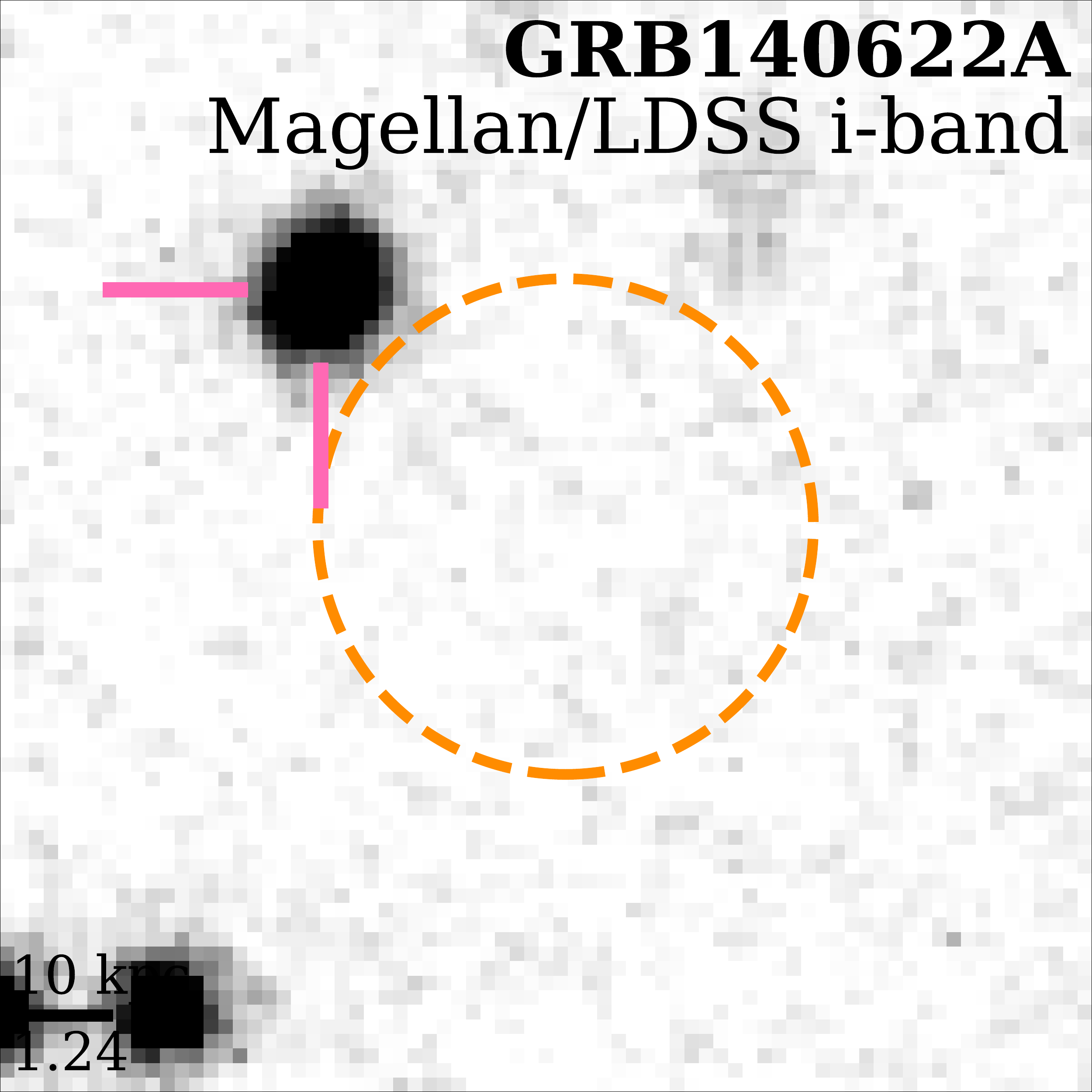}
\includegraphics[width=0.245\textwidth]{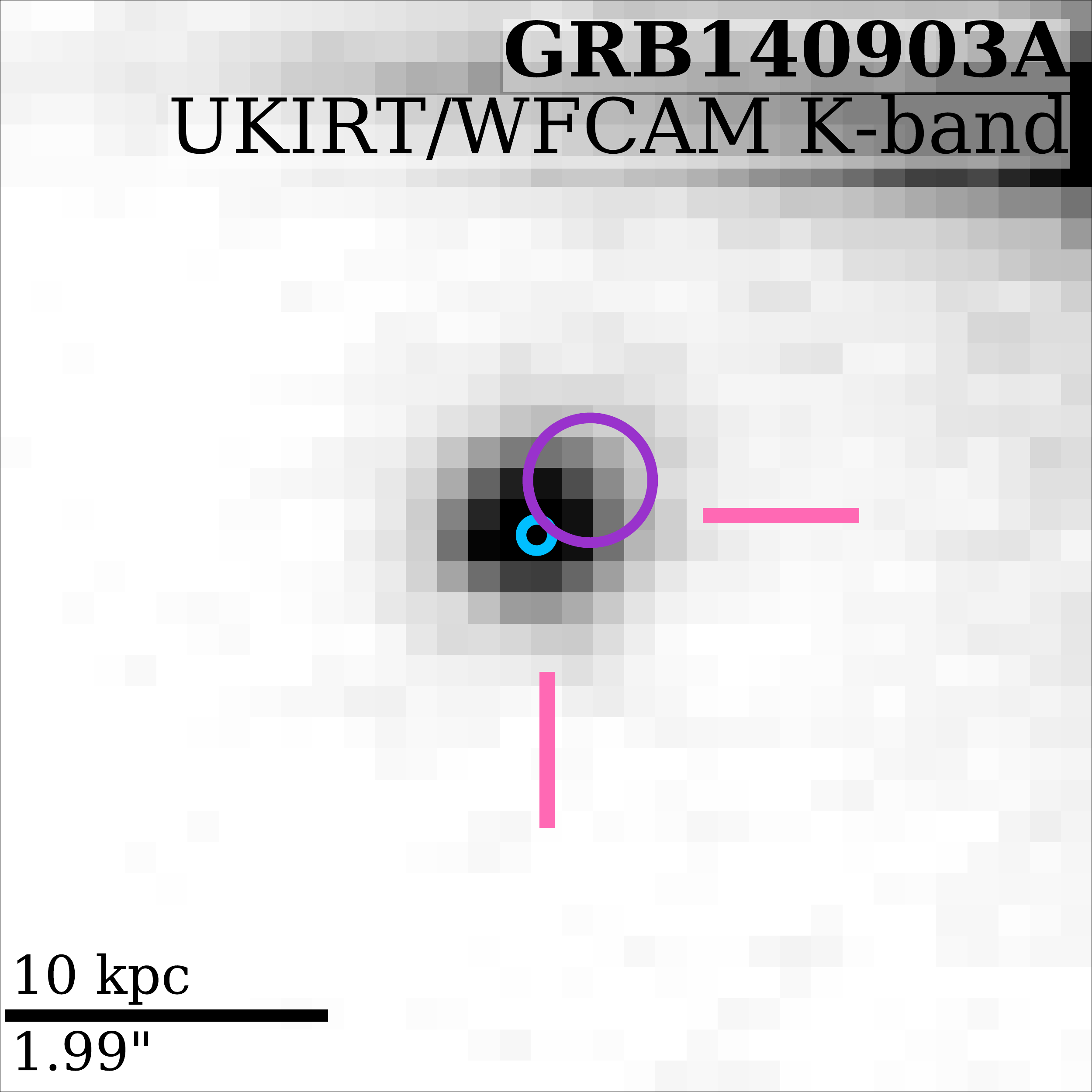}
\includegraphics[width=0.245\textwidth]{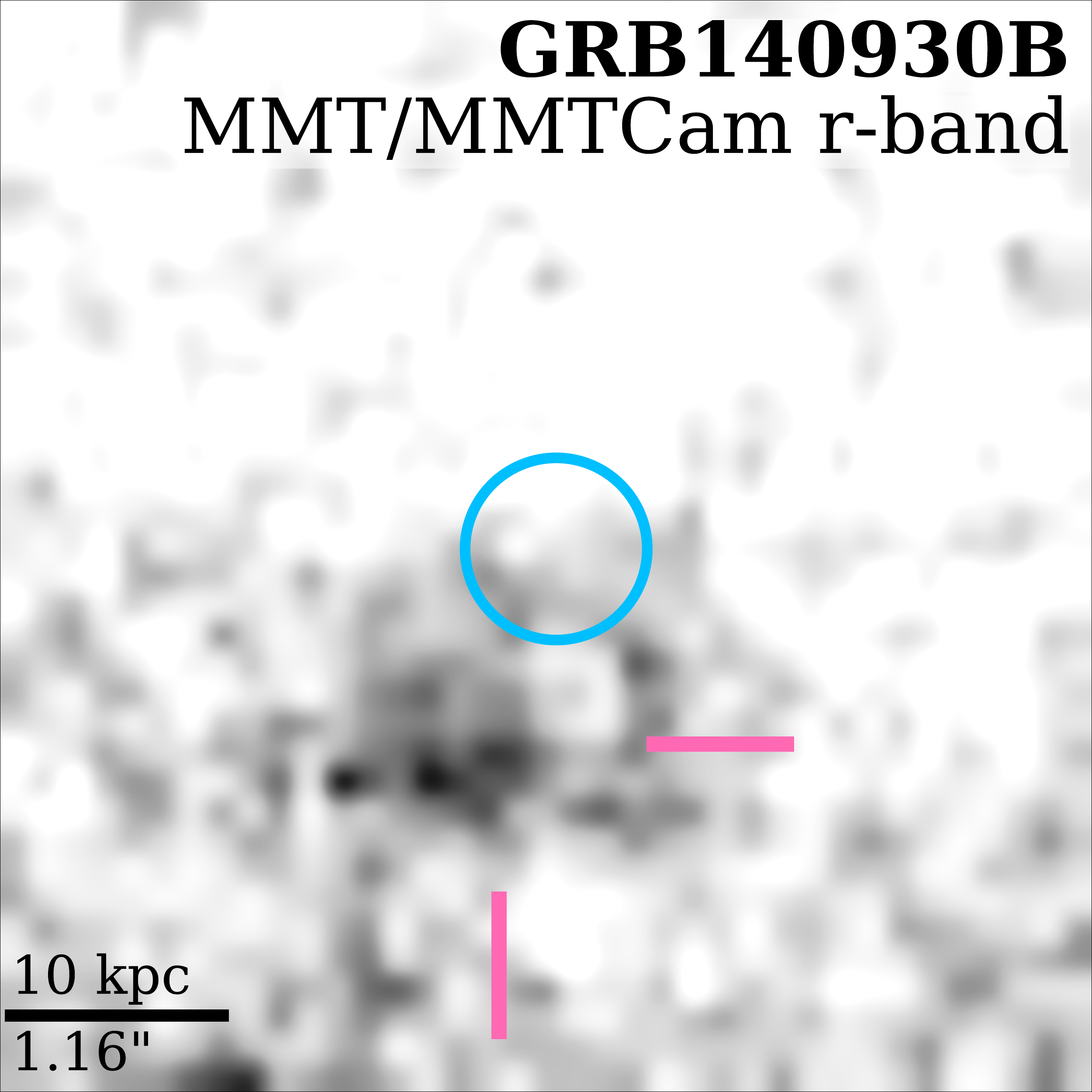}
\includegraphics[width=0.245\textwidth]{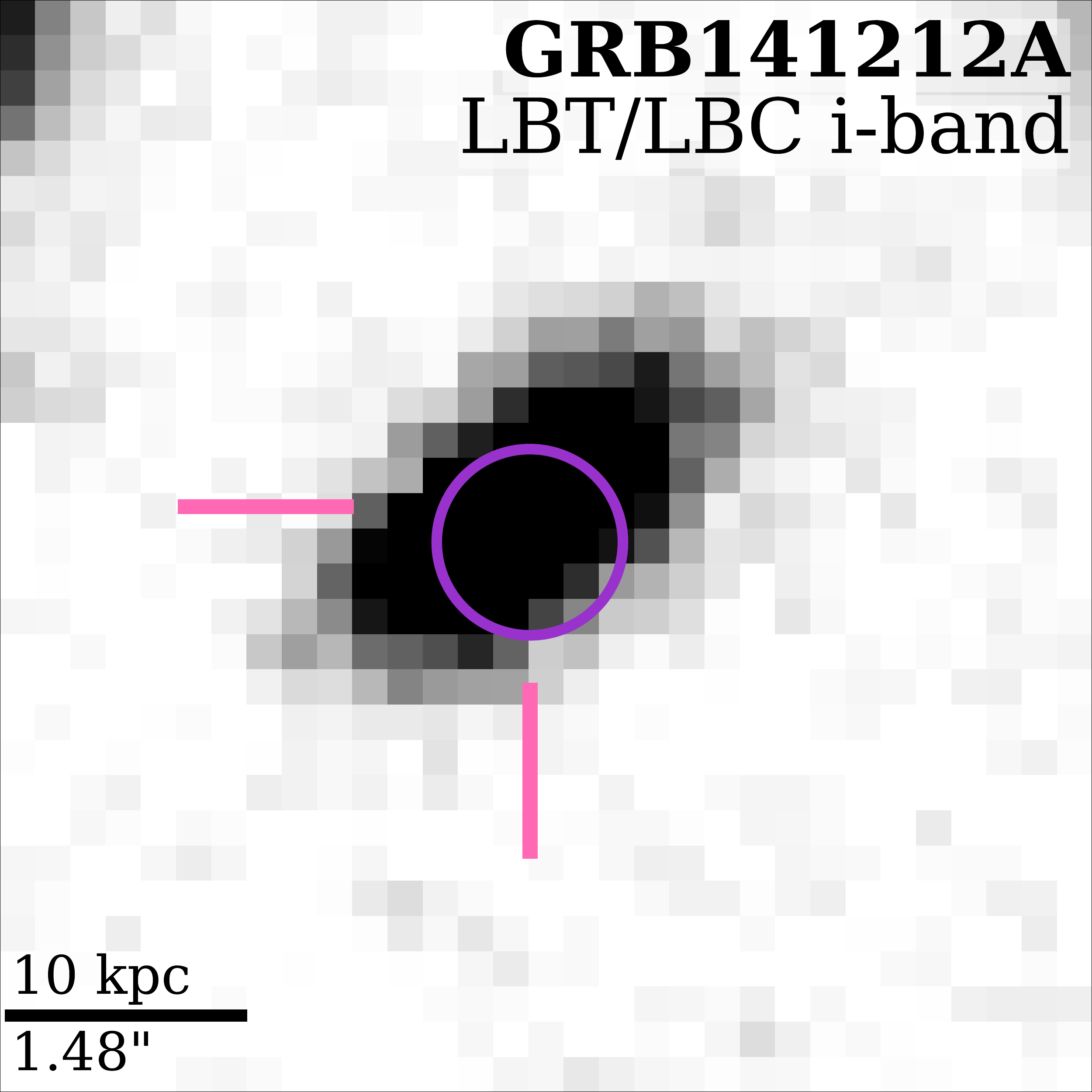}
\includegraphics[width=0.245\textwidth]{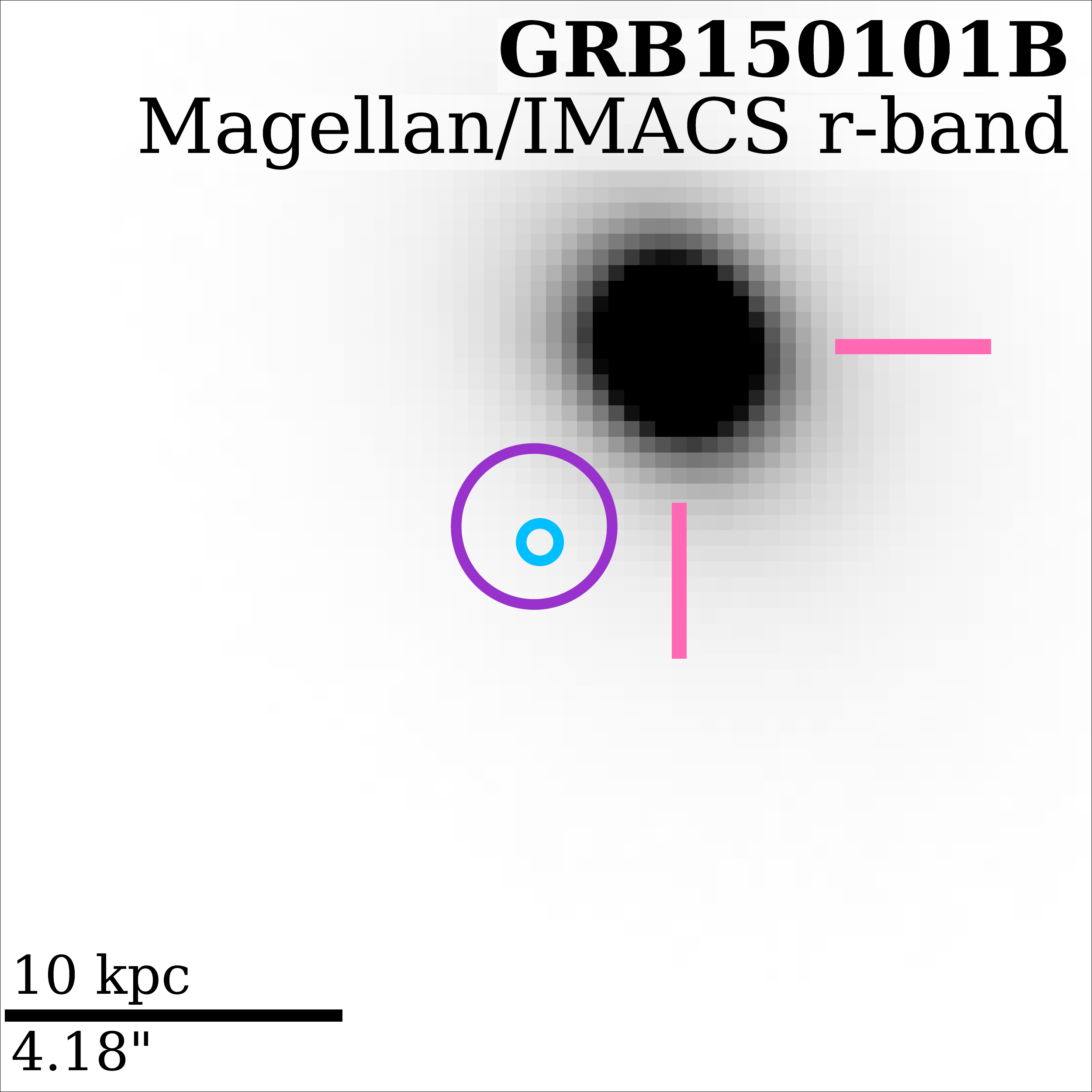}
\includegraphics[width=0.245\textwidth]{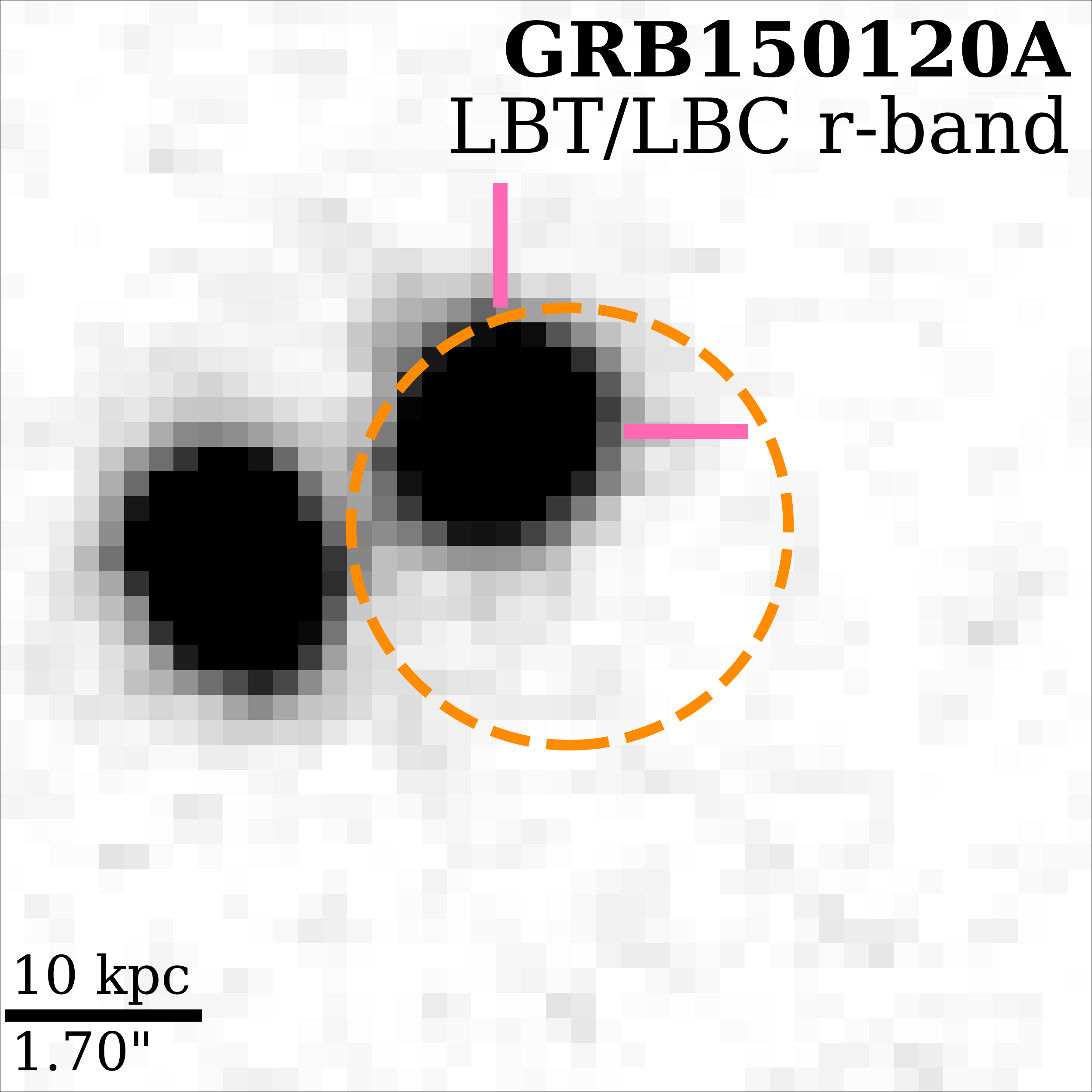}
\includegraphics[width=0.245\textwidth]{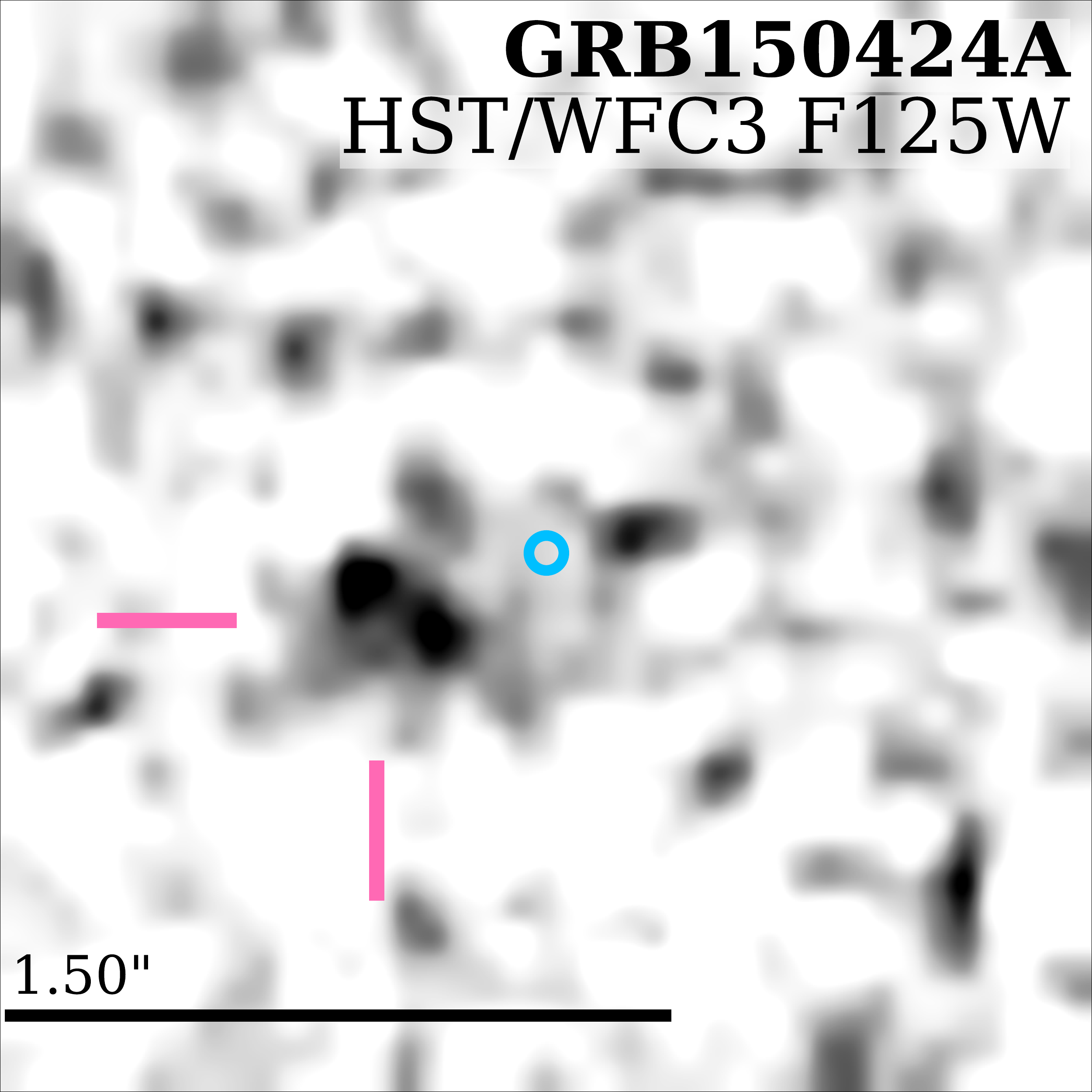}
\includegraphics[width=0.245\textwidth]{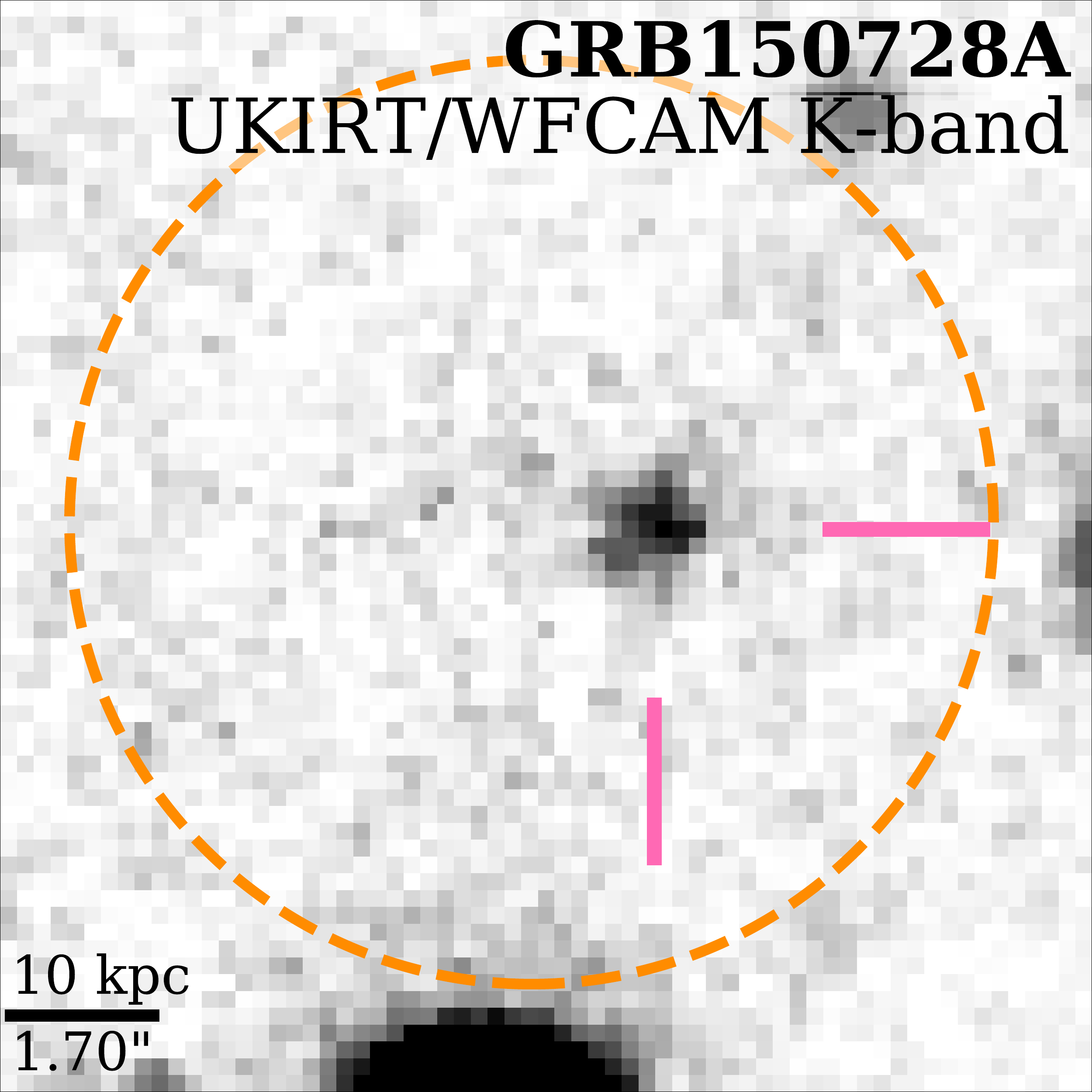}
\includegraphics[width=0.245\textwidth]{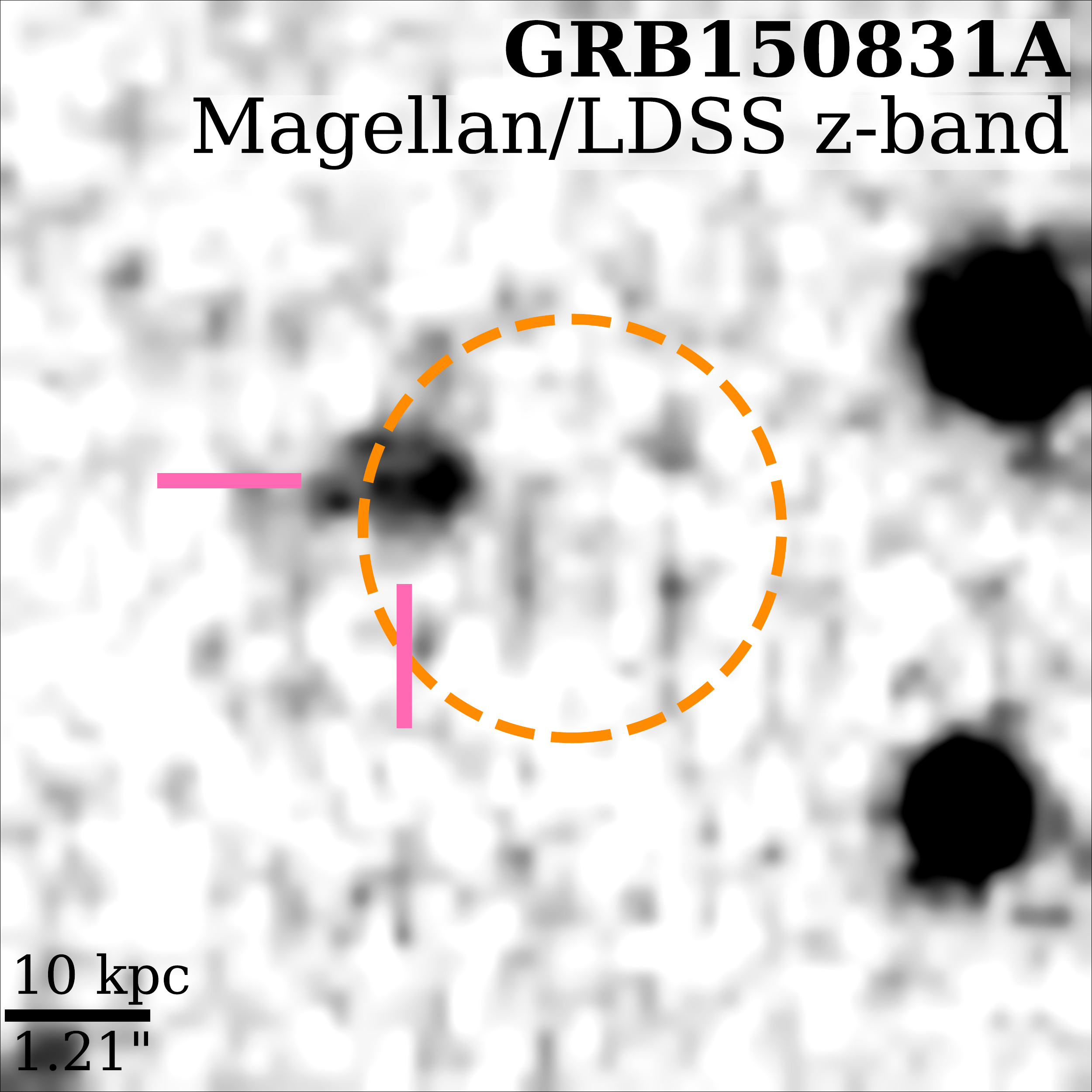}
\includegraphics[width=0.245\textwidth]{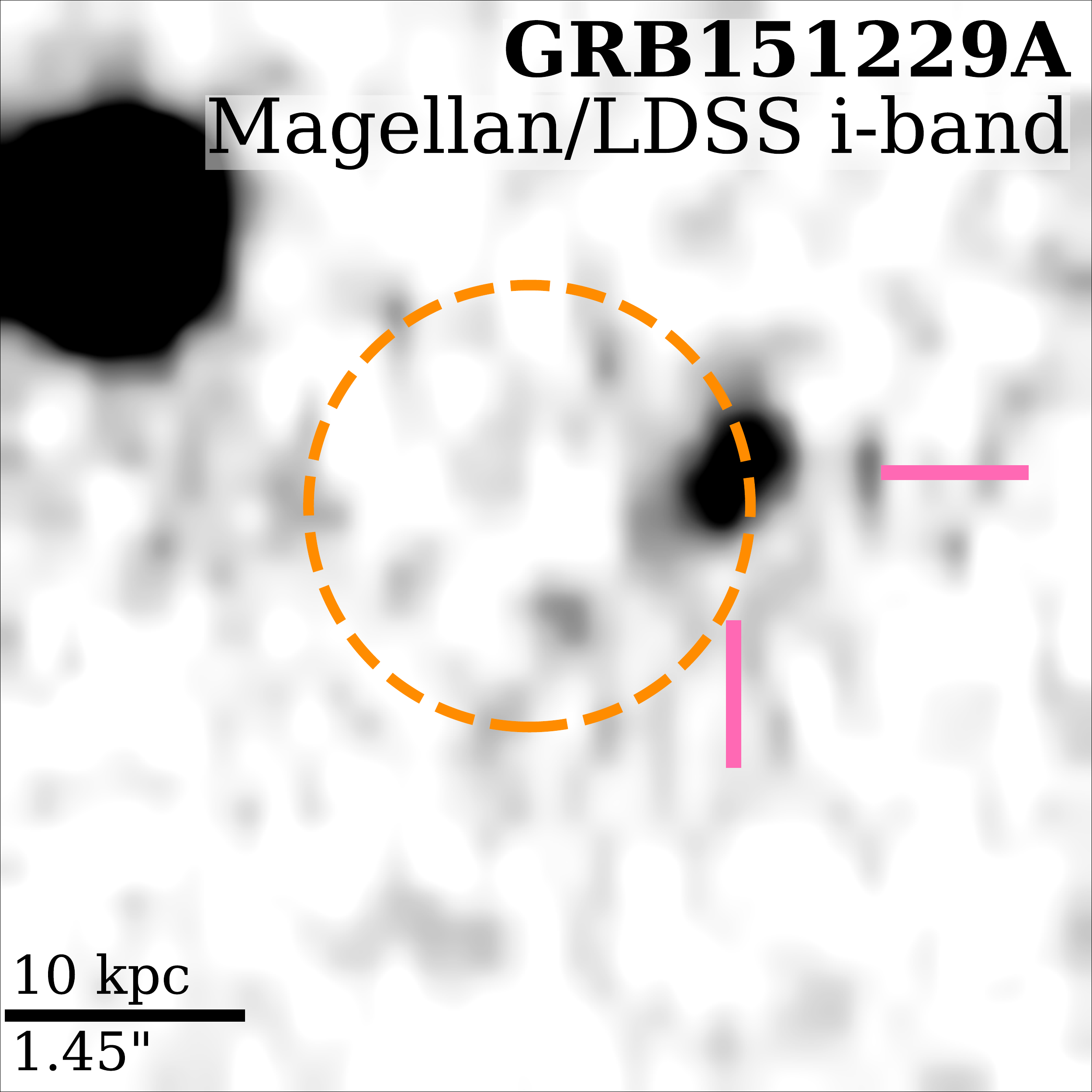}
\includegraphics[width=0.245\textwidth]{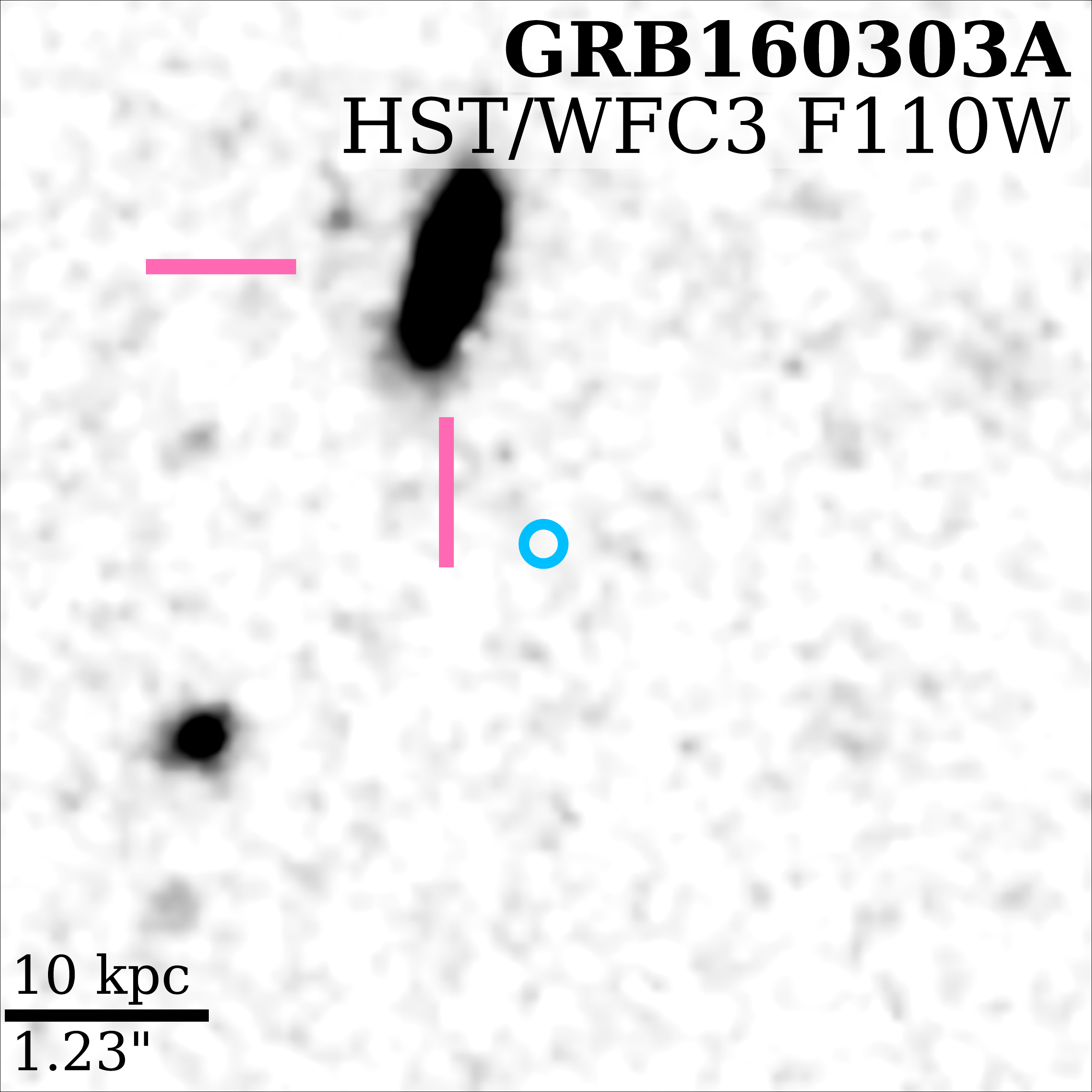}
\includegraphics[width=0.245\textwidth]{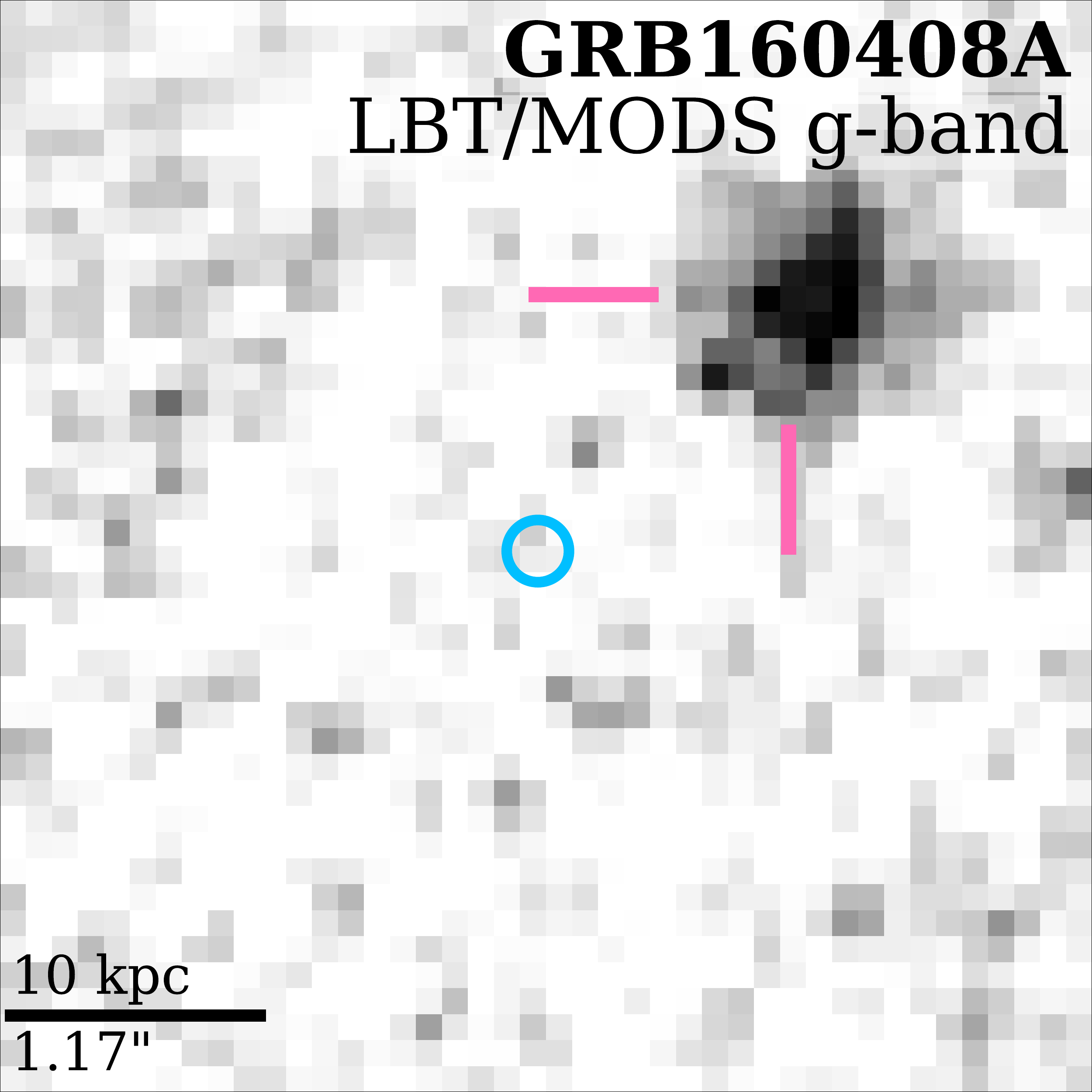}
\includegraphics[width=0.245\textwidth]{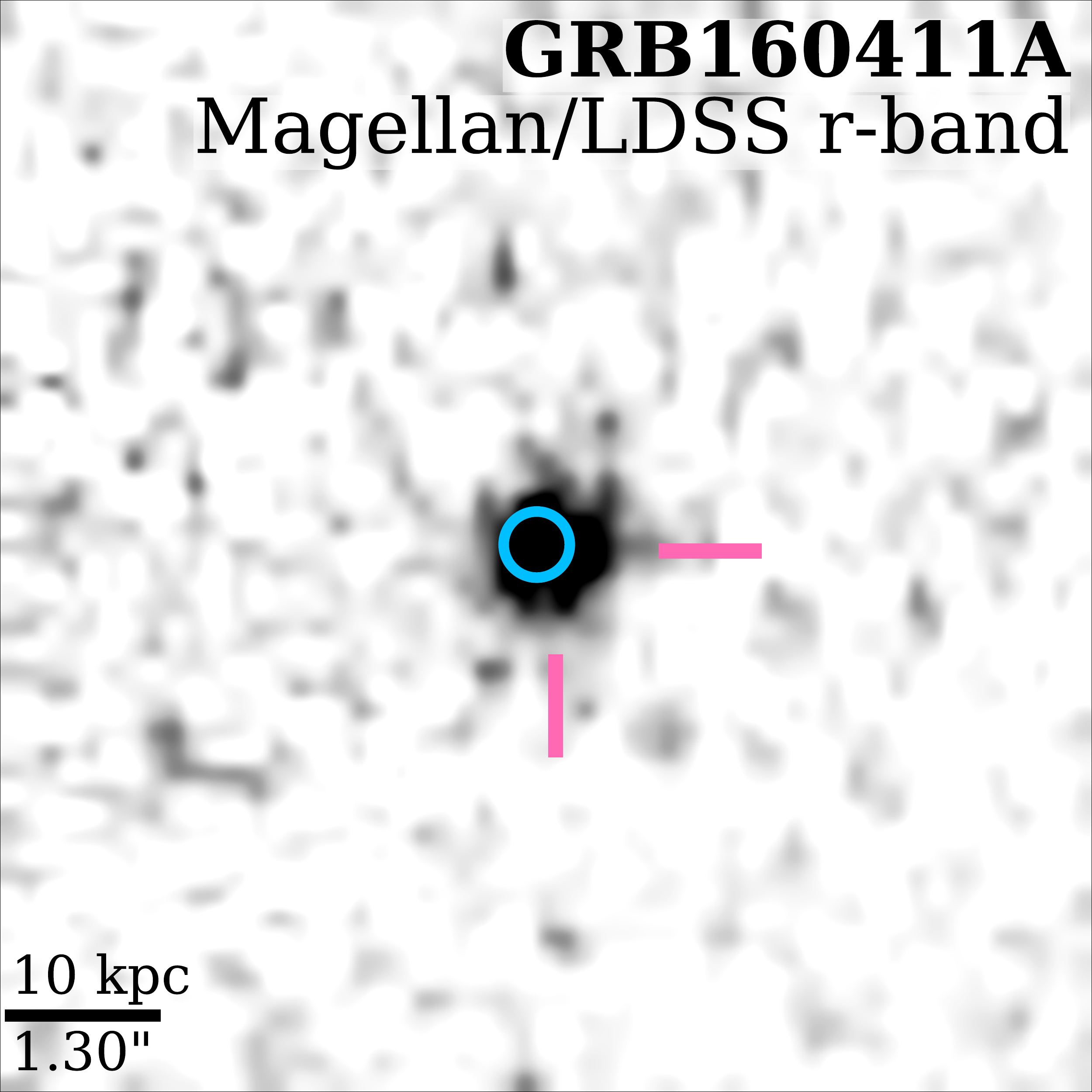}
\includegraphics[width=0.245\textwidth]{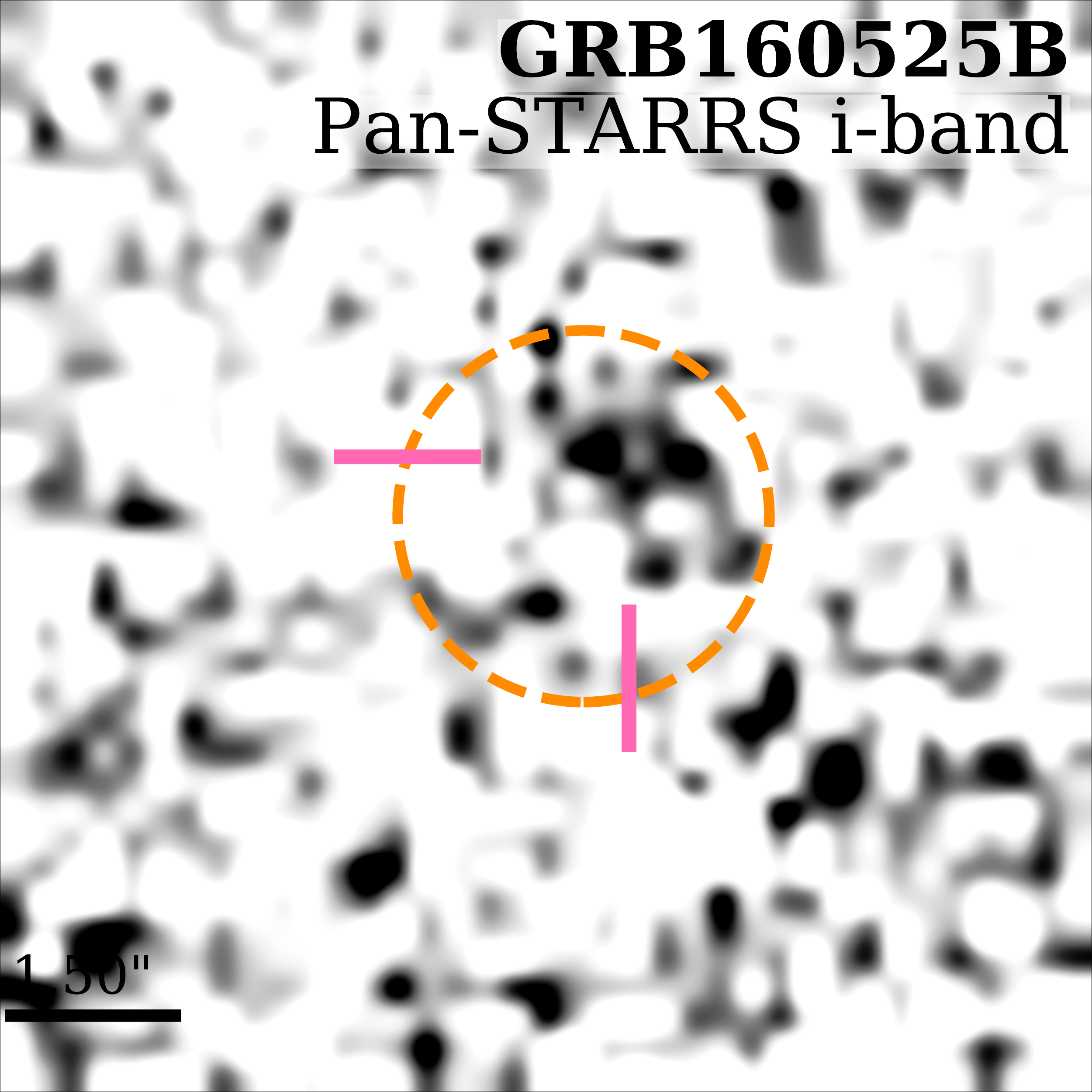}
\includegraphics[width=0.245\textwidth]{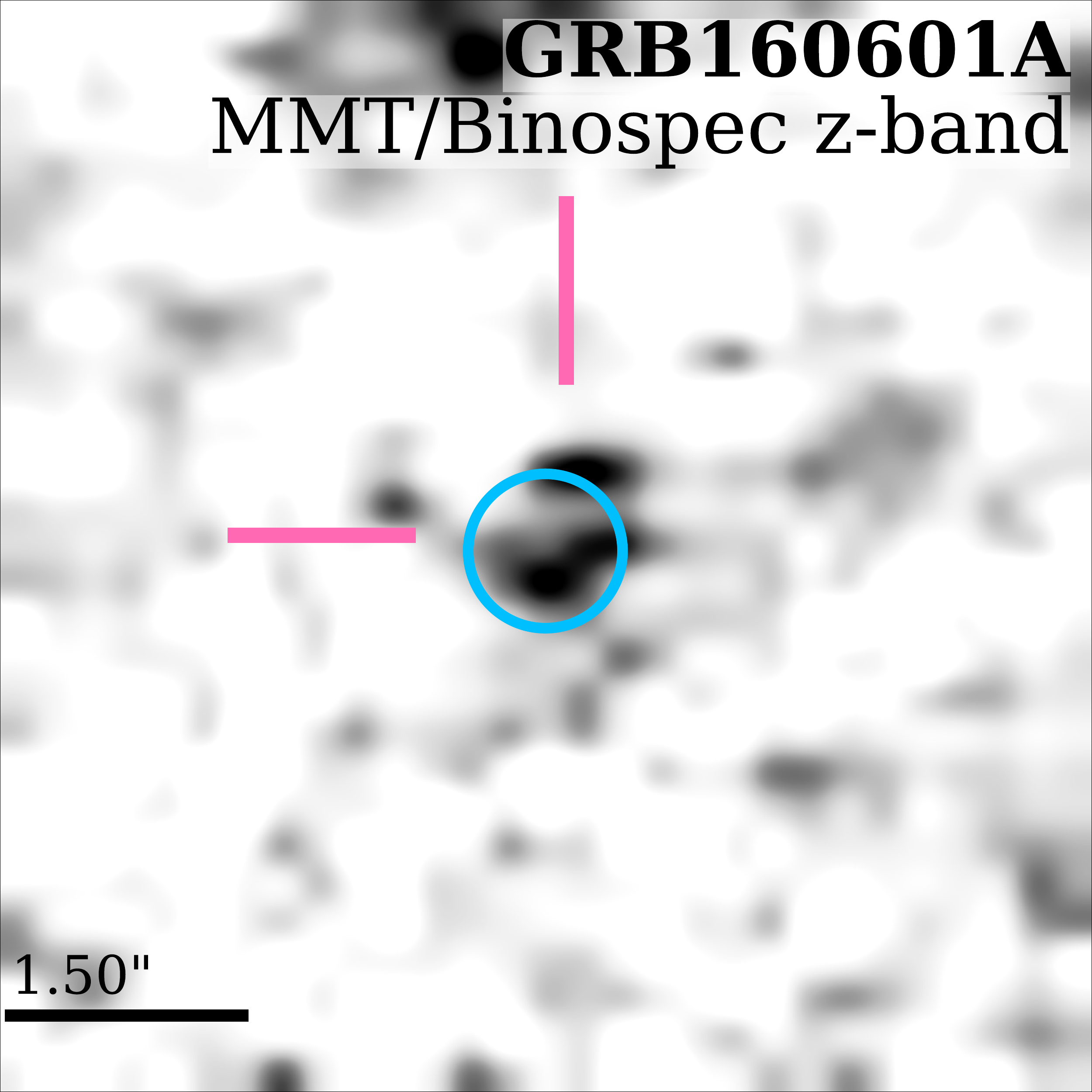}
\includegraphics[width=0.245\textwidth]{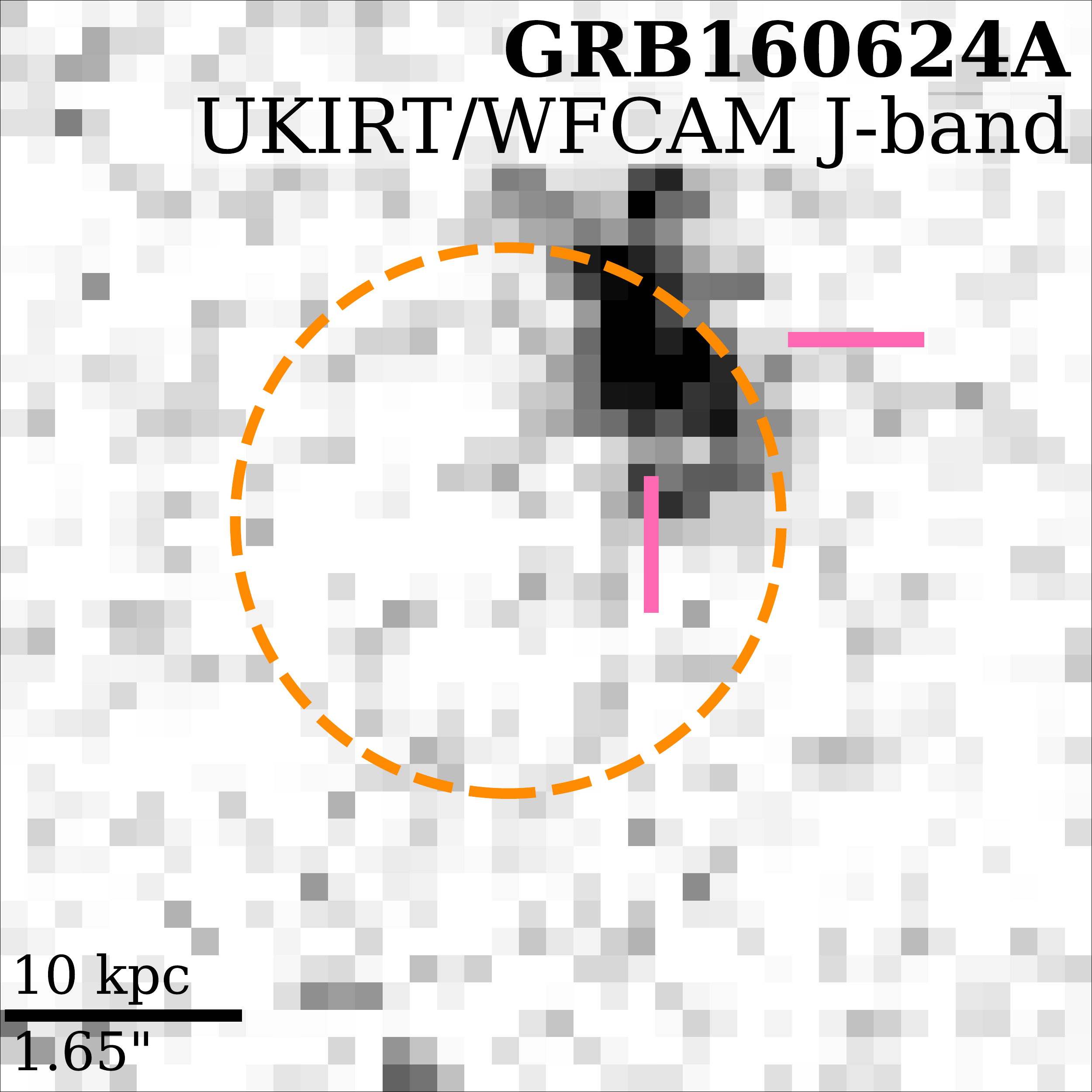}
\includegraphics[width=0.245\textwidth]{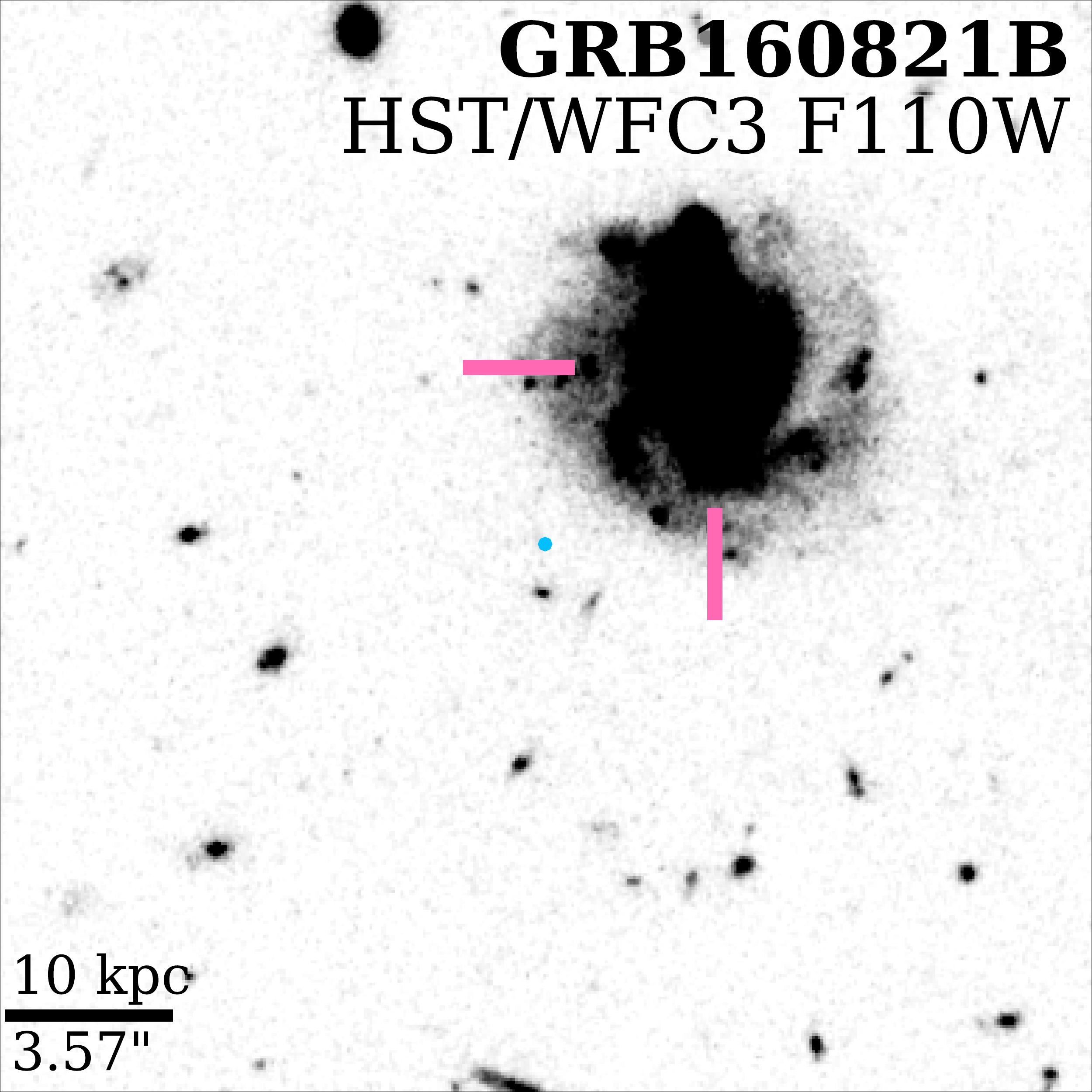}
\caption{Representative images of the host galaxies of the short GRBs in our catalog. In each panel, the most precise afterglow localization(s) for each burst is/are plotted (XRT 90\%: orange dashed, optical $1\sigma$: blue, {\it Chandra} or VLA $1\sigma$: purple). The putative host galaxy is denoted by the pink cross-hairs. All images are oriented North up and East to the left.}
\end{figure*}

\addtocounter{figure}{-1}

\begin{figure*}[t]
\centering
\includegraphics[width=0.245\textwidth]{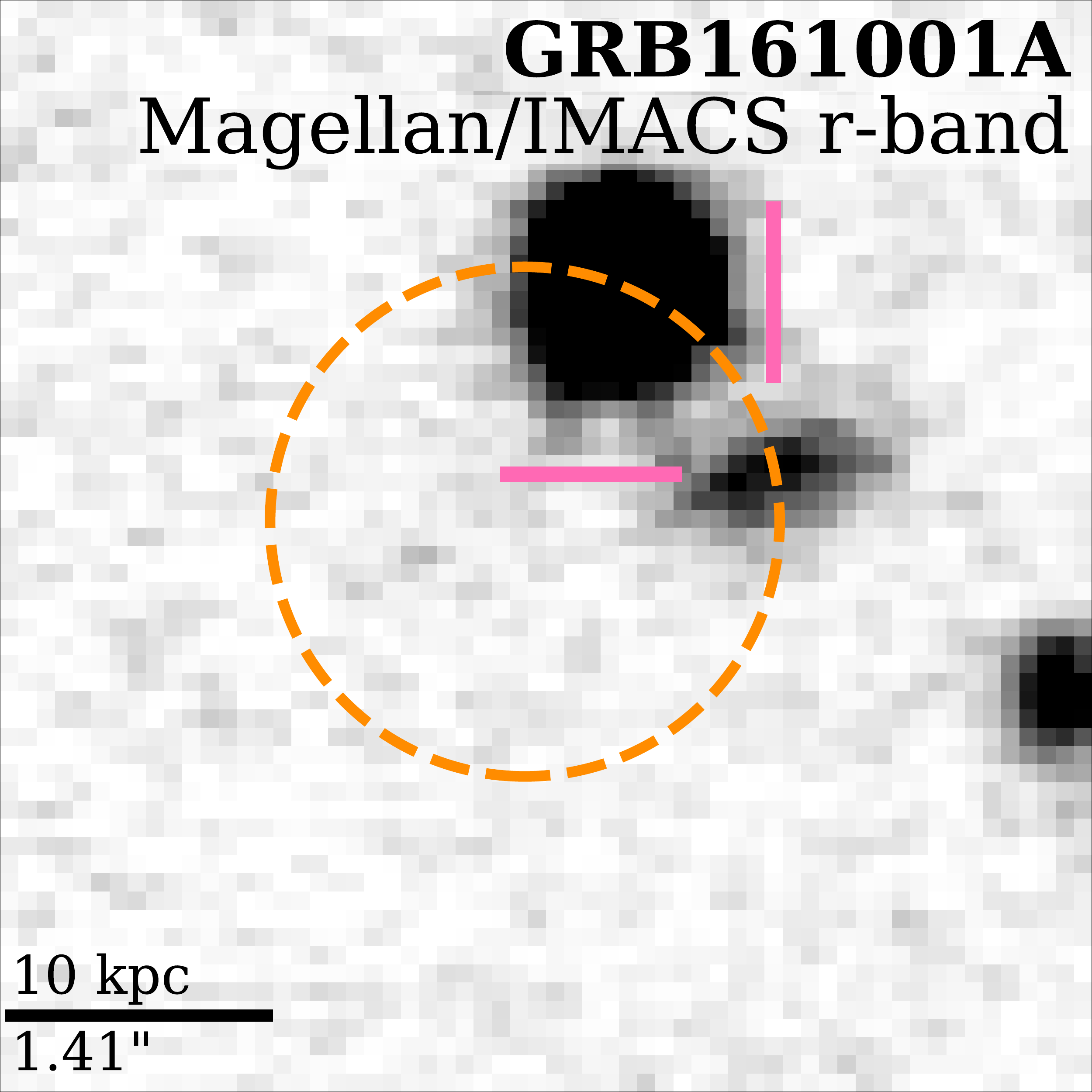}
\includegraphics[width=0.245\textwidth]{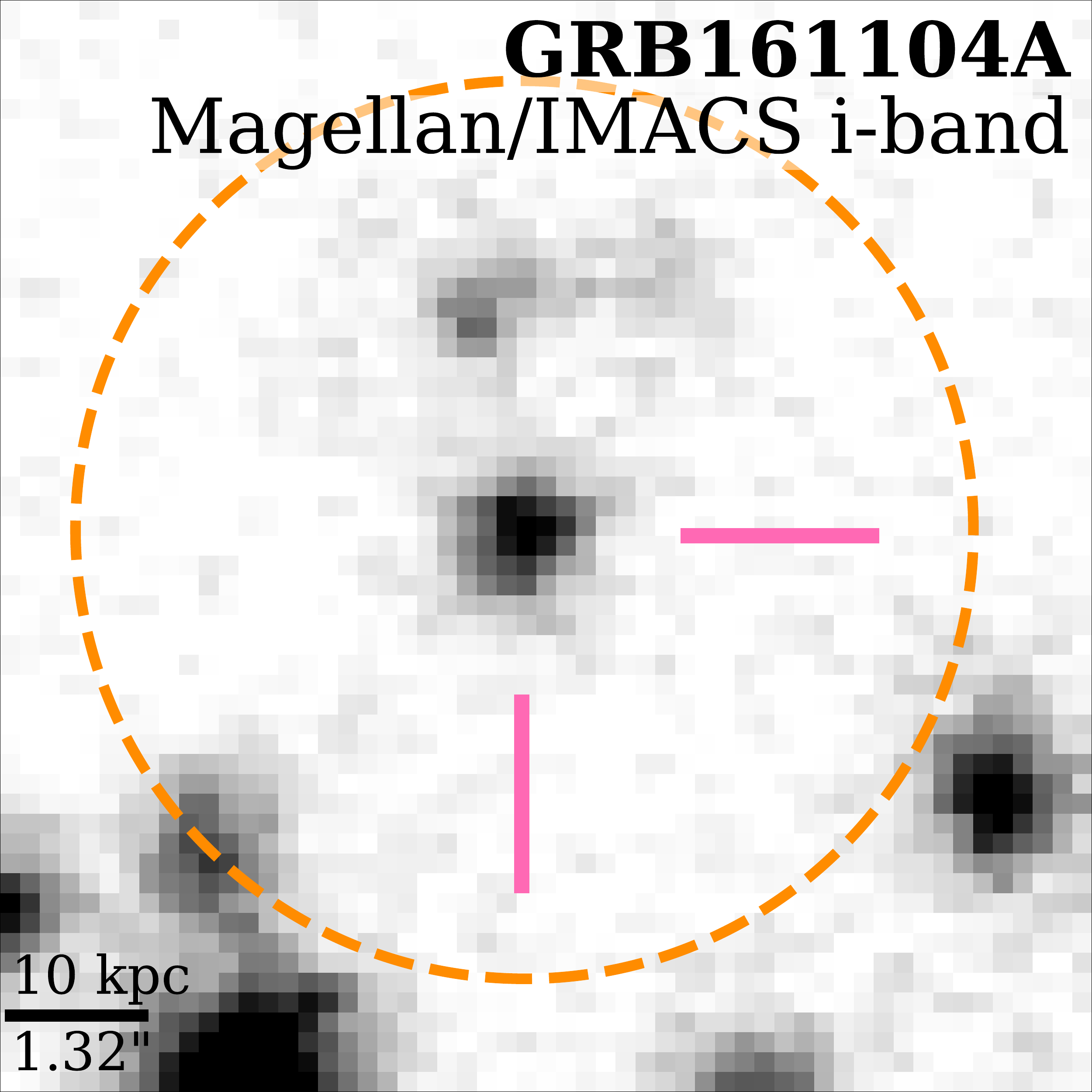}
\includegraphics[width=0.245\textwidth]{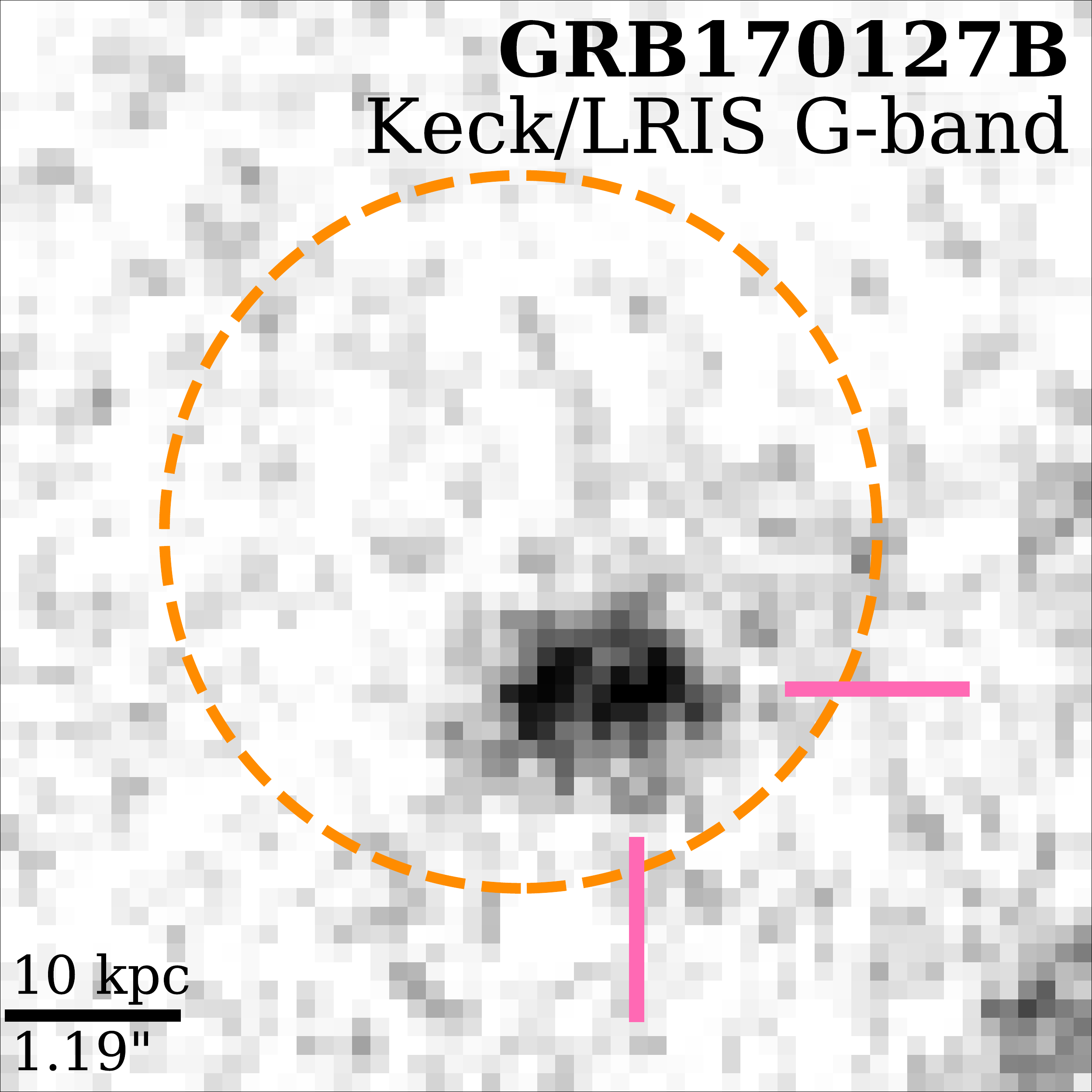}
\includegraphics[width=0.245\textwidth]{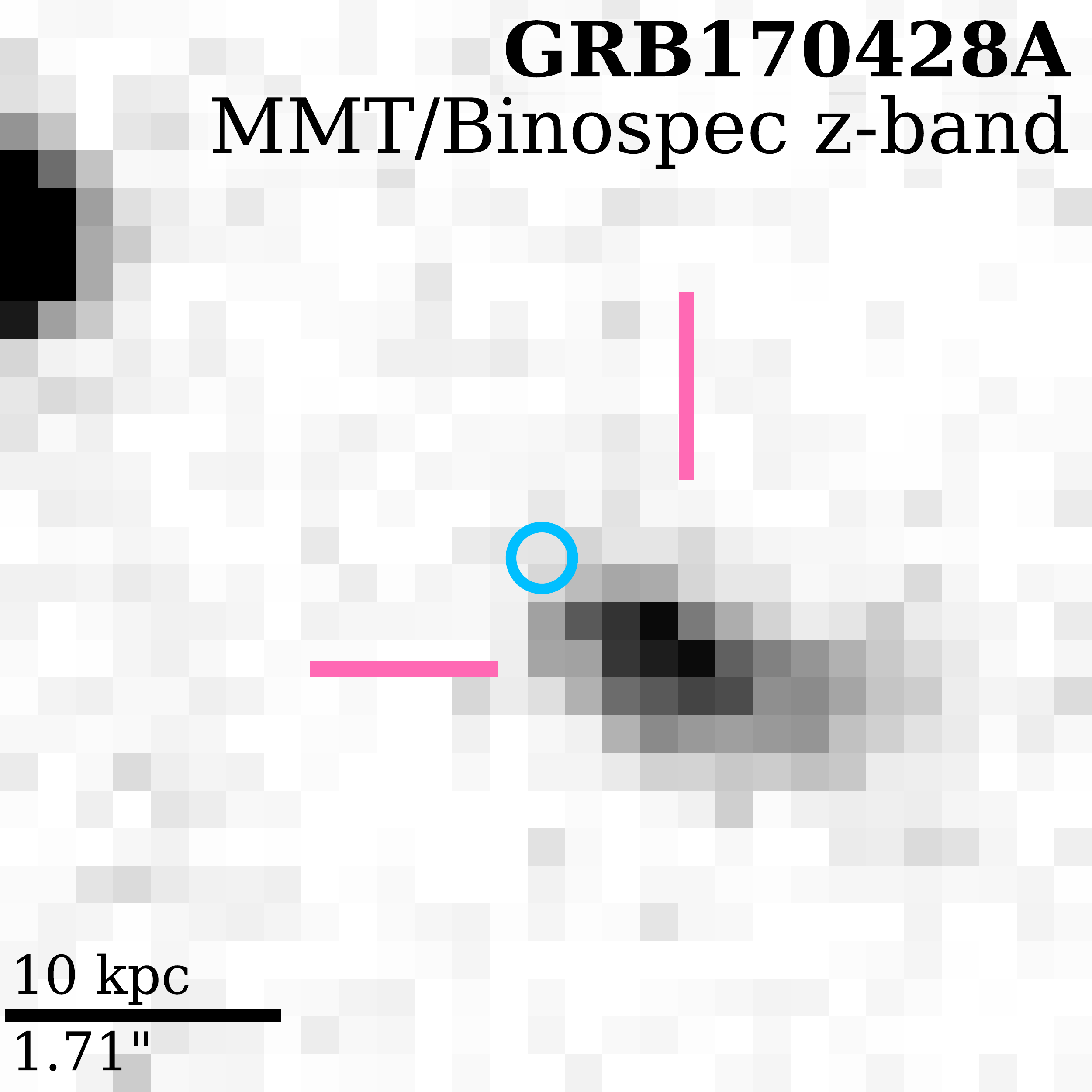}
\includegraphics[width=0.245\textwidth]{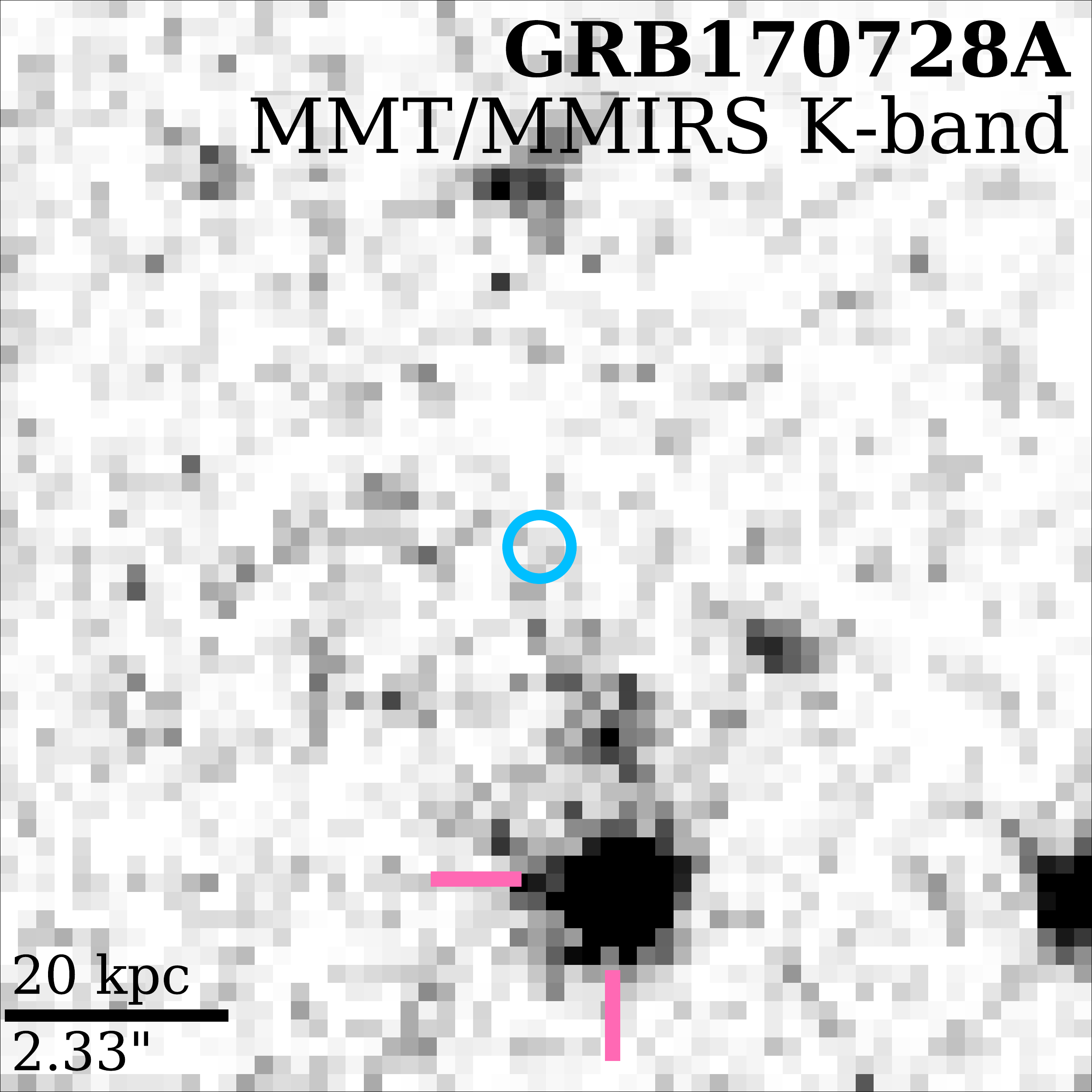}
\includegraphics[width=0.245\textwidth]{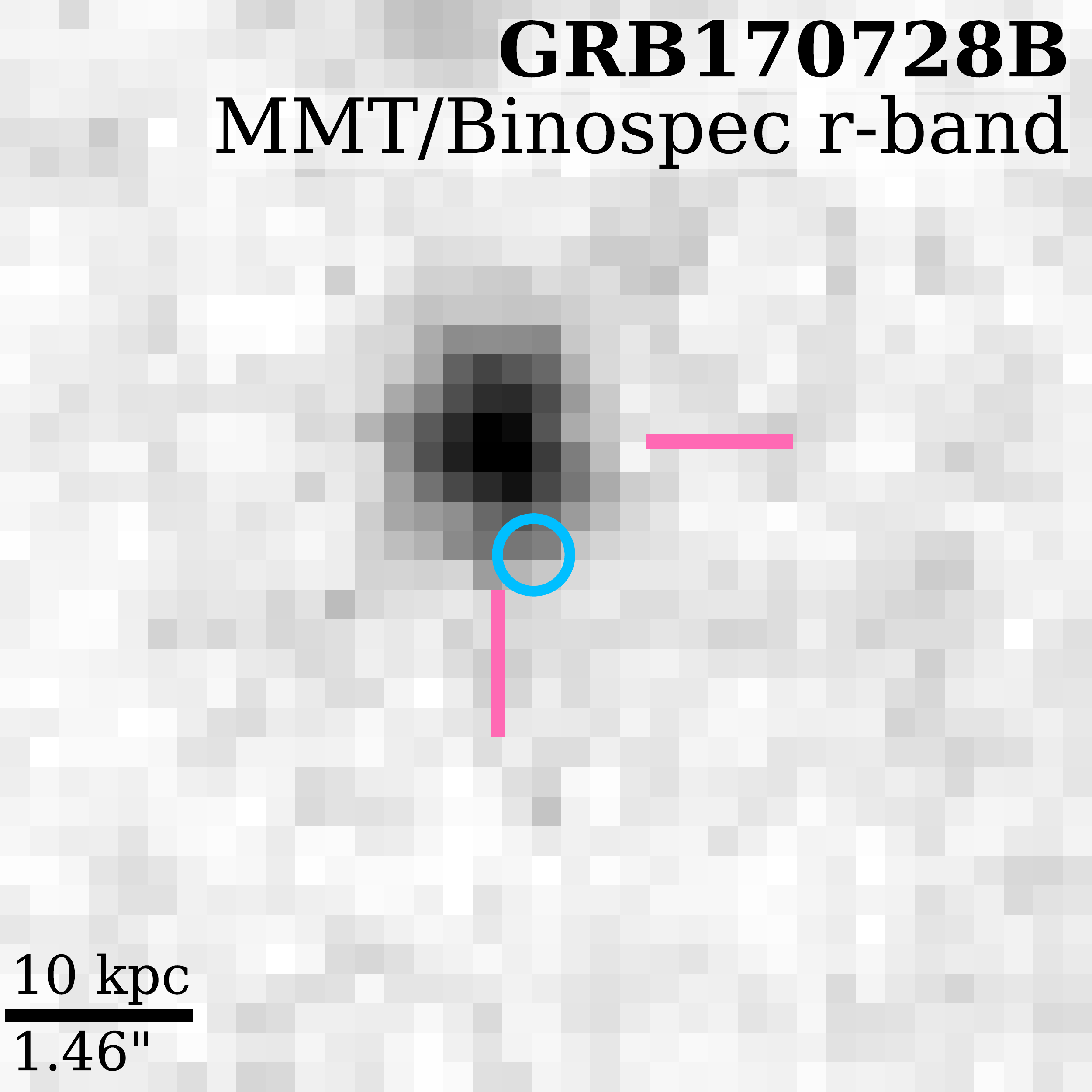}
\includegraphics[width=0.245\textwidth]{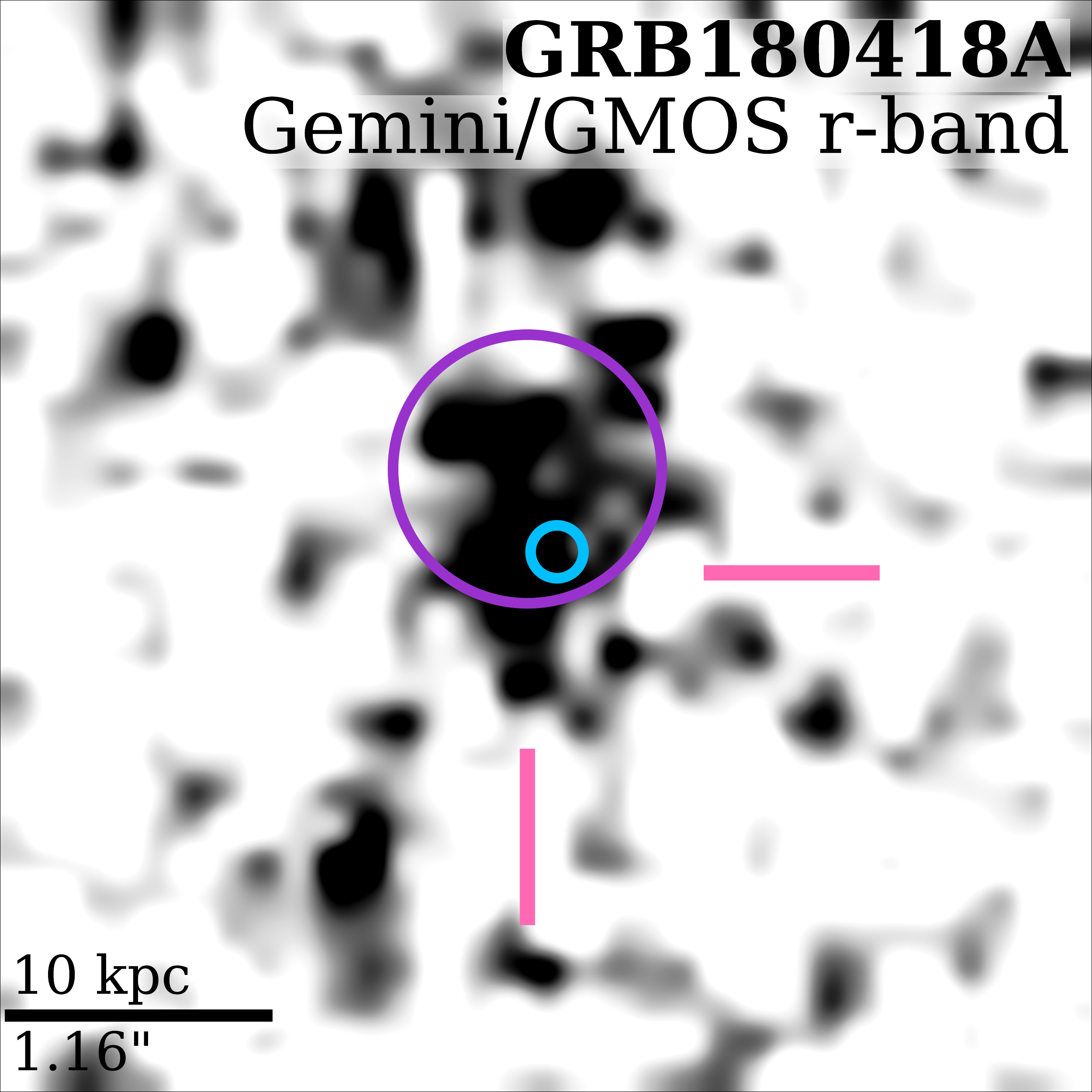}
\includegraphics[width=0.245\textwidth]{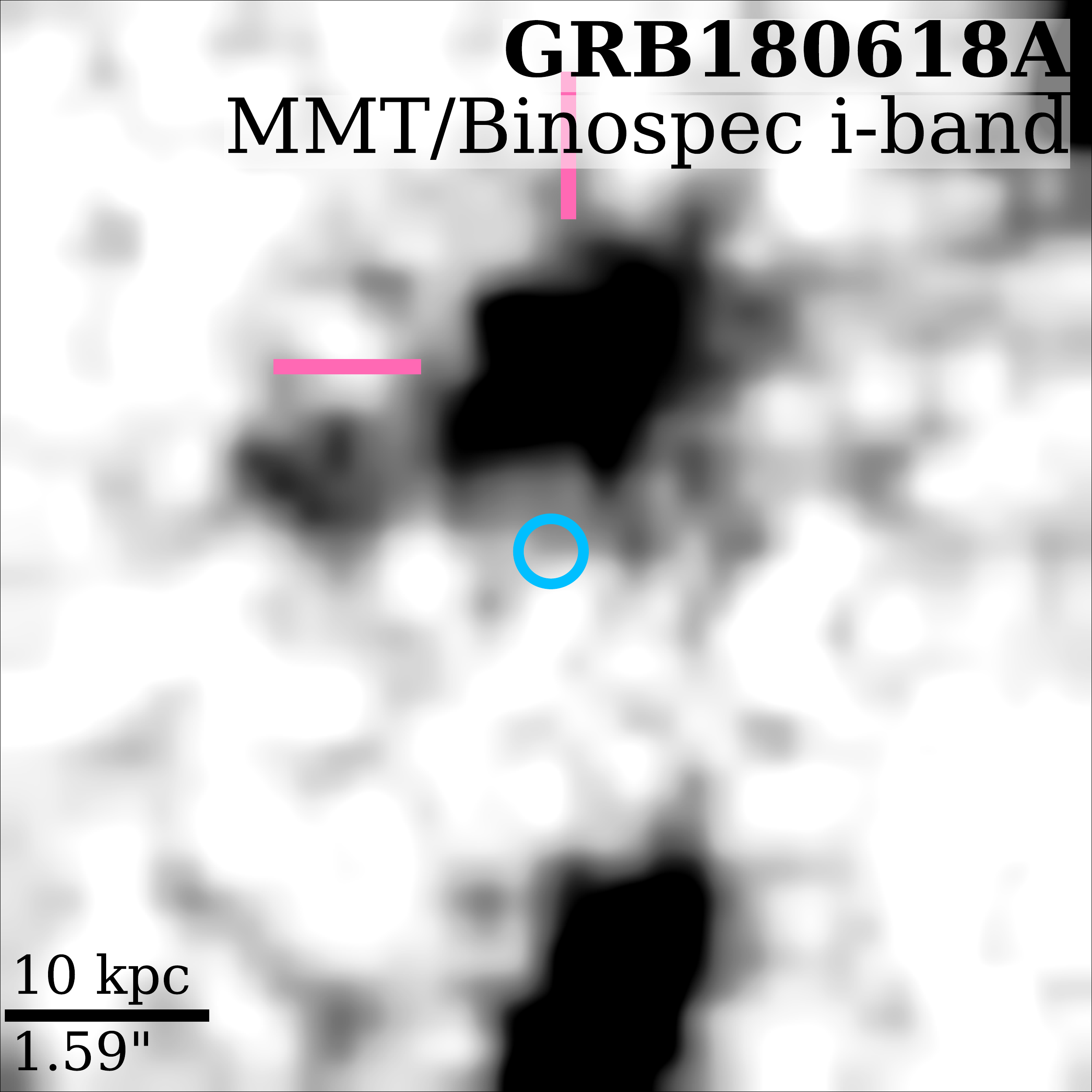}
\includegraphics[width=0.245\textwidth]{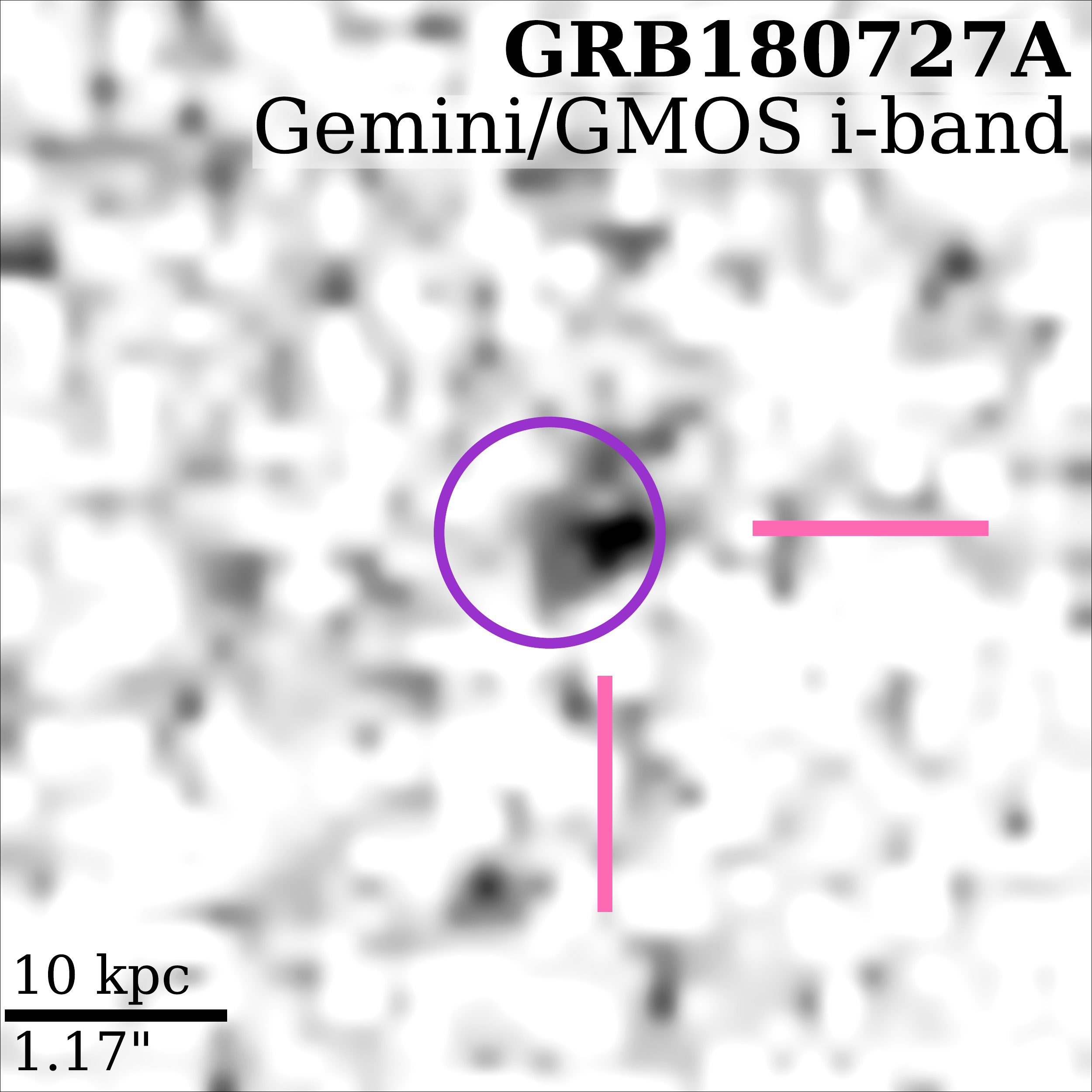}
\includegraphics[width=0.245\textwidth]{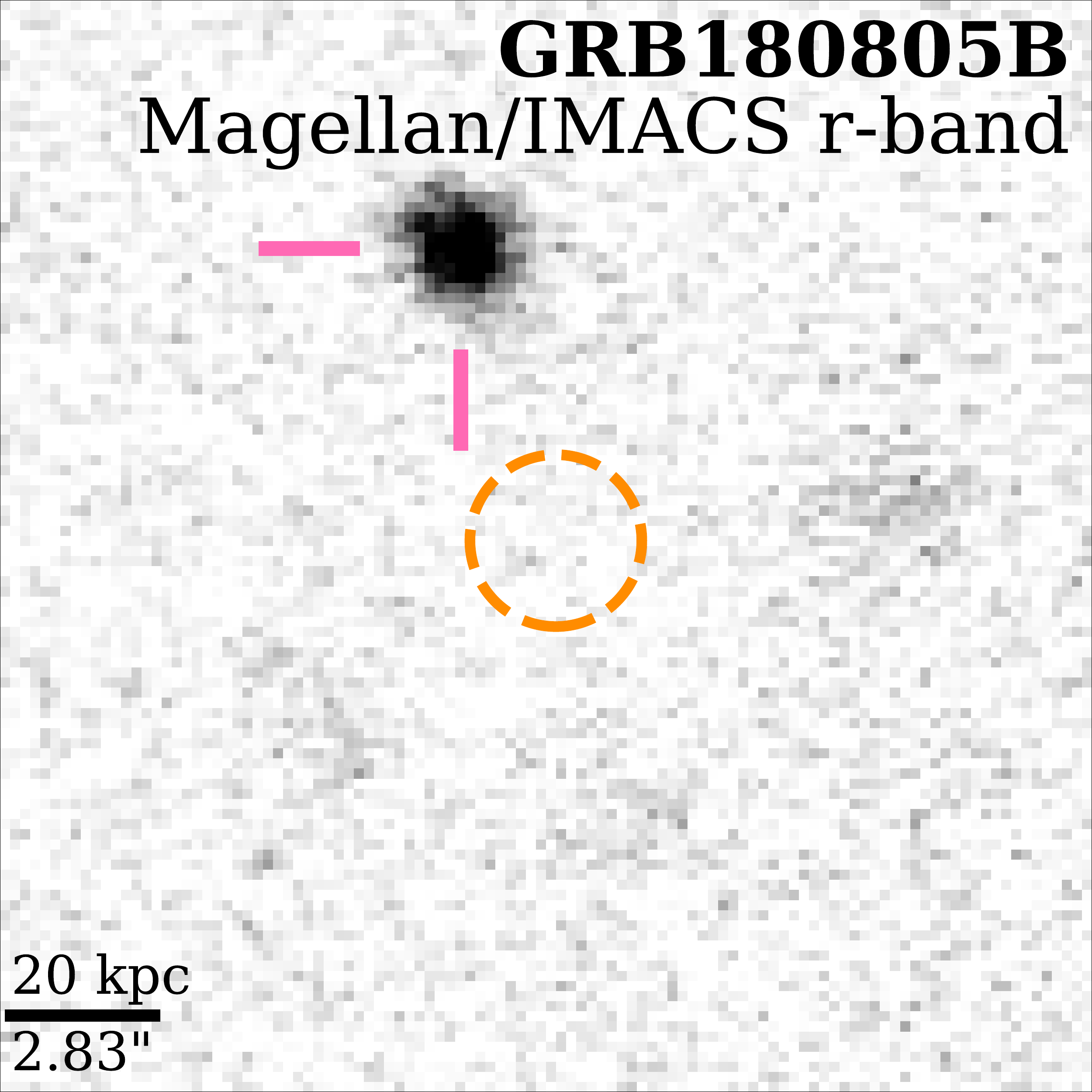}
\includegraphics[width=0.245\textwidth]{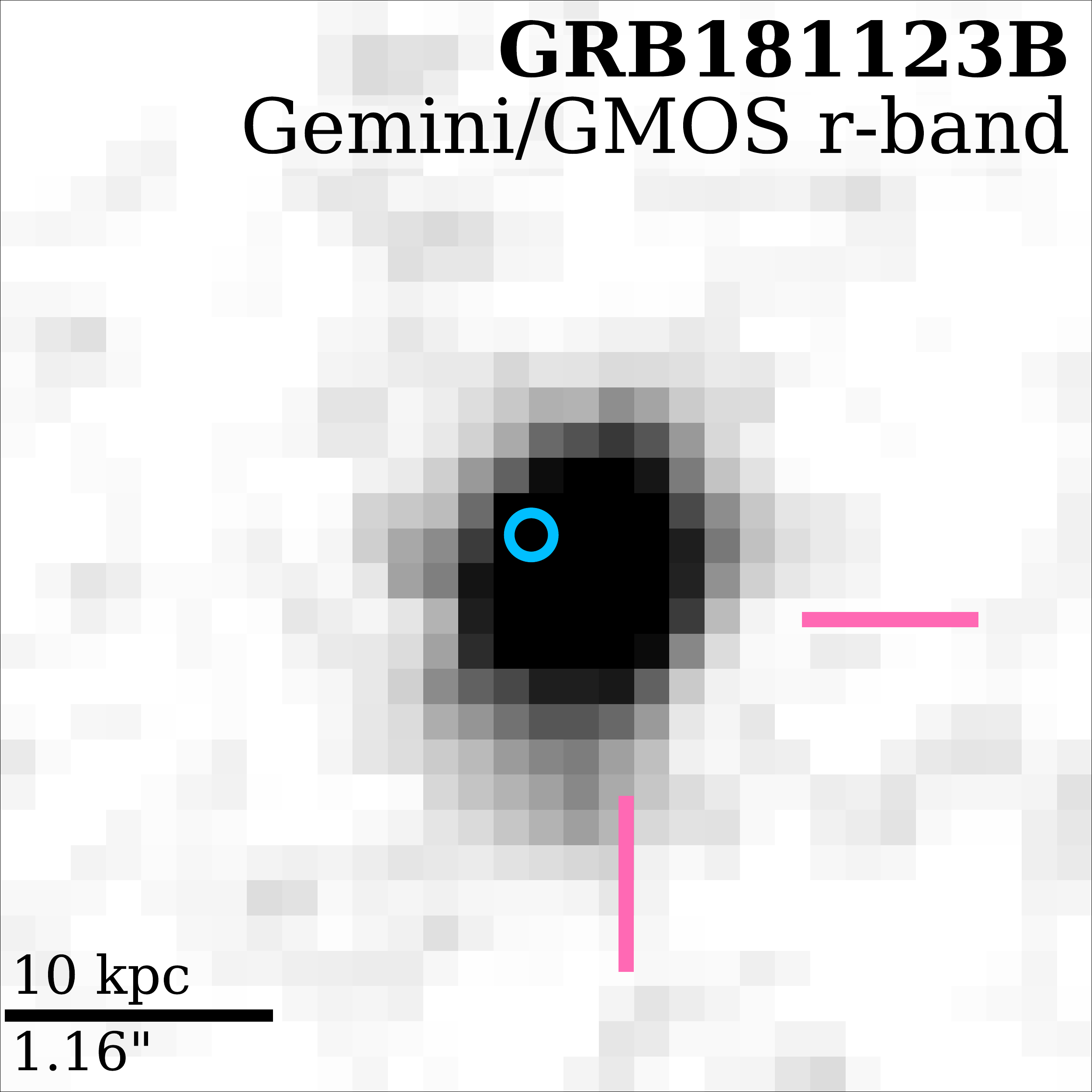}
\includegraphics[width=0.245\textwidth]{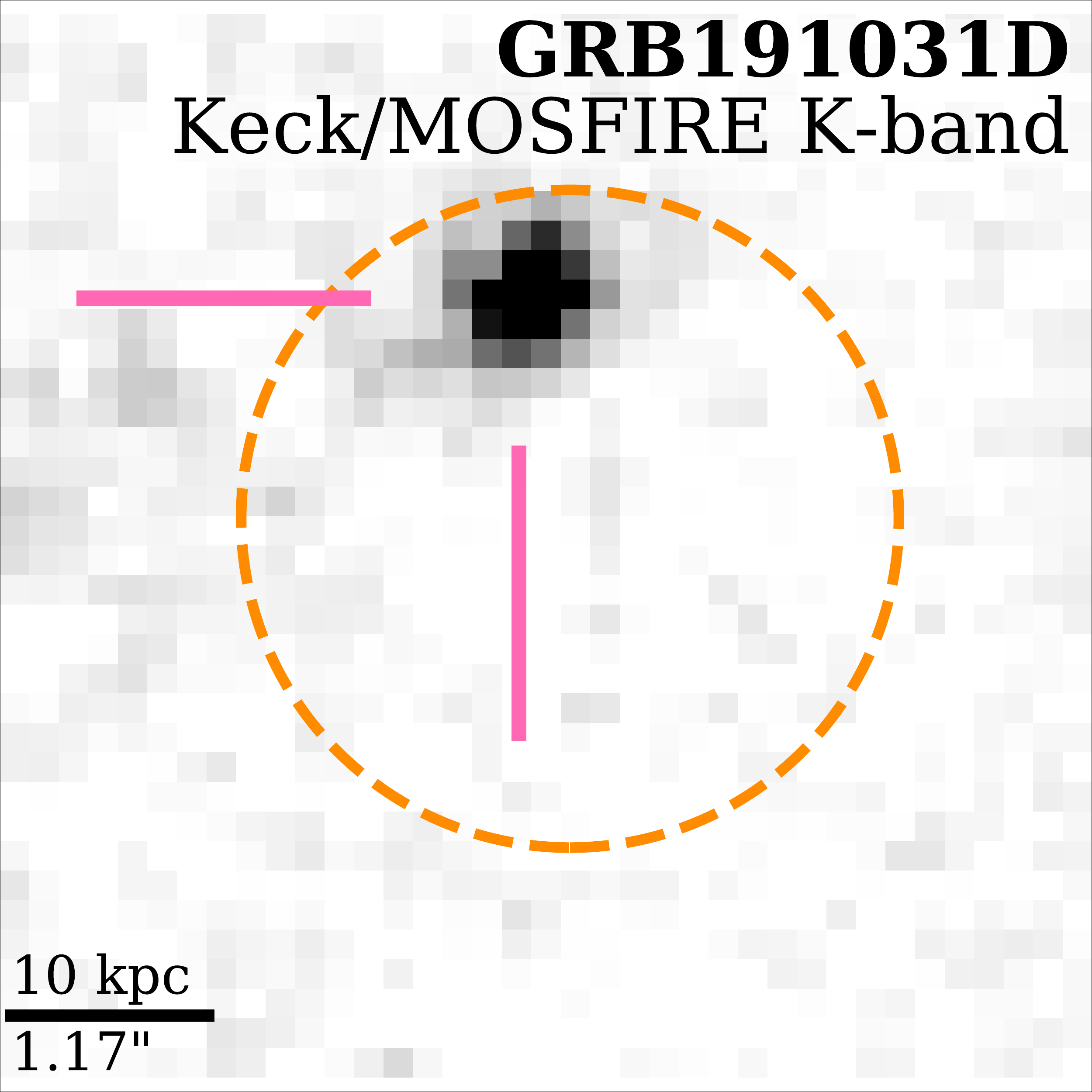}
\includegraphics[width=0.245\textwidth]{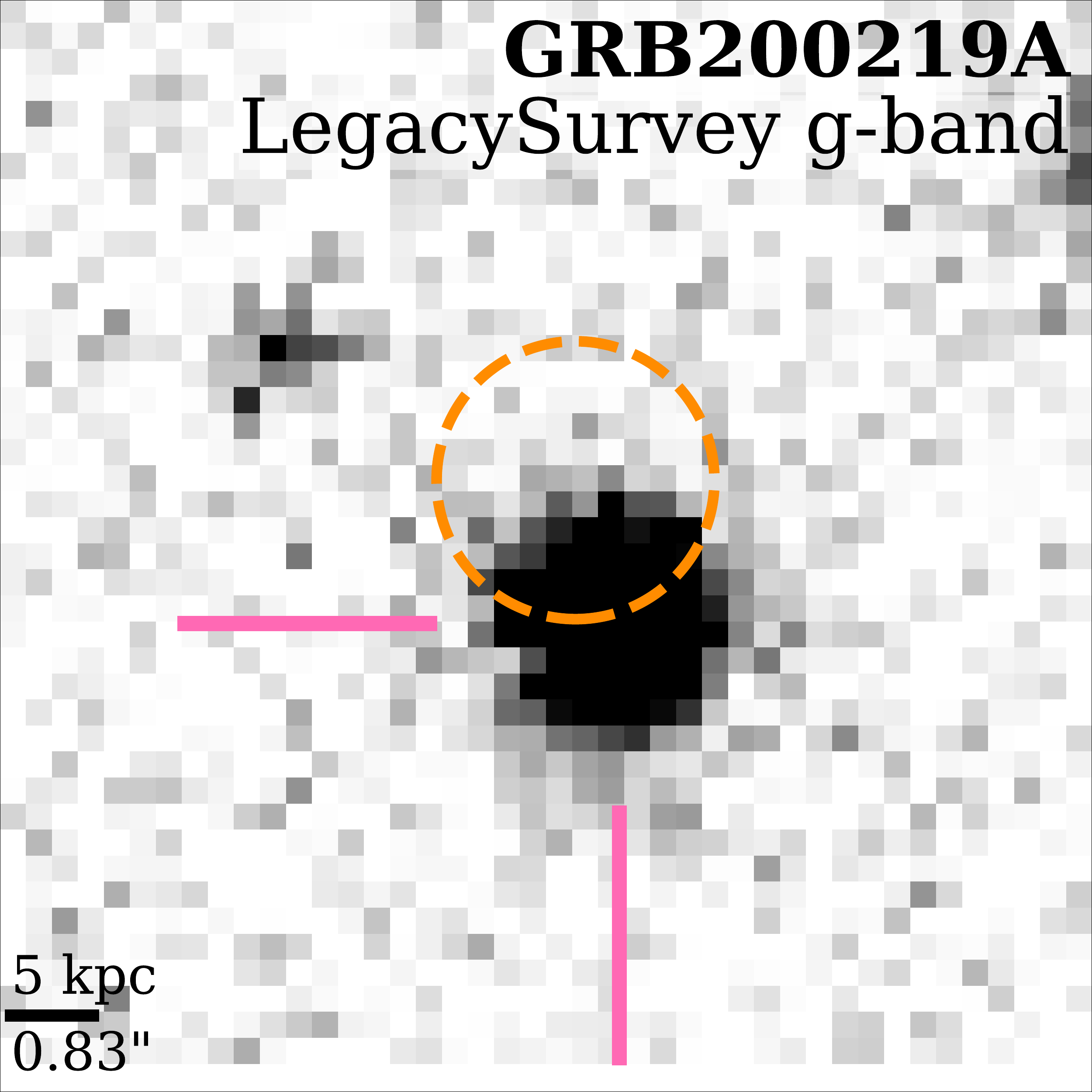}
\includegraphics[width=0.245\textwidth]{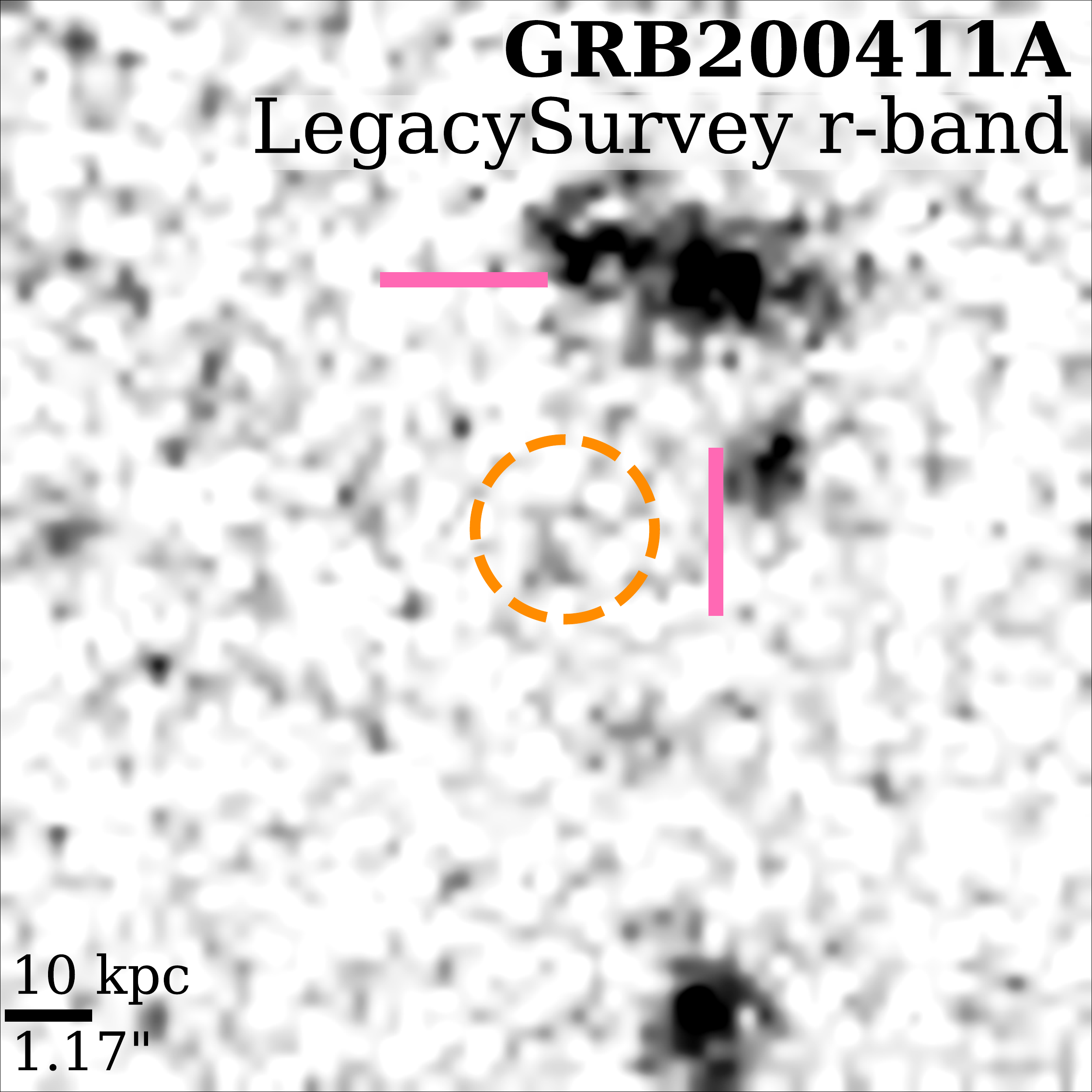}
\includegraphics[width=0.245\textwidth]{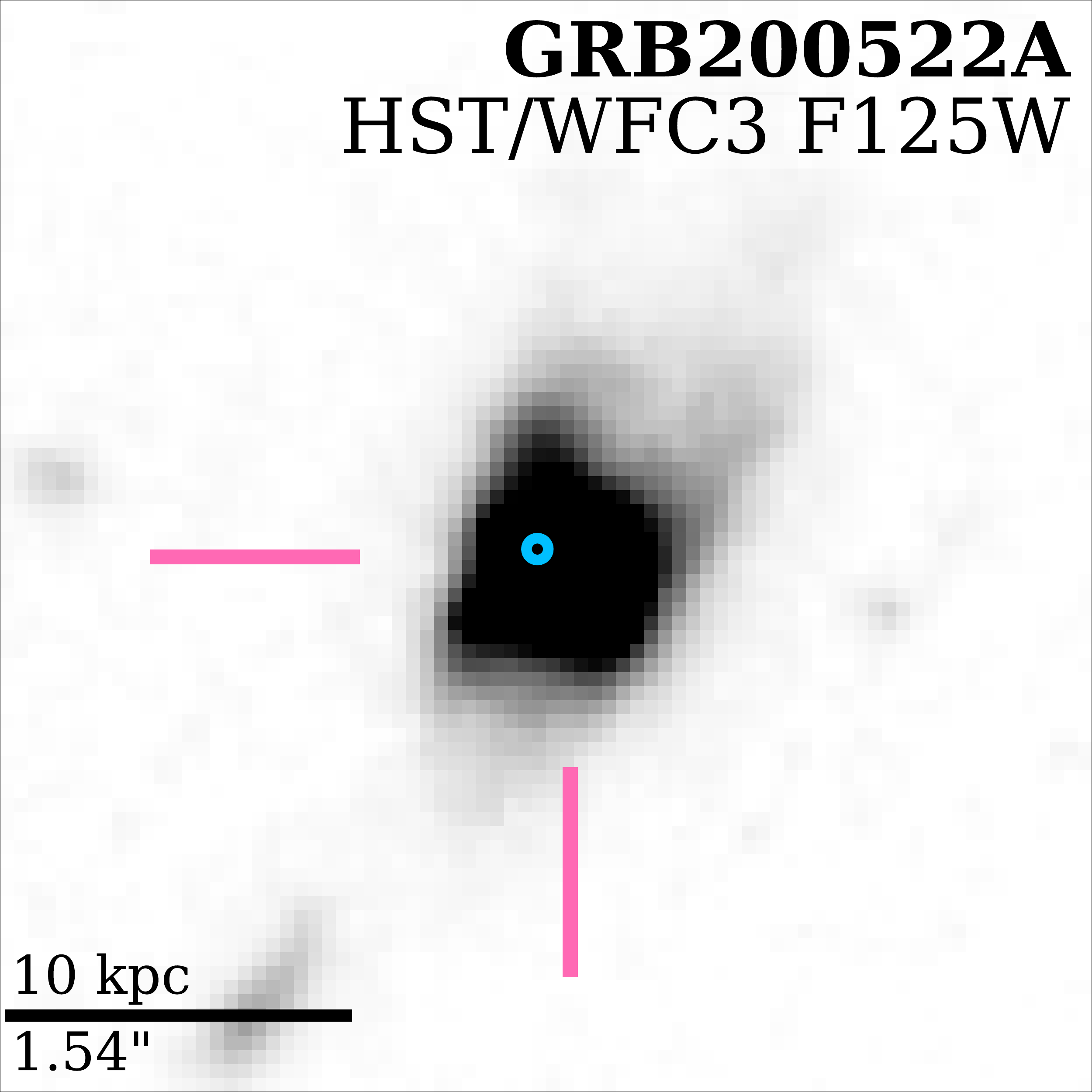}
\includegraphics[width=0.245\textwidth]{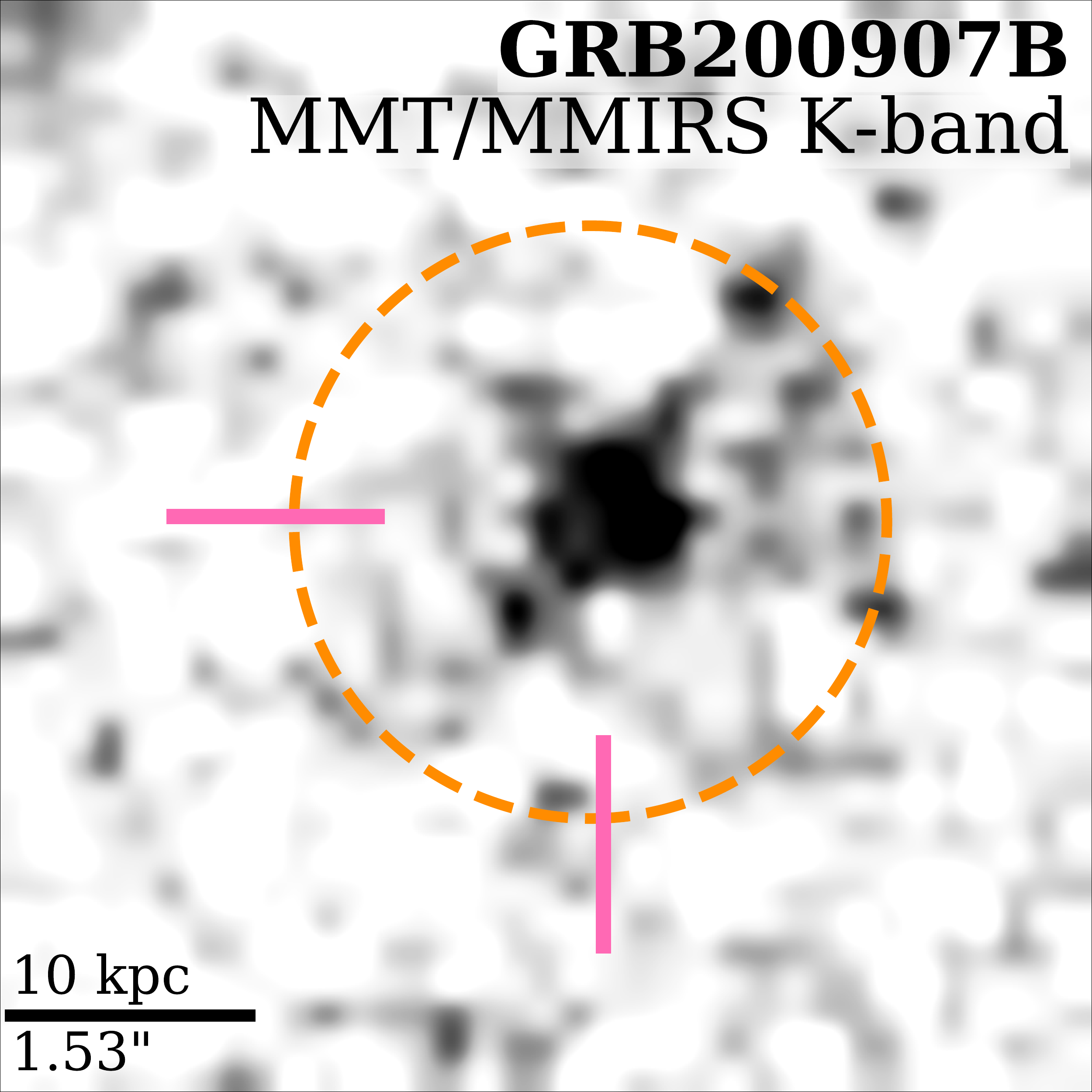}
\includegraphics[width=0.245\textwidth]{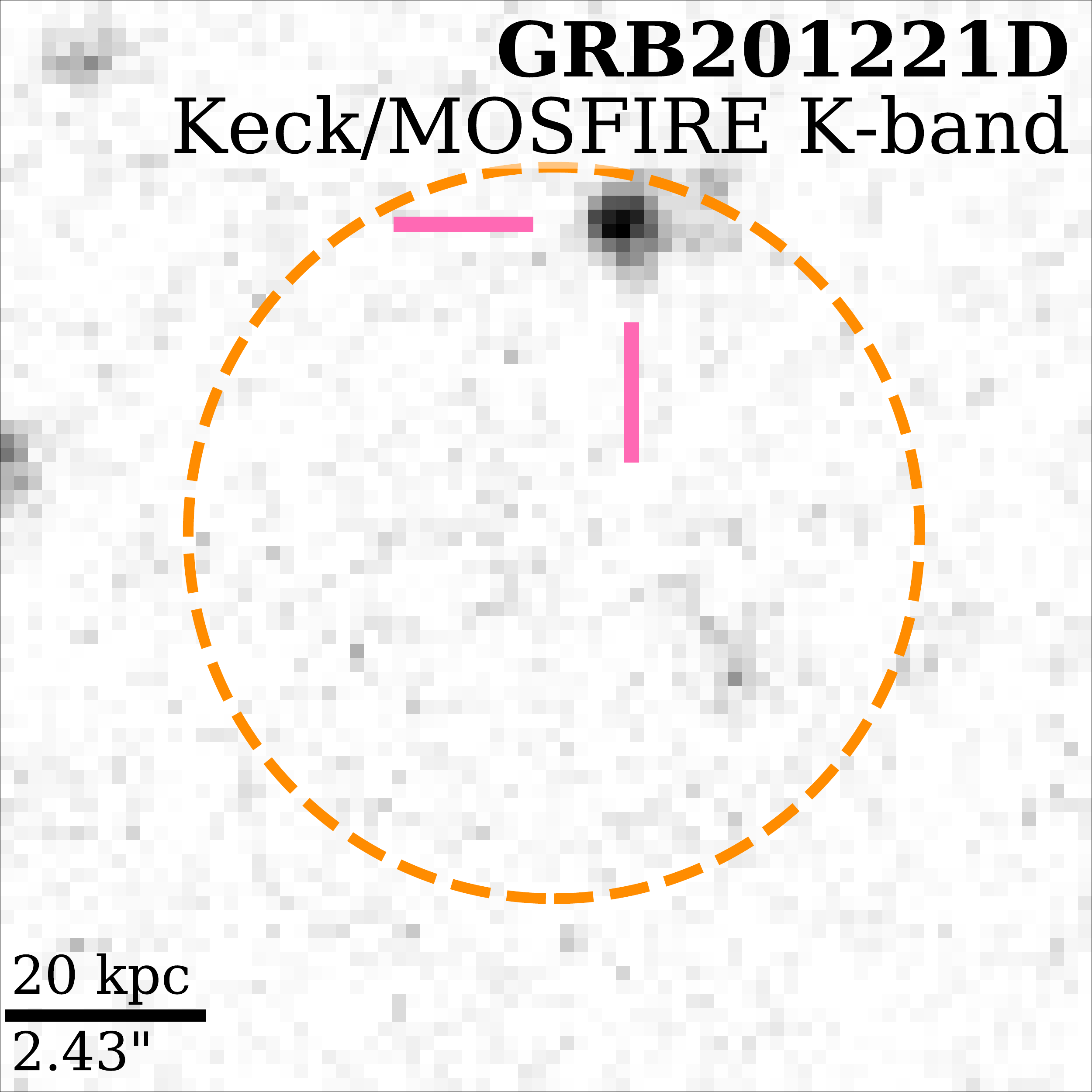}
\includegraphics[width=0.245\textwidth]{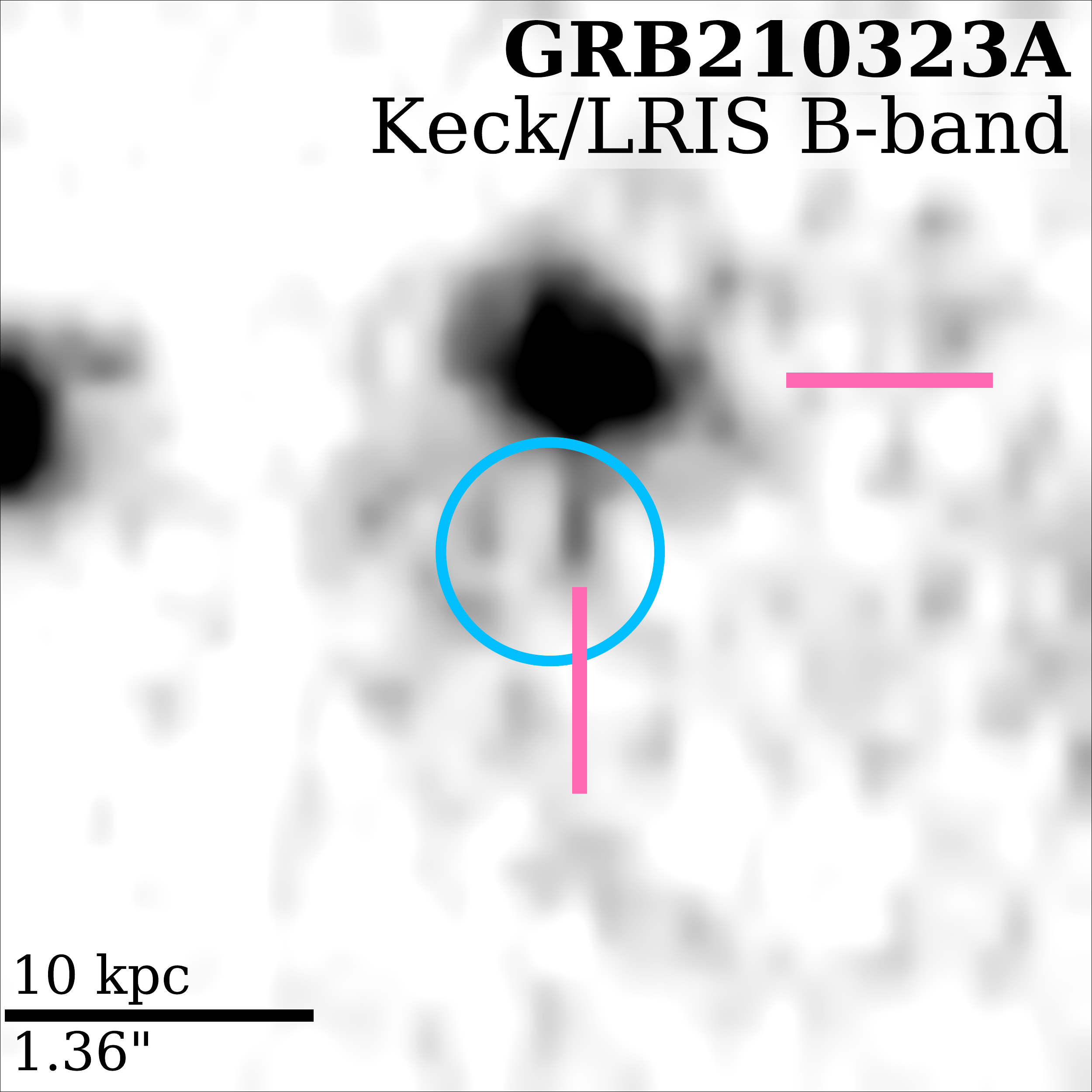}
\includegraphics[width=0.245\textwidth]{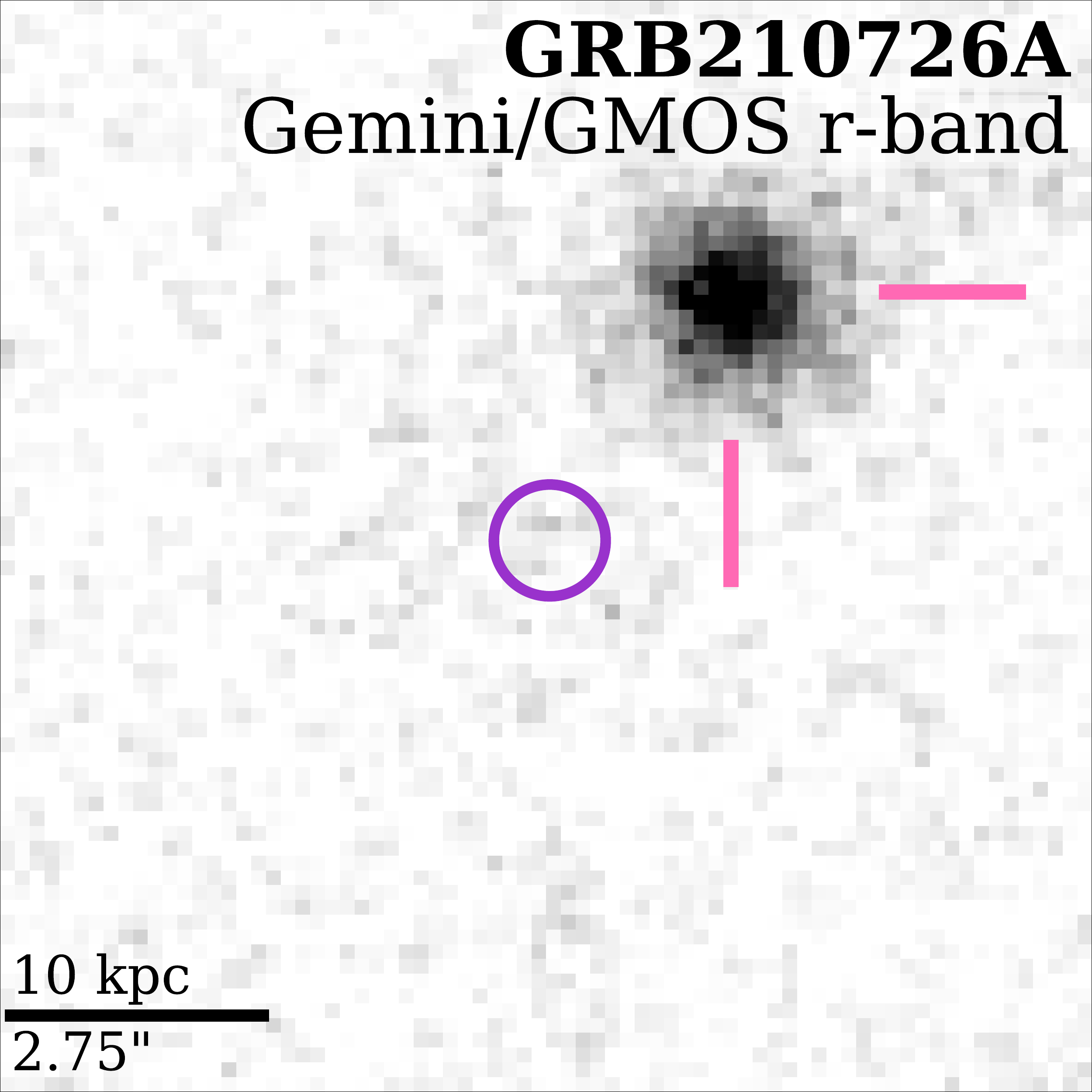}
\includegraphics[width=0.245\textwidth]{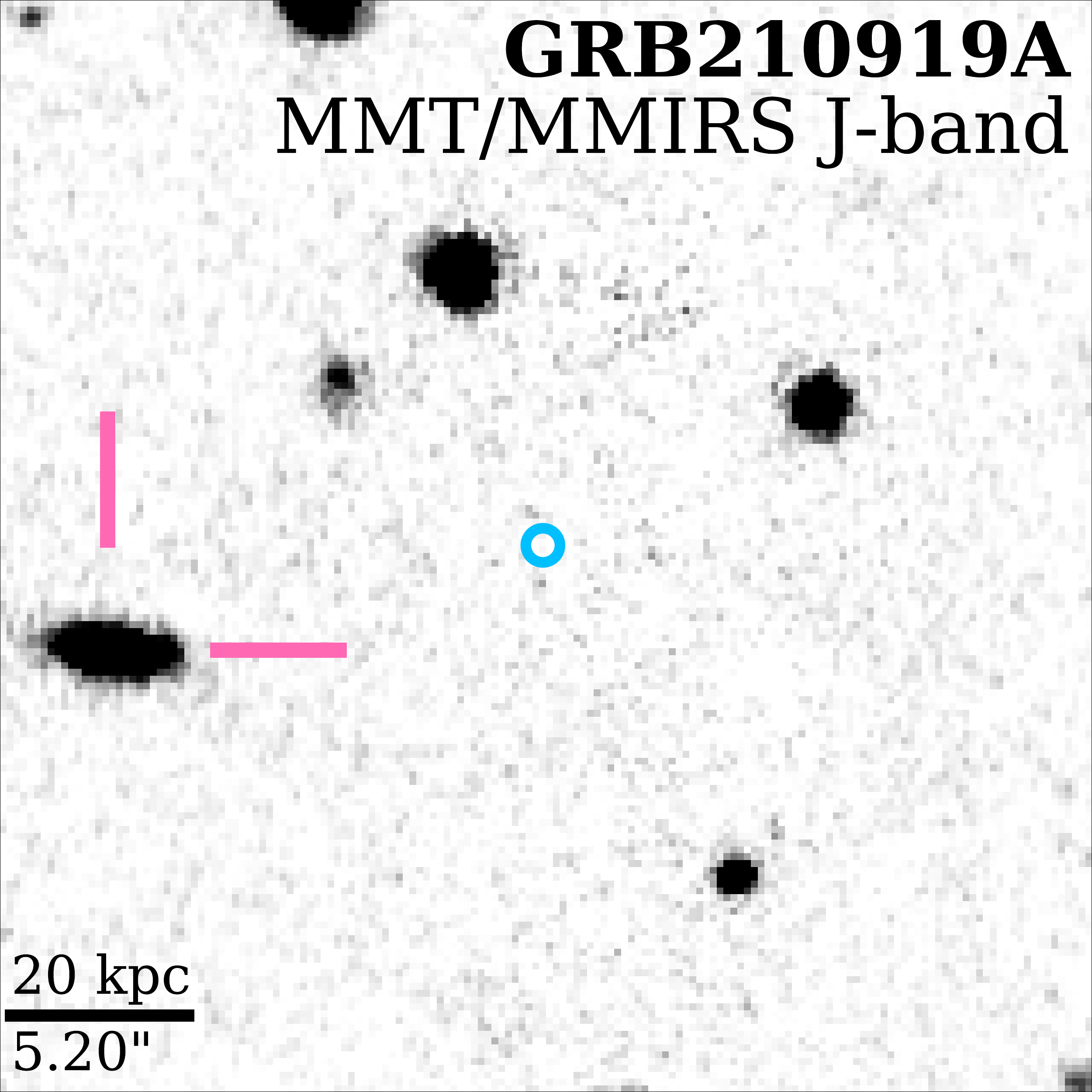}
\caption{Representative images of the host galaxies of the short GRBs in our catalog. In each panel, the most precise afterglow localization(s) for each burst is/are plotted (XRT 90\%: orange dashed, optical $1\sigma$: blue, {\it Chandra} or VLA $1\sigma$: purple). The putative host galaxy is denoted by the pink cross-hairs. All images are oriented North up and East to the left.}
\end{figure*}

\addtocounter{figure}{-1}

\begin{figure*}[!t]
\centering
\includegraphics[width=0.245\textwidth]{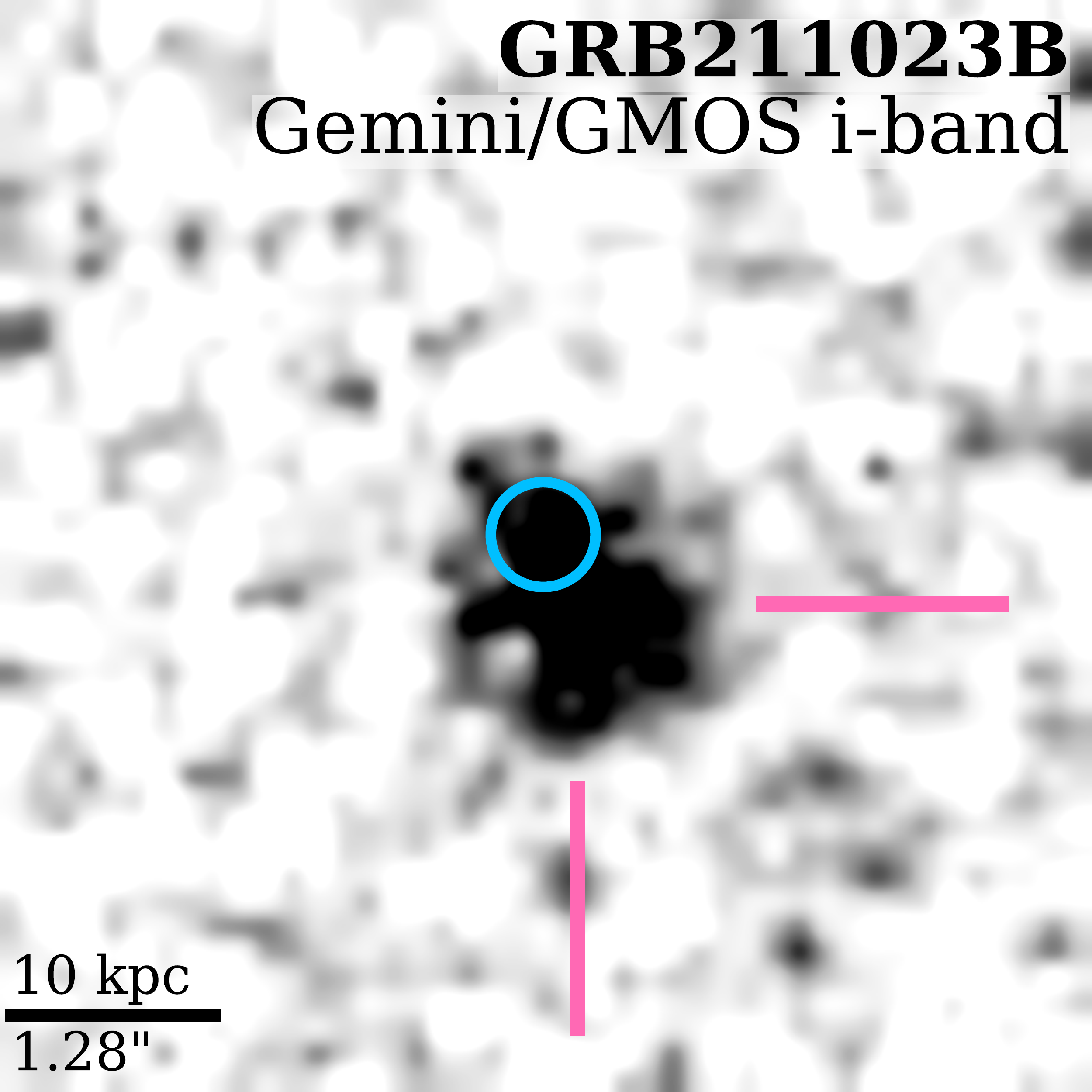}
\includegraphics[width=0.245\textwidth]{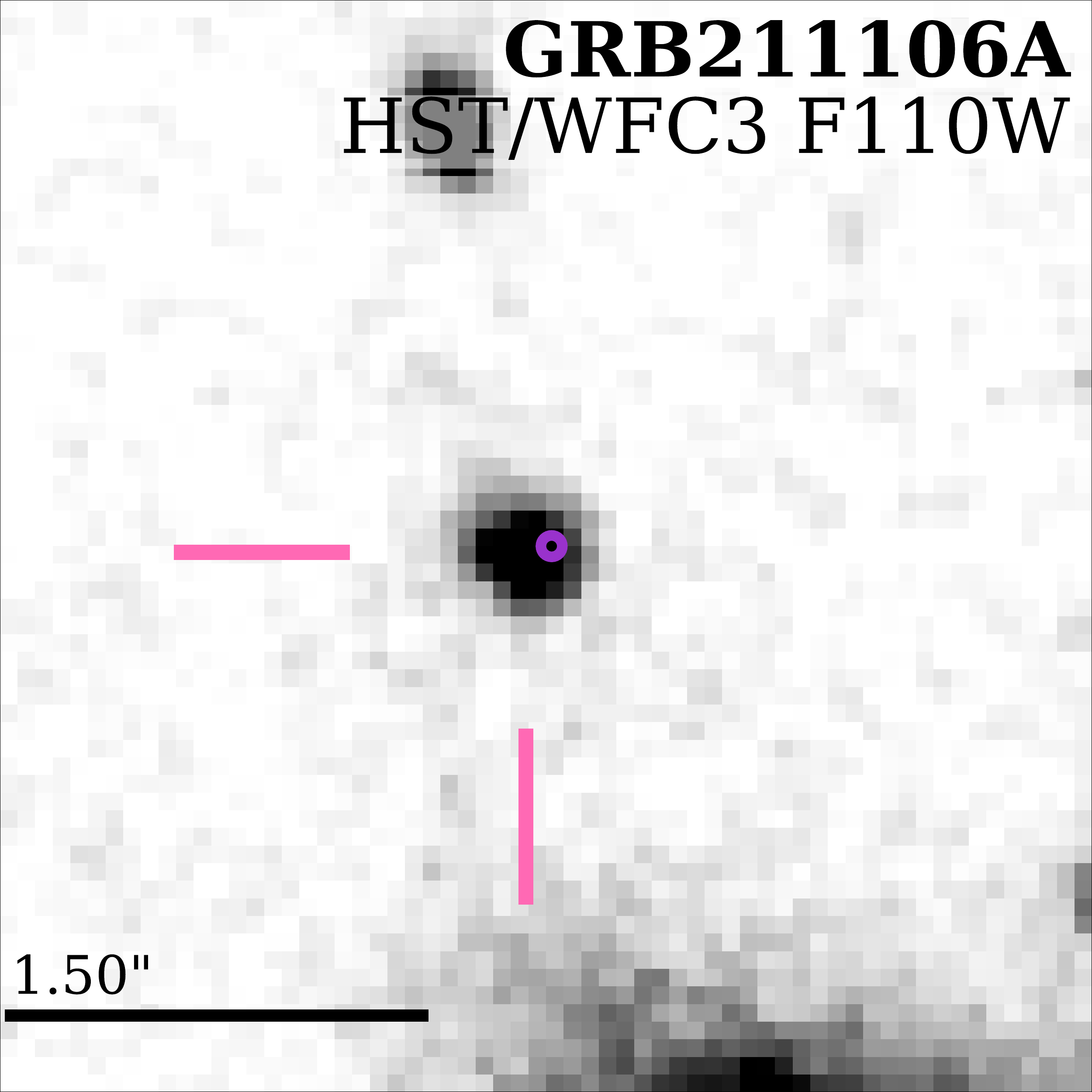}
\includegraphics[width=0.245\textwidth]{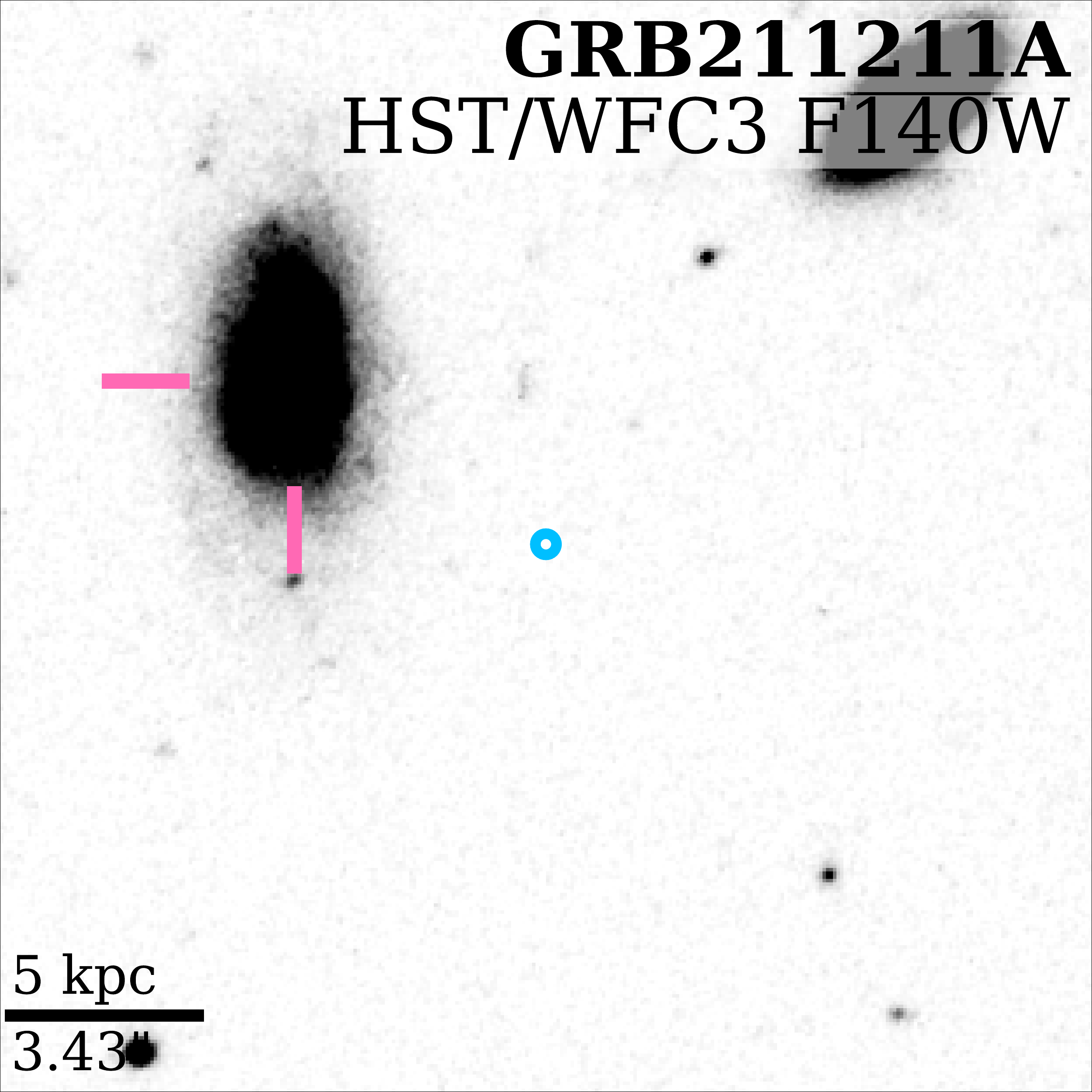}
\caption{Representative images of the host galaxies of the short GRBs in our catalog. In each panel, the most precise afterglow localization(s) for each burst is/are plotted (XRT 90\%: orange dashed, optical $1\sigma$: blue, {\it Chandra} or VLA $1\sigma$: purple). The putative host galaxy is denoted by the pink cross-hairs. All images are oriented North up and East to the left.}
\end{figure*}

\subsection{{\it Hubble Space Telescope} Observations}
\label{sec:hst}

We supplement the ground-based imaging with {\it HST} observations for ten events, consisting of events that we have not yet analyzed in our previous works. Six GRBs were imaged in a single band with the F110W filter using the infrared channel on the Wide-Field Camera~3 (WFC3/IR) under Program 14685 (PI:~Fong). The purpose of this program was to search for redder host galaxies that may have gone undetected from ground-based optical observations. Indeed, at the time of observation, five of these events had no reported or detected host galaxy at their sub-arcsecond optical afterglow positions to the limits of ground-based imaging (GRBs\,091109B, 110112A, 130912A, 131004A, and 150423A). We recover plausible host galaxies for three of these events (Section~\ref{sec:pcc}). The sixth event observed in F110W, GRB\,160303A, had a ground-based GTC detection of a possible host galaxy \citep{gcn19160}, and this is well-detected in our {\it HST} imaging (Figure~\ref{fig:imagepanel}). These {\it HST} data also newly appear in \citet{otd+22}.

The four additional events have multiple epochs and filters (GRBs\,060614, 150424A, 160624A and 160821B) under Programs 10917 (PI:~Fox), 13830, 14237 (PI:~Tanvir) and 14357 (PI:~Troja), and were previously published \citep{gfp+06,kgv+17,jlw+18,otd+21,ltl+19,tcb+19,rfk+21}. For each of these latter events, we select the filter which yields the highest signal-to-noise ratio (SNR) for sources in the image to perform our subsequent analysis.

For data reduction, we retrieved pre-processed images from the {\it HST} archive\footnote{https://archive.stsci.edu/hst/} for observations of these nine short GRBs. We used the {\tt astrodrizzle} routine as part of the {\tt drizzlepac} package in PyRAF to apply standard distortion corrections and combine the exposures for each event and filter. For WFC3/IR imaging, we use pixfrac=0.8 and pixscale=$0.0642''$~pixel$^{-1}$, half of the native pixel scale. For WFC3/UVIS images, we first apply a charge transfer efficiency (CTE) correction using  the standalone Fortran program\footnote{http://www.stsci.edu/hst/wfc3/tools/cte\_tools}, and then use {\tt astrodrizzle} to combine the CTE-corrected exposures using  pixscale=$0.033''$ pixel$^{-1}$. For the ACS images, we use pixfrac=1.0 and pixscale=$0.05''$ pixel$^{-1}$. The images are shown in Figure~\ref{fig:imagepanel}.

We also compile available photometric data and reduced imaging for all remaining short GRBs with {\it HST} observations from our previous works. This results in the addition of 26 short GRBs with {\it HST} data. These bursts and their references are listed in Table~\ref{tab:sample} and Appendix Table~\ref{tab:phot}.

\clearpage
\startlongtable
\begin{deluxetable*}{lcccccccc}
\tabletypesize{\scriptsize}
\linespread{1.1}
\tablecolumns{9}
\tablewidth{0pt}
\tablecaption{Short Gamma-ray Burst Host Galaxy Sample
\label{tab:sample}}
\tablehead{
\colhead{GRB} &
\colhead{RA} &
\colhead{Decl.} &
\colhead{Redshift} &
\colhead{Filter} &
\colhead{$m_{\rm AB}$} & 
\colhead{$P_{\rm cc}$} & 
\colhead{Class} &
\colhead{References} \\
\colhead{} &
\colhead{(J2000)} &
\colhead{(J2000)} &
\colhead{} &
\colhead{} &
\colhead{(AB mag)} & 
\colhead{} &
\colhead{} &
\colhead{}
}
\startdata
050509B &	\ra{12}{36}{12.875} &	\dec{+28}{58}{58.84 } & $0.2248 \pm 0.0002$ & $r$ & $17.123 \pm 0.01$ & $5 \times 10^{-3}$ & Gold & 1-4 \\
050709  &	\ra{23}{01}{26.765} &	\dec{-38}{58}{40.422} & $0.1607 \pm 0.0004$ & $R$ & $21.258 \pm 0.07$ & $3 \times 10^{-3}$ & Gold & 2, 5-8 \\
050724  &	\ra{16}{24}{44.410} &	\dec{-27}{32}{26.393} & $0.254 \pm 0.001$ & $R$ & $19.829 \pm 0.03$ & $2 \times 10^{-5}$ & Gold	& 2, 9-10 \\
050813 &    \ra{16}{07}{57.200} & 	\dec{+11}{14}{53.09} & $0.719 \pm 0.001$ & $R$ & $23.43 \pm 0.07$ & 0.2 & Bronze & 11-13  \\
051210  &	\ra{22}{00}{40.942} &	\dec{-57}{36}{47.063} & $2.58_{-0.17}^{+0.11}$ & $r$ & $24.043 \pm 0.15$ & $0.04$ & Silver	& 2, 8, This work \\
051221A &	\ra{21}{54}{48.653} &	\dec{+16}{53}{27.335} & $0.5464 \pm 0.0001$ & $r$ & $22.178 \pm 0.09$ & $5 \times 10^{-5}$ & Gold	& 2, 8, 14 \\
\hline 
060121  &	\ra{09}{09}5{2.026} &	\dec{+45}{39}{45.538} & \nod & F606W & $26.27$ & $2 \times 10^{-3}$ & Gold & 2	 \\
060313  &	\ra{04}{26}{28.402} &	\dec{-10}{50}{39.901} & \nod & F475W & $26.68$ & $3 \times 10^{-3}$ & Gold & 2	 \\
060614$^a$  &   \ra{21}{23}{32.102} &   \dec{-53}{01}{36.436} & $0.125 \pm 0.002$ & F814W & $21.92 \pm 0.1$ & $3 \times 10^{-4}$ & Gold & 15 \\
060801  &	\ra{14}{12}{01.262} &	\dec{+16}{58}{55.97 } & $1.131 \pm 0.001$ & $r$ & $23.202 \pm 0.11$ & $0.02$ & Gold	& 8, 16 \\
061006  &	\ra{07}{24}{07.808} &	\dec{-79}{11}{55.188} & $0.461 \pm 0.0007$ & $r$ & $24.153 \pm 0.09$ & $4 \times 10^{-4}$ & Gold & 2, 8, 17, This work \\
061201  &	\nod &	\nod			     & \nod & F160W & $\gtrsim 26.4$ & \nod & Inconclusive & 18  	 \\
061210  &	\ra{09}{38}{05.362} &	\dec{+15}{37}{18.877} & $0.4095 \pm 0.0001$ & $r$ & $21.396 \pm 0.05$ & $0.02$ & Gold & 8, 16 \\
\hline
070429B &	\ra{21}{52}{03.691} &	\dec{-38}{49}{42.82 } & $0.902 \pm 0.001$ & $r$ & $23.283 \pm 0.04$ & $3 \times 10^{-3}$ & Gold	& 8, 19 \\
070707  &	\ra{17}{50}{58.555} &	\dec{-68}{55}{27.60 } & \nod & F606W & $26.857 \pm 0.12$ & $7.0 \times 10^{-3}$ & Gold	 & 18 \\
070714B &	\ra{03}{51}{22.272} &	\dec{+28}{17}{50.943} & $0.923 \pm 0.001$ & $r$ & $24.889 \pm 0.21$ & $5 \times 10^{-3}$ & Gold & 18-20 \\
070724  &	\ra{01}{51}{14.068} &	\dec{-18}{35}{38.47 } & $0.457 \pm 0.0007$ & $r$ & $20.776 \pm 0.03$ & $8 \times 10^{-4}$ & Gold & 8, 18, This work \\
070729  &	\ra{3}{45}{15.808} &	\dec{-39}{19}{18.590} & $0.52_{-0.28}^{+1.17}$ & $r$ & $23.019 \pm 0.263$ & $0.036$ & Silver & This work	 \\
070809  &	\ra{13}{35}{04.177} &	\dec{-22}{08}{33.01 } & $0.473$ & $r$ & $20.142 \pm 0.02$ & $6 \times 10^{-3}$ & Gold & 8, 18, 21 \\
071227  &	\ra{3}{52}{31.026} &	\dec{-55}{59}{00.89 } & $0.381$ & $r$ & $20.635 \pm 0.05$ &  $0.01$	 & Gold & 8, 17-18 \\
\hline
080123  &	\ra{22}{35}{46.943} &	\dec{-64}{53}{54.973} & $0.495$ & $r$ & $20.96 \pm 0.05$ & $0.11$ & Bronze & 8 \\
080503  &	\ra{19}{06}{28.901} &	\dec{+68}{47}{34.78 } & \nod & F606W & $27.151 \pm 0.2$ & $0.05$ & Silver & 18, 22 \\
080905A &	\ra{19}{10}{42.045} &	\dec{-18}{52}{54.51 } & $0.1218 \pm 0.0003$ & $R$ & $18.0 \pm 0.5$ & $0.01$ & Gold	& 18, 23 \\
081226A & \ra{08}{22}{00.45} & \dec{-69}{01}{49.5} & \nod & $r$ & $26.029 \pm 0.34$ &  0.01 & Gold & 24 \\
\hline
090305A &	\ra{16}{07}{07.596} &	\dec{-31}{33}{22.54 } & \nod & F160W & $25.292 \pm 0.10$ & $7 \times 10^{-3}$ & Gold & 18 \\
090426A & \ra{12}{36}{18.047} & \dec{+32}{59}{09.46} & 2.609$^{\ddagger}$ & F160W & $25.57 \pm 0.07$ & $1.5 \times 10^{-4}$ & Gold & 18, 25 \\
090510  &	\ra{22}{14}{12.623} &	\dec{-26}{34}{58.55 } & $0.903 \pm 0.001$ & $i$ & $22.452 \pm 0.14$ & $8 \times 10^{-3}$ & Gold	& 8, 18, 26 \\
090515  &	\ra{10}{56}{35.847} &	\dec{+14}{26}{42.84 } & 0.403 & $r$ & $20.268 \pm 0.05$ & $0.05$ & Silver	& 8, 18, 21 \\
091109B &	\ra{07}{30}{56.55} &	\dec{-54}{05}{23.22 } & \nod & F110W & $27.808 \pm 0.24$ &  $0.11$ & Bronze	& This work \\
\hline
100117A &	\ra{00}{45}{04.661} &	\dec{-01}{35}{42.02 } & $0.914 \pm 0.0004$ & $r$ & $24.40 \pm 0.10$ & $7 \times 10^{-5}$ & Gold	& 18, 27 \\
100206A &	\ra{3}{08}{39.142} &	\dec{+13}{09}{29.34 } & $0.407 \pm 0.002$ & $R$ & $21.53 \pm 0.09$ & $0.02$ & Gold	& 28, 29 \\
100625A &	\ra{1}{03}{10.918} &	\dec{-39}{05}{18.44 } & $0.452 \pm 0.002$ & $r$ & $22.659 \pm 0.09$ & $0.04$ & Silver	  & 30 \\ 
101219A &	\ra{4}{58}{20.497} &	\dec{-02}{32}{22.45 } & $0.7179 \pm 0.0008$ & $r$ & $24.083 \pm 0.05$ & $0.06$ & Silver	 & 30 \\
101224A &	\ra{19}{03}{41.919} &	\dec{45}{42}{48.86  } & $0.454 \pm 0.0007$ & $r$ & $22.071 \pm 0.052$  & $0.015$ & Gold	& This work \\
\hline
110112A &	\nod	 &	\nod		       		 & \nod & F110W & $\gtrsim 28.0$ & $0.44$ & Inconclusive & This work \\
111117A &	\ra{0}{50}{46.268} &	\dec{+23}{00}{41.41 } & $2.211 \pm 0.001$ & $r$ & $23.789 \pm 0.11$ & $0.024$ & Silver & 31-32 \\
\hline
120305A &	\ra{03}{10}{08.754} &	\dec{28}{29}{35.87  } & $0.225 \pm 0.001$ & $r$ & $22.398 \pm 0.050$ &  $0.053$ & Silver & This work \\
120804A &	\ra{15}{35}{47.510} &	\dec{-28}{46}{56.11 } & $0.74_{-0.33}^{+0.79}$ & $r$ & $26.406 \pm 0.200$ & $0.02$ & Gold & 33, This work  \\
121226A &	\ra{11}{14}{34.121} &	\dec{-30}{24}{22.84 } & $1.37_{-0.06}^{+0.05}$ & $r$ & $24.309 \pm 0.06$ & $0.019$ & Gold	& This work \\
\hline
130515A &	\ra{18}{53}{45.021} &	\dec{-54}{16}{50.72 } & $0.8 \pm 0.01$ & $r$ & $22.651 \pm 0.040$ & $0.081$	 & Silver & This work \\
130603B &	\ra{11}{28}{48.231} &	\dec{+17}{04}{18.61 } & $0.3568 \pm 0.0005$ & $r$ & $21.06 \pm 0.06$ & $2 \times 10^{-3}$ & Gold	& 18, 34-35 \\
130716A &	\ra{11}{58}{17.862} &	\dec{+63}{03}{15.35} & $2.2_{-0.37}^{+0.35}$ & $r$ & $24.894 \pm 0.344$ & $0.36$ & Bronze & This work	 \\
130822A &	\ra{1}{51}{42.708} &	\dec{-3}{12}{25.447} & $0.154 \pm 0.001$ & $r$ & $18.248 \pm 0.063$ & $0.086$ & Silver	 & This work \\
130912A &	\ra{03}{10}{22.2} &	\dec{13}{59}{48.74} & \nod & F110W & $27.471 \pm 0.23$ & $0.12$ & Bronze & This work \\
131004A &	\ra{19}{44}{27.064} &	\dec{-2}{57}{30.429} & 0.717$^{\ddagger}$ & F110W & $25.464 \pm 0.09$ & $0.055$	& Silver & 36, This work \\
\hline
140129B &	\ra{21}{47}{01.649} &	\dec{26}{12}{23.270 } & $0.43 \pm 0.003$ & $r$ & $23.55 \pm 0.07$ & $8.7 \times 10^{-4}$ & Gold & This work	 \\
140516A & \nod & \nod & \nod & $i$ & $\gtrsim 26.1$ & \nod & Inconclusive & This work \\
        &      &      &      & $K$ & $\gtrsim 23.6$ & \nod & & This work \\
140622A &	\ra{21}{08}{41.744} &	\dec{-14}{25}{06.166} & $0.959 \pm 0.001$ & $r$ & $22.703 \pm 0.042$ & $0.10$ & Bronze & This work \\
140903A &	\ra{15}{52}{03.265} &	\dec{+27}{36}{10.71 } & $0.3529 \pm 0.0002$ & $r$ & $21.367 \pm 0.194$ & $6.2 \times 10^{-5}$ & Gold	 & 37, This work \\
140930B &	\ra{0}{25}{23.473} &	\dec{+24}{17}{37.93 } & $1.465 \pm 0.001$ & $r$ & $24.206 \pm 0.248$ & $0.021$ &  Silver & This work \\
141212A &	\ra{2}{36}{29.957} &	\dec{+18}{08}{47.228} & $0.596 \pm 0.001$ & $r$ & $22.945 \pm 0.056$ & $2.9 \times 10^{-4}$ & Gold	& This work \\
\hline
150101B &	\ra{12}{32}{04.973} &	\dec{-10}{56}{00.50 } & $0.134 \pm 0.003$ & $r$ & $16.604 \pm 0.04$ & $4.8 \times 10^{-4}$ & Gold	 & 38 \\
150120A &	\ra{0}{41}{16.563} &	\dec{+33}{59}{42.598} & $0.4604 \pm 0.0004$ & $r$ & $22.051 \pm 0.063$ & $1.9 \times 10^{-3}$ & Gold	& This work \\
150423A &	\nod &	\nod & $1.394^{\ddagger}$ & F110W & $\gtrsim 28.1$ & \nod & Inconclusive & 39, This work \\
150424A &	\ra{10}{09}{13.406} &	\dec{-26}{37}{51.745} & \nod & F125W & $26.293 \pm 0.15$ & $0.06$ & Bronze$\dagger$	& This work \\
150728A &	\ra{19}{28}{54.808} &	\dec{+33}{54}{58.22 } & $0.461 \pm 0.0005$ & $i$ & $21.420 \pm 0.054$ &  $0.018$ & Gold	 & This work \\
150831A &	\ra{14}{44}{05.939} &	\dec{-25}{38}{05.78 } & $1.09_{-0.19}^{+0.1}$ & $r$ & $24.434 \pm 0.446$ &  $0.037$ & Silver & This work	 \\
151229A &	\ra{21}{57}{28.701} &	\dec{-20}{43}{54.80 } & $0.63_{-0.35}^{+0.49}$ & $i$ & $24.924 \pm 0.134$ &  $0.040$ & Silver & This work	 \\
\hline
160303A &	\ra{11}{14}{48.119} &	\dec{+22}{44}{33.420} & $1.01_{-0.4}^{+0.19}$ & F110W & $23.774 \pm 0.02$ & $0.096$ & Silver & 40, This work \\
160408A &	\ra{8}{10}{29.580} &	\dec{+71}{07}{45.03 } & $1.90_{-0.53}^{+0.38}$ & $r$ & $25.736 \pm 0.162$ & $0.14$	& Bronze & This work \\
160410A &	\nod	 & \nod	& $1.7177 \pm 0.0001^{\ddagger}$ & $r$ & $>27.2$ & \nod & Inconclusive &  41, This work \\
160411A &	\ra{23}{17}{25.355} &	\dec{-40}{14}{30.56 } & $0.82_{-0.45}^{+0.64}$	& $r$ & $24.532 \pm 0.134$ & $7.2 \times 10^{-4}$ & Gold & This work \\
160525B & \ra{9}{57}{32.227} &	\dec{+51}{12}{24.813} & \nod & $i$ & $24.08 \pm 0.30$ & $0.018$ & Gold & 42 \\
160601A &	\ra{15}{39}{43.949} &	\dec{+64}{32}{30.604} & \nod & $z$ & $24.947 \pm 0.344$ & $8.9 \times 10^{-4}$ & Gold & This work	 \\
160624A &	\ra{22}{00}{46.145} &	\dec{+29}{38}{39.336} & $0.4842 \pm 0.0005$	& $r$ & $21.960 \pm 0.047$ &  $0.037$ & Silver & This work \\
160821B &	\ra{18}{39}{53.994} &	\dec{+62}{23}{34.427} & $0.1619 \pm 0.0002$ & $r$ & $19.548 \pm 0.004$ & $0.044$ & Silver	 & This work \\
160927A & \nod & \nod & \nod & $G$ & $\gtrsim 25.7$ & \nod & Inconclusive & This work \\
        &      &      &      & $J$ & $\gtrsim 24.4$ & \nod &  & This work \\
161001A &	\ra{4}{47}{40.530} &	\dec{-57}{15}{39.184} & $0.67 \pm 0.02$ & $r$ & $22.968 \pm 0.046$ & $0.045$ & Silver	 & 39, This work \\
161104A &	\ra{05}{11}{34.37} &	\dec{-51}{27}{36.29 } & $0.793 \pm 0.003$ & $r$ & $23.847 \pm 0.10$ & $0.06$ & Bronze$^\dagger$ & 43  \\
\hline
170127B &	\ra{1}{19}{54.415} &	\dec{-30}{21}{29.615} & $2.28 \pm 0.14$ & $r$ & $25.320 \pm  0.290$ & $0.098$ & Silver & This work	\\
170428A &	\ra{22}{00}{18.710} &	\dec{+26}{54}{56.280} & $0.453 \pm 0.001$ & $r$ & $22.346 \pm 0.100$ & $6.7 \times 10^{-3}$ & Gold & This work	 \\
170728A &	\ra{3}{55}{33.116 } &	\dec{+12}{10}{51.04 } & $1.493 \pm 0.009$ & $R$ & $24.735 \pm 0.136$ &  $0.22$ & Bronze	& This work \\
170728B &	\ra{15}{51}{55.529} &	\dec{+70}{07}{22.038} & $0.62_{-0.37}^{+1.34}$ & $r$ & $23.313 \pm 0.096$ & $8.3 \times 10^{-3}$ & Gold	& This work \\
\hline
180418A &	\ra{11}{20}{29.21} &	\dec{24}{55}5{8.734 } & $1.56_{-0.43}^{+0.21}$ & $r$ & $25.729 \pm 0.21$ & $1.5 \times 10^{-3}$ & Gold	 & 44, This work \\
180618A &	\ra{11}{19}{45.801} &	\dec{+73}{50}{15.03 } & $0.52_{-0.11}^{+0.09}$ & $i$ & $22.183 \pm 0.081$ & $8.2 \times 10^{-3}$ & Gold  & This work  \\
180727A &	\ra{23}{06}{40.038} &	\dec{-63}{03}{07.088} & $1.95_{-0.58}^{+0.5}$ & $r$ & $26.486 \pm 0.277$ & $8.6 \times 10^{-3}$ & Gold	 & This work \\
180805B &	\ra{1}{43}{07.655} &	\dec{-17}{29}{33.091} & $0.6612 \pm 0.002$ & $r$ & $22.153 \pm 0.063$ & $0.042$ & Silver & This work  \\
181123B &	\ra{12}{17}{27.91} &	\dec{+14}{35}{52.27 } & $1.754 \pm 0.001$ & $r$ & $23.92 \pm 0.19$ & $4.4 \times 10^{-3}$ & Gold & 45 \\
\hline
191031D &	\ra{18}{53}{09.522} &	\dec{+47}{38}{40.13 } & $1.93_{-1.44}^{+0.22}$ & $r$ & $24.462 \pm 0.263$ & $0.043$ & Silver & This work \\
\hline
200219A &	\ra{22}{50}{33.108} &	\dec{-59}{07}{11.579} & $0.48 \pm 0.02$ & $r$ & $20.661 \pm 0.05$ & $2.2 \times 10^{-3}$ & Gold	& 13, This work \\
200411A &	\ra{03}{10}{39.135} &	\dec{-52}{18}{59.545} & $1.93_{-0.25}^{+0.15}$ & $r$ & $22.564 \pm 0.042$ &  $0.11$ & Bronze	& 13, This work \\
200522A &	\ra{00}{22}{43.717} &	\dec{-00}{16}{57.466} & $0.5536 \pm 0.0003$ & $r$ & $21.196 \pm 0.02$ & $3.5 \times 10^{-5}$ & Gold & 42, 46-48 \\
200907B &	\ra{05}{56}{06.951 } &	\dec{+6}{54}{22.637 } & $0.56_{-0.32}^{+1.39}$ & $i$ & $23.936 \pm 0.108$ &  $9 \times 10^{-3}$ & Gold & This work \\
201006A$^b$ &	\nod	 &	\nod & \nod & $K$ & $\gtrsim 23.6$ & \nod & Constraint & This work		       	 \\
201221D &	\ra{11}{24}{14.064} &	\dec{+42}{08}{40.047} & $1.055 \pm 0.001$ & $r$ & $23.418 \pm 0.076$ & $0.12$ & Bronze	& 49, This work \\
\hline
210323A &	\ra{21}{11}{47.320} &	\dec{+25}{22}{09.989} & $0.733 \pm 0.001$ & $r$ & $24.972 \pm 0.252$ & $0.013$ & Gold & This work \\
210726A &	\ra{12}{53}{09.638} &	\dec{+19}{11}{27.319} & $0.2244 \pm 0.0002$ & $r$ & $22.027 \pm 0.200$ & $0.036$ & Silver & 13, This work \\
210919A &	\ra{05}{21}{01.954} &	\dec{+1}{18}{40.022 } & $0.2415 \pm 0.001$ & $r$ & $20.50 \pm 0.05$ & $0.13$ & Bronze & 13, This work \\
211023B &	\ra{11}{21}{14.311} &	\dec{+39}{08}{08.36 } & $0.862 \pm 0.001$ & $r$ & $24.361 \pm 0.377$ &  $4.7 \times 10^{-3}$ & Gold	& 13, This work \\
211106A &	\ra{22}{54}{20.541} &	\dec{-53}{13}{50.548} & \nod & F814W & $25.791 \pm 0.069$ & $5.5 \times 10^{-4}$ & Gold & This work, 50 \\
211211A$^a$ & \ra{14}{09}{10.467} & \dec{+27}{53}{21.050} & $0.0763 \pm 0.0002$ & F606W & $19.57 \pm 0.01$ & 0.0136 & Gold & 51 \\
\hline
170817A$^{c}$ & \ra{13}{09}{47.70} & \dec{-23}{23}{02.0} & $0.009787 \pm 0.000057$ & $r$ & $12.44 \pm 0.01$ & $4.9 \times 10^{-4}$ & \nod &  52-53 \\
\enddata
\tablecomments{Magnitudes $m_{\lambda}$ are not corrected for Galactic extinction, $A_{\lambda}$, in the direction of the burst. Photometric redshift uncertainties correspond to 68\% confidence and the methods to determine them are described in \citet{BRIGHT-II}. \\
$\dagger$ Hosts reclassified as Bronze due to at least one other galaxies with comparably low $P_{\rm cc}$ value in the field. \\
$\ddagger$ Redshift determined from afterglow. \\
$^a$ Long-duration GRBs thought to be associated with NS merger origins. \\
$^b$ This burst has a high Galactic extinction of $A_V=3.5$~mag and is therefore considered an observing constraint burst with a sightline that precludes a meaningful host galaxy search in the optical bands. However, we report a $K$-band limit here on a galaxy within the XRT position for completeness. \\
$^c$ This host is only considered as a point of comparison to the cosmological short GRB sample. \\
{\bf References:} (1) \citealt{bpp+06},
(2) \citealt{fbf10},
(3) \citealt{SDSS-DR13}, 
(4) \citealt{scs+06}, 
(5) \citealt{cmi+06},
(6) \citealt{ffp+05},
(7) \citealt{hwf+05},
(8) \citealt{lb10},
(9) \citealt{bpc+05},
(10) \citealt{gcg+06},
(11) \citealt{fsk+07},
(12) \citealt{pbc+06},
(13) \citealt{sdh+21}, 
(14) \citealt{sbk+06},
(15) \citealt{gfp+06},
(16) \citealt{bfp+07},
(17) \citealt{dmc+09},
(18) \citealt{fb13},
(19) \citealt{cbn+08},
(20) \citealt{gfl+09},
(21) \citealt{ber10a},
(22) \citealt{pmg+09},
(23) \citealt{rwl+10},
(24) \citet{nkg+12},
(25) \citealt{lbb+10},
(26) \citealt{mkr+10},
(27) \citealt{fbc+11},
(28) \citealt{pmm+12},
(29) \citealt{WISE},
(30) \citealt{fbc+13},
(31) \citealt{mbf+12},
(32) \citealt{skm+18},
(33) \citealt{bzl+13},
(34) \citealt{dtr+14},
(35) \citealt{cpp+13},
(36) \citealt{gcn15307},
(37) \citealt{tsc+16},
(38) \citealt{fmc+16},
(39) \citealt{smg+19},
(40) \citealt{gcn19160},
(41) \citealt{atk+21},
(42) \citealt{cmm+16}, 
(43) \citealt{nfd+20},
(44) \citealt{rfv+21},
(45) \citealt{pfn+20},
(46) \citealt{flr+21},
(47) \citealt{Papovich2016}, 
(48) \citealt{Timlin2016}, 
(49) \citealt{gcn29133},
(50) \citealt{GRB211106A-Laskar},
(51) \citealt{rgl+22},
(52) \citealt{bbf+17},
(53) \citealt{SIMBAD}}
\end{deluxetable*}

\subsection{Literature or Archival Survey Photometry}

To supplement these observations, we draw from published ground-based data in the literature, focusing on (i) previously published, well-characterized hosts, or (ii) hosts which lack imaging in a given filter in our catalog. In this vein, we collect literature photometry for \totlitphotnum\ bursts. We list the references for all of these data in Table~\ref{tab:sample}. When made available by corresponding authors, we also provide the reduced stacks from these works on the BRIGHT website. We emphasize that the literature data set is comprehensive for a given host in that we attempt to fill out the SED, but does not include all literature photometry that exists for every host galaxy.

Finally, for \totarchivephotnum\ bursts, we include archival photometric survey data. We draw from Two Micron All Sky Survey All-Sky (2MASS, \citealt{scs+06}), Legacy Surveys Data Release 9 (DR 9, \citealt{sdh+21}), Pan-STARRS Data Release (DR 2, \citealt{fmc+20}), Sloan Digital Sky Survey Data Release 12 (SDSS DR12, \citealt{aaa+15}), SDSS DR13, \citep{SDSS-DR13}, Spitzer \citep{Papovich2016, Timlin2016}, and Wide-field Infrared Survey Explorer (WISE, \citealt{WISE}).

\subsection{Afterglow Observations}
\label{sec:afterglow}

To obtain afterglow positions and thus burst locations with respect to putative host galaxies, we first use ground-based optical discovery images when available. In particular, we utilize a combination of our Target-of-Opportunity programs on the twin 6.5-m Magellan telescopes, the 6.5-m MMT, the twin Gemini telescopes, and the 60-in Palomar Observatory P60 telescope (PIs: Berger, Fong, Cenko, Cucchiara), as well as publicly-available ground-based imaging from the 4.2-m William Herschel Telescope (WHT) and the $8.2$-m Very Large Telescope (VLT). For all ground-based observations, we use the same procedures for data reduction as for our host galaxy imaging, described in Section~\ref{sec:obs}.

For the subset of bursts for which there exists {\it HST} afterglow discovery imaging, we retrieve and process the images as described in Section~\ref{sec:hst}. For both ground-based and {\it HST} imaging, if the position of the afterglow is contaminated by host galaxy light in the discovery image, we use the HOTPANTS software package \citep{bec15} to perform image subtraction between the afterglow images and late-time templates to produce residual images for accurate afterglow centroiding. We use {\tt Source Extractor} ({\tt SExtractor}; \citealt{SExtractor}) to determine the positional uncertainties of the afterglows, $\sigma_{\rm GRB}$. We calculate a range of $\sigma_{\rm GRB} \approx 10-120$~mas for ground-based discoveries, and $\sigma_{\rm GRB}\approx 1-4$~mas for bursts with {\it HST}-detected afterglows.

For bursts for which the most precise afterglow localization is from the {\it Chandra X-ray Observatory}, we retrieve Level II files from the {\it Chandra} archive, and we use CIAO/{\tt wavdetect} to determine their positions and uncertainties. We describe how relative astrometry is performed to the host galaxy images in Section~\ref{sec:astrometry}. For afterglows detected with the Karl G. Jansky Very Large Array (VLA) or Atacama Large Millimeter Array (ALMA), we use CASA/{\tt jmfit} to fit a 2D Gaussian to the afterglow. Finally, for 33 bursts, the most precise afterglow localization is from {\it Swift}/XRT. In these cases, we use the published positions and uncertainties\footnote{We draw the positions and uncertainties from \url{https://www.swift.ac.uk/xrt_positions/}, using values as of June 2022.}, with the methods described in \citet{ebp+09}.

\section{Host Galaxy Associations}
\label{sec:assoc}

\subsection{Astrometry}
\label{sec:astrometry}

For each host galaxy stack, we perform absolute astrometry using common sources between the host galaxy imaging and available source catalogs: Gaia DR2 \citep{Gaia-DR2}, Pan-STARRS (PS1; \citealt{cmm+16}), SDSS DR12 \citep{aaa+15}, or 2MASS. We use {\tt SExtractor} to determine the centroids of common sources, and IRAF/{\tt ccred} and {\tt ccsetwcs} to determine the astrometric solution from each image to the catalog. We find that a fourth-order polynomial with six free parameters corresponding to a shift, scale, and rotation in each coordinate provides robust solutions. The $1\sigma$ absolute astrometric uncertainties have a range of $\sigma_{\rm abs}\approx 0.1-0.3''$. We report the host galaxy positions in Appendix Table~\ref{tab:phot}.

In order to make host galaxy identifications, it is necessary to align the afterglow and host galaxy imaging to the same frame. Thus, we additionally perform relative astrometry from the available afterglow imaging (from ground-based facilities, {\it HST}, or {\it Chandra} X-ray Observatory) to the host galaxy images, which themselves are tied to an absolute astrometic system. We again use a combination of {\tt SExtractor} and IRAF using common sources. The $1\sigma$ uncertainty on the afterglow position includes the afterglow positional uncertainty (see Section~\ref{sec:afterglow}) and the relative astrometic tie uncertainty ($\sigma_{\rm rel}\approx 10-100$~mas) added in quadrature. If we do not have access to afterglow discovery images, we use published positions from the literature or GCNs. If uncertainties are not provided with those positions, we assume a conservative $1\sigma$ uncertainty of $0.5''$. For the 33 bursts with no sub-arcsecond localization, and X-ray positions only from {\it Swift}/XRT, we use the published XRT positions directly (typically $\approx 1.5-5\arcsec$; \citealt{ebp+09}). For 89 of the \totsamp\ fields, we show a representative, deep optical or NIR image centered on the most precise afterglow position for each burst in Figures~\ref{fig:imagepanel}-\ref{fig:hostless}\footnote{We only lack imaging for GRB\,081226A and use the results reported in \citet{nkg+12} for our subsequent analysis.}.

\subsection{Probability of Chance Coincidence}
\label{sec:pcc}

We use the available imaging data to determine the most probable host galaxy for each burst by calculating the probability of chance coincidence ($P_{cc}$) for nearby galaxies in the field of view. The fields of view are typically $>2.5'$ in radius, corresponding to $\approx\!1$~Mpc at $z=0.5$. For context, the largest observed projected physical offsets for short GRBs are only $\lesssim 100$~kpc \citep{ber10,tlt+14}. The $P_{cc}$ method requires two ingredients: angular offsets between a host galaxy candidate and burst location, $\delta R$, and putative host galaxy magnitudes, $m_i$. The methods to calculate $\delta R$ are described in Section~\ref{sec:offsets}.

Using our deepest available image for each burst, which is typically in the $r$-band or $JH$-bands, we start with the extended source which has the smallest angular offset to the most precise afterglow position. For this source, we perform aperture photometry using the IRAF/{\tt apphot} package. We begin with default apertures of 2.5$\theta_{\rm FWHM}$, but often use larger apertures to fully encompass the galaxy light. For background regions, we use annuli immediately surrounding the putative hosts, and adjust the radii as needed to avoid any contaminating sources. We determine zeropoints either by using stars with catalogued magnitudes in the field of the host galaxy, or by using standard star fields taken on the same night at similar airmasses. For the determination of optical zeropoints, we use SDSS DR12, Pan-STARRS (applying transformations to the SDSS system; \citealt{tsl+12}), or the USNO-B catalogs. For NIR zeropoints, we use the 2MASS catalog. When relevant, we convert from the Vega to the AB system, using standard transformations or instrument-specific conversions. For {\it HST} data, we use the relevant tabulated zeropoint for each instrument and filter\footnote{\url{http://www.stsci.edu/hst/wfc3/ir\_phot\_zpt} and \url{https://acszeropoints.stsci.edu/}}, again varying the apertures and background regions based on the size of each galaxy.

We follow the methodology of \citet{bkd02} to calculate $P_{cc}$ based on the surface density of galaxies brighter than a given magnitude, $\sigma(\leq m_i)$ within a radius, $R_i$. To determine $\sigma(\leq m_i)$, we interpolate $r$-band or $H$-band number counts from galaxy surveys (compiled in \citealt{hpm+97,bsk+06} and \citealt{msw+06}) and integrate the relevant function for $m\leq m_i$ depending on the filter of the observation. For the value of $R_i$, we use the maximum of $\delta R$ or $\sigma_{\rm GRB}$, which is an approximation on the methods described in \citet{bkd02} and \citet{bbf16}, in the absence of effective radii measurements for all putative GRB hosts. If the most proximal extended source has $P_{cc}\lesssim 0.01$, we consider this to be the host galaxy of that burst.

If the most proximal source to the GRB has $P_{cc}\gtrsim 0.01$, we continue to perform photometry of all extended sources in the image using the IRAF/{\tt apphot} package. We discard noticeably fainter galaxies with increasing angular distance $\delta R$ from the burst since these objects will have a lower probability of being the host galaxy based on chance alignment arguments. We then calculate $P_{cc}$ for each of these sources. For each burst, the minimum in the probability function, $P_{\rm cc, min}$, corresponds to the most probable host galaxy. In many cases, this still ends up being the most proximal host galaxy, although in some cases a galaxy at a larger separation is favored as the host.

\subsection{Gold, Silver and Bronze Samples}
\label{sec:classifications}

In general, the $P_{cc}$ method can recover clear host galaxies for most bursts. However, it favors apparently brighter galaxies at a given angular separation, and it is difficult to interpret if two putative hosts for a given burst have similarly low $P_{cc}$ values. To reflect the varying robustness of associations and the nuances of the method, we divide our sample into four categories based on the minimum probability of chance coincidence value, $P_{\rm cc, min}$ as follows:

\begin{itemize}
    \item {\it Gold:} Bursts with putative hosts that have  $P_{\rm cc, min} \leq 0.02$. There are 49 events in the Gold sample. These represent short GRBs with the most robust host associations, although they are likely to be biased toward bursts with sub-arcsecond localizations (e.g., optical afterglows), smaller offsets, and apparently brighter hosts.
    \item {\it Silver:} Bursts with putative hosts that have $0.02 < P_{\rm cc, min} \leq 0.10$. There are 22 events in the Silver sample. These represent short GRBs with moderately robust host associations. This sample is subject to less of the biases outlined for the Gold sample.
    \item {\it Bronze:} Bursts with putative hosts that have $0.10 < P_{\rm cc, min} \leq 0.20$. There are 13 events in the Bronze sample. These represent short GRBs with the least robust host associations, but for which more probable alternatives do not exist. There is likely a small loss of integrity in individual host assignments (addressed in Section~\ref{sec:disc}). However this is an important sample to include as it is least subject to the biases of the $P_{cc}$ assignment method.
    \item {\it Inconclusive:} Bursts for which the lowest value for an extended source in the field is $P_{\rm cc, min}>0.20$. There are 6 events in the Inconclusive sample. In all cases, there are deep optical or NIR limits on a coincident host galaxy to $\gtrsim 26-28$~mag.
\end{itemize}

\noindent In general, we follow the above guidelines to associate each short GRB to its host galaxy. However, we make three exceptions and modifications to the above scheme based on information from their afterglows, or an analysis that yields multiple putative hosts with similarly low $P_{cc}$ values.

For GRB\,150424A, the first and second closest galaxies (in angular separation) have similar values of $P_{cc}=0.06$ and $0.04$, respectively, both of which would place the burst in the Silver sample. The second closest galaxy has a spectroscopic redshift of $z=0.3$, but this is at odds with the inferred value from the afterglow SED, which implies $z=1.0^{+0.3}_{-0.2}$ \citep{gcn17758,kgv+17}. Instead, the closest galaxy, which was first reported in early {\it HST} imaging \citep{gcn18100} has a probable redshift range of $z \approx 0.9-1.6$ based on the likely location of the 4000\AA\ break \citep{jlw+18}. Thus, despite having a slightly higher $P_{cc}$ value, we conclude that the closest galaxy is the most likely host of GRB\,150424A. We downgrade this burst to the Bronze sample given the more ambiguous nature of this association.

For GRB\,161104A, the galaxies with the two lowest $P_{cc}$ values have $P_{cc}=0.06$ and $P_{cc}=0.08$ \citep{nfd+20}, which would nominally place this burst in the Silver sample. The closest galaxy, with $P_{cc}=0.06$, is fully encompassed in the XRT position and is part of a galaxy cluster at $z\approx 0.79$. We consider this to be the host galaxy, but we also demote this burst to the Bronze sample.

Finally, for GRB\,061201, there are two galaxies at very different offsets that have identical $P_{cc}$ values of 0.07, which would be a Silver sample burst. While one of the putative hosts is at $z=0.111$ and has previously been considered as a tentative association \citep{sdp+07,fb13}, we cannot conclude on an individual host association and consider this to be an Inconclusive burst.

Overall, we make associations for \totassoc\ events. The resulting minimum $P_{cc}$ value, corresponding to the most probable host galaxy, for each short GRB is listed in Table~\ref{tab:sample}. We also list the optical or NIR magnitude, and the host classification (Gold, Silver, Bronze, Inconclusive). We confirm many associations that were previously made in the literature, as well as make 26 new identifications\footnote{In addition, we corroborate the host associations of eight short GRBs recently reported in \citet{otd+22}.}. We also revise associations for three bursts: GRBs\,070729, 161001A and 191031D. For GRB\,070729, the XRT position shifted significantly compared to the initial published position on which the host association was made. The updated position coincides with a faint source on the edge of the XRT position (90\% confidence) while the original host published in \citet{lb10} now has a substantially larger value of $P_{cc}$. For GRB\,161001A, \citet{smg+19} notes a faint extended source in the wings of an M-star that overlaps with the XRT position with $z=0.891$, that is taken to be the host galaxy. However, this galaxy is not apparent in our imaging and instead we find a (presumably) brighter $r \approx 22.9$~mag galaxy coincident with the XRT position (Figure~\ref{fig:imagepanel}). We consider this to be the host galaxy and determine a photometric redshift of $z \approx 0.67$. Finally, for GRB\,191031D, \citet{otd+22} identified a galaxy outside of the XRT position at $z\approx0.5$ as the host. However, our deep Keck NIR imaging reveals a red, extended source within the XRT position with $P_{cc}=0.04$ (compared to the larger value of $P_{cc}\approx 0.1$ for the galaxy identified by \citealt{otd+22}). Thus, we revise this host identification and photometric redshift to $z \approx 1.9$. We use the categories in our downstream analysis and in \citet{BRIGHT-II}.

\subsection{Inconclusive Host Associations}

\renewcommand{\thefigure}{\arabic{figure}}
\begin{figure*}[t]
\centering
\includegraphics[width=0.325\textwidth]{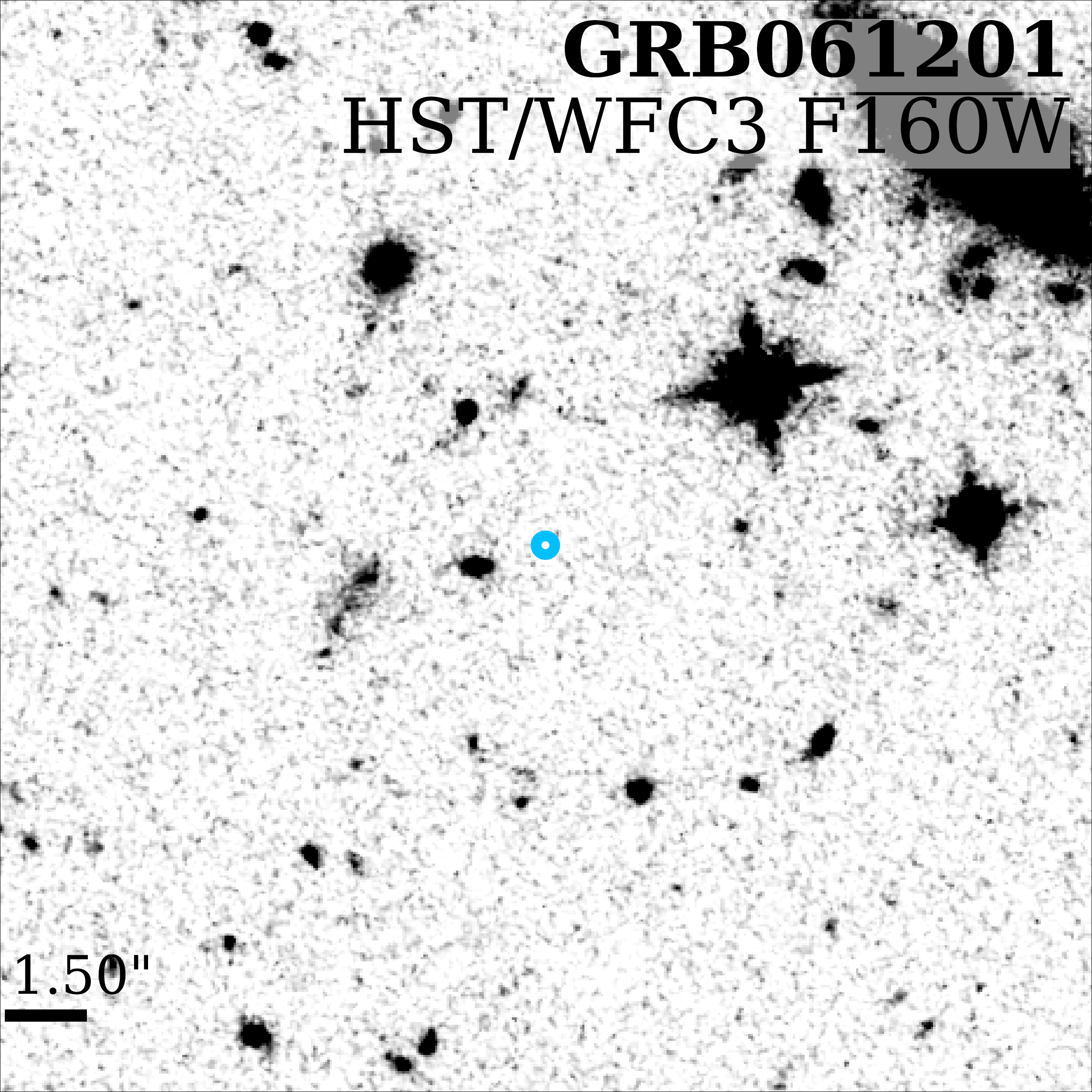}
\includegraphics[width=0.325\textwidth]{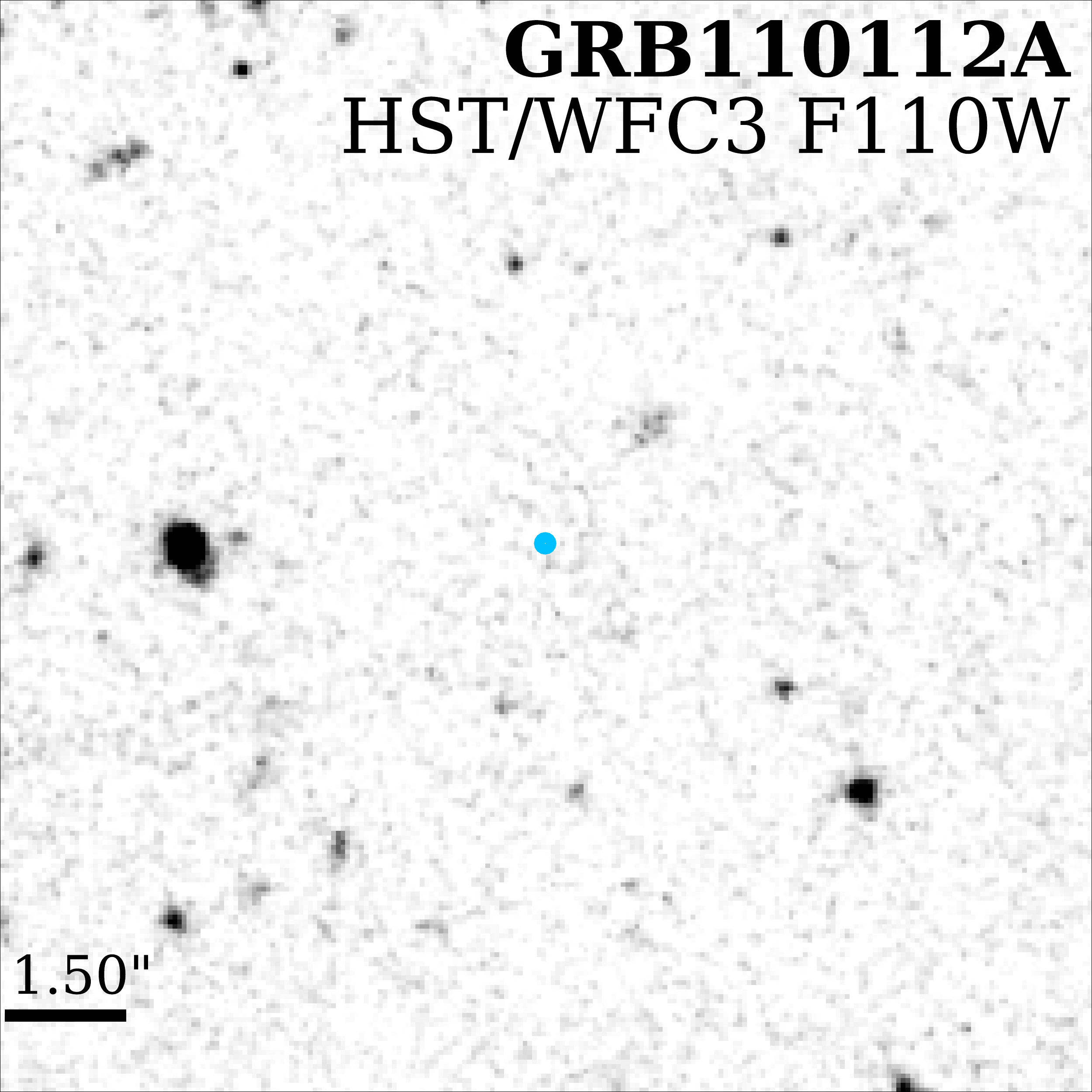}
\includegraphics[width=0.325\textwidth]{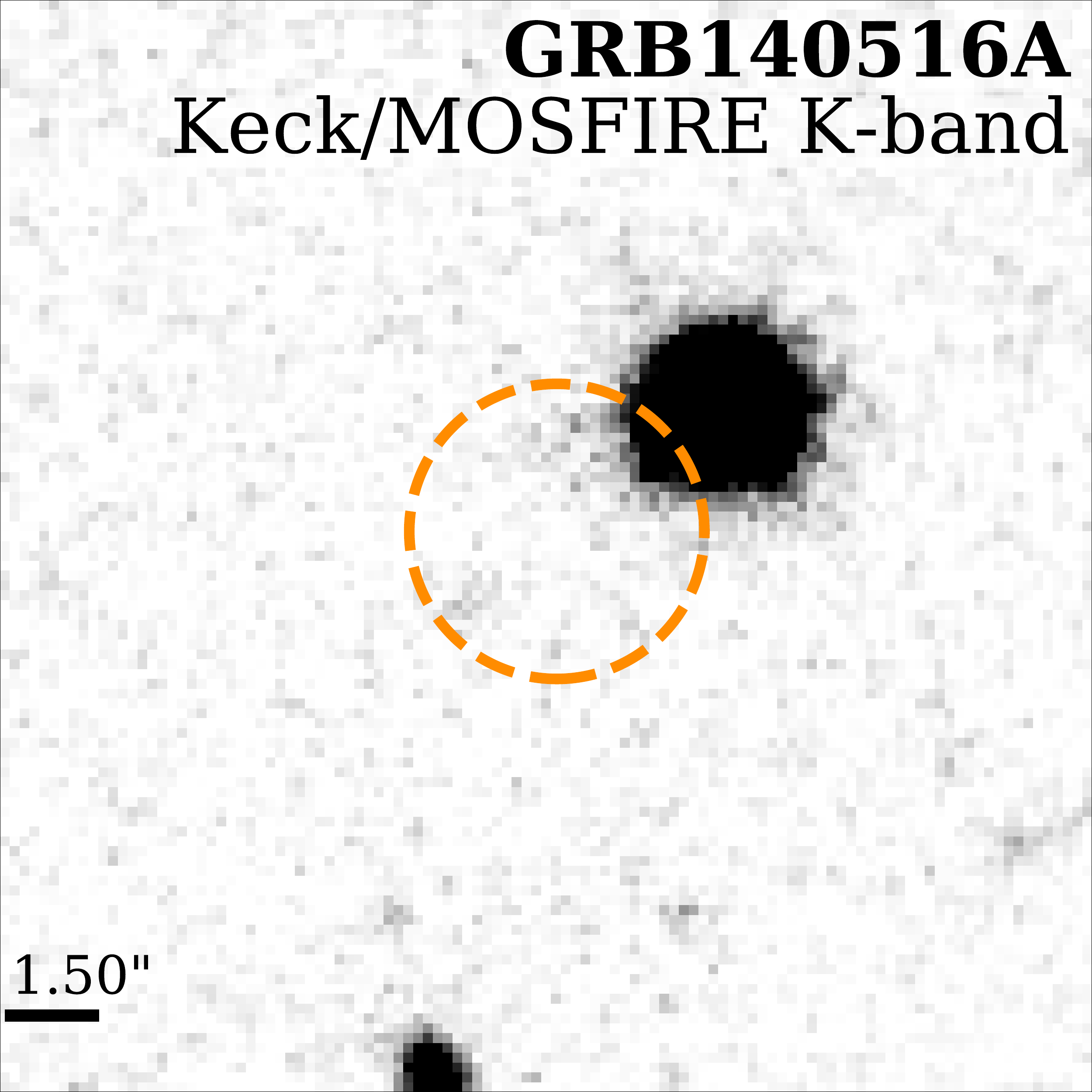}
\includegraphics[width=0.325\textwidth]{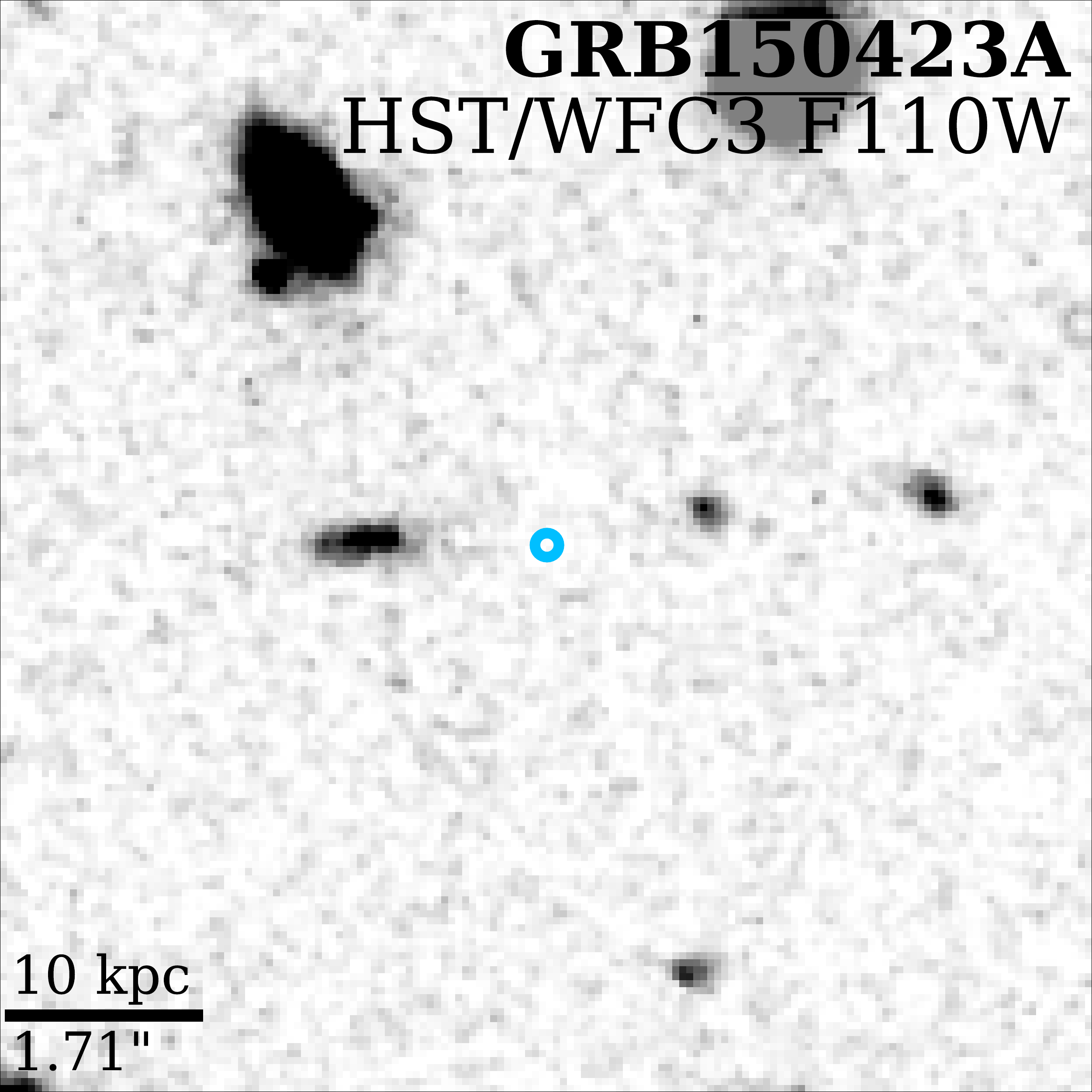}
\includegraphics[width=0.325\textwidth]{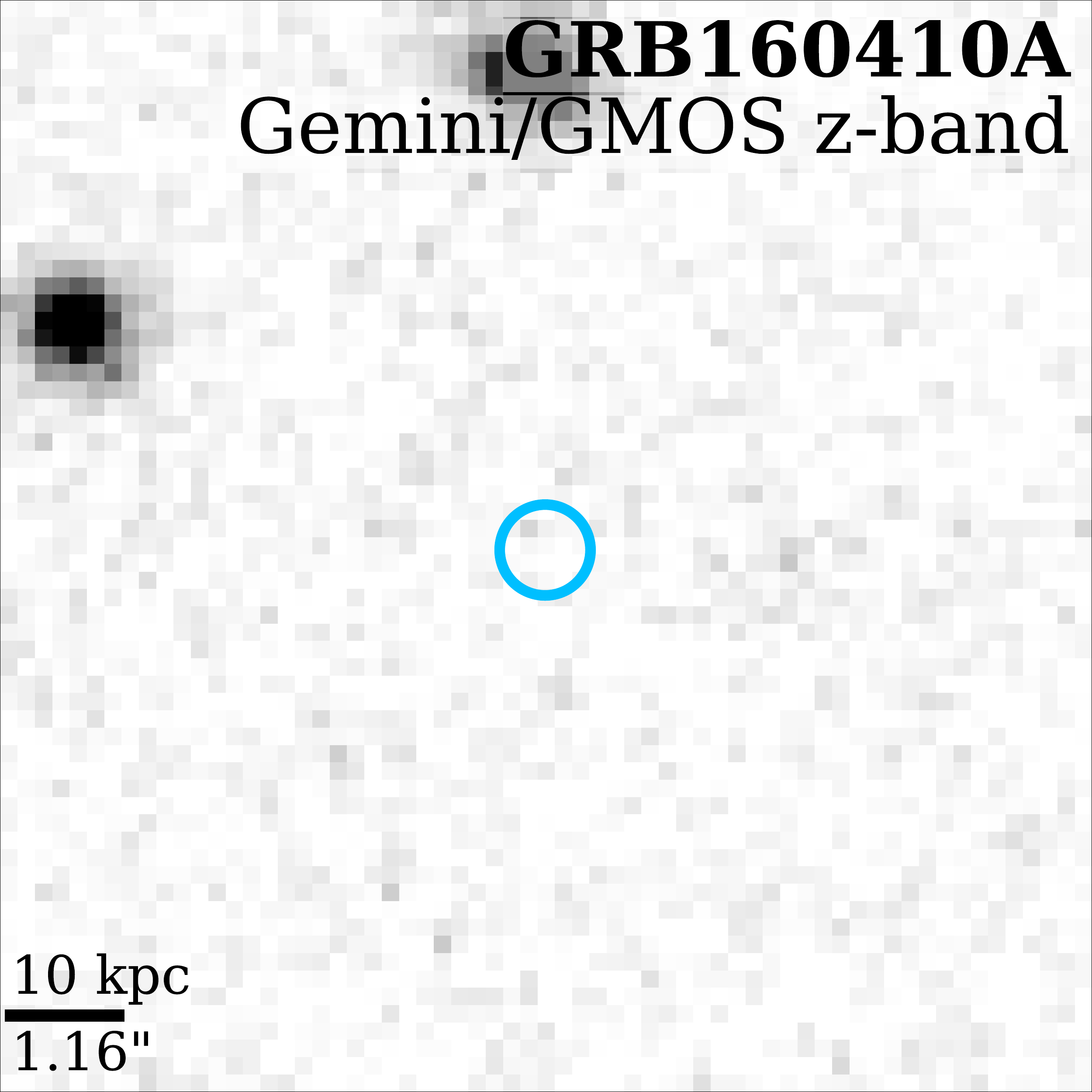}
\includegraphics[width=0.325\textwidth]{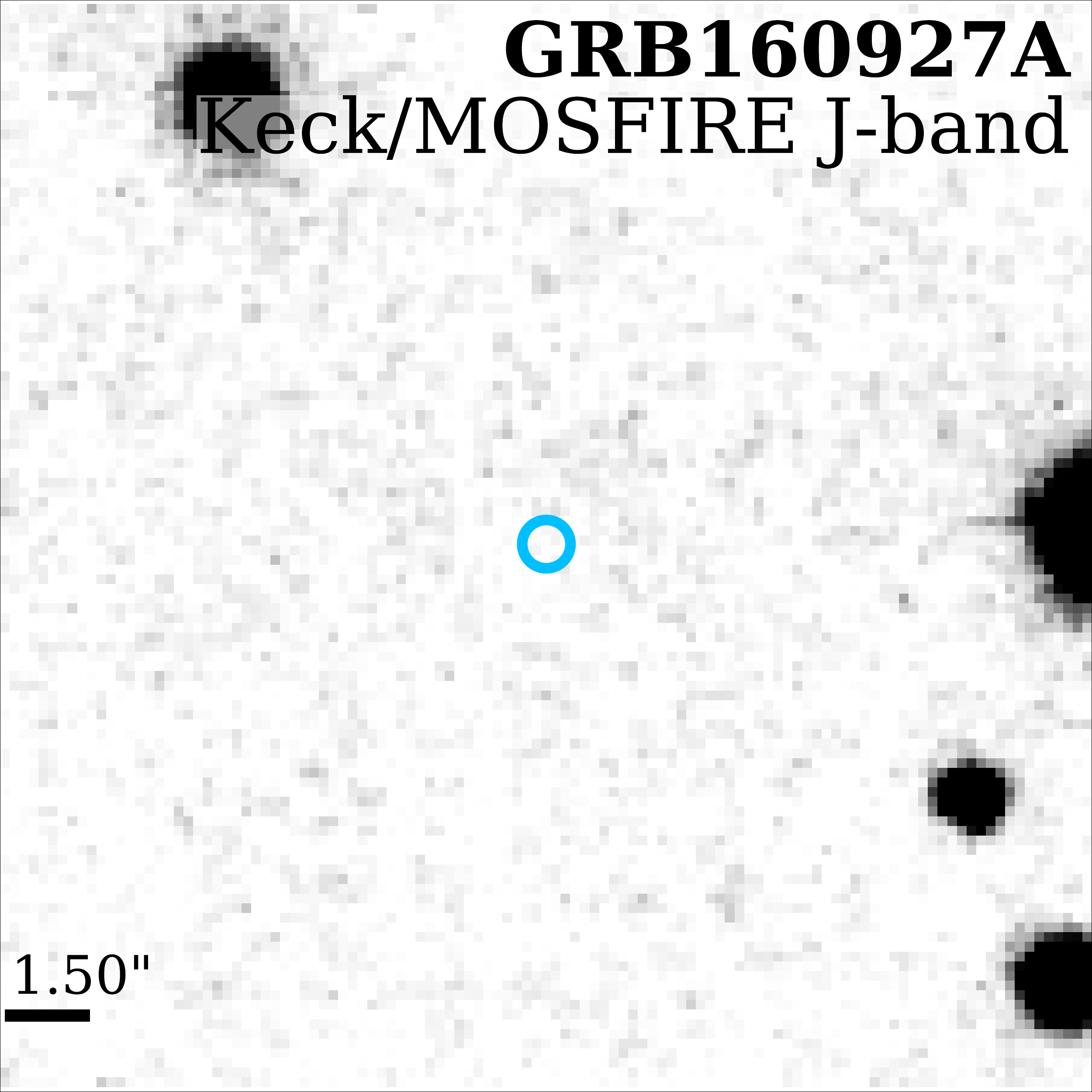}
\includegraphics[width=0.325\textwidth]{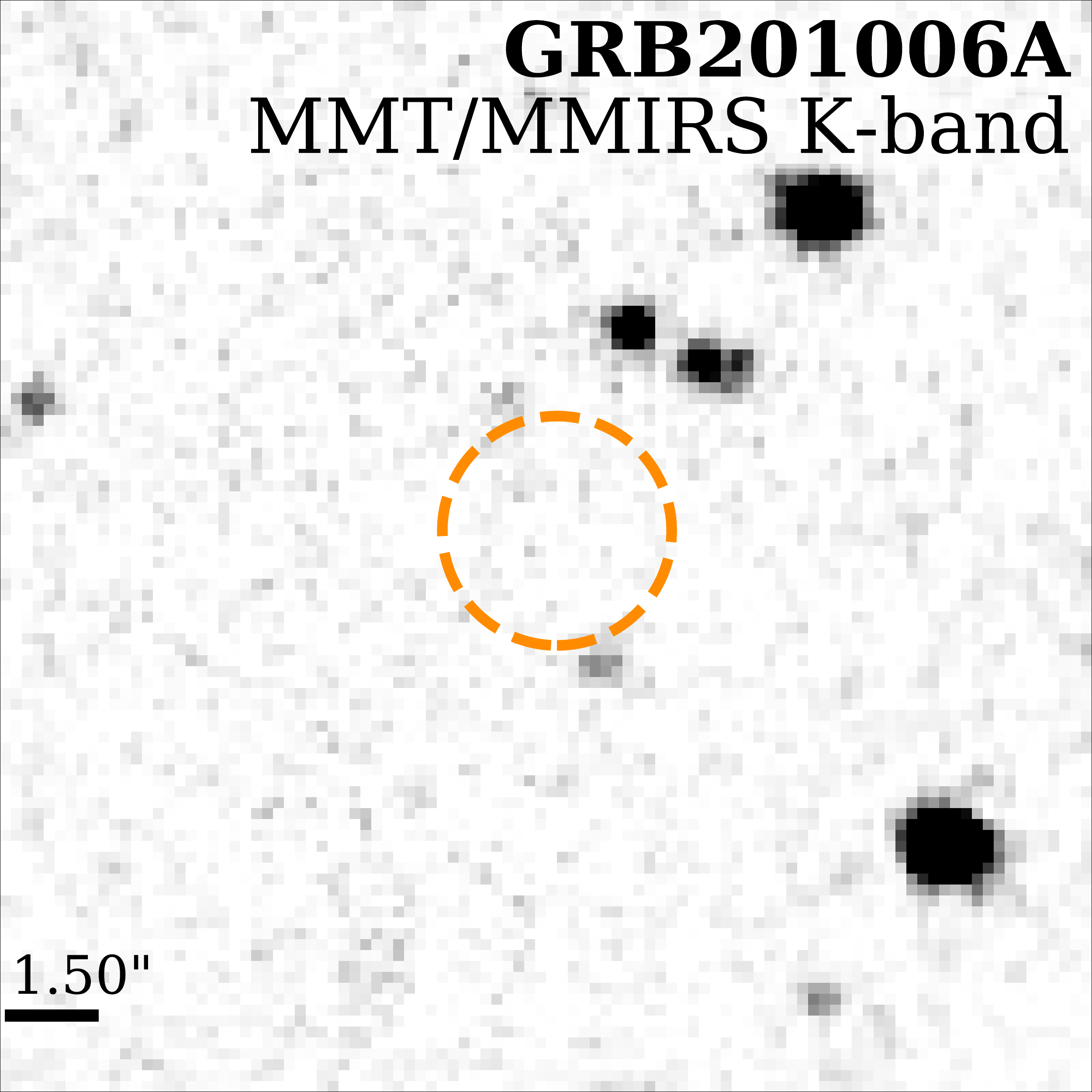}
\caption{The fields of six short GRBs with no clear host association ($P_{\rm cc,min} \gtrsim 0.25$), classified as ``Inconclusive'' associations. Orange dashed circles are XRT positions, while solid blue circles are optical positions. Also displayed is the field of GRB\,201006A which has a high extinction siteline of $A_V=3.5$~mag (falling in the observing constraint category), but for which we have deep $K$-band imaging to place limits on a redder host. These images reach $3\sigma$ depths of $\gtrsim 26$-$28$~mag on coincident host galaxies.
\label{fig:hostless}}
\end{figure*}

We show the fields of the six bursts in the Inconclusive host association class in Figure~\ref{fig:hostless}. We also show a seventh burst, GRB\,201006A, which has a high extinction sightline through the Galaxy of $A_V=3.5$~mag but for which we have deep $K$-band imaging in which the extinction is less severe. Thus, we consider this burst to have meaningful information at NIR wavelengths only. Five have sub-arcsecond localizations from optical afterglows while two (GRBs\,140516A and 201006A) only have XRT positions. 

This class of bursts was formerly termed ``host-less'' \citep{ber10}. The fields of GRBs\,061201, 110112A, and 160410A were previously studied and classified as such \citep{ber10,fb13,atk+21} (although deeper {\it HST} imaging for GRB\,110112A is presented here). We introduce four new cases. Both GRB\,140516A\footnote{We note that the brighter object on the outskirts of the XRT position is a star.} and 201006A have faint extended sources in the vicinity of their XRT positions (90\% confidence) although no galaxy reaches the threshold of $P_{\rm cc, min}\leq 0.2$. 

Two of these bursts, GRBs\,150423A and 160410A have known redshifts from their afterglows (\citealt{smg+19,atk+21}; Table~\ref{tab:sample}). From {\it HST} imaging, we find that GRB\,150423A has no host to $m_{\rm F110W} \gtrsim 28.1$~mag at the position of its optical afterglow, constraining any spatially coincident galaxy to $L_r \lesssim 3 \times 10^{8}L_{\odot}$ at $z=1.394$. The most probable host galaxy using the $P_{cc}$ method has $z=0.456$ \citep{gcn17744} with $P_{cc} \approx 0.2$, but is inconsistent with the afterglow redshift. Meanwhile, the next most probable host galaxy (and the one at smallest angular separation to the afterglow) has $P_{cc}=0.23$, too high to meet the Bronze class threshold. For GRB\,160410A at $z=1.7177$, the reported limit is $r\gtrsim 27.2$~mag \citep{atk+21}, which constrains any coincident host to $L_r \lesssim 1.6 \times 10^{9}L_{\odot}$.

Overall, while this population may represent bursts with the largest offsets, and is thus an important population to include, it comprises only a small fraction of the total short GRBs in this sample. We also note that several formerly ``host-less'' bursts have associations which meet the threshold to fall in one of the Gold, Silver, or Bronze categories, thus highlighting the importance of using a uniform method across all bursts for association.

\section{Spectroscopic Catalog \& Data Analysis}
\label{sec:specobs}

\begin{deluxetable*}{lcccccc}
\linespread{1.2}
\tabletypesize{\scriptsize}
\tablecaption{Log of Spectroscopic Observations \label{tab:specobs}}
\tablecolumns{7}
\tablewidth{0pt}
\tablehead{
\colhead{GRB} &
\colhead{Facility/Instrument} &
\colhead{Exposures} &
\colhead{Lines Identified} & 
\colhead{Previously Published?} &
\colhead{Re-reduced?} &
\colhead{Reduction Method} \\
}
\startdata
050509B & Keck/DEIMOS & $3 \times 300$ & Ca~II H\&K & \citet{bpp+06} & yes & PypeIt \\
050709 & Gemini/GMOS & $2 \times 1200$ & H$\beta$, [O~III]$\lambda \lambda 4959, 5007$ & \citet{ffp+2005} & yes & IRAF \\
050724 & Gemini/GMOS & $4 \times 1800$ & Ca~II H\&K & \citet{bpc+05} & yes & IRAF \\
051221A & Gemini/GMOS & $2 \times 1800$ & [O~II]$\lambda3727$, H$\beta$, [O~III]$\lambda \lambda 4959, 5007$ & \citet{sbk+06} & yes & IRAF \\
060614 & Gemini/GMOS & 4x1200 & [O~III]$\lambda \lambda 4959, 5007$, H$\alpha$ & \citet{nah+17} & yes & PypeIt \\
060801 & Gemini/GMOS & $2 \times 900$ & [O~II]$\lambda 3727$ & \citet{bfp+07} & yes & IRAF \\
061006 & Gemini/GMOS & $2 \times 1830$ & [O~II]$\lambda3727$, [O~III]$\lambda \lambda 4959, 5007$ & \citet{bfp+07} & yes & IRAF \\
070429B & Keck/LRIS & $2 \times 1500$ & [O~II]$\lambda 3727$ & \citet{cbn+08} & yes & PypeIt \\
070714B & Keck/LRIS & $1 \times 2100$ & [O~II]$\lambda 3727$ & \citet{cbn+08} & yes & PypeIt \\
070724A$^*$ & Keck/DEIMOS & $2 \times 1800$ & [O~II]$\lambda3727$, [Ca~II]H\&K, H$\beta$, [O~III]$\lambda \lambda 4959, 5007$ & no & yes & PypeIt \\
090510 & VLT/FORS2 & $1 \times 1800$ & [O~II]$\lambda3727$, [O~III]$\lambda 5007$ & \citet{mkr+10} & yes & PypeIt \\
100117A & Gemini/GMOS & $4 \times 1460$ & Ca~II H\&K & \citet{fbc+11} & no & \nod \\
100206A & Keck/LRIS & $2 \times 600$ & [O~II]$\lambda 3727$, Ca~II H\&K, H$\beta$, [O~III]$\lambda\lambda 4959, 5007$, & \citet{pmm+12} & no & \nod \\
 & & & H$\alpha$, [N~II]$\lambda \lambda 6549,6584$, [S~II]$\lambda \lambda 6717, 6731$ \\
100625A & Magellan/LDSS & $2 \times 2700$ & Ca~II H\&K, H$\beta$, H$\delta$ & \citet{fbc+13} &  no & \nod \\
101219A & Gemini/GMOS & $4 \times 1800$ & [O~II]$\lambda3727$, H$\beta$, [O~III]$\lambda \lambda 4959, 5007$ & \citet{fbc+13} & no & \nod \\
101224A & LBT/MODS & $8 \times 600$ & H$\gamma$, H$\beta$, [O~III]$\lambda \lambda 4959, 5007$, H$\alpha$ & no & yes & IRAF \\
120305A & LBT/MODS & $2 \times 900$ & [O~III]$\lambda 5007$, H$\alpha$ & no & yes & IRAF \\
130515A & Magellan/IMACS & $2 \times 1200$ & None & no & yes & IRAF \\
130603B & Gemini/GMOS & $2 \times 900$ & [O~II]$\lambda3727$, H$\beta$, [O~III]$\lambda \lambda 4959, 5007$ & \citet{cpp+13} & no & \nod \\
130822A & LBT/MODS & $3 \times 600$ & [O~II]$\lambda 3727$, $H\beta$, H$\alpha$ & no & yes & IRAF \\
140129B & Keck/DEIMOS & $3 \times 1800$ & [O~II]$\lambda 3727$, $H\beta$, [O~III]$\lambda 5007$ & no & yes & PypeIt \\
140622A & Magellan/LDSS & $2x1800$ & [O~II]$\lambda 3727$, $H\beta$ & no & yes & IRAF \\
140903A$^*$ & Keck/LRIS & $2x1200$ & [O~III]$\lambda 5007$, $H\alpha$ & no & yes & IRAF \\
140930B & LBT/MODS & $4 \times 1200$ & [O~II]$\lambda 3727$ & no & yes & IRAF \\
141212A & Gemini/GMOS & $3 \times 900$ & H$\beta$, [O~III]$\lambda \lambda 4959, 5007$ & no & yes & PypeIt \\
150101B & Magellan/IMACS & $2 \times 600$ & H$\beta$, Mg$\lambda5175$, NaD$\lambda5892$, TiO$\lambda7050$ & \citet{fmc+16} & no & \nod \\
150120A & Gemini/GMOS & $1 \times 900$ & [O~II]$\lambda 3727$, H$\beta$ & no & yes & PypeIt \\
150728A & Keck/DEIMOS & $3 \times 1800$ & [O~II]$\lambda3727$, H$\beta$, [O~III]$\lambda \lambda 4959, 5007$ & no & yes & PypeIt \\
151229A & Keck/LRIS & $3 \times 1200$ & None & no & yes & PypeIt \\
160411A & Magellan/LDSS & $3 \times 1800$ & None & no & yes & IRAF \\
160624A$^*$ & LBT/MODS & $8 \times 600$ & $H\beta$, [O~III]$\lambda 5007$ & no & yes & IRAF \\
160821B$^*$ & Keck/DEIMOS & $2 \times 900$ & H$\gamma$, H$\beta$, [O~III]$\lambda \lambda 4959, 5007$ & no & yes & PypeIt \\
161104A & Magellan/IMACS & $3 \times 1800$ & Ca~II H\&K & \citet{nfd+20} & no & \nod \\
170428A & MMT/Binospec & $4 \times 900$ & [O~II]$\lambda 3727$,  [O~III]$\lambda 5007$ & no & yes & MMT/IRAF \\
170728A & Keck/DEIMOS & $3 \times 1800$ & [O~II]$\lambda 3727$ & no & yes & PypeIt \\
180805B & Keck/LRIS & $2 \times 1200$ & [O~II]$\lambda3727$, H$\gamma$, H$\beta$, [O~III]$\lambda \lambda 4959, 5007$ & no & yes & PypeIt \\
180618A & MMT/Binospec & $2 \times 1800$ & None & no & yes & MMT/IRAF \\
181123B & Gemini/FLAMINGOS & $30 \times 120$ & H$\beta$ & \citet{pfn+20} & no & \nod \\
200522A & Keck/LRIS & $3 \times 900$ & [O~II]$\lambda3727$, H$\gamma$, H$\beta$, [O~III]$\lambda \lambda 4959, 5007$ & \citet{flr+20} & no & \nod \\
201221D & Keck/LRIS & $3 \times 1240$ & [O~II]$\lambda 3727$ & no & yes & IRAF \\
210323A & Keck/DEIMOS & $3 \times 1800$ & [O~II]$\lambda3727$, [O~III]$\lambda \lambda 4959, 5007$ & no & yes & PypeIt \\
210919A & Keck/DEIMOS & $3 \times 1800$ & [O~II]$\lambda 3727$,[O~III]$\lambda 4959$, [O~III]$\lambda$, H$\alpha$ & no & yes & PypeIt \\
211023B & Keck/DEIMOS & $2 \times 1800$ & [O~II]$\lambda3727$, [O~III]$\lambda \lambda 4959, 5007$ & no & yes & PypeIt \\
211211A & Keck/DEIMOS & $2 \times 1500$ & H$\beta$, [O~III]$\lambda \lambda 4959, 5007$, H$\alpha$, & \citet{rgl+22} & no & \nod \\
& & & [N~II]$\lambda \lambda 6549,6584$, [S~II]$\lambda \lambda 6717, 6731$ \\
\enddata
\tablecomments{Spectroscopic observations of short GRB host galaxies. Reduction methods are stated for bursts analyzed or re-analyzed in this work. Spectra that were not re-reduced were donated by the corresponding authors to this work for the BRIGHT database. \\
$^*$ These hosts have different previously-published spectra than the ones presented here. Our new spectra taken with different instruments of the same objects are consistent with the literature findings (GRBs 070724A: \citealt{bcf+09}; 140903A: \citealt{tsc+16}, 160624A: \citealt{otd+21}, and 160821B: \citealt{ltl+19}).}
\end{deluxetable*}

The second major goal of this study is to build a spectroscopic catalog of short GRB host galaxies. We draw from new spectroscopic observations, archival data that were previously published in the literature, and donated reduced spectra from corresponding authors. 

\subsection{Spectroscopic Observations}

We obtained spectroscopic observations for 21 short GRB hosts with unpublished spectra: GRBs\,101224A, 120305A, 130822A, 140129B, 140622A, 140930B, 141212A, 150120A, 150728A, 151229A, 161001A, 160411A, 170428A 170728A, 180618A, 180805B, 201221D, 210323A, 210726A, 210919A, and 211023B. For these observations, we used the twin 6.5-m Magellan/Baade and Clay telescopes, 8-m Gemini-North and Gemini-South telescopes, 6.5-m MMT, twin 10-m Keck I and II telescopes, and the 10.2-m LBT. Additionally, we obtained our own spectroscopic observations for four short GRBs which have previously published spectra from other telescopes and have known redshifts: GRBs 070724A \citep{lb10}, 140903A \citep{tsc+16}, 160624A \citep{otd+21}, and 160821B \citep{kkl17}. We list the telescope and instruments used for these new observations in Table \ref{tab:telescopes}, as well as the details of the spectra in Table~\ref{tab:specobs}.

We also draw from archival and literature sources to obtain spectroscopy for the remaining host galaxies. Our aim is to build as complete a spectroscopic catalog as possible to enable the uniform stellar population modeling analysis in \citet{BRIGHT-II}. Thus, we first retrieved the raw 2D spectra and calibration files of 10 short GRB hosts from observatory archives for re-reduction and analysis. We note that these same spectra were previously published in the following works: GRBs 050509B \citep{bpp+06}, 050709 \citep{ffp+05}, 050724 \citep{bpc+05}, 051221A \citep{sbk+06}, 060614 \citep{nah+17}, 060801 \citep{bfp+07}, 061006 \citep{bfp+07}, 070429B \citep{cbn+08}, 070714B \citep{cbn+08}, and 090510 \citep{mkr+10}. From the literature, we also obtained reduced 1D object and error spectra for ten events from the corresponding authors: GRBs 100117A \citep{fbc+11}, 100206A \citep{pmm+12}, 100625A, 101219A \citep{fbc+13}, 130603B \citep{cpp+13}, 150101B \citep{fmc+16}, 161104A \citep{nfd+20}, 181123B \citep{pfn+20}, 200522A \citep{flr+21}, and 211211A \citep{rgl+22}. This archival and literature sample uses Magellan, Gemini, Keck, and the Very Large Telescope (VLT). We list the details of these spectra, when available, in Table~\ref{tab:specobs}.

\subsection{Spectroscopic Reduction \& Analysis}

\begin{figure*}
\centering
\includegraphics[width=0.52\textwidth]{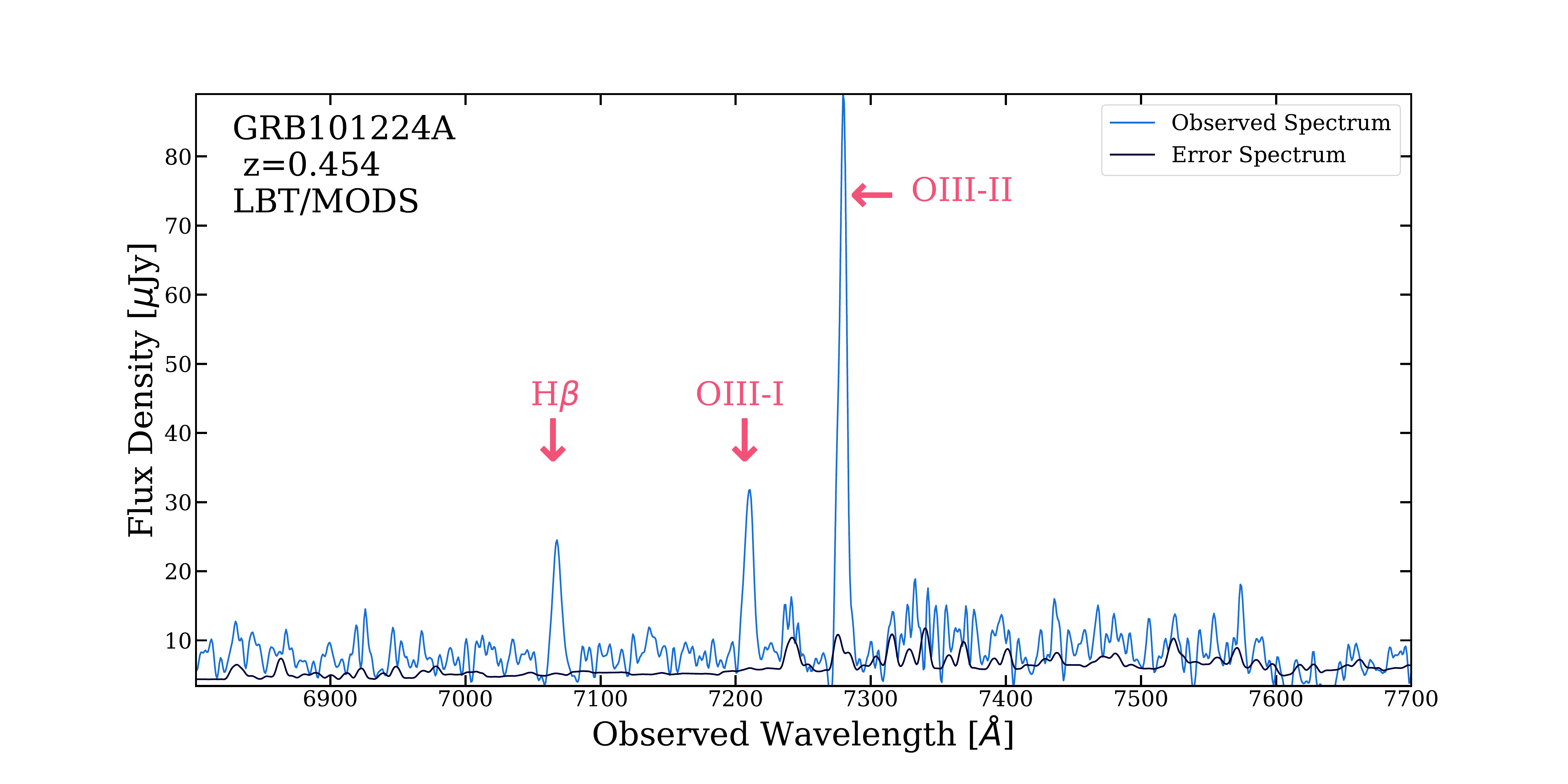}
\hspace{-0.5in}
\includegraphics[width=0.52\textwidth]{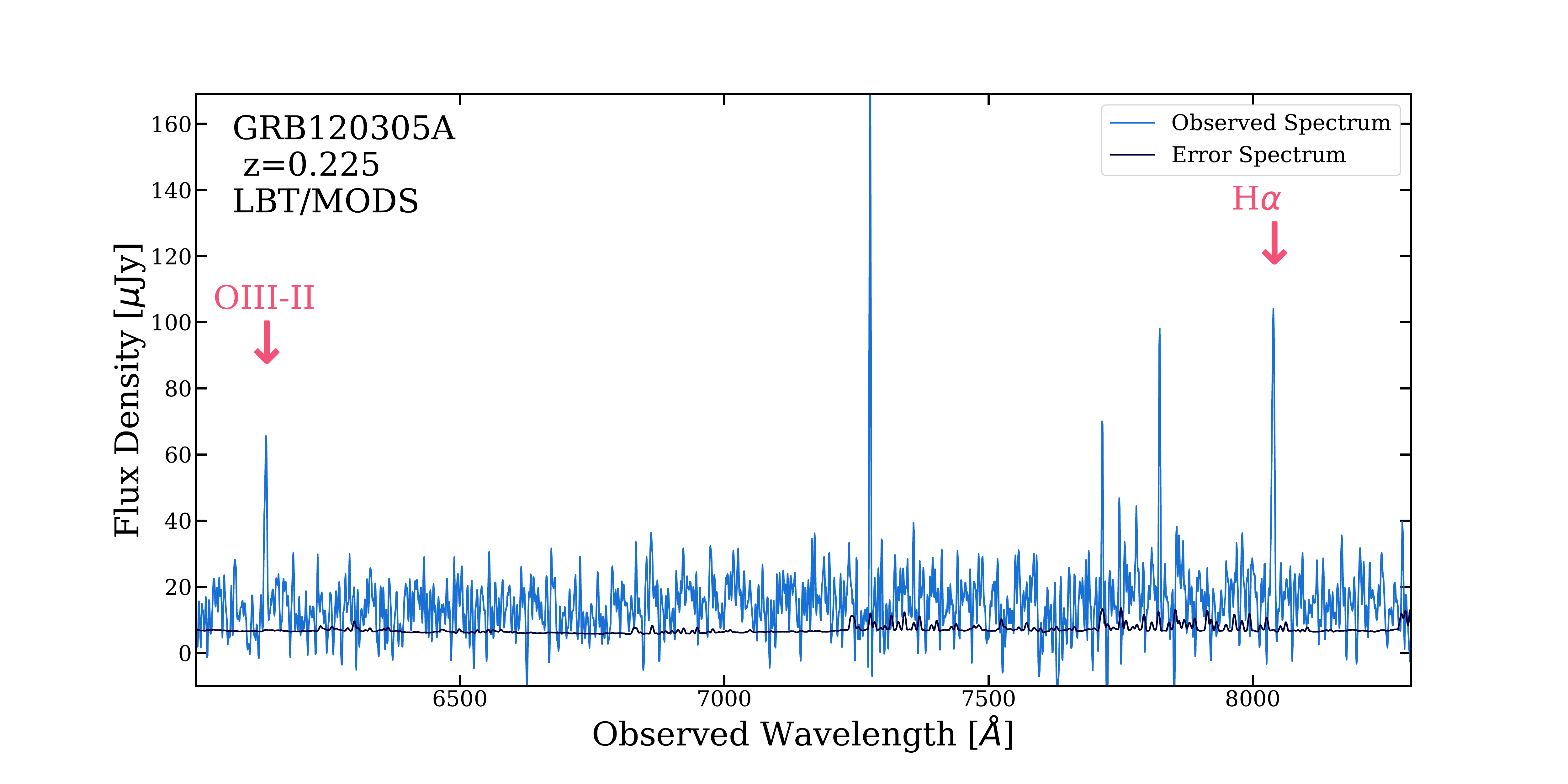}
\hspace{-0.5in}
\includegraphics[width=0.52\textwidth]{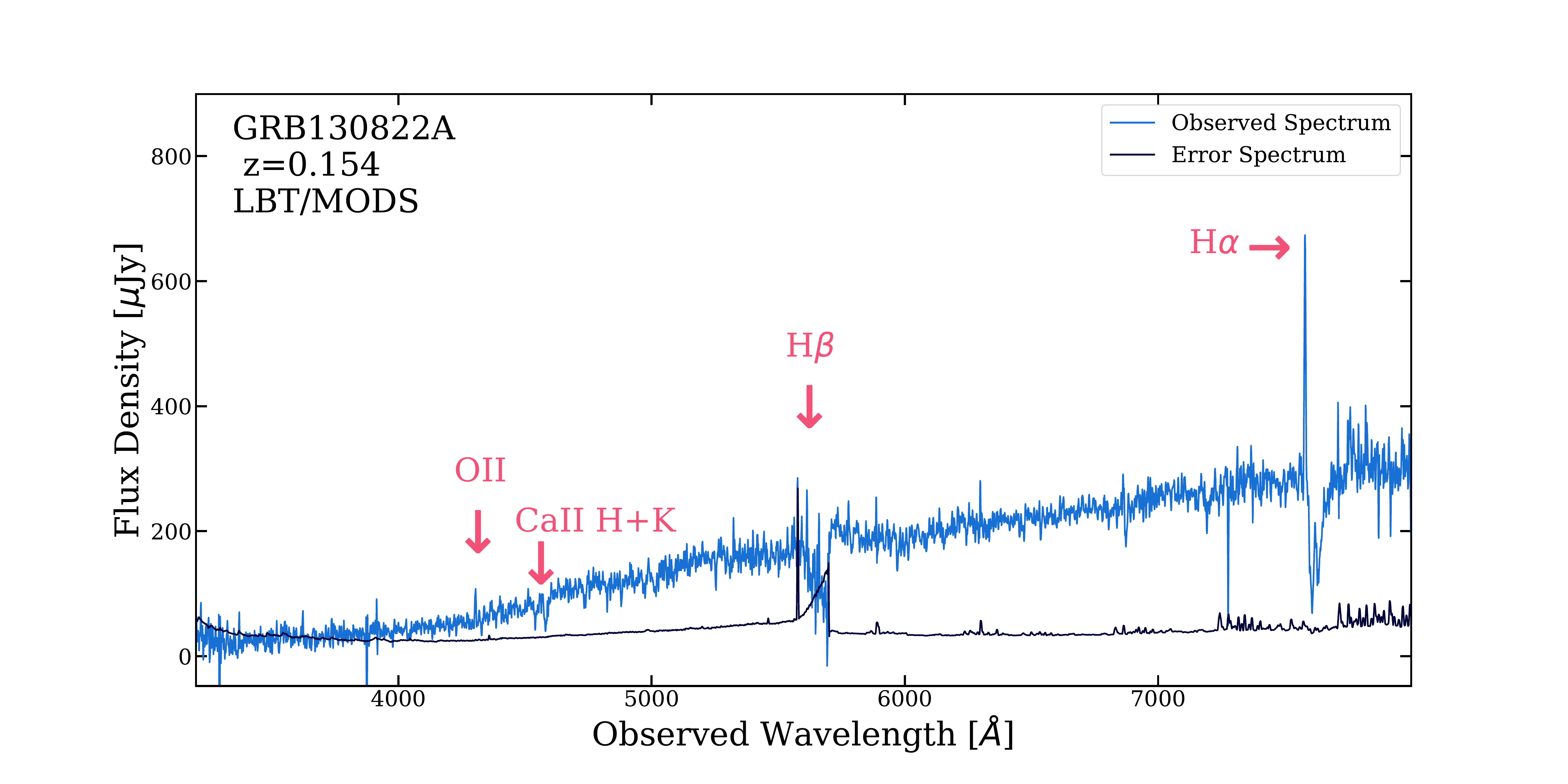}
\hspace{-0.5in}
\includegraphics[width=0.52\textwidth]{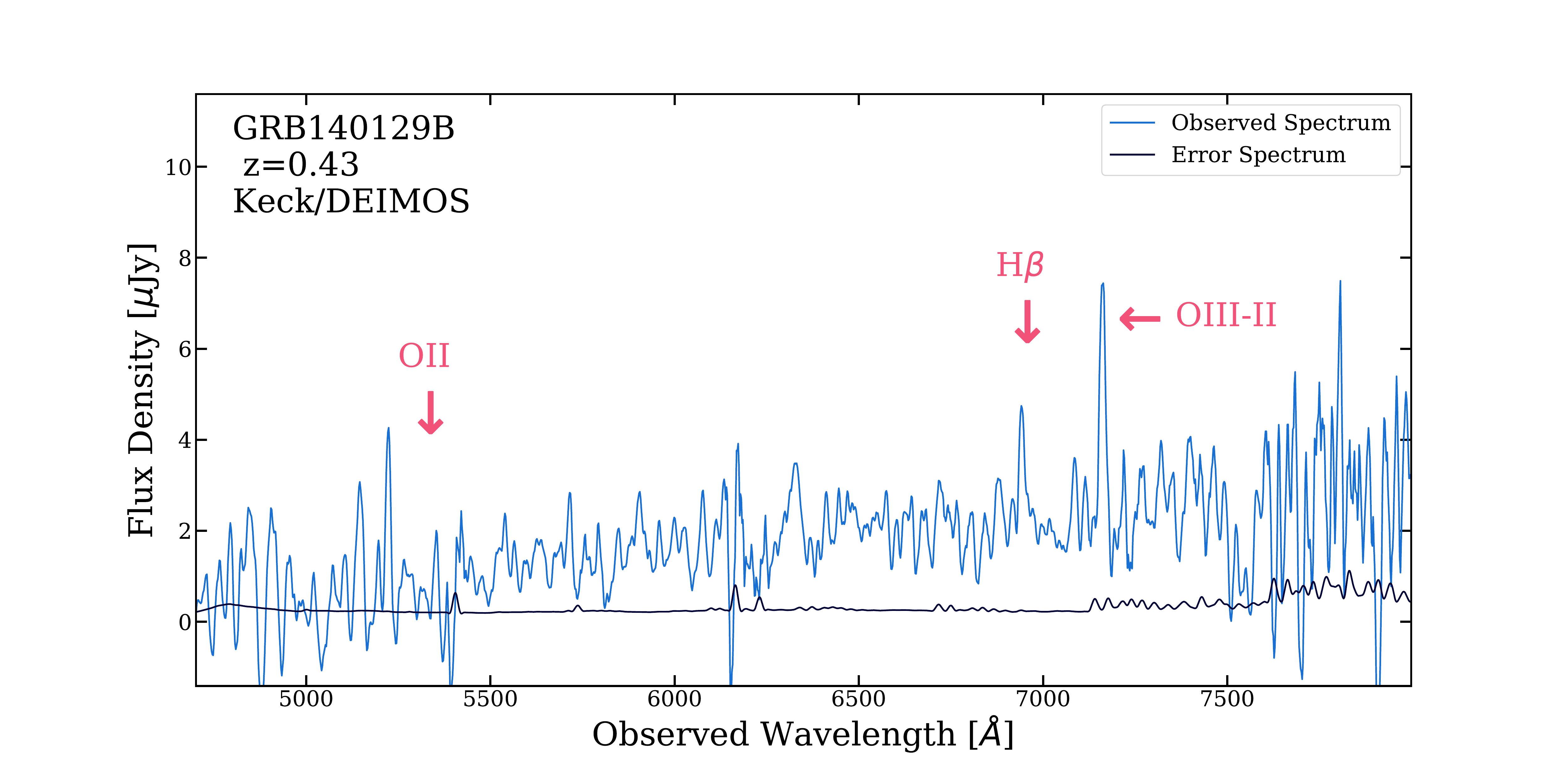}
\hspace{-0.5in}
\includegraphics[width=0.52\textwidth]{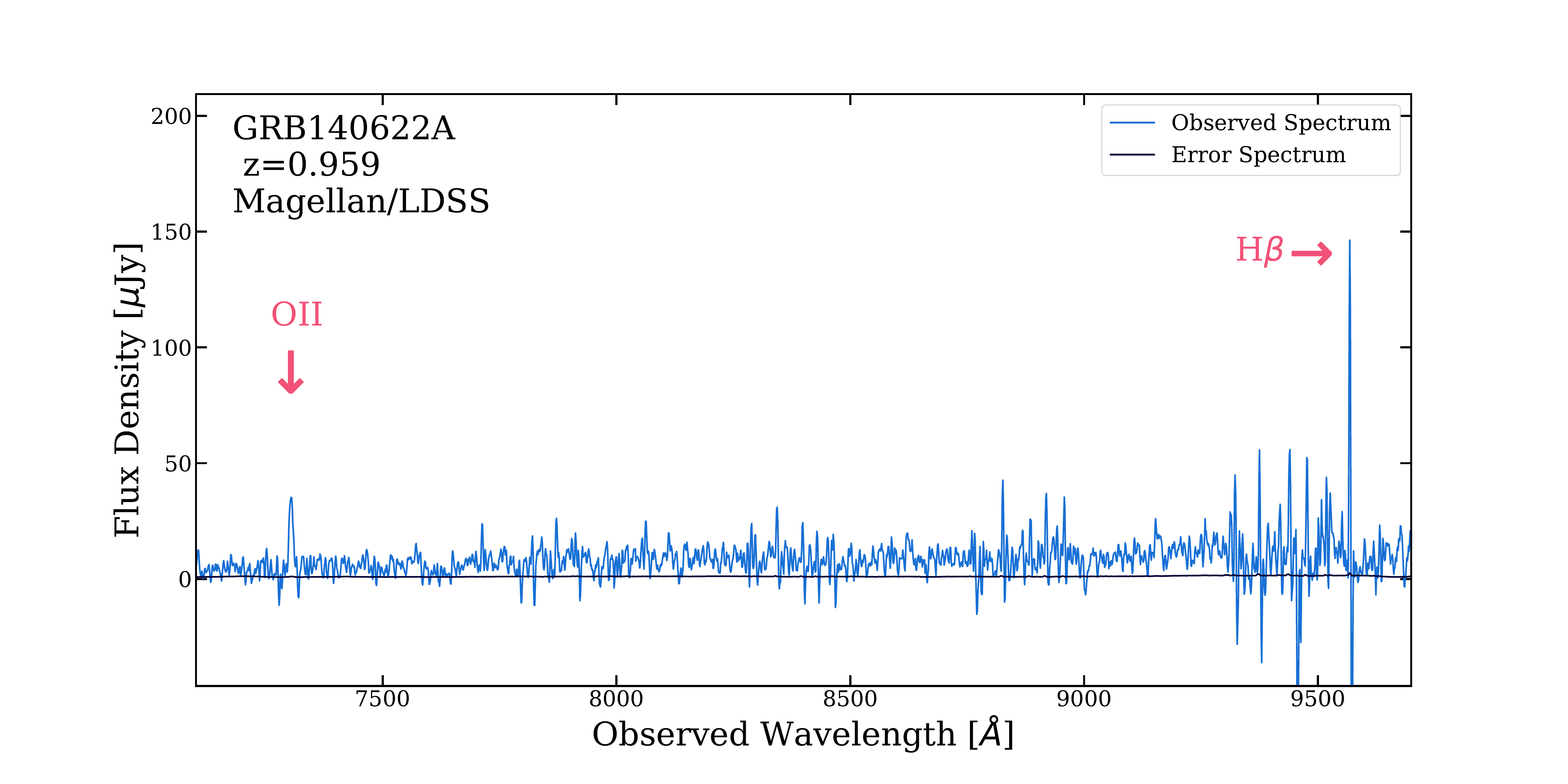}
\hspace{-0.5in}
\includegraphics[width=0.52\textwidth]{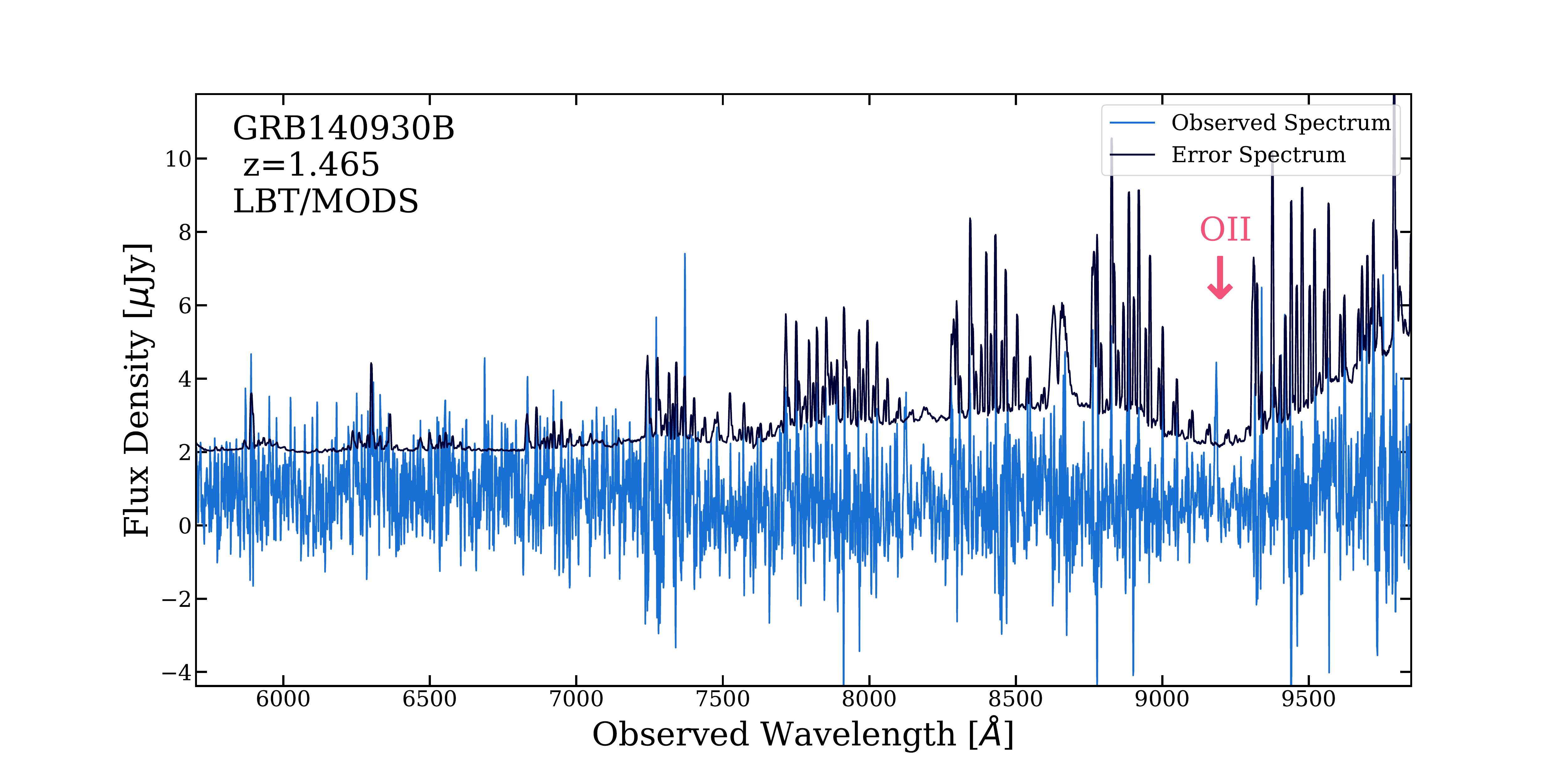}
\hspace{-0.5in}
\includegraphics[width=0.52\textwidth]{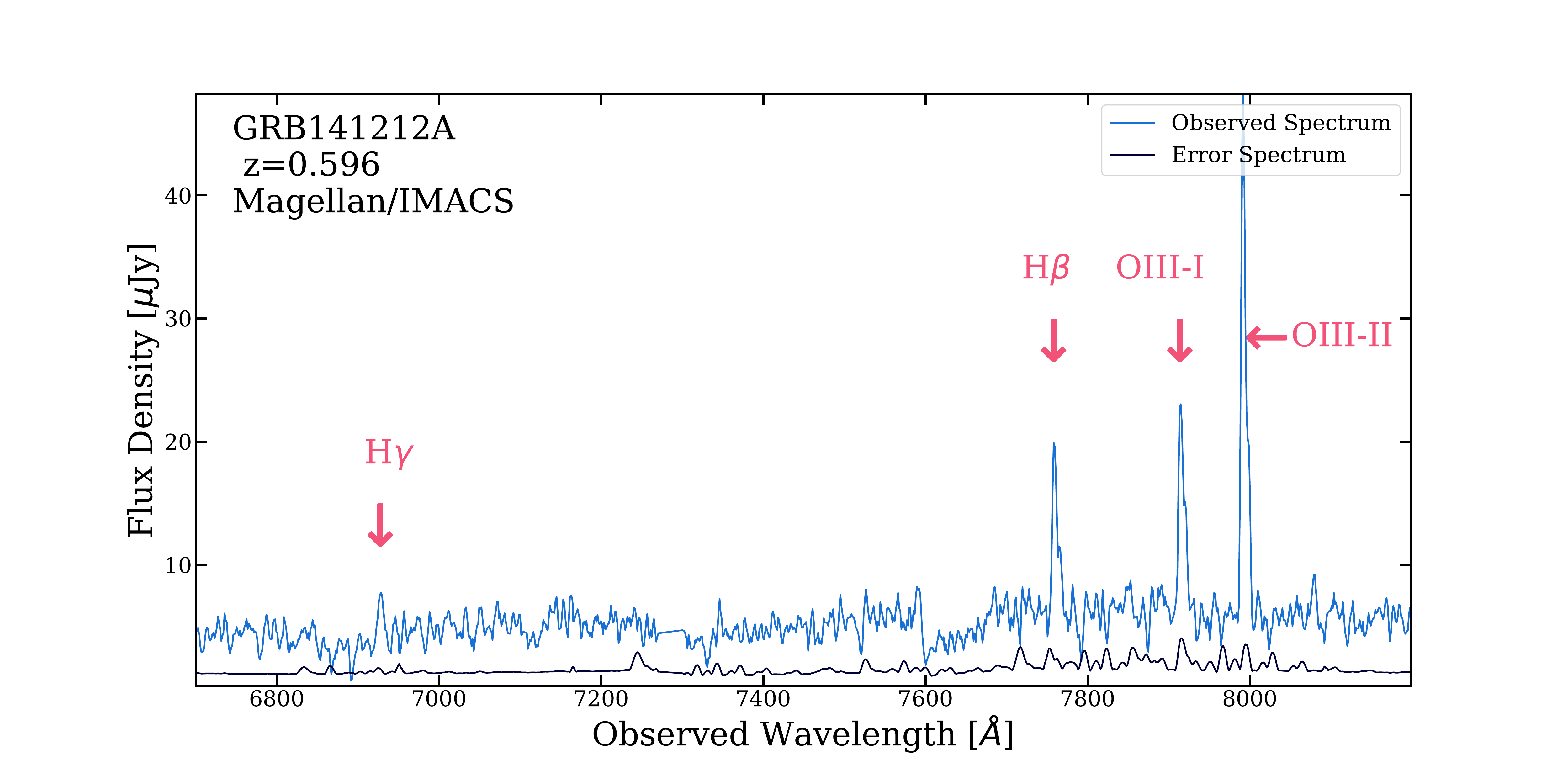}
\hspace{-0.5in}
\includegraphics[width=0.52\textwidth]{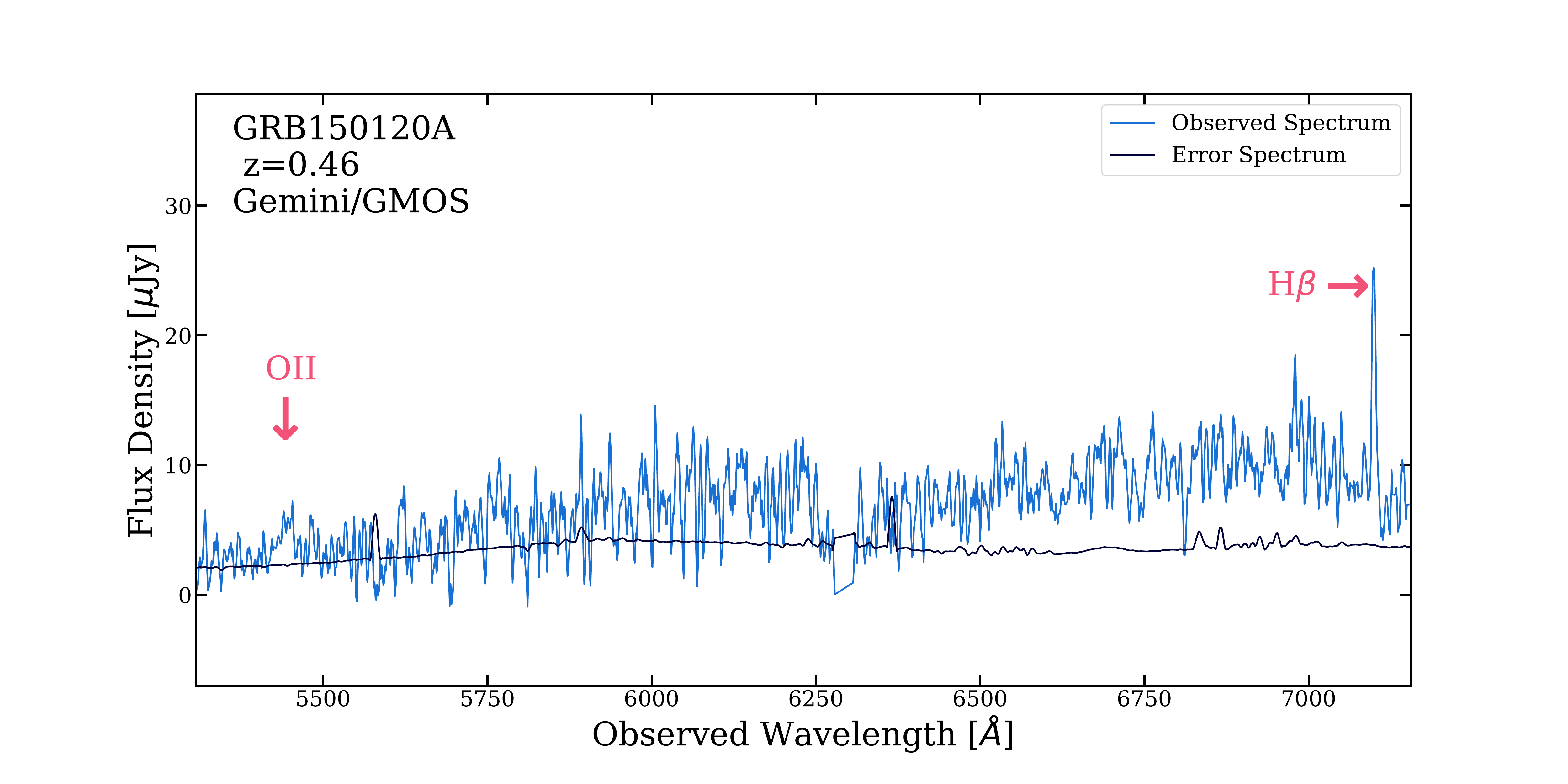}
\hspace{-0.5in}
\includegraphics[width=0.52\textwidth]{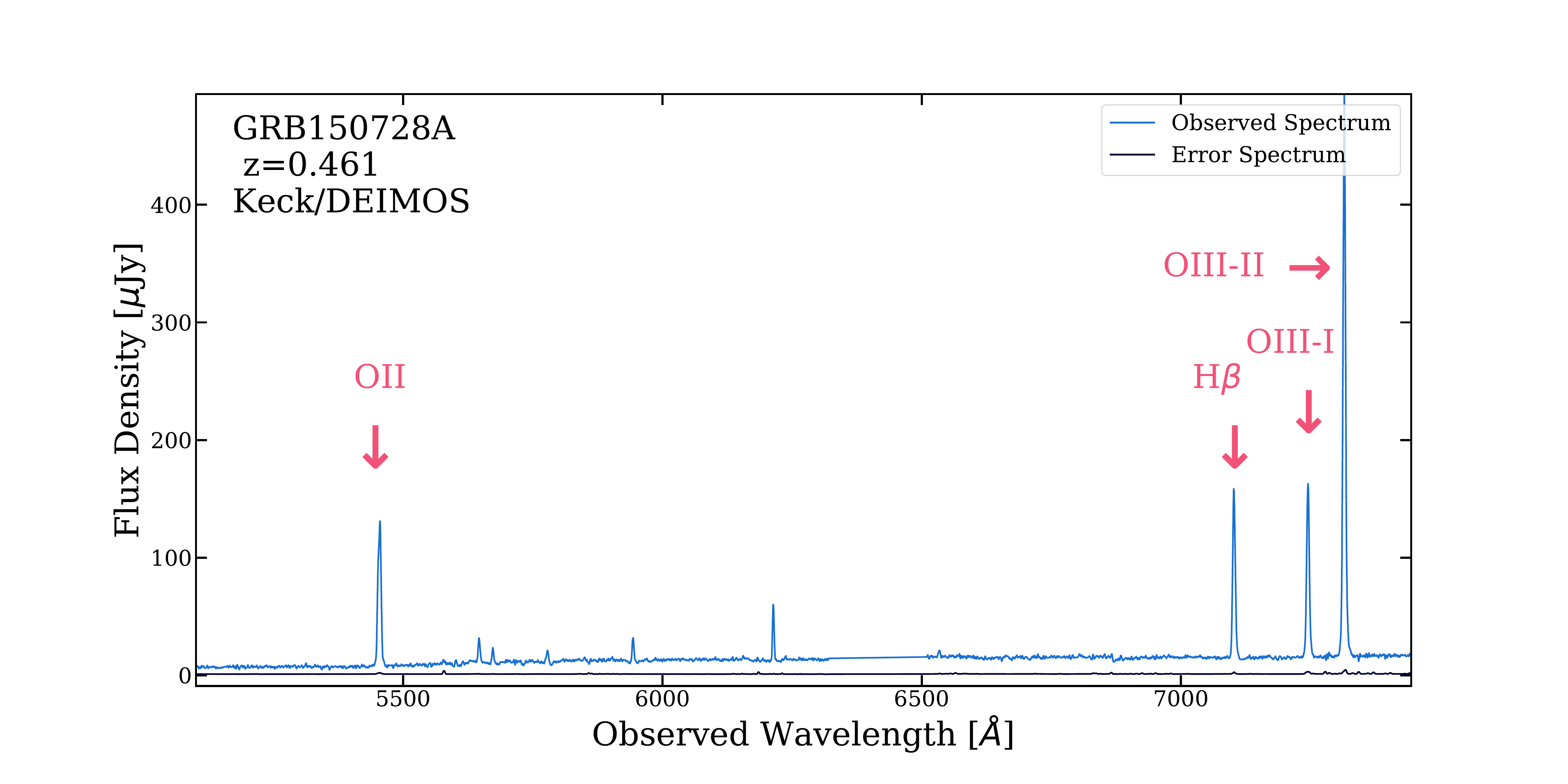}
\hspace{-0.5in}
\includegraphics[width=0.52\textwidth]{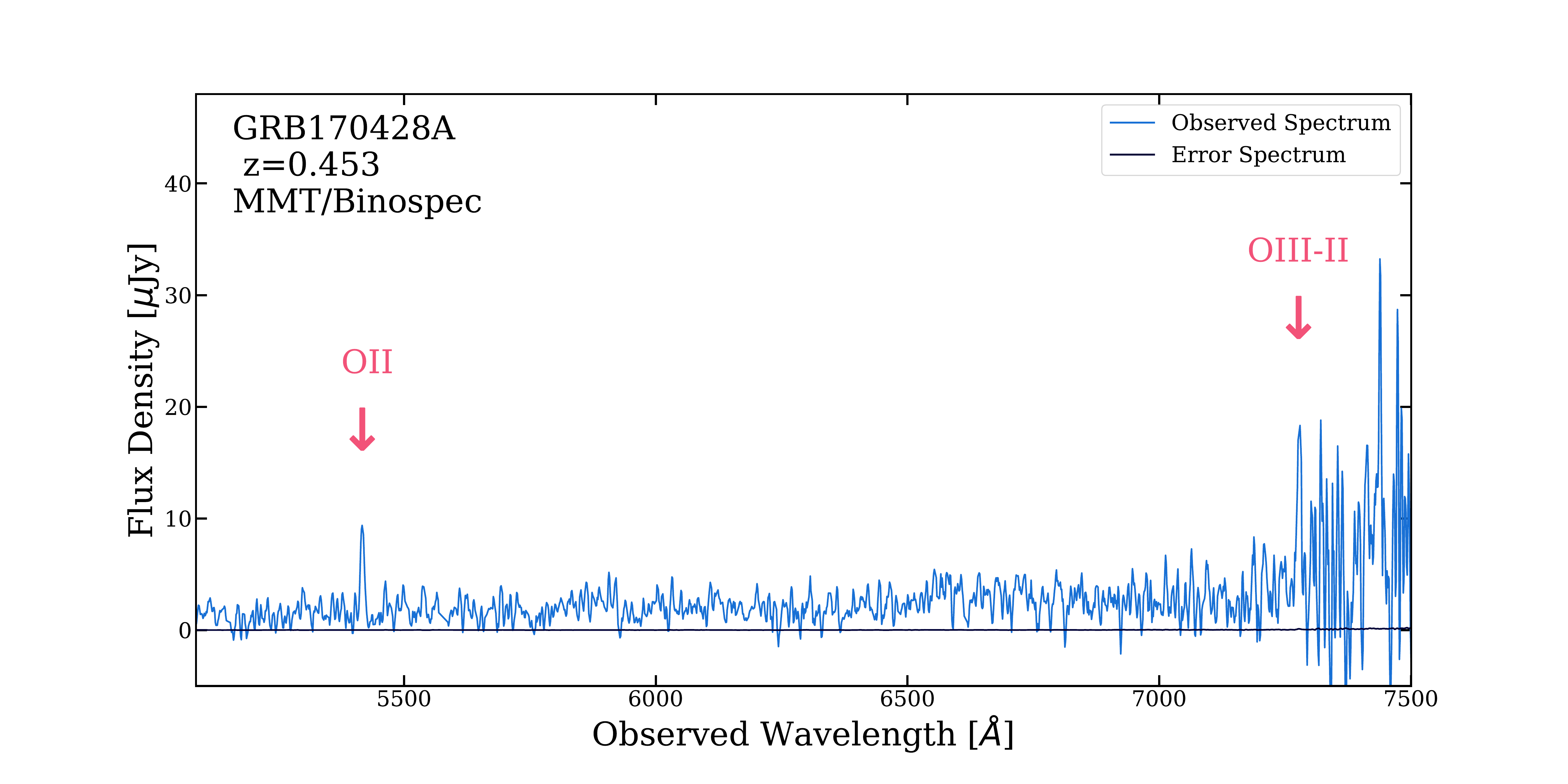}
\caption{Optical spectroscopy of 17 short GRB host galaxies that are newly presented in this work. In each panel, the spectral lines which enable redshift determination are denoted. Four additional new spectra have consistent results with previous works and thus are not shown.
\label{fig:specpanel}}
\end{figure*}

\addtocounter{figure}{-1}
\renewcommand{\thefigure}{\arabic{figure} (Cont.)}

\begin{figure*}
\centering
\includegraphics[width=0.52\textwidth]{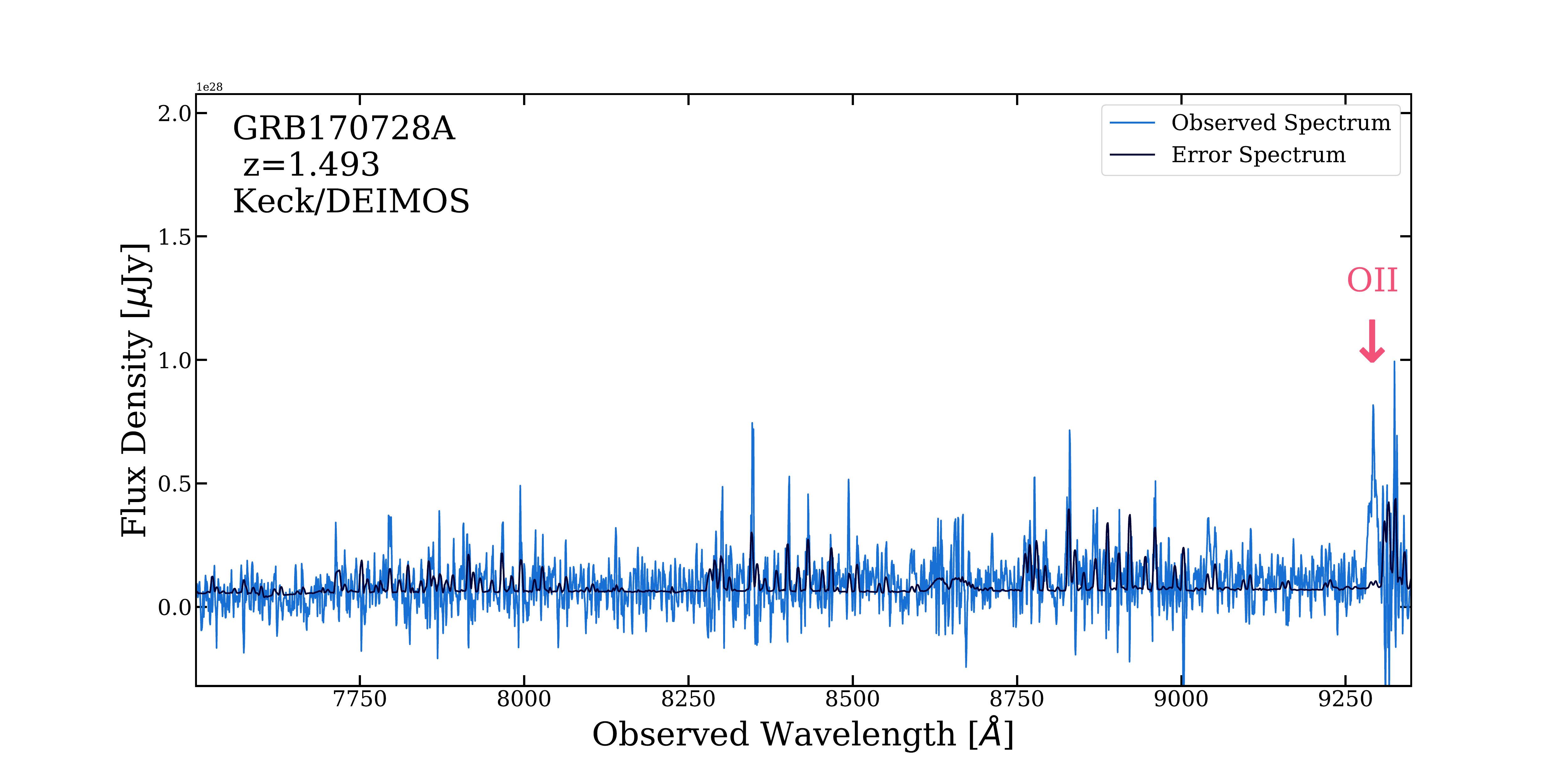}
\hspace{-0.5in}
\includegraphics[width=0.52\textwidth]{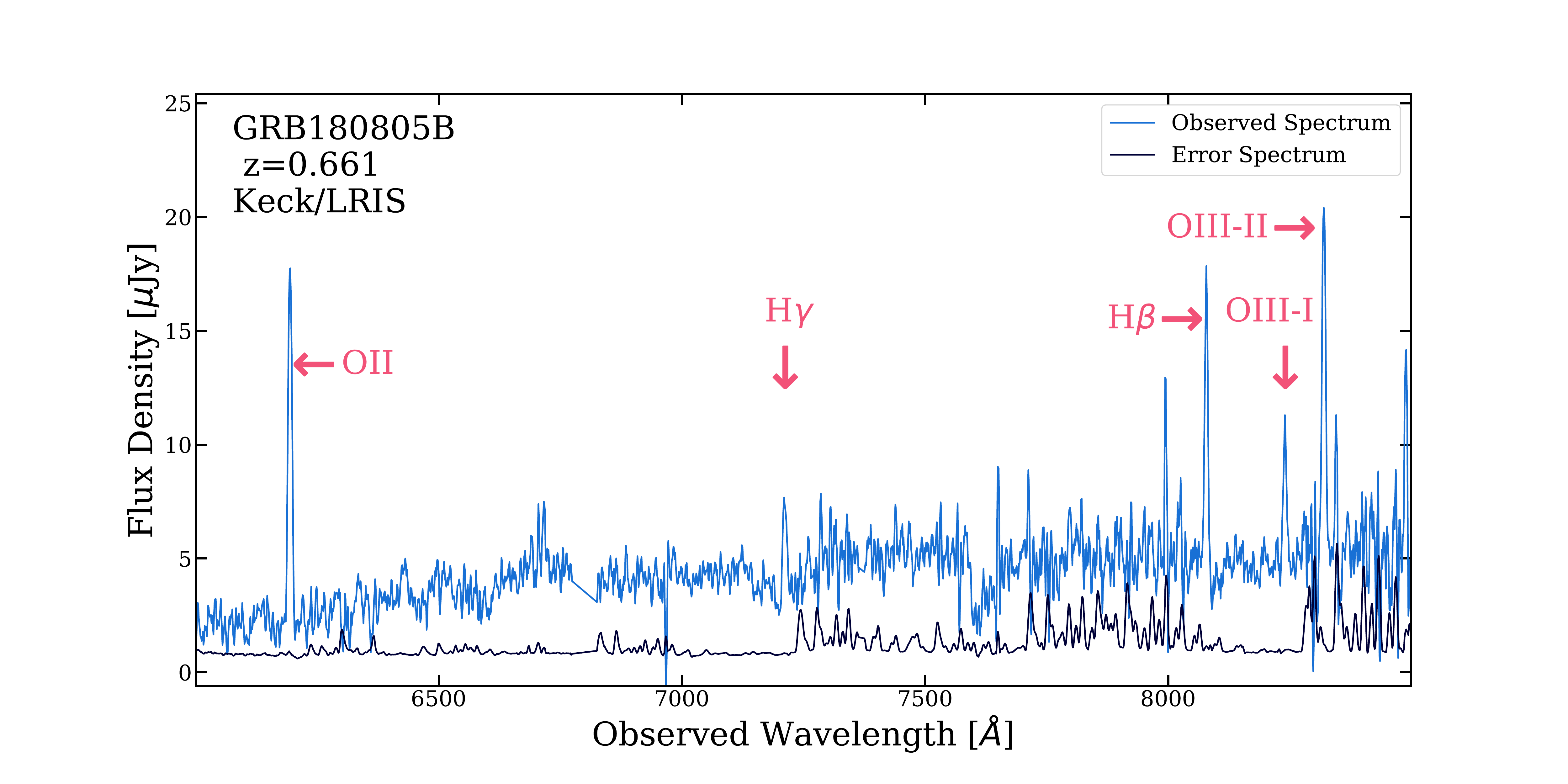}
\hspace{-0.5in}
\includegraphics[width=0.52\textwidth]{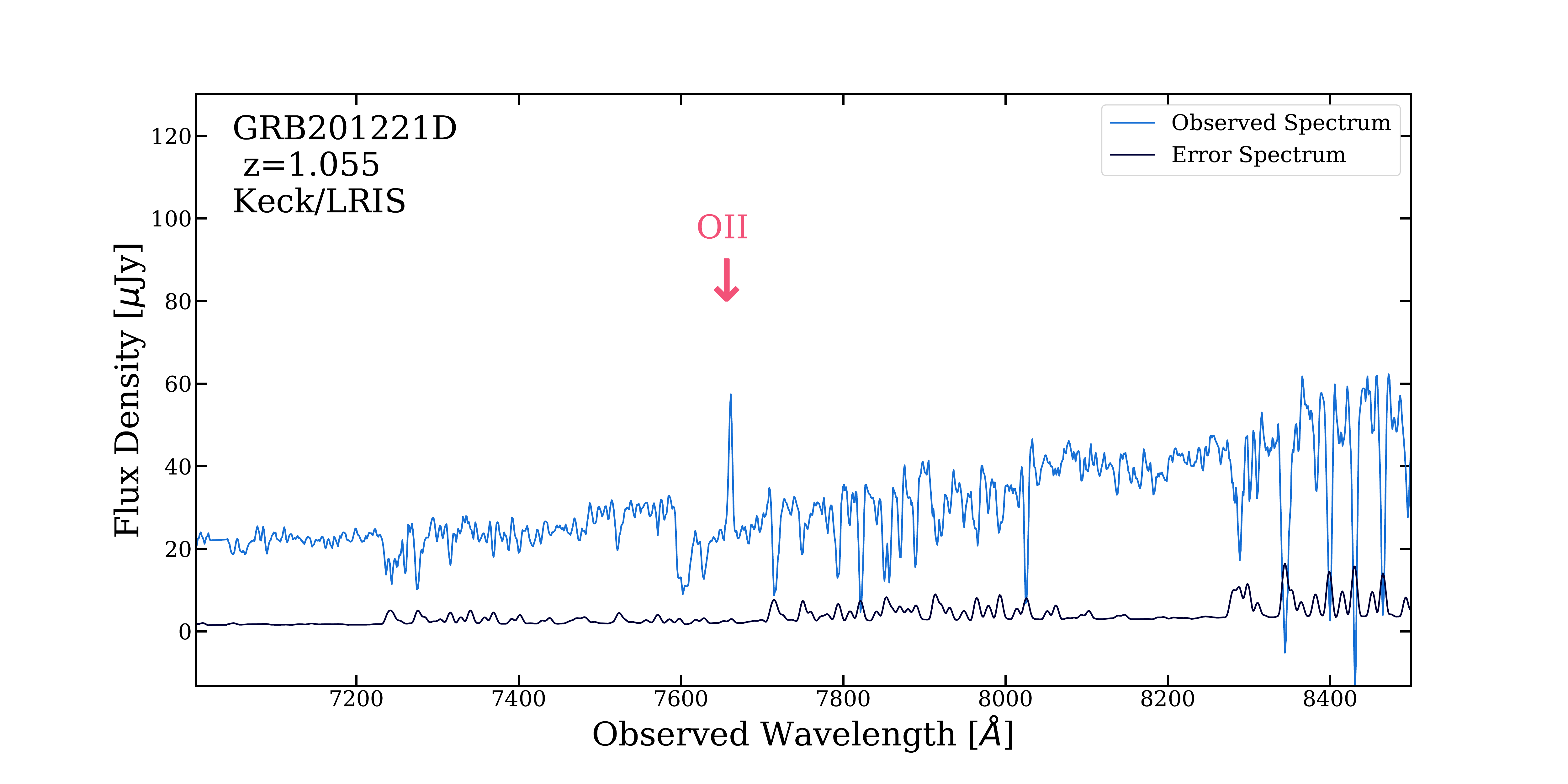}
\hspace{-0.5in}
\includegraphics[width=0.52\textwidth]{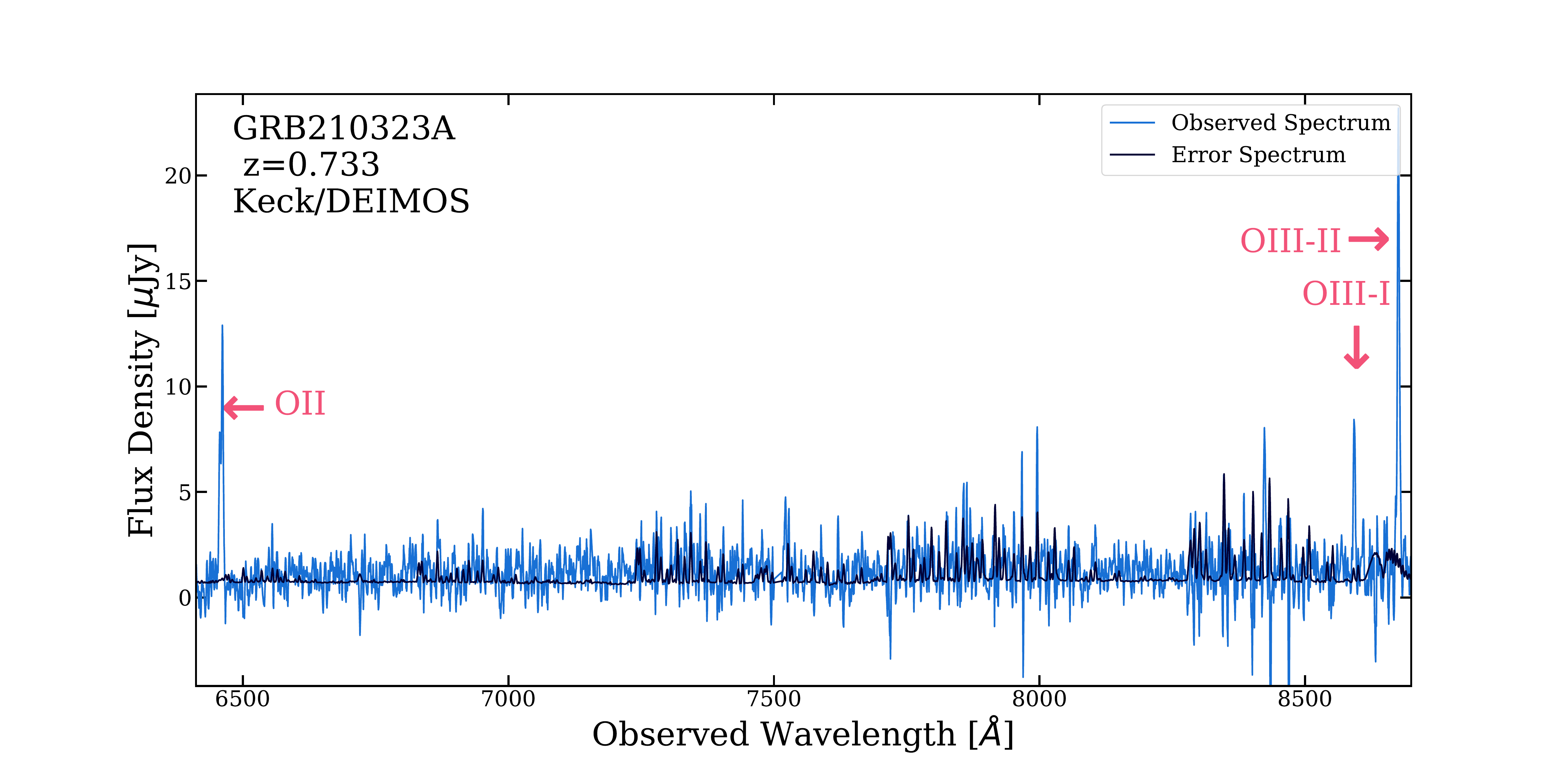}
\hspace{-0.5in}
\includegraphics[width=0.52\textwidth]{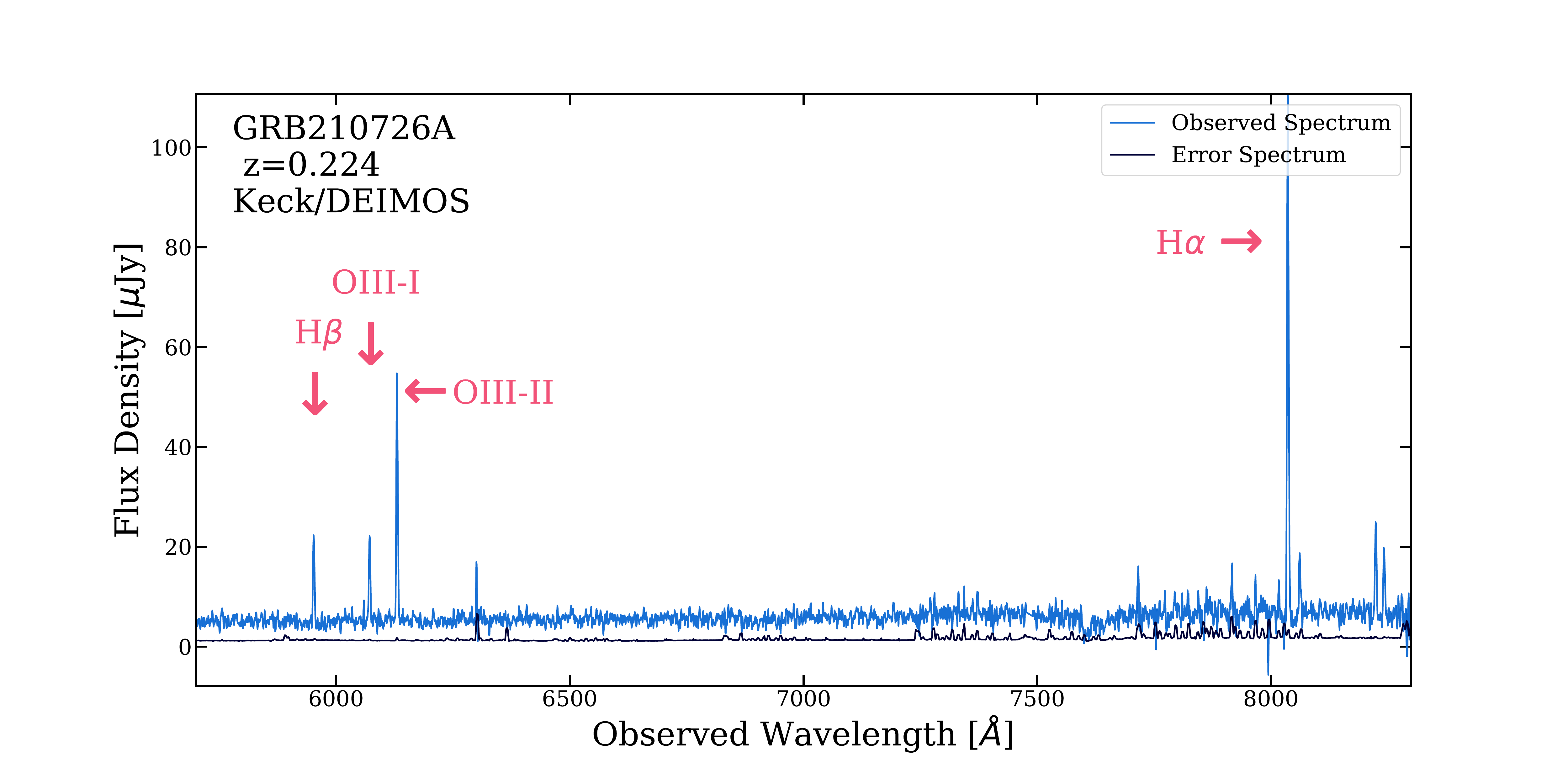}
\hspace{-0.5in}
\includegraphics[width=0.52\textwidth]{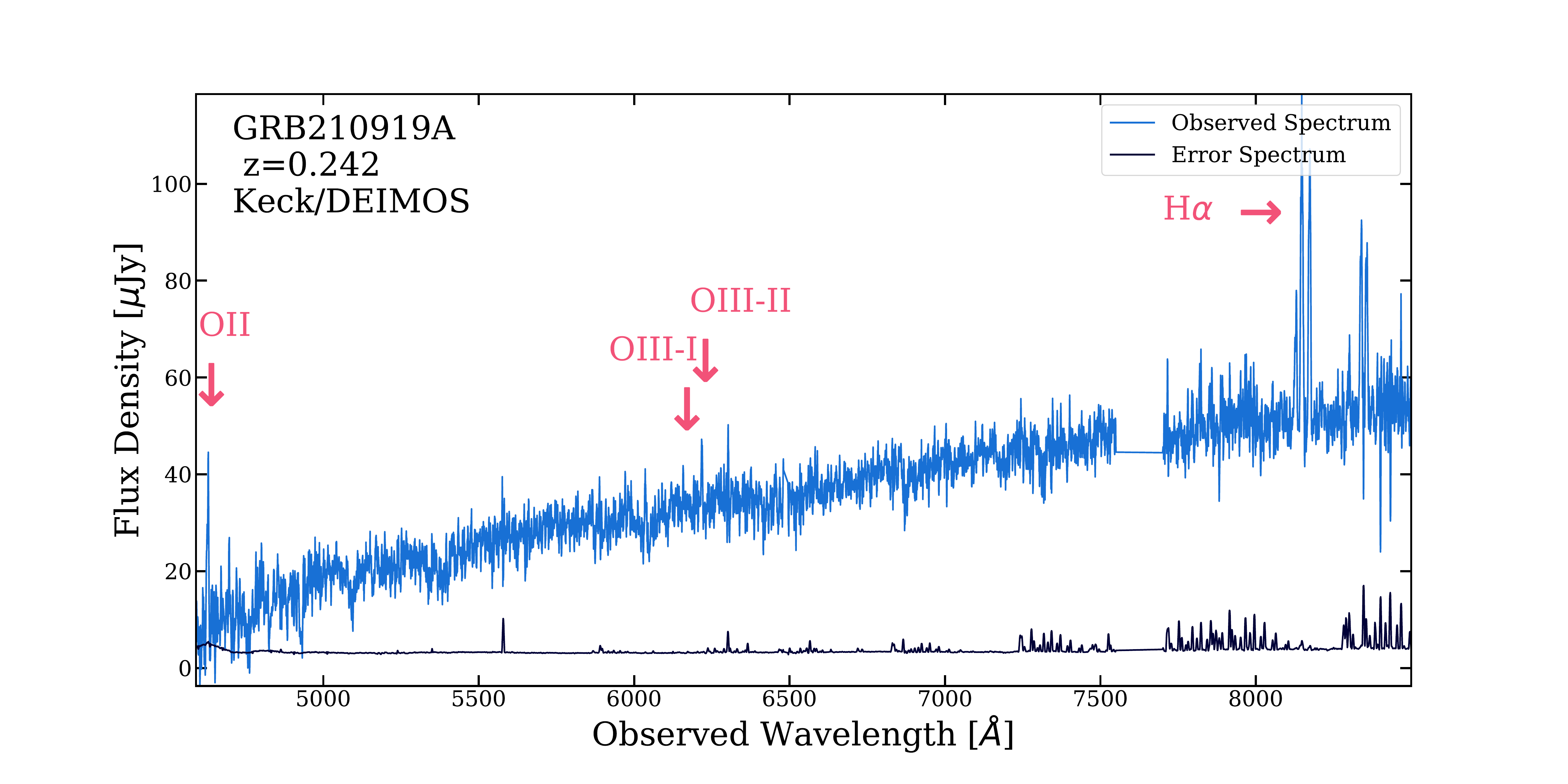}
\hspace{-0.5in}
\includegraphics[width=0.52\textwidth]{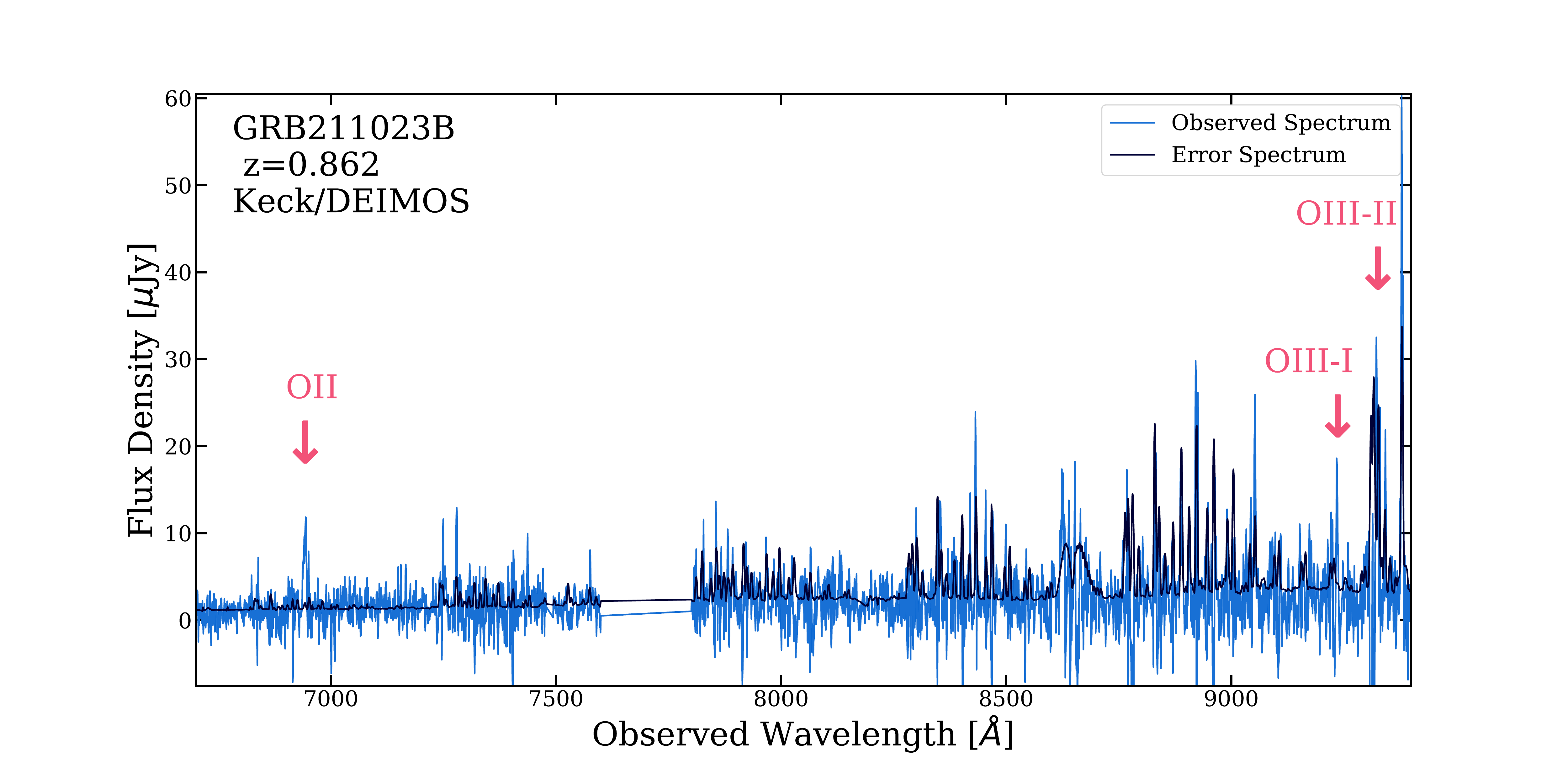}
\caption{Optical spectroscopy of 17 short GRB host galaxies that are newly presented in this work. In each panel, the spectral lines which enable redshift determination are denoted. Four additional new spectra have consistent results with previous works and thus are not shown.}
\end{figure*}

We use a combination of reduction tasks and methods depending on the origin of the spectra. For the subset of 10 host galaxies with spectra taken with LBT/MODS, Magellan (IMACS and LDSS), and Keck/LRIS (GRBs 101224A, 120305A, 130515A, 130822A, 140622A, 140903A, 140930B, 160411A, 160624A, and 201221D), we use standard {\tt IRAF} tasks in the \texttt{ccdred}, \texttt{longslit}, \texttt{immatch} packages to process and co-add the data \citep{iraf1, iraf2}. For each set of data, we subtract the overscan regions or apply bias corrections, apply flat-field corrections, model the sky background, and subtract this background from the individual frames. We co-add individual background-subtracted 2D frames and then use IRAF/{\tt apall} to extract the 1D spectra. We perform wavelength calibration using standard arc lamp spectra (HgNeArKrXe for MODS, HeNeAr for LDSS, NeArKrXe for IMACS, and HgNeArCdZn for LRIS). We apply spectrophotometric flux calibration using standard stars taken at a similar airmass on the same night in the same spectral set-up as the host spectra. We determine the error spectrum by performing the aforementioned reduction steps, but on the 2D spectra without sky subtraction. We perform standard error propagation in the combination. The spectra for six of these short GRB hosts are shown in Figure~\ref{fig:specpanel}.

For five hosts (GRBs 050709, 050724, 051221A, 060801, and 061006) with Gemini/GMOS (EEV detector) observations, we use the {\tt gemini/gmos} IRAF package. We apply bias subtraction, flat-field corrections, and model the sky background. We determine the wavelength solutions using CuAr arc lamps and calibrate the individual 2D science frames with the \texttt{gswavelength} and \texttt{gssubtract} tasks. We apply flux calibration to the spectra with a standard star taken within the same observing semester. We extract the 1D spectra with \texttt{gsextract} and the combine these with \texttt{gscombine}. The Gemini IRAF package propagates variance in traces through each task, which we use to determine the final error spectra.

For two hosts (GRB 170428A and 180618A), we obtain 1D, coadded, flux and wavelength calibrated spectra from the MMT/Binospec observatory products. This data was reduced with the instrument's spectroscopic reduction software, which is based in IDL\footnote{https://bitbucket.org/chil\_sai/binospec/wiki/Home}. The software automatically applies a flat-field and sky background correction. It uses a barycentric wavelength calibration and flux calibrates based on the spectrophotometric standard taken on the same night at a similar airmass. It extracts a 1D spectrum from co-added 2D frames, using a $1 \arcsec$ radius, and provides an uncertainty, which we use as the error spectrum.

Finally, for 19 hosts with data from Keck (DEIMOS and LRIS), Gemini/GMOS (E2V and Hamamatsu detectors), and VLT/FORS2, we used the Python Spectroscopic Data Reduction Pipeline ({\tt PypeIt}; \citealt{PypeIt}) for data processing and spectral extraction. These hosts are GRBs\,050509B, 060614, 070429B, 070714B, 070724A, 090510, 140129B, 141212A, 150120A, 150728A,  151229A, 160821B, 170728A, 180805B, 191031D, 210323A, 210726A, and 210919A, and 211023B (Table~\ref{tab:specobs}). In {\tt PypeIt}, we apply an overscan and/or bias subraction, flat-field correction and perform wavelength calibration and spectral extraction (using the {\tt boxcar} method with a $1.5-2.5\arcsec$ radius, in order to include all of the emission line flux). We apply flux calibrations using appropriate spectrophotometric standards.  We co-add the flux-calibrated 1D spectra and apply a telluric correction using an atmospheric model. {\tt PypeIt} determines the variance on each trace, which we use to determine the error spectra.

For all 25 new and unpublished spectra in our sample as well as the 10 re-reduced spectra, we use the \citet{ccg+1989} extinction law and the $A_V$ in the direction of each burst \citep{sf11} to correct for Galactic extinction. We then normalize the host spectra to their extinction-corrected photometry. The final 1D spectra of 17 hosts with determined redshifts and their spectral line identifications are displayed in Figure~\ref{fig:specpanel}. Not shown are the four new spectra which have consistent results with previous works.

\subsection{Feature Identification and Redshift Determinations}

For 21 hosts of the 25 hosts with new, unpublished spectroscopic observations, we determine redshifts through feature identification. The most common features in our spectra are [O~II]$\lambda$3727, H$\beta$, [O~III]$\lambda$4959, [O~III]$\lambda$5007, and H$\alpha$ for star-forming galaxies, and the Ca~II H\&K for quiescent or transitioning galaxies. We search for high S/N spectral lines (S/N > 5) with Gaussian like structures in both the 1D and 2D frames. When multiple spectral lines are found, we use the mean of the Gaussian lines to determine ratios between each pair of lines. We compare these ratios to those of the rest-frame wavelengths of spectral lines at redshifts between $0 \leq z \leq 3.0$. We require that the ratios of observed lines are within 0.1\% of the rest-frame spectral line ratios to maintain accuracy in line determination. From there, we can determine what each observed line is and the redshift. We determine error on the redshift by fitting the spectral lines and their direct background ($\sim \pm 100$ \AA) with a Gaussian profile and determining the mean of $1\sigma$ uncertainties on each line.

For 17 hosts, we have at least two spectral lines with S/N $> 5$ above the continuum from which we can determine a common redshift. Out of these 17, 7 are completely new redshifts, unconfirmed in GCNs or other works. The redshifts of GRBs 101224A, 130822A, 140622A, 170428A, and 180805B are consistent with those reported in \citet{otd+22}. A few redshifts were reported in GCNs and are also consistent with our findings: GRBs 141212A \citep{GCN141212A_redshift} and 210919A \citep{GCN210919A_redshift}. The GCN redshift of 211023B \citep{GCN211023B_redshift} is slightly inconsistent with our result and only based on one detected emission line; thus we consider this a new redshift. 

For three hosts (GRBs 140930B, 170728A, and 201221D), we can only identify a single emission line. However, in all cases the width of this line is double-peaked in nature, suggesting it is likely a doublet, specifically the [O~II]$\lambda\lambda$3727, 3729 doublet. In addition, these lines were all found at higher wavelengths ($> 9100$\AA for GRBs 140930B and 170728A, and $\approx 7659$\AA for GRB 201221D).  If these lines were instead H$\beta$ or [O~III], we would expect to detect the [O~II], H$\beta$, and [O~III] in all three spectra, or if these lines were H$\alpha$, we would be able to detect all the common emission lines in the spectra of GRBs 140930B and 201221D, and everything but [O~II] in the spectrum of 170728A. Given that these additional lines are not detected, despite the wavelength coverage of these spectra, it is most likely that the identified lines are [O~II].  Therefore, in all three cases, we identify the line as the [O~II]$\lambda$3727 doublet. GRB\,140930B and 170728A do not have published spectroscopy or redshifts so we report these redshifts for the first time. Our redshift of $z=1.055 \pm 0.001$ is close, although not formally consistent with, the previously reported redshift in the GCNs ($z = 1.045$; \citealt{GCN201221D_redshift}). We present all spectroscopic redshifts and uncertainties in Table \ref{tab:sample}. 

For short GRBs with photometric data only, we use the stellar population inference code {\tt Prospector} \citep{bdj21, Leja_2017} to model their SEDs and determine their photometric redshifts. The full stellar population modeling methods and analysis are described in \citet{BRIGHT-II}. We report 20 new photometric redshifts (see \citealt{BRIGHT-II}, Appendix for fits). We note that our photometric redshifts for 5 GRBs (070729, 120804A, 151229A, 191031D, and 200411A) differ from the literature, due to a combination of modeling assumptions (such as truncated redshift priors in other works), less complete data sets, or incorrect host associations. Combined with the literature sample, we find a median redshift for the full population of $0.6$ with a 68\% credible interval on the distribution of [-0.25,+0.90], and a higher median and credible interval of $1.08$ [-0.61,+1.0] for the photometric redshift population; this is discussed and explored in more detail in \citet{BRIGHT-II}.

\renewcommand{\thefigure}{\arabic{figure}}
\begin{figure*}[t]
\centering
\includegraphics[width=0.45\textwidth]{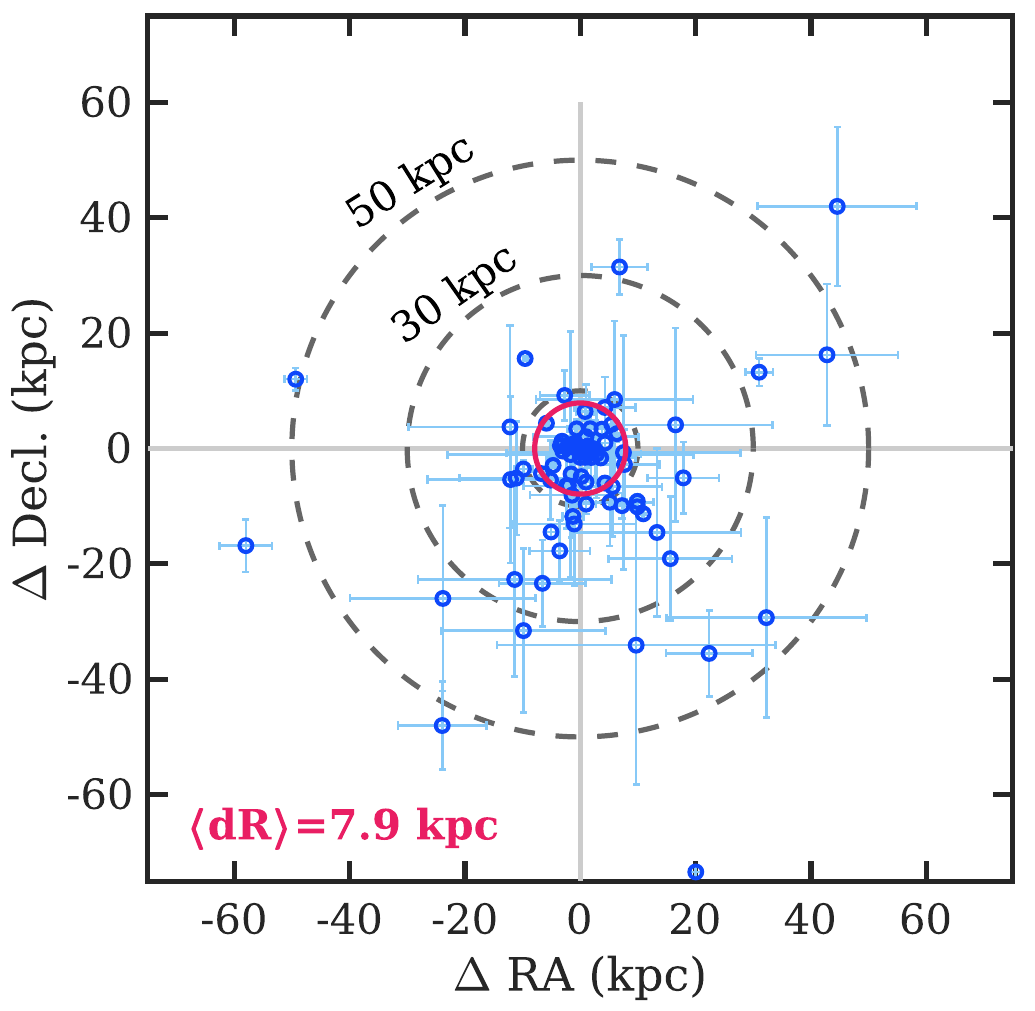}
\includegraphics[width=0.45\textwidth]{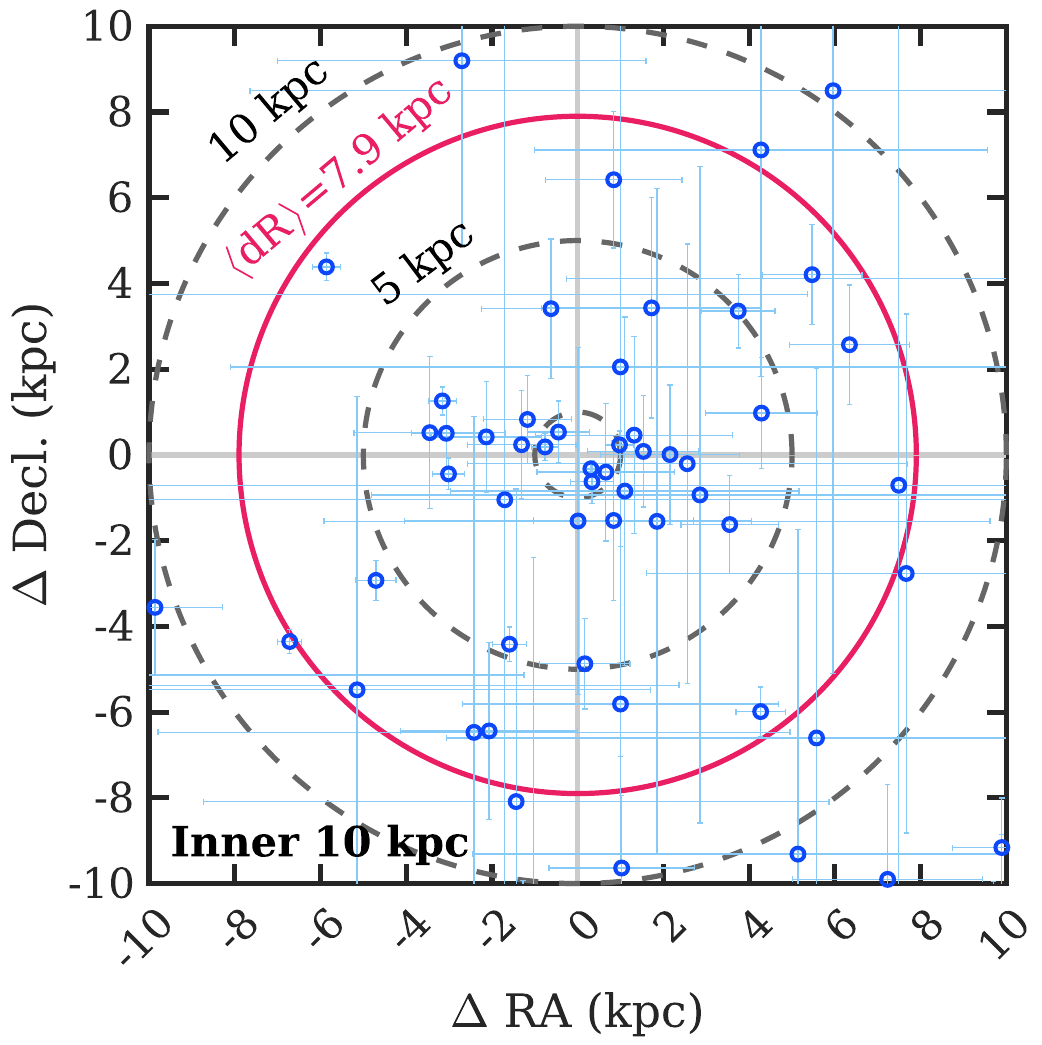}
\caption{{\it Left:} The locations of \totoffset\ short GRBs with respect to their host galaxy centers (represented by the origin), where uncertainties on individual measurements correspond to $1\sigma$ confidence. The axes are oriented with North up and East to the left, and the scale is in the frame of the host galaxy (in which negative values correspond to East and/or South of the galaxy). Concentric dashed circles denote 10, 30, and 50 kpc offsets, while the red circle denotes the median on the full distribution of $7.9$~kpc. {\it Right:} Same as left panel, but for only the inner 10~kpc from the host galaxy centers. The panels demonstrate that while most short GRBs reside at $\lesssim 10$~kpc from their host galaxies, a significant fraction lie outside of this galactocentric radius.
\label{fig:offset_circular}}
\end{figure*}

\section{Galactocentric Offsets}
\label{sec:offsets}

\subsection{Angular, Physical, and Host-Normalized Offsets}

With imaging, host identifications, and redshifts in-hand, we now turn to the locations of short GRBs with respect to their hosts. In the context of their NS mergr progenitors, offsets are an observable diagnostic for a combination of progenitor kicks and delay times (e.g., \citealt{zkn+20}. To determine the position of each GRB relative to its host galaxy, and thus measure precise offsets, we perform relative astrometry by aligning each of the afterglow discovery images to the host galaxy imaging. We consider three sources of uncertainty in the offsets: the afterglow centroid ($\sigma_{\rm GRB}$), the astrometric tie uncertainty between the afterglow discovery and the host images ($\sigma_{\rm GRB \rightarrow host}$), and the host galaxy centroid uncertainty ($\sigma_{\rm host}$). We perform the astrometric tie in the same manner as described in Section~\ref{sec:astrometry}. To determine the host centroid uncertainty, we again use {\tt SExtractor}, and find a range of values, $\sigma_{\rm gal} \approx 1-50$~mas. This is generally the smallest source of uncertainty.

For each galaxy/filter combination, we first use the afterglow and host position to measure angular offsets ($\delta R$). The offsets and accompanying combined uncertainties are listed in Table~\ref{tab:offsets}. To convert to physical offsets, we use the redshifts in Table~\ref{tab:sample}. The values of angular and physical offsets for \totoffset\ short GRBs are listed in Table~\ref{tab:offsets}. For bursts with unknown redshift or no redshift information, we assume $z=1$. While the median redshift for the entire population is lower ($z\approx 0.6$), we assume that host galaxies which lack redshift information are at higher redshifts than the median. We also note that the angular diameter distance at $z \gtrsim 0.5$ is relatively flat, so the exact choice of redshift beyond this value will not have a large effect on the physical offset distribution.

The high angular resolution of {\it HST} data enables us to calculate the effective radii, $r_e$, and thus host-normalized offsets, which we determine from surface brightness profile fitting. When given the choice, we select the filter which corresponds to the rest-frame optical band of the host, as there can be small size differences between filters. We use the IRAF/{\tt ellipse} routine to generate elliptical intensity isophotes and construct one-dimensional radial surface brightness profiles for the most probable host galaxy for each burst. For each observation, we allow the center, ellipticity, and position angle of each isophote to vary. Using a $\chi^2$-minimization grid search, we fit each profile with a S\'{e}rsic model with three free parameters: S\'{e}rsic index $n$ \citep{cb99}, the effective radius ($r_e$, also known as the half-light radius), and the effective surface brightness ($\Sigma_e$). A single S\'{e}rsic component provides an adequate fit ($\chi^{2}_{\nu} \approx 0.4-1.5$) for most of the host galaxies. We perform this analysis for the hosts of 10 short GRBs with {\it HST} data that do not already have half-light radii measurements determined in this same manner.

Finally, we compile offset measurements for 32 short GRBs with ground-based or {\it HST} data \citep{bpc+07,fbf10,nkg+12,fb13,fmc+16,nfd+20,pfn+20,flr+21,rfv+21,GRB211106A-Laskar,rgl+22}. We re-calculate the physical and host-normalized offsets and uncertainties using the best-fit redshifts and same cosmological parameters as used in this work. We also assume $z=1$ for bursts with unknown redshift, and use updated, enhanced XRT positions for bursts in which the most precise position comes from the X-ray afterglow \citep{gtb+07,ebp+09}. These combined corrections result in minor revisions to the originally published values and are listed in Table~\ref{tab:offsets}. 

We show all of the physical offsets of \totoffset\ short GRBs in Figure~\ref{fig:offset_circular}, with each short GRB's host galaxy center represented by the origin. While the majority of short GRBs occur at $\lesssim 10$~kpc and drive the median of $\approx 7.9$~kpc (Section~\ref{sec:offset_distribution}), a substantial fraction occur outside of this galactocentric radius.

\subsection{Offset Distributions}
\label{sec:offset_distribution}

\begin{figure*}[t]
\centering
\includegraphics[width=0.45\textwidth]{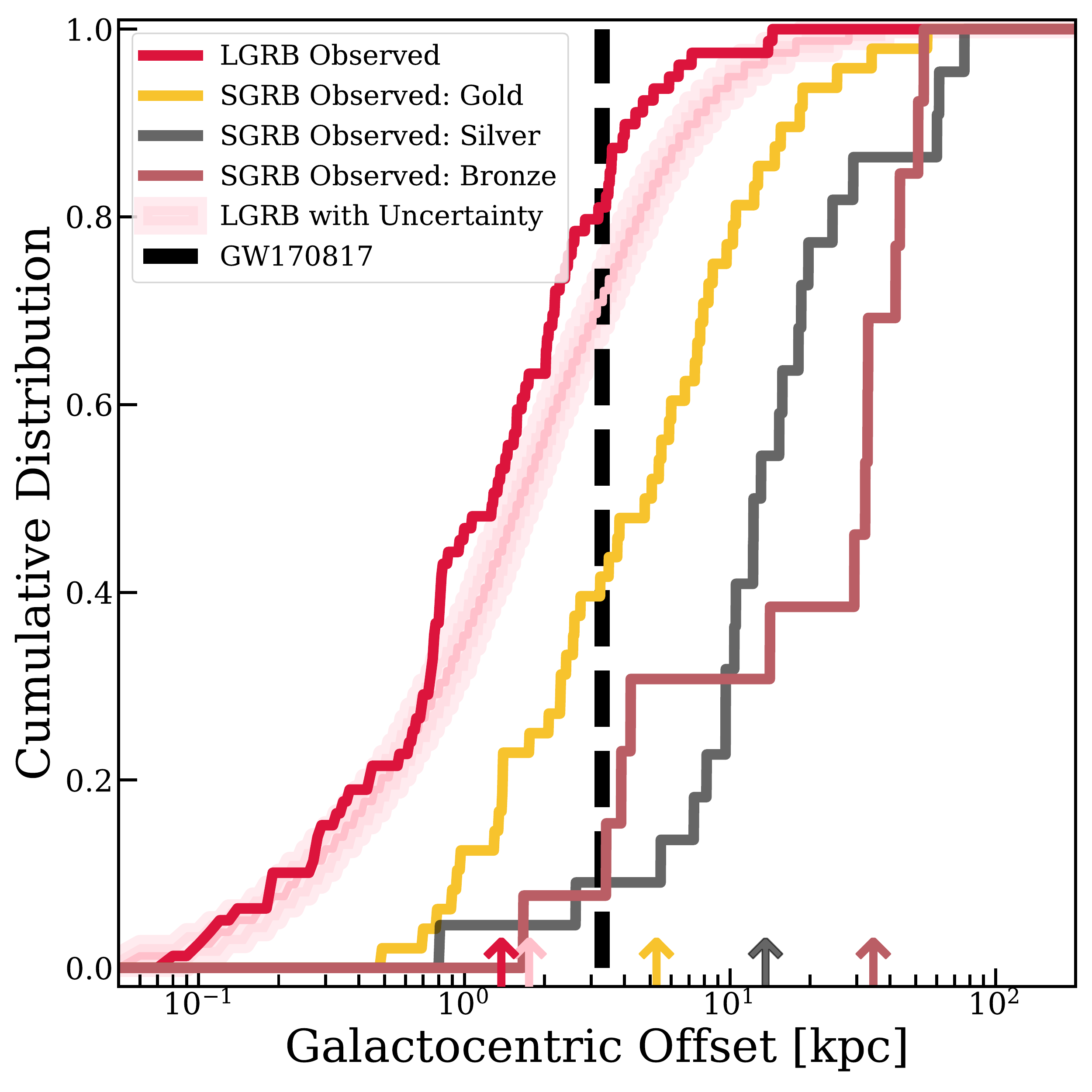}
\includegraphics[width=0.49\textwidth]{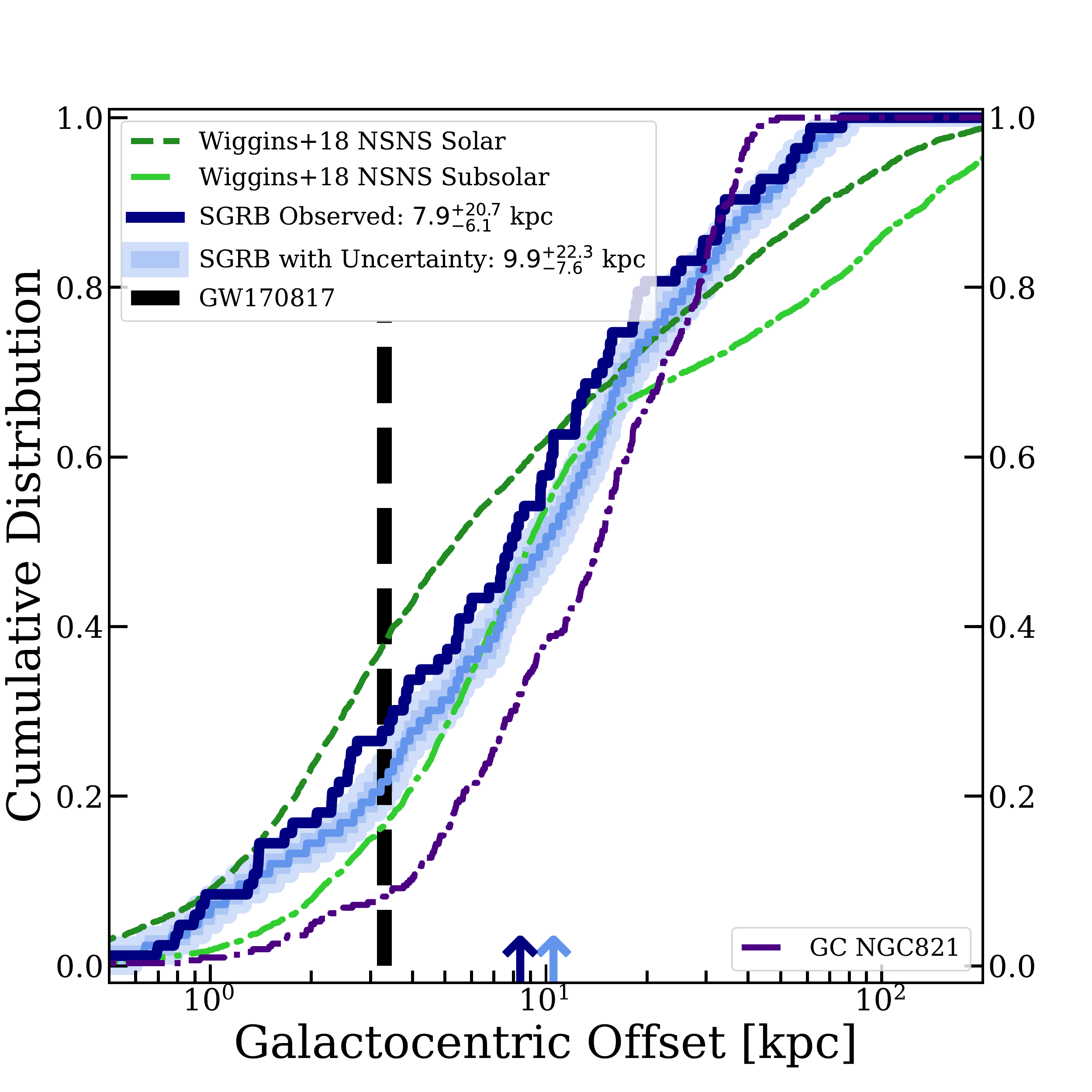}
\caption{{\it Left:} The observed physical offset distribution of short GRBs, divided into the Gold, Silver and Bronze samples; arrows from the bottom denote medians of each distribution. As expected, the bursts with the most tenuous associations (Bronze sample) are farther from their hosts than the Gold and Silver samples. Also shown are the observed offsets of long GRBs and the sampled distributions taking into account uncertainties (red; \citealt{bbf16}). {\it Right:} The observed physical offset distribution of short GRBs (navy blue) and the sampled distributions taking into account measurement uncertainties (shaded light blue). A comparison to representative NS-NS merger models for differing metallicities (dotted and dash-dotted green lines; \citealt{wfs+18}) shows that the observed and model distributions are overall consistent, although there is a relative dearth of observed high-offset short GRBs. Also shown is the observed distribution of globular clusters in the elliptical galaxy NGC821 (dash-dotted black line; \citealt{sfs+08}); overall, short GRBs are clearly not as extended as this population.
\label{fig:offset_physical}}
\end{figure*}

\begin{figure*}
\centering
\includegraphics[width=0.45\textwidth]{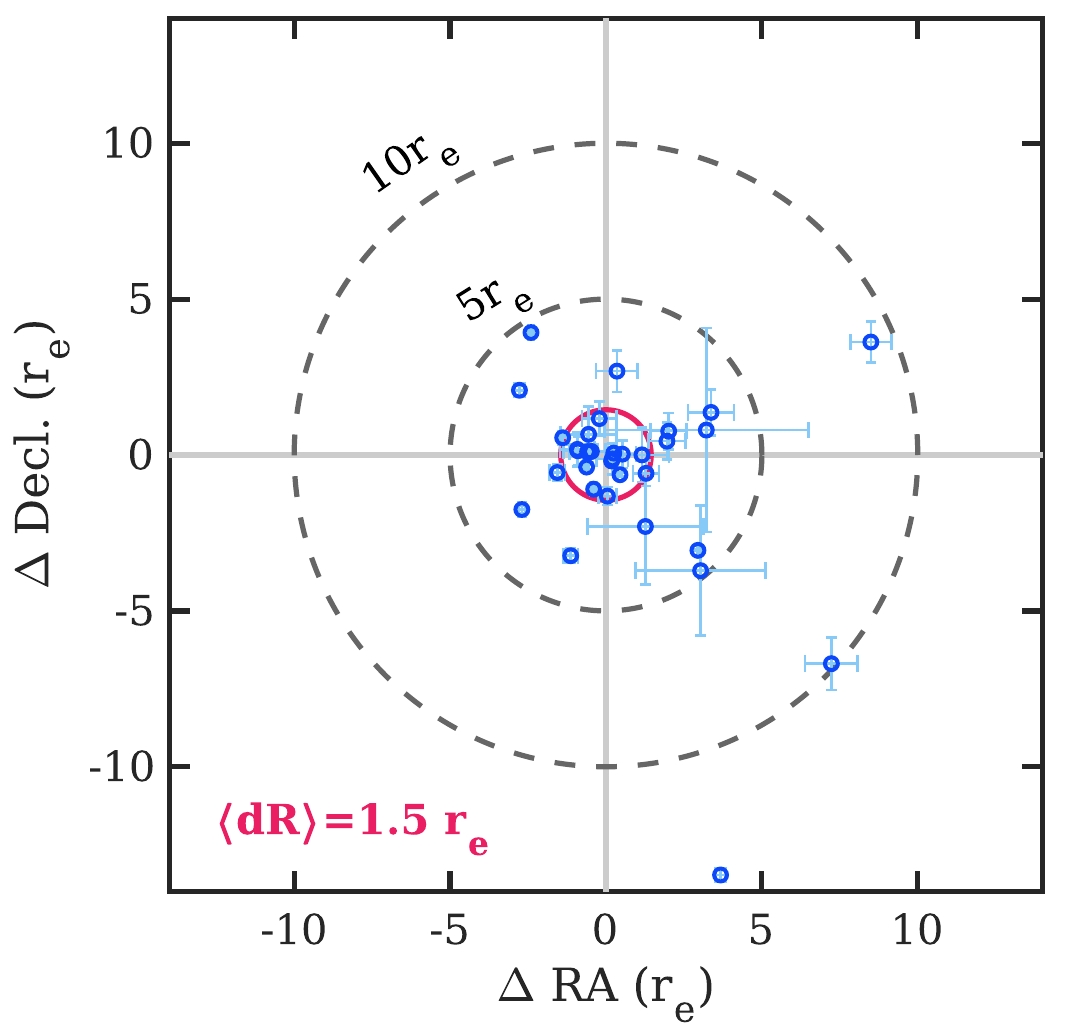}
\includegraphics[width=0.45\textwidth]{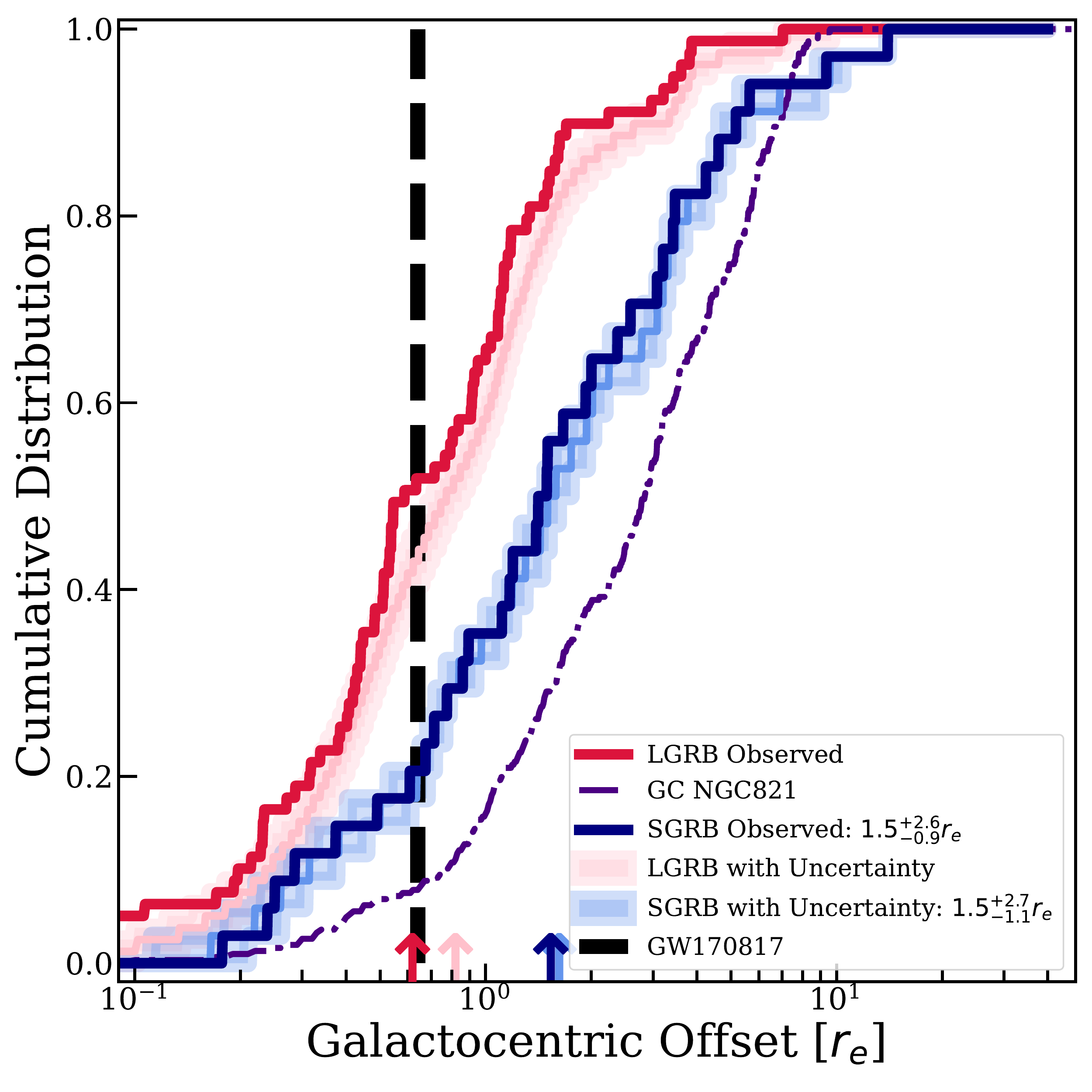}
\caption{{\it Left:} The host-normalized offsets of 34 short GRBs with respect to their host galaxy centers (represented by the origin), where uncertainties on individual measurements correspond to $1\sigma$ confidence. The axes are oriented with North up and East to the left, and the scale is in the frame of the host galaxy (in which negative values correspond to East and/or South of the galaxy). Concentric dashed circles denote $1r_e$, $5r_e$, and $10r_e$ offsets, while the red circle denotes the median on the full distribution of $1.5r_e$. {\it Right:} The observed distribution of host-normalized offsets for short GRBs (blue solid line) and the sampled distributions with uncertainties (blue shaded region). Also shown are the corresponding distributions of long GRBs (red; \citealt{bbf16}) and globular clusters in the elliptical galaxy NGC821 (\citealt{sfs+08}; dash-dotted black line). Overall, short GRBs are farther from their hosts than long GRBs but closer than globular clusters, even when normalized by their host galaxy sizes.
\label{fig:offset_hostnorm}}
\end{figure*}

We now combine the distribution for \totoffset\ short GRBs, which are all events for which offset measurements are available. This sample includes 34 short GRBs which have enough information for host-normalized offsets. Due to the inclusion of both XRT-localized and sub-arcsecond localized bursts, the measurement uncertainties vary significantly across the population and have an impact on the offset distribution. Thus, it is imperative to take these uncertainties into account in the final distribution. Indeed, driven largely by the few arcsecond-size XRT positions, 15 short GRBs are in the regime in which the measurement uncertainty is larger than the offset.

To account for these uncertainties and the fact that offsets must be a positive definite value, we use the Rice distribution for each short GRB given their offset and $1\sigma$ uncertainty. We randomly sample from the resulting distributions 500 times, using the method described in \citet{bbf16}. In particular, we use Equation~2 in \citet{bbf16} for the Rice distribution, and note that when $\delta R/\sigma_{\delta R} > 5$, the Rice distribution is the same as a Gaussian distribution. We then build 500 cumulative distribution functions (CDF) and compare against the observed offsets. Since the CDFs are built from the random samples, the sampled distributions overall have larger physical offsets than the observed distribution since they are driven by the larger uncertainties. We show the distributions of angular, physical and host-normalized offsets in Figures~\ref{fig:offset_physical}-\ref{fig:offset_hostnorm}.

For the observed distribution, we calculate a median of $1.24 \arcsec$ [$-0.97\arcsec, +3.29\arcsec$] (16th and 84th percentiles on the full distribution) or $1.50\arcsec$ [$-1.15\arcsec, +3.94\arcsec$] for the sampled distributions. For the physical offset distribution, we find an observed median of $7.92$~kpc with an interval on the full distribution of [-6.13, +20.71] kpc, or $9.82$ [-7.50, +22.86]~kpc for the sampled distributions. We note that the physical offset distribution includes 13~events with an assumed $z=1$, although the distribution and median minimally change when excluding these events. We find that the short GRBs in the Bronze sample are overall farther from their host galaxies than the Gold or Silver samples (Figure~\ref{fig:offset_physical}); this is to be expected given their less robust associations. Thus, the Gold sample median of $\approx 4.92$~kpc can be interpreted as a minimum on the short GRB median offset.

Finally, for 34 short GRBs with effective radii measurements, we find medians of $1.45r_e$ [$-0.93r_e, +2.57r_e$] ($1.54r_e$ [$-1.06r_e, +2.71r_e$]) for the observed (sampled) distributions. Overall, we find that the observed physical offset distribution here is more extended than determined in previous literature, with a median that is $\approx 2-3$~kpc higher \citep{fbf10,fb13,otd+22}.

We compare the observed distributions of short GRBs to those of long GRBs \citep{bbf16}, the predicted distributions of field BNS mergers for solar and sub-solar metallicities (\citealt{wfs+18}; Figure~\ref{fig:offset_physical}), and the observed distributions of globular clusters in the elliptical galaxy NGC821 (\citealt{sfs+08}; Figures~\ref{fig:offset_physical}-\ref{fig:offset_hostnorm}). Overall, short GRBs occur at larger offsets than long GRBs, are in reasonable agreement with the expected locations of BNS mergers, and are occur closer to their host galaxies than the observed distributions of globular clusters, in terms of both physical and host-normalized offsets. We further explore the relationship between offset and host galaxy type in \citet{BRIGHT-II}.

\startlongtable
\tabletypesize{\small}
\begin{deluxetable*}{lccccccccc}
\tablecolumns{9}
\tablewidth{0pc}
\tablecaption{Short GRB Angular, Physical, and Host-Normalized Offsets
\label{tab:offsets}}
\tablehead {
\colhead {GRB}		&
\colhead {$z$}		&
\colhead {Offset}	&
\colhead {$\sigma$}	&
\colhead {Offset$^a$}	&
\colhead {$\sigma$}	&
\colhead {Offset}	&
\colhead {$\sigma$}	&
\colhead {Reference$^b$} \\
\colhead {}		&
\colhead {}		&
\colhead {($''$)}	&
\colhead {($''$)}	&
\colhead {(kpc)}        &
\colhead {(kpc)}        &
\colhead {($r_e$)} &
\colhead {($r_e$)} &
\colhead {} 
}
\startdata
050509B & 0.2248 & 15.10 & 3.40 & 55.19 & 12.43 & 2.59 &	0.58 & 1 \\
050709 & 0.1607 & 1.35 & 0.020 & 3.76 & 0.056  & 2.00 &	0.030 & 1 \\
050724 & 0.254 & 0.68 & 0.020 & 2.74 & 0.080 & 0.67 &	0.020 & 1  \\
050813 & 0.719 & 5.96 & 2.34 & 43.57 & 17.37 & \nod & \nod & This work \\
051210 & 2.58 & 3.56 & 2.00 & 29.08 & 16.34  & 5.65 & 	3.17 & 1 \\
051221A & 0.5464 & 0.32 & 0.030 & 2.08 & 0.19  & 0.89 &	0.083 & 1 \\
060121 & \nod & 0.119 & 0.046 & 0.97 & 0.37 & 0.18 &	0.069 & 1  \\
060313 & \nod & 0.32 & 0.068 & 2.60 & 0.55  & 1.39 &	0.30 & 1 \\
060614 & 0.125 & 0.31 & 0.35 & 0.70 & 0.79 & 0.86 & 0.97 & This work \\
060801 & 1.131 & 1.23 & 1.31 & 10.25 & 10.92 & \nod & \nod & 2 \\
061006 & 0.461 & 0.24 & 0.05 & 1.39 & 0.29  & 0.37 &	0.077 & 1 \\
061210 & 0.4095 & 2.82 & 2.61 & 15.51 & 14.36 & \nod & \nod & 3 \\
070429B & 0.902 & 0.76 & 1.7 & 6.00 & 13.44 & 1.17 &	2.62 & 3 \\
070707 & \nod & 0.4 & 0.03 & 3.25 & 0.24 & 1.11 &	0.083 & 3 \\
070714B & 0.923 & 1.55 & 0.11 & 12.33 & 0.87 & 5.17 &	0.37 & 3 \\
070724 & 0.457 & 0.94 & 0.03 & 5.52 & 0.18 & 1.49 &	0.048 & 3 \\
070729 & 0.52 & 3.13 & 2.3	& 19.72 & 14.49 & \nod & \nod & This work \\
070809 & 0.473 & 5.70 & 0.46 & 34.11 & 2.75 & 9.34 &	0.75 & 3 \\
071227 & 0.381 & 2.80 & 0.05 & 14.74 & 0.26 & 3.08 &	0.055 & 3 \\
080123	& 0.495	& 8.74 & 1.25 & 53.63 &	7.67 & \nod & \nod & This work \\
080503 & \nod & 0.9 & 0.03 & 7.31 & 0.24 & 3.46 &	0.12 & 3 \\
080905 & 0.1218 & 8.29 & 0.08 & 18.30 & 0.18 & 4.61 & 	0.044 & 3 \\
081226A$^{c}$ & \nod & $<0.5$ & \nod & $<4.06$ & \nod & \nod & \nod & 4 \\
090305 & \nod & 0.43 & 0.030 & 3.49 & 0.24 & 1.19 &	0.083 & 3 \\
090426 & 2.609 & 0.060 & 0.030 & 0.49 & 0.24 & 0.29 &	0.14 & 3 \\
090510 & 0.903 & 1.33 & 0.37 & 10.51 & 2.92 & 1.66 &	0.46 & 3 \\
090515 & 0.403 & 13.98 & 0.03 & 76.19 & 0.16 & 13.98 & 0.03 & 3 \\
091109 & \nod & 0.52 & 0.05 & 4.22 & 0.41 & 1.93 & 0.19  & This work \\
100117A & 0.914 & 0.17 & 0.04 & 1.35 & 0.32  & 0.61 & 0.14 & 2 \\
100206 & 0.407 & 4.59 & 2.37 & 25.28 & 13.05 & \nod & \nod & This work \\
100625A & 0.452 & 0.45 & 1.16 & 2.63 & 6.77 & \nod & \nod & This work \\
101219 & 0.718 & 0.75 & 0.91 & 5.48 & 6.65 & \nod & \nod & This work \\
101224 & 0.454 & 2.18 & 2.31 & 12.75 & 13.51 & \nod & \nod & This work \\
111117 & 2.211 & 1.25 & 0.2 & 10.52 & 1.68 & \nod & \nod & This work \\
120305 & 0.225 & 4.967 & 1.44 &	18.09 &	5.25 & \nod & \nod & This work \\
120804 & 0.74 & 0.27 & 0.15 & 2.30 & 1.28 & \nod & \nod & This work \\
121226A & 1.37 & 0.27 & 1.07	& 2.31 & 9.15 &	\nod & \nod & This work \\
130515A	& 0.8 &	8.05 &	1.81 &	61.22 &	13.77 &	\nod & \nod & This work \\
130603B & 0.3568 & 1.07 & 0.04 & 5.40 & 0.20 & 0.71 &	0.027 & 3 \\
130716A & 2.2 &	3.93 & 1.69	& 33.08 &	14.23 & \nod & \nod & This work \\
130822A & 0.154 & 22.32 & 1.82 & 60.09 & 4.90 & \nod & \nod & This work \\
130912A & \nod & 0.48 & 0.13 & 3.90 & 1.06 & 1.41 & 	0.38 & This work \\
131004A & 0.717 & 0.11 & 0.030 & 0.80 & 0.22 & 0.25 &	0.068 & This work \\
140129B & 0.43 & 0.31 & 0.31 & 1.76 & 1.76 & \nod & \nod & This work \\
140622A & 0.959 & 4.1 & 1.4 & 32.95 & 11.25 & \nod & \nod & This work \\
140903A & 0.351 & 0.18 & 0.02 & 0.90 & 0.10 & \nod & \nod & This work \\
140930B	& 1.465	& 1.12 &	0.5 &	9.62 &	4.30 &	\nod & \nod & This work \\
141212A & 0.596 & 2.782 & 1.823 & 18.75 & 12.29 & \nod & \nod & This work \\
150101B & 0.134 & 3.07 & 0.030 & 7.36 & 0.072 & 0.78 & 	0.0076 & 5, This work \\
150120A & 0.46 & 0.81 & 1.094 & 4.77 & 6.44 & \nod & \nod & This work  \\
150424A	& \nod &	0.42 &	0.04 &	3.41 &	0.32 &	1.5 &	0.14 & This work \\
150728A &	0.461 &	1.28 &	3.44 &	7.52 &	20.29 & \nod & \nod & This work  \\
150831A	& 1.09	& 1.48 &	1.18 &	12.21 &	9.77 & \nod & \nod & This work  \\
151229A	& 0.63 &	1.18 &	0.88 &	8.16 &	6.05 & \nod & \nod & This work  \\
160303A &	1.01 &	1.88 &	0.11 &	15.31 &	0.90 &	3.42 &	0.2 & This work \\
160408A &	1.9	& 1.65 & 0.15	& 14.13 & 1.25 & \nod & \nod & This work  \\
160411A &	0.82 &	0.18 &	0.3 &	1.40 &	2.30 & \nod & \nod & This work  \\
160525B & \nod	&	1.06 &	1.06 &	8.61 &	8.61 & \nod & \nod & This work  \\
160601A & \nod	&	0.17 &	0.5 &	1.38 &	4.06 & \nod & \nod & This work  \\
160624A & 0.4842 & 1.59 & 1.03 & 9.63 & 6.24 & 2.37 &	1.54 & This work \\
160821B	& 0.1619 &	5.61 &	0.01 &	15.74 &	0.03 &	4.24 &	0.008 & This work \\
161001A	& 0.67	& 2.61	& 0.88 &	18.54 &	6.22 & \nod & \nod & This work  \\
161104A & 0.793 & 0.219 & 2.19 & 1.66 & 16.60 & \nod & \nod & 6 \\
170127B	& 2.28	& 1.24 & 1.63 &	10.37 &	13.60 & \nod & \nod & This work \\
170428A & 0.453 & 1.32 & 0.58 & 7.72 & 3.39 & \nod & \nod & This work \\
170728A & 1.493 & 3.75 & 0.35 & 32.25 & 3.01 & \nod & \nod & This work \\
170728B & 0.62 & 0.99 & 0.30 & 6.76 & 2.06 & \nod & \nod & This work \\
180418A & 1.56 & 0.16 & 0.04 & 1.30 & 0.32 & \nod & \nod & 7 \\
180618A	& 0.52 &	1.54 &	0.27 &	9.70 &	1.69 & 	\nod & \nod & This work \\
180727A &	1.95 &	0.3 &	0.6 &	2.56 &	5.13 &	\nod & \nod & This work \\
180805B &	0.6612 &	3.44 &	1.06 &	24.30 &	7.49 & 	\nod & \nod & This work \\
181123B & 1.754 & 0.59 & 0.16	 & 5.08 & 1.38 & \nod & \nod & 8 \\
191031D	& 1.93 &	1.53 &	1.25 &	13.08 &	10.69 &	\nod & \nod & This work \\
200219A &	0.48 &	1.38 &	0.88 &	8.30 &	5.28 & \nod & \nod & This work \\
200411A &	1.93 &	4.91 &	0.88 &	41.98 &	7.48 & 	\nod & \nod & This work \\
200522A &	0.5536 &	0.143 &	0.029 &	0.93 &	0.19 &	0.24 &	0.048 & 9 \\
200907B &	0.56 &	0.37 &	1.19 &	2.41 &	7.78 & 	\nod & \nod & This work \\
201221D &	1.055	& 3.57	& 2.93	& 29.35 &	24.09 & \nod & \nod & This work \\
210323A &	0.733 &	0.8	& 0.5 &	5.89 &	3.68 & 	\nod & \nod & This work \\
210726A &	0.2244 &	3.37 &	0.61 &	12.26 &	2.22 & \nod & \nod & This work \\
210919A &	0.2415 &	13.28 &	0.5 &	51.05 &	1.92 & 	\nod & \nod & This work \\
211023B &	0.862 &	0.49 &	0.33 &	3.84 &	2.57 & \nod & \nod & This work \\
211106A &	\nod &	0.097 &	0.036 &	0.79 &	0.29 &	0.49 &	0.18 & 10, This work \\
211211A &	0.0763 &	5.44 &	0.02 &	7.92 &	0.029 & 3.20 & 0.01 & 11, This work \\
\enddata
\tablecomments{Galactocentric offsets for \totoffset\ bursts, and one upper limit on the offset (for GRB\,081226A). Physical offsets are calculated using the same cosmological parameters across all bursts.
For bursts with optical afterglow detections and no publicly-available afterglow imaging, we assume an astrometric tie error of $0.5''$ in our calculation of the offset uncertainty. The positions for bursts with only XRT positions are based on the enhanced XRT positions as of May 2022 \citep{ebp+09}. \\
$^{a}$ For bursts with unknown redshift, physical offsets are calculated for an assumed $z=1$. \\
$^{b}$ Angular offset measurement references: (1) \citealt{fbf10}, (2) \citealt{bfp+07}, (3) \citealt{fb13}, (4) \citealt{nkg+12}, (5) \citealt{fmc+16}, (6) \citealt{nfd+20}, (7) \citealt{rfv+21}, (8) \citealt{pfn+20}, (9) \citealt{flr+21}, (10) \citealt{GRB211106A-Laskar}, (11) \citealt{rgl+22}. \\
$^{c}$ An angular offset of $<0.5''$ is reported in \citet{nkg+12}, but an afterglow position is not.
}
\end{deluxetable*}


\section{Discussion}
\label{sec:disc}

\begin{figure}[t]
\centering
\includegraphics[width=0.5\textwidth]{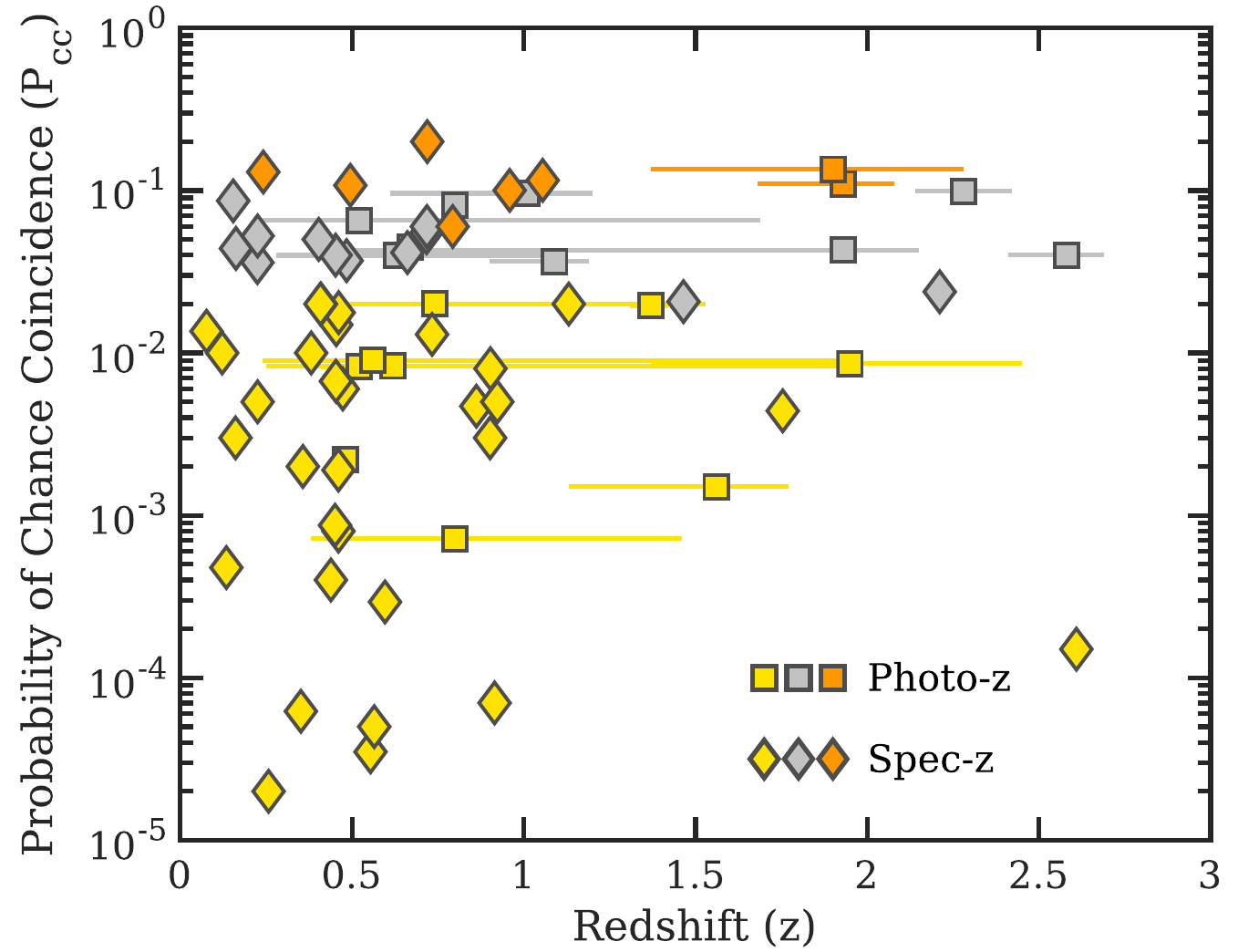}
\caption{Probability of chance coincidence versus redshift for the host associations in our sample, color coded by classification as Gold (yellow), Silver (gray) or Bronze (orange). Diamonds denote spectroscopic redshifts while squares represent photometric redshifts. Error bars correspond to $1\sigma$ confidence. As expected, the average $P_{cc}$ increases with redshift, especially beyond $z \gtrsim 1$. Put another way, the robustness of association decreases with increasing redshift. We also note the larger prevalence of photometric redshifts at $z\gtrsim 1$. Not shown are host galaxies with unknown redshift (which are largely Bronze classifications and are likely to be low-luminosity hosts or at $z \gtrsim 1$).
\label{fig:pccz}}
\end{figure}

\begin{figure*}[t]
\centering
\includegraphics[width=0.6\textwidth]{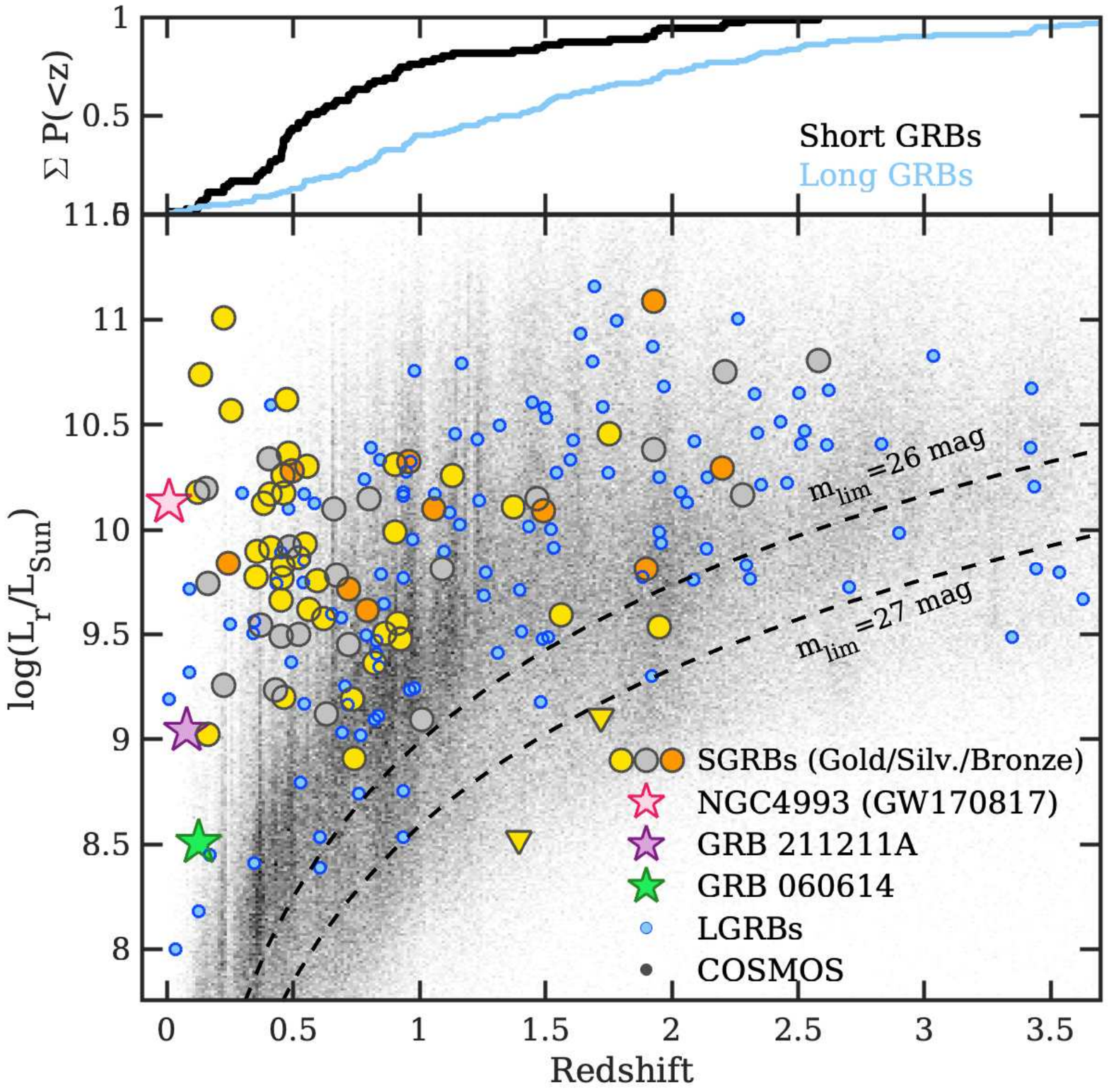}
\vspace{-0.2in}
\caption{{\it Bottom:} $r$-band host luminosities versus redshift for 73 short GRBs with redshifts and $r$-band magnitudes color coded by gold, silver and bronze samples (circles). Two bursts with limits on a coincident host and known redshifts from their afterglows are also shown (gold triangles). Approximate limits of the ground-based catalog ($m_{\rm lim} \approx 26$~mag) and {\it HST} sample ($m_{\rm lim}\gtrsim 27$~mag) are plotted as dashed lines. For the subset of bursts which lack a coincident host to $\gtrsim 27$~mag, these searches are sensitive enough to rule out the faint end of the observed short GRB host luminosity function to $z \approx 2$, as well as the fainter end of the field galaxy population (grayscale; COSMOS; \citealt{lmi+16}). Also shown are 85 long GRB host luminosities over the same redshift range (black circles; \citealt{sgl09,slt+10,hmj+12,vsj+15,bbf16}), the host galaxy of the long GRB\,211211A which has a likely kilonova (purple star; \citealt{rgl+22}), the host galaxy of the possible short GRB-EE, GRB\,060614 (green star; \citealt{gfp+06}), and the host of GW170817, NGC4993 (red star). Overall, short GRB hosts exist in more luminous galaxies than long GRBs at the same redshifts (although there is substantial overlap in their luminosity functions). {\it Top:} Cumulative distribution of redshifts for short (black) and long (blue) GRBs.
\label{fig:lumz}}
\end{figure*}

\subsection{Diversifying the Population of Short GRB Hosts: Redshifts, Luminosities \& Offsets}
We first address the effects of our methods in defining host galaxy associations. Here, we have adopted a uniform, generous criteria of association, in which every burst with a galaxy in the field that has $P_{\rm cc,min} \lesssim 0.2$ is assigned to a host. Realistically, with this method we inevitably inherit some incorrect host assignments when imposing a high $P_{cc}$ threshold of 20\% for any given association. In particular, we expect $\lesssim 2\%$ of Gold, $\lesssim 10\%$ of Silver, and $\lesssim 20\%$ of Bronze associations to be spurious. This results in an expected total of $\lesssim 5.8$~incorrect identifications, which is only $\approx 6.9\%$ of our total sample of associations. On the other hand, by only including the most robust associations (Gold sample, $P_{\rm cc, min} \leq 0.02$) in short GRB studies, we subject the sample to biases and draw conclusions that may not be reflective of the entire population. Moreover, since the compact object binary progenitors of short GRBs experience natal kicks and thus systemic velocities from their birth sites (e.g., \citealt{fk97,fwh99,bpb+06,vns+18}), it is inevitable that a fraction of events will explode far from their host galaxies, largely irrespective of host brightness. Thus, it is expected that due to the nature of their progenitors and the nature of the $P_{cc}$ method, some short GRBs will still have correctly assigned host galaxies with relatively high $P_{cc}$ values.

Indeed, we find that by including the less robust host associations (Silver and Bronze), we capture a substantial number of bursts at $z\gtrsim 1$ (Figure~\ref{fig:pccz}). Similarly, we find that the inclusion of the Silver and Bronze hosts results in additional lower luminosity hosts with $L_R \lesssim 10^{10}L_{\odot}$ (Figure~\ref{fig:lumz}). Finally, the locations of the most robustly associated short GRBs (Gold) are on average 3~kpc closer to their host galaxies than the sample including all associations ($\approx 4.9$~kpc versus $\approx 7.9$~kpc; Figure~\ref{fig:offset_physical}). To summarize, we find that establishing a generous association criteria helps to capture more bursts at larger offsets, as well as higher-redshift ($z \gtrsim 1$) and lower-luminosity hosts. While this may come at a cost of incorrect host assignments at the level of $\lesssim 7\%$, we note that the Gold sample still comprises over half of the host associations. It is also the sub-sample with the most information (e.g., redshifts, host-normalized offsets) so these bursts still dominate the distributions in every property. Overall, it is clear that including more associations results in a diversification of the known population of short GRB hosts. In terms of drawing physical conclusions for their progenitors, we further explore the effect of host associations on stellar population properties in \citet{BRIGHT-II}.

We next explore the effects of including photometric redshifts (the full modeling methods go hand-in-hand with the stellar population properties, and are thus described in detail in \citealt{BRIGHT-II}). Figure~\ref{fig:pccz} shows that the inclusion of a large sample of photometric redshifts also captures higher-redshift events. Indeed, for $z\gtrsim 1$, we find that there is a steep drop-off in spectroscopic redshifts, and in turn, a higher frequency of bursts with photometric redshifts (Figure~\ref{fig:pccz}). This is in part due to the so-called `redshift desert' in which the most prevalent optical spectral features for redshift identification are shifted into the NIR band where ground-based spectroscopy is less sensitive. At these redshifts, the apparent faintness of the host galaxies also preclude high S/N spectroscopy and only photometric redshifts are possible. In short, including short GRB hosts with photometric redshifts helps to fill out the short GRB redshift distribution at $z\gtrsim 1$.

This population is particularly important for constraining the functional form of the delay time distribution (DTD). Indeed, \citet{wp15} and \citet{pfn+20} found that even a few short GRBs at $z\gtrsim 1.5$ could rule out log-normal DTD models to high confidence, whereas power-law DTD models could accommodate tens of events within the {\it Swift} short GRB population at $z\gtrsim 1$. Here, we find that $\approx 25\%$ of bursts with known redshift have $z \gtrsim 1$, while $\approx 16\%$ have $z\gtrsim 1.5$. As not all bursts have associations or redshifts, these likely represent lower limits on the fractions. Indeed, ten bursts with identified hosts are too faint to have determined redshifts, while 6 events have inconclusive host associations. In the extreme case that all 16 of these events are at $z>1$ or $z>1.5$ in similar proportions to the sample with known redshifts, then as many as $\approx 44\%$ (28\%) could reside at $z>1$ ($z>1.5$). In summary, we find that $\approx 25-44\%$ of short GRBs originate at $z>1$ whereas $\approx 16-28\%$ originate at $z>1.5$. Our results are in broad agreement with the finding that $\approx 20-50\%$ of short GRBs could reside at $z>1$ based on bursts with no identified host galaxies \citep{otd+22}. In \citet{Zevin+DTD}, we use the full sample of host galaxies with star formation histories, stellar population ages and stellar masses derived in \citet{BRIGHT-II} to calculate a power-law DTD with a fairly steep slope of $\propto t^{-1.8}$ (e.g., more short delay-time systems compared to a canonical $\propto t^{-1}$ model), commensurate with the larger fraction of bursts at higher redshift reported here.

Using optical magnitudes and redshifts for 73~short GRBs, we find a range of host optical luminosities, $L_r\approx 3.2 \times 10^{8}L_{\odot}$ to $\approx 1.2 \times 10^{11}L_{\odot}$ with a median of $\langle L_r \rangle \approx 8 \times 10^{9}L_{\odot}$ (Figure~\ref{fig:lumz}). This is a factor of $\approx 2$ lower than the previous medians based on smaller samples of events \citep{ber09,ber14,fbb+17}. This difference can naturally be explained by the inclusion of a larger number of lower-luminosity hosts, some of which are less robust associations compared to the focus of earlier works.  The short GRB photometric catalog presented here generally reaches $m_{\rm lim} \approx 26$~mag ($3\sigma$), which is sensitive enough to detect $L_r\lesssim 10^{9}L_{\odot}$ galaxies out to $z\approx 1$. Despite this, there is a lack of short GRB hosts with $L_r\lesssim 10^{9}L_{\odot}$ when compared to the field galaxy sample from the COSMOS2015 survey \citep{lmi+16}. Although such galaxies are more common in the universe, this is likely due to the fewer numbers of stars, and thus BNS progenitors, in these galaxies, and goes hand in hand with the lack of low-mass short GRB hosts \citep{BRIGHT-II}.

We also compare the short GRB population to a sample of 85 long GRB hosts at $z \lesssim 3$ \citep{sgl09,slt+10,hmj+12,vsj+15,bbf16}. Long GRBs are known to originate from massive star progenitors and overall explode in lower luminosity host galaxies \citep{sgl09,slt+10}. Figure~\ref{fig:lumz} shows that while there is substantial overlap in the observed host luminosity functions, there are more long GRB hosts with $L_r\lesssim 10^{9}L_{\odot}$ compared to almost none for the short GRB host population. Indeed, there is only one short GRB host with $L_r\lesssim 10^{9}L_{\odot}$ compared to ten long GRB hosts. When compared to the short GRB host population, the host galaxies of long-duration GRB\,211211A and the possible short GRB-EE GRB\,060614 are among the least luminous galaxies compared to the short GRB host sample. On the other hand, NGC4993 (the host galaxy of GW170817) is more typical of short GRB hosts, albeit older and more quiescent (cf., \citealt{BRIGHT-II}).

For the six hosts with Inconclusive associations, two have known redshifts from their afterglows (GRBs\,150423A and 160410A). The limits on coincident hosts for these bursts are deep enough to rule out hosts down to $L_{r}\approx 10^{8.5}-10^{9}L_{\odot}$ (Figure~\ref{fig:lumz}). For the remaining four Inconclusive bursts, the limits are deep enough to detect the faintest known short GRB hosts to $z\approx 2.5$. Thus, these bursts either originate from hosts with particularly low luminosities or are highly-offset from hosts that are closer to the median.

In terms of galactocentric offsets, the hosts with the most robust associations can be taken as a minimum distribution. Compared to other studies which determined a median of $\approx 5-6$~kpc \citep{fbf10,fb13,otd+22}, our median value for the entire population is larger, with $\approx 7.9$~kpc. This is largely because previous studies focused on the most robust associations (e.g., the Gold sample); indeed we find a smaller median of $\approx 4.9$~kpc for this sample alone. In keeping with previous work, we find that short GRBs are also $\approx 6$ times larger in physical offsets than long GRBs, and $\approx 2.5$ times larger in host-normalized offsets. This is consistent with the migratory ability of short GRB progenitors and their long delay times, compared to the young and relatively stationary massive star progenitors of long GRBs. Compared to NS merger models, we also find a dearth of observed highly-offset ($\gtrsim 30$~kpc) SGRBs (Figure~\ref{fig:offset_physical}). This can be reconciled if most of the Inconclusive associations originate from hosts at large offsets (as opposed to from low-luminosity, spatially coincident hosts). 

Early work suggested that dynamical channels in globular clusters could form BNS mergers and be responsible for $\approx 30\%$ of observed short GRBs \citep{gpm+06}. However, more recent globular cluster simulations have shown the rate of BNS and NSBH mergers to be negligible in clusters compared to the field \citep{yfk+20}. This is corroborated by the lack of a globular cluster to deep limits at or proximal to the position of the BNS merger GW170817 \citep{fba+19,lll+19,kfb+22}. Indeed, when we compare to the globular cluster distributions, we find that at most $\lesssim 10\%$ of observed short GRBs could originate from globular clusters {\it in situ}.

We finally examine whether or not short GRB-EE or possible short GRB-EE events are distinct in their properties. The mechanism powering the extended emission is uncertain, although has been proposed to be linked to protomagnetar winds \citep{bmt+12}, two-component jets \citep{bp11}, or a progenitor-specific phenomenon such as NS-BH merger origins \citep{tko+08,glt20}. Equipped with a large sample of short GRBs, we now examine this question in further detail, here and in \citet{BRIGHT-II}. For the 14 short GRB-EE bursts with optical luminosities and magnitudes, we find that these bursts span the full range in host luminosity with a median of $\langle L_{r,EE}\rangle \approx 1.0 \times 10^{10}L_{\odot}$, virtually indistinguishable from the classical short GRB sample. We also find a clear lack of correlation with any galaxy-scale environmental property (e.g., stellar population age, stellar mass;  \citealt{BRIGHT-II}). In terms of locations, \citet{tko+08} claimed that short GRB-EE events lie closer to their host galaxies than classical short GRBs, and likely arise from NS-BH mergers; this was tentatively supported in \citet{glt20} although the latter works finds their host-normalized offsets to be indistinguishable. Here, we perform an Anderson-Darling test between these two populations to test the null hypothesis that their projected physical (host-normalized) offsets are drawn from the same underlying distribution. We calculate a $p$-value of $p=0.25$ ($p=0.25$). Thus, we cannot rule out the null hypothesis in both cases, and find that these two populations are statistically indistinguishable in terms of the locations with respect to their hosts. We note that if the observed population of short GRB-EE events arise from different progenitors than classical short GRBs, these differences are not manifested as a distinct set of environmental properties. 

\subsection{Selection Effects and Assessing Potential Contamination to Sample}
Here we address the selection effects of our sample born out of the required inclusion criteria, and assess the potential contamination from events that originate from collapsars (e.g., ``traditional'' long GRBs). First and foremost, robust host galaxy identifications require localization via the detection of afterglows. The brightness of the afterglow depends on a combination of kinetic energy and circumburst density (e.g., \citealt{gs02}). Thus, the requirement of a detected afterglow might translate to a missing population of bursts at large offsets in the galaxy halos or IGM, or in quiescent galaxies for which the average ISM densities are lower \citep{Perna22}. Indeed, Figure~\ref{fig:offset_physical} shows that the predicted spatial distributions of BNS mergers expect $\approx 20\%$ of the population to reside at $\gtrsim 50$~kpc \citep{wfs+18}. We briefly quantify how many bursts we could be missing at larger offsets. In our study, the requirement for the detection of an afterglow reduces the total available sample by $\approx 28\%$. However, the majority ($\approx 70\%$) of the events which lack X-ray afterglows have delayed XRT follow-up observations, either due to observatory constraints or discoveries in BAT ground analysis, whereas only a couple of events in our well localized sample had delayed X-ray follow-up. Thus, most of the events which lack X-ray afterglows are not likely to have systematically fainter X-ray afterglows than the rest of the population. This leaves $\approx 12$~events with presumably fainter X-ray afterglows that resulted in afterglow non-detections. If these events indeed went undetected as a result of lower circumburst densities and larger offsets, their inclusion in this sample would have an effect of $\lesssim 10\%$. While the optical afterglows of highly-offset bursts have been found to be fainter \citep{ber10}, a more recent study exploring the X-ray afterglows did not find any correlation between offset and X-ray afterglow brightness \citep{otd+22}. In summary, we do not expect the population missing at large offsets to be substantial.

Our second major criteria for selection is based on the observed $\gamma$-ray duration ($T_{90}$). The duration-based classification has been shown to be a detector-specific and imperfect delineation between NS merger and collapsar events \citep{bnp+13,jss+20}, and thus we expect to inherit some contamination in our sample. This is rooted in a few examples in which the duration does not provide a one-to-one mapping to the progenitor. For instance, the {\it Fermi} short GRB\,200826A had a duration of $\approx 1.1$~sec (30-500~keV), but has a photometric excess fully consistent with an associated SN, indicative of a massive star (non-NS merger) origin \citep{asa+21}. Similarly, the {\it Swift} and {\it Fermi} long GRB\,211211A was found to have a photometrically-identified kilonova, consistent with an NS merger origin \citep{rgl+22}. To assess the degree of potential contamination by true collapsar events, we apply both the \cite{bnp+13} and \cite{jss+20} criteria to our sample, and subsequently search for any systematic variations between bursts classified as collapsars via the various schemes. \cite{bnp+13} determine the probability of a given burst arising from a collapsar based on the number counts of bursts at a given duration, and the expectation of a plateau in durations corresponding to the break-out time from a collapsar. They critically conclude that spectral hardness is, if anything, a more important tool for distinction than duration. They subsequently provide a route for the determination of the collapsar probability as a function of duration, as well as in different spectral hardness bins ($f_{\rm NC}$). Since the fit parameters for the spectral hardness bins are not provided directly in \cite{bnp+13} we re-calculate them based on the provided distributions. Furthermore, we calculate the probabilities based on single power-law spectral fits to the bursts in our sample from the more recent BAT catalog, resulting in some small (typically not important) discrepancies between our values and those in \cite{bnp+13}. For the \cite{jss+20} we directly determine bursts which are either in the short or long category. In this case, \cite{jss+20} used a machine learning approach to identify features which distinguish short and long GRBs, and these appear to provide a cleaner separation than duration alone. However, they also cannot perform this analysis on bursts of very short duration, and thus we assign these bursts to the short class. It should also be noted that this classification scheme places some short GRB-EE in the long category, perhaps most notably GRB 060614. Our results are listed in Table~\ref{tab:class}. 

We can then confront the outcomes of these 
classification schemes based on $\gamma$-ray properties and examine trends with offsets. The baseline expectation is that true NS merger events will have larger offsets than collapsar contaminants. \citet{bnp+13} found that $\approx 35\%$ of {\it Swift} GRBs with $T_{90}\leq2$~s could be true collapsars. If these probabilities provide an accurate indication of progenitor, we would overall expect that bursts with larger values of $f_{\rm NC}$ would have larger offsets. However, we find that both the potential contaminants and the pure NS merger populations under this scheme have indistinguishable offset distributions and span the full range.  

In addition to offsets, we can test if the populations of bursts identified as likely collapsars differ from those identified as mergers (shorts) in other properties. Perhaps most notably these include the redshift and the host specific star formation rate (the latter of which are reported in \citealt{BRIGHT-II}). Since the long GRB population is typically found at higher redshift than the short burst population, we may expect that the collapsar contamination may give rise to the apparent high redshift short GRBs. Similarly, while short GRBs do arise predominantly from star-forming host galaxies, their specific star formation rates are lower. However, AD tests do not reveal any significant differences between the redshifts or specific star formation rates between our sample split according to the two alternative schemes. Moreover, the classification of \citet{bnp+13} predicts that $\approx 1/3$ of $z<0.5$ events in our sample are contaminated by true collapsars but this is at odds with the fact that the large majority of events at these redshifts have strong constraints on associated SN emission. We further find that the distributions between our sample and the classical long GRB sample are quite distinct in several properties, including host luminosities, physical offsets (this work), stellar masses and star formation rates \citep{BRIGHT-II}. These vastly different distributions are also at odds with a large contamination fraction by true collapsars, and if true, would require extreme host properties in the contaminating population to reconcile the differences.

\cite{zzv+09} suggest that a full characterization of the population should include the consideration of multiple physical criteria, including host type, offset, location in high energy correlations and the presence of supernovae as a route of distinction between the two classes of burst which they term type I (mergers) and type II (collapsars). Not all of the ideal information is available for each burst. However, based on the available information, the majority of our sample would be classed as type I, or would be  inconclusive.

The lack of apparent differences between the samples when employing different inclusion criteria demonstrates that our duration cut is as efficient as others in identifying true, merger-driven short GRBs. While it is impossible to precisely quantify the contamination, since alternative cuts do not result in significantly different distributions in the core parameters, our physical conclusions are not sensitive to our choice of method of determining what constitutes a short GRB.

\section{Conclusions}
\label{sec:conc}

We have presented photometric, spectroscopic, and galactocentric offset catalogs which describe the host galaxies of short GRBs and their locations within them. Our sample comprises \totsamp\ events discovered in 2005--2021 primarily discovered by {\it Swift}. We come to the following main conclusions:

\begin{itemize}
    \item With 1--11 imaging filters per host galaxy, we newly contribute \totbrightphot\ photometric data points in the optical and NIR bands, reaching depths of $\approx 24-27$~mag and $\approx 23-26$~mag, respectively. We also present 25 new host galaxy spectra and determine 17 spectroscopic redshifts, spanning $z \approx 0.15-1.6$.
    \item Including associations previously made in the literature, we report host galaxy associations for \totassoc\ events, including 26 new associations. This comprises $\approx 56\%$ of the total {\it Swift} population of short GRBs. For the remaining $\approx 44\%$ of events for which host associations cannot be made with present data, the large majority have observing constraints or lack X-ray afterglows, precluding meaningful observations for host identifications.
    \item Taking into account individual measurement uncertainties, we determine a median projected physical offset of $\approx 7.9$~kpc (16th--84th percentile on the full distribution of 1.8--28.8~kpc) for \totoffset\ bursts which is $\approx\!2-3$~kpc larger than previously found. For 34 short GRBs with effective radii measurements, we find a median host-normalized offset of $\approx 1.5r_e$ (0.57--4.1$r_e$), although we note that this population is largely comprised of bursts with the most robust host associations. The physical and host-normalized offset distributions are a factor of $\approx 6$ and $\approx 2.5$~times larger than those of long GRBs, respectively.
    \item The most robust associations account for over half of host identifications (the Gold sample). The Gold sample bursts generally have more luminous host galaxies, lower redshifts, and smaller offsets than those with less robust host associations (Silver and Bronze samples). Thus, the inclusion of less robust associations, even if risking a small potential loss of integrity at the level of $\lesssim 7\%$, is important when drawing conclusions on their progenitors.
    \item We find that likely $\approx 25-44\%$ of observed {\it Swift} short GRBs originate at $z>1$, whereas $16-28\%$ originate at $z>1.5$. The true frequency of this population relative to the low-redshift sample provides discriminating power among DTD models, and in particular, the prevalence of shorter delay-time systems.
    \item In terms of optical luminosity, NGC4993 (the host galaxy of GW170817) has similar properties to the rest of the host population. However, the host galaxies of possible short GRB-EE\,060614 and the potentially merger-driven long GRB\,211211A are on the low-luminosity end of the population. Overall the short GRB host population exhibits diversity in terms of intrinsic luminosities and locations.
    \item As a population, we find that short GRBs with extended emission (including those tentatively classified as such) and classical short GRBs are statistically indistinguishable in terms of their host galaxy luminosities, projected physical offsets, and host-normalized offsets from their hosts. Thus, if these two populations arise from different progenitors, the progenitors do not select for a distinct set of environmental properties.
\end{itemize}

\noindent The launch of {\it Swift} enabled the discovery of the first short GRB afterglows in 2005. This crucially paved the way for the first few host galaxy associations \citep{gso+05,vlr+05,ffp+05,bpc+05,bpp+06}. Thanks to the continued longevity of {\it Swift} and concerted efforts over 17 years to identify as many host galaxies as possible, we now have a legacy sample of \totassoc\ host galaxies. This sample serves as a cosmological anchor against which future multi-messenger BNS and NSBH merger environments can be compared. In particular, the advent of 3G gravitational wave detectors in the next two decades, which will be sensitive to BNS mergers to $z \approx 1$ and beyond \citep{DAWNReport,CEHorizons,Kalogera-3G}, will provide a direct comparison to the short GRB host population. Those bursts without clear host associations with present facilities can be pursued with the {\it James Webb} Space Telescope or {\it Nancy Grace Roman} Space Telescope to potentially unveil a population of high-redshift ($z\gtrsim 2$ short GRBs). Using the largest and broadest possible data set currently available, this series of legacy catalogs paves the way for interpretation of short GRB progenitors, such as their inference on their stellar population properties and delay time distributions, and fundamental properties of their compact object binary progenitors.

\section*{Acknowledgements}
The Fong Group at Northwestern acknowledges support by the National Science Foundation under grant Nos. AST-1814782, AST-1909358 and CAREER grant No. AST-2047919. W.F. gratefully acknowledges support by the David and Lucile Packard Foundation, the Alfred P. Sloan Foundation, and the Research Corporation for Science Advancement through Cottrell Scholar Award \#28284. A.E.N. acknowledges support from the Henry Luce Foundation through a Graduate Fellowship in Physics and Astronomy. Y.D. is supported by the National Science Foundation Graduate Research Fellowship under Grant No. DGE-1842165. M.N. is supported by the European Research Council (ERC) under the European Union’s Horizon 2020 research and innovation programme (grant agreement No.~948381) and by a Fellowship from the Alan Turing Institute. R.M. acknowledges support by National Science Foundation under Award Nos.~AST-1909796 and AST-1944985.

This research is based on observations made with the NASA/ESA Hubble Space Telescope obtained from the Space Telescope Science Institute, which is operated by the Association of Universities for Research in Astronomy, Inc., under NASA contract NAS 5–26555. These observations are associated with program \#14685, 13830, 14237, and 14357. W. M. Keck Observatory and MMT Observatory access was supported by Northwestern University and the Center for Interdisciplinary Exploration and Research in Astrophysics (CIERA). Some of the data presented herein were obtained at the W. M. Keck Observatory, which is operated as a scientific partnership among the California Institute of Technology, the University of California and the National Aeronautics and Space Administration. The Observatory was made possible by the generous financial support of the W. M. Keck Foundation. Observations reported here were obtained at the MMT Observatory, a joint facility of the University of Arizona and the Smithsonian Institution.

Based on observations obtained at the international Gemini Observatory, a program of NOIRLab, which is managed by the Association of Universities for Research in Astronomy (AURA) under a cooperative agreement with the National Science Foundation on behalf of the Gemini Observatory partnership: the National Science Foundation (United States), National Research Council (Canada), Agencia Nacional de Investigaci\'{o}n y Desarrollo (Chile), Ministerio de Ciencia, Tecnolog\'{i}a e Innovaci\'{o}n (Argentina), Minist\'{e}rio da Ci\^{e}ncia, Tecnologia, Inova\c{c}\~{o}es e Comunica\c{c}\~{o}es (Brazil), and Korea Astronomy and Space Science Institute (Republic of Korea).

This paper includes data gathered with the 6.5 meter Magellan Telescopes located at Las Campanas Observatory, Chile

The United Kingdom Infrared Telescope (UKIRT) was supported by NASA and operated under an agreement among the University of Hawaii, the University of Arizona, and Lockheed Martin Advanced Technology Center; operations are enabled through the cooperation of the East Asian Observatory. We thank the Cambridge Astronomical Survey Unit (CASU) for processing the WFCAM data and the WFCAM Science Archive (WSA) for making the data available.

This work made use of data supplied by the UK Swift Science Data Centre at the University of Leicester.

The LBT is an international collaboration among institutions in the United States, Italy and Germany. The LBT Corporation partners are: The University of Arizona on behalf of the Arizona university system; Istituto Nazionale di Astrofisica, Italy;  LBT Beteiligungsgesellschaft, Germany, representing the Max Planck Society, the Astrophysical Institute Potsdam, and Heidelberg University; The Ohio State University; The Research Corporation, on behalf of The University of Notre Dame, University of Minnesota and University of Virginia.

This work is based in part on observations made with the Spitzer Space Telescope, which was operated by the Jet Propulsion Laboratory, California Institute of Technology under a contract with NASA.

This research has made use of the HST-COSMOS database, operated at CeSAM/LAM, Marseille, France.
 
\vspace{5mm}
\facilities{MMT (Binospec, MMIRS), Magellan:Clay (LDSS3), Magellan:Baade (IMACS, FourStar), UKIRT (WFCAM, UFTI), Keck:I (LRIS, MOSFIRE), Keck:II (DEIMOS), HST (WFC3), LBT (LBC, MODS), Gemini:South (GMOS, FLAMINGOS-2), Gemini:North (GMOS, NIRI), Swift (XRT)}

\software{PypeIt \citep{PypeIt}, IRAF \citep{iraf1,iraf2}, SExtractor \citep{SExtractor}}

\appendix 
\restartappendixnumbering

\section{Photometric Catalog} \label{appendix_photometry}

Here we present the \totfits\ photometric data points that make up the photometric catalog. We also list derived host galaxy positions from the available imaging, as described in Section~\ref{sec:astrometry}.

\startlongtable 
\tabletypesize{\footnotesize}
\begin{deluxetable*}{lccCcc}
\tablecolumns{6}
\tablewidth{0pc}
\tablecaption{Photometric Catalog
\label{tab:phot}}
\tablehead{
\colhead {GRB}	&
\colhead {RA}	&
\colhead {Decl.}	 &
\colhead {Filter} 		&
\colhead {$m_{\rm AB}$} &
\colhead{Ref.} \\
\colhead {}	&
\colhead {(J2000)}	 &
\colhead {(J2000)} 		&
\colhead {} 		&
\colhead {(AB mag)}	&
\colhead{}
}
\startdata
050509B & \ra{12}{36}{12.875} & \dec{+28}{58}{58.84} & $u$ & $20.32 \pm 0.13$ & \citet{SDSS-DR13} \\
        &                     &                      & $g$ & $18.52 \pm 0.02$ & \citet{SDSS-DR13} \\
        &                     &                      & $r$ & $17.12 \pm 0.01$ & \citet{SDSS-DR13} \\
        &                     &                      & $i$ & $16.60 \pm 0.01$ & \citet{SDSS-DR13} \\
        &                     &                      & $z$ & $16.25 \pm 0.01$ & \citet{SDSS-DR13} \\
        &                     &                      & $J$ & $16.16 \pm 0.14$ & \citealt{scs+06} \\
        &                     &                      & $H$ & $15.84 \pm 0.18$ & \citealt{scs+06} \\
        &                     &                      & $K$ & $15.98 \pm 0.16$ & \citealt{scs+06} \\
        &                     &                      & $F814W$ & $16.28 \pm 0.05$ & \citet{fbf10} \\
050709 & \ra{23}{01}{26.765} & \dec{-38}{58}{40.422} & B & 22.05 $\pm$ 0.10 & \citet{hwf+05} \\
 &  &  & V & 21.34 $\pm$ 0.07 & \citet{cmi+06} \\
 &  &  & R & 21.26 $\pm$ 0.07 & \citet{cmi+06} \\
 &  &  & I & 21.01 $\pm$ 0.08 & \citet{cmi+06} \\
 &  &  & J & 20.76 $\pm$ 0.08 & \citet{lb10} \\
 &  &  & K & 21.04 $\pm$ 0.16 & \citet{lb10} \\
 &  &  & F450W & 21.48 $\pm$ 0.05 & \citet{fbf10} \\
 &  &  & F814W & 21.11 $\pm$ 0.05 & \citet{fbf10} \\
050724 & \ra{16}{24}{44.41} & \dec{-27}{32}{26.393} & U & $>$ 26.42 & \citet{gcg+06} \\
 &  &  & B & 22.34 $\pm$ 0.12 & \citet{gcg+06} \\
 &  &  & V & 20.69 $\pm$ 0.05 & \citet{gcg+06} \\
 &  &  & R & 19.83 $\pm$ 0.03 & \citet{gcg+06} \\
 &  &  & I & 19.01 $\pm$ 0.20 & \citet{bpc+05} \\
 &  &  & J & 17.83 $\pm$ 0.04 & \citet{gcg+06} \\
 &  &  & H & 17.24 $\pm$ 0.05 & \citet{gcg+06} \\
 &  &  & K & 16.82 $\pm$ 0.05 & \citet{gcg+06} \\
 &  &  & F450W & 22.63 $\pm$ 0.05 & \citet{fbf10} \\
 &  &  & F814W & 19.93 $\pm$ 0.05 & \citet{fbf10} \\
050813 & \ra{16}{07}{57.2} & \dec{+11}{14}{53.09} & R & 23.43 $\pm$ 0.07 & \citet{pbc+06} \\
 &  &  & g & 25.45 $\pm$ 0.44 & \citet{sdh+21} \\
 &  &  & r & 23.03 $\pm$ 0.07 & \citet{sdh+21} \\
 &  &  & i & 22.52 $\pm$ 0.12 & \citet{fsk+07} \\
 &  &  & z & 21.77 $\pm$ 0.05 & \citet{sdh+21} \\
051210 & \ra{22}{00}{40.942} & \dec{-57}{36}{47.063} & g & 24.29 $\pm$ 0.34 & \citet{lb10} \\
 &  &  & r & 24.04 $\pm$ 0.15 & \citet{lb10} \\
 &  &  & i & 24.90 $\pm$ 0.22 & \citet{lb10} \\
 &  &  & z & 24.06 $\pm$ 0.21 & \citet{lb10} \\
 &  &  & K & $>$ 20.91 & \citet{lb10} \\
 &  &  & F675W & 21.19 $\pm$ 0.05 & \citet{fbf10} \\
051221A & \ra{21}{54}{48.653} & \dec{+16}{53}{27.335} & g & 23.74 $\pm$ 0.07 & \citet{lb10} \\
 &  &  & r & 22.18 $\pm$ 0.09 & \citet{sbk+06} \\
 &  &  & i & 22.13 $\pm$ 0.17 & \citet{sbk+06} \\
 &  &  & J & 22.01 $\pm$ 0.20 & \citet{lb10} \\
 &  &  & K & 22.30 $\pm$ 0.15 & \citet{lb10} \\
 &  &  & F555W & 22.09 $\pm$ 0.05 & \citet{fbf10} \\
 &  &  & F814W & 21.55 $\pm$ 0.05 & \citet{fbf10} \\
060121 & \ra{09}{09}{52.026} & \dec{+45}{39}{45.538} & F606W & 26.27 $\pm$ 1.00 & \citet{fbf10} \\
060313 & \ra{04}{26}{28.402} & \dec{-10}{50}{39.901} & F475W & 26.68 $\pm$ 1.00 & \citet{fbf10} \\
 &  &  & F775W & 25.75 $\pm$ 1.00 & \citet{fbf10} \\
060614 & \ra{21}{23}{32.102} & \dec{-53}{01}{36.436} & U & 22.38 $\pm$ 0.10 & \citet{gfp+06} \\
 &  &  & B & 23.02 $\pm$ 0.10 & \citet{gfp+06} \\
 &  &  & V & 22.77 $\pm$ 0.10 & \citet{gfp+06} \\
 &  &  & R & 22.57 $\pm$ 0.10 & \citet{gfp+06} \\
 &  &  & I & 21.95 $\pm$ 0.10 & \citet{gfp+06} \\
 &  &  & F606W & 22.76 $\pm$ 0.10 & \citet{gfp+06} \\
 &  &  & F814W & 21.95 $\pm$ 0.10 & \citet{gfp+06} \\
060801 & \ra{14}{12}{01.262} & \dec{+16}{58}{55.97} & g & 23.44 $\pm$ 0.09 & \citet{lb10} \\
 &  &  & r & 23.20 $\pm$ 0.11 & \citet{lb10} \\
 &  &  & i & 23.05 $\pm$ 0.19 & \citet{lb10} \\
 &  &  & z & 22.88 $\pm$ 0.10 & \citet{lb10} \\
 &  &  & J & $>$ 21.52 & \citet{lb10} \\
 &  &  & K & $>$ 19.91 & \citet{lb10} \\
061006 & \ra{07}{24}{07.808} & \dec{-79}{11}{55.188} & r & 24.15 $\pm$ 0.09 & \citet{lb10} \\
 &  &  & z & 23.28 $\pm$ 0.25 & \citet{lb10} \\
 &  &  & B & 25.75 $\pm$ 0.12 & \citet{dmc+09} \\
 &  &  & V & 24.56 $\pm$ 0.07 & \citet{dmc+09} \\
 &  &  & R & 24.14 $\pm$ 0.12 & \citet{dmc+09} \\
 &  &  & I & 23.44 $\pm$ 0.12 & \citet{dmc+09} \\
 &  &  & J & 22.91 $\pm$ 0.20 & \citet{dmc+09} \\
 &  &  & K & 22.60 $\pm$ 0.25 & \citet{lb10} \\
 &  &  & F814W & 22.29 $\pm$ 0.05 & \citet{fbf10} \\
 &  &  & F555W & 24.95 $\pm$ 0.05 & \citet{fbf10} \\
061201 & \nod & \nod & F160W & $>$ 26.40 & \citet{fb13} \\
061210 & \ra{09}{38}{05.362} & \dec{+15}{37}{18.877} & g & 23.28 $\pm$ 0.10 & \citet{lb10} \\
 &  &  & r & 21.40 $\pm$ 0.05 & \citet{lb10} \\
 &  &  & i & 21.63 $\pm$ 0.10 & \citet{lb10} \\
 &  &  & z & 21.32 $\pm$ 0.14 & \citet{lb10} \\
 &  &  & J & 21.32 $\pm$ 0.15 & \citet{lb10} \\
 &  &  & K & 20.33 $\pm$ 0.10 & \citet{lb10} \\
070429B & \ra{21}{52}{03.691} & \dec{-38}{49}{42.82} & g & 24.40 $\pm$ 0.20 & \citet{lb10} \\
 &  &  & r & 23.28 $\pm$ 0.04 & \citet{lb10} \\
 &  &  & i & 21.89 $\pm$ 0.09 & \citet{lb10} \\
 &  &  & z & 21.76 $\pm$ 0.12 & \citet{lb10} \\
 &  &  & F160W & 20.60 $\pm$ 0.03 & \citet{cbn+08} \\
070707 & \ra{17}{50}{58.555} & \dec{-68}{55}{27.6} & F160W & 26.04 $\pm$ 0.24 & \citet{fb13} \\
 &  &  & F606W & 26.86 $\pm$ 0.12 & \citet{fb13} \\
070714B & \ra{03}{51}{22.272} & \dec{+28}{17}{50.943} & g & 25.77 $\pm$ 0.34 & \citet{gfl+09} \\
 &  &  & r & 24.89 $\pm$ 0.21 & \citet{gfl+09} \\
 &  &  & i & 23.97 $\pm$ 0.12 & \citet{gfl+09} \\
 &  &  & z & 23.98 $\pm$ 0.13 & \citet{gfl+09} \\
 &  &  & J & 23.18 $\pm$ 0.12 & \citet{gfl+09} \\
 &  &  & H & 23.66 $\pm$ 0.20 & \citet{gfl+09} \\
 &  &  & K & 23.02 $\pm$ 0.13 & \citet{gfl+09} \\
 &  &  & F475W & 25.36 $\pm$ 0.06 & \citet{fb13} \\
 &  &  & F160W & 23.06 $\pm$ 0.02 & \citet{fb13} \\
070724 & \ra{01}{51}{14.068} & \dec{-18}{35}{38.47} & g & 21.56 $\pm$ 0.06 & \citet{lb10} \\
 &  &  & r & 20.78 $\pm$ 0.03 & \citet{lb10} \\
 &  &  & i & 20.46 $\pm$ 0.03 & \citet{lb10} \\
 &  &  & z & 20.28 $\pm$ 0.04 & \citet{lb10} \\
 &  &  & J & 20.02 $\pm$ 0.02 & \citet{lb10} \\
 &  &  & H & 19.79 $\pm$ 0.02 & \citet{lb10} \\
 &  &  & K & 19.71 $\pm$ 0.04 & \citet{lb10} \\
 &  &  & F160W & 19.90 $\pm$ 0.02 & \citet{fb13} \\
070729 & \ra{03}{45}{15.808} & \dec{-39}{19}{18.59} & r & 23.02 $\pm$ 0.26 & This work \\
 &  &  & i & 22.74 $\pm$ 0.36 & This work \\
 &  &  & J & 22.68 $\pm$ 0.18 & This work \\
 &  &  & K & 22.77 $\pm$ 0.37 & This work \\
070809 & \ra{13}{35}{04.177} & \dec{-22}{08}{33.01} & g & 22.15 $\pm$ 0.05 & \citet{lb10} \\
 &  &  & r & 20.14 $\pm$ 0.02 & \citet{lb10} \\
 &  &  & i & 19.46 $\pm$ 0.05 & \citet{lb10} \\
 &  &  & K & 17.99 $\pm$ 0.04 & \citet{lb10} \\
 &  &  & F160W & 18.26 $\pm$ 0.01 & \citet{fb13} \\
 &  &  & F606W & 20.68 $\pm$ 0.03 & \citet{fb13} \\
071227 & \ra{03}{52}{31.026} & \dec{-55}{59}{00.89} & g & 22.87 $\pm$ 0.13 & \citet{lb10} \\
 &  &  & r & 20.64 $\pm$ 0.05 & \citet{lb10} \\
 &  &  & i & 20.50 $\pm$ 0.04 & \citet{lb10} \\
 &  &  & z & 19.79 $\pm$ 0.03 & \citet{lb10} \\
 &  &  & J & 19.17 $\pm$ 0.06 & \citet{lb10} \\
 &  &  & K & 18.16 $\pm$ 0.06 & \citet{lb10} \\
 &  &  & F160W & 18.74 $\pm$ 0.01 & \citet{fb13} \\
080123 & \ra{22}{35}{46.943} & \dec{-64}{53}{54.973} & g & 22.16 $\pm$ 0.06 & \citet{lb10} \\
 &  &  & r & 20.96 $\pm$ 0.05 & \citet{lb10} \\
 &  &  & i & 20.54 $\pm$ 0.07 & \citet{lb10} \\
 &  &  & z & 20.16 $\pm$ 0.20 & \citet{lb10} \\
 &  &  & J & 20.32 $\pm$ 0.05 & \citet{lb10} \\
 &  &  & K & 19.59 $\pm$ 0.06 & \citet{lb10} \\
080503 & \ra{19}{06}{28.901} & \dec{+68}{47}{34.78} & F160W & 25.88 $\pm$ 0.07 & \citet{fb13} \\
 &  &  & F606W & 27.15 $\pm$ 0.20 & \citet{pmg+09} \\
080905A & \ra{19}{10}{42.045} & \dec{-18}{52}{54.51} & R & 18.00 $\pm$ 0.50 & \citet{rwl+10} \\
 &  &  & F160W & 25.97 $\pm$ 0.11 & \citet{fb13} \\
081226A & \ra{08}{22}{00.45} & \dec{-69}{01}{49.5} & g & 26.25 $\pm$ 0.24 & \citet{nkg+12} \\
 &  &  & r & 26.03 $\pm$ 0.34 & \citet{nkg+12} \\
 &  &  & i & $>$ 25.00 & \citet{nkg+12} \\
 &  &  & z & $>$ 24.50 & \citet{nkg+12} \\
 &  &  & J & $>$ 22.20 & \citet{nkg+12} \\
 &  &  & H & $>$ 21.60 & \citet{nkg+12} \\
 &  &  & K & $>$ 20.60 & \citet{nkg+12} \\
090305A & \ra{16}{07}{07.596} & \dec{-31}{33}{22.54} & F160W & 25.29 $\pm$ 0.10 & \citet{fb13} \\
090426A & \ra{12}{36}{18.05} & \dec{+32}{59}{09.42} & F160W & 25.57 $\pm$ 0.07 & \citet{fb13} \\
090510 & \ra{22}{14}{12.623} & \dec{-26}{34}{58.55} & g & 23.86 $\pm$ 0.08 & \citet{lb10} \\
 &  &  & i & 22.45 $\pm$ 0.14 & \citet{lb10} \\
 &  &  & z & 22.69 $\pm$ 0.17 & \citet{lb10} \\
 &  &  & J & 21.81 $\pm$ 0.15 & \citet{lb10} \\
 &  &  & F160W & 21.80 $\pm$ 0.01 & \citet{fb13} \\
090515 & \ra{10}{56}{35.847} & \dec{+14}{26}{42.84} & g & 21.97 $\pm$ 0.02 & \citet{lb10} \\
 &  &  & r & 20.27 $\pm$ 0.05 & \citet{lb10} \\
 &  &  & i & 19.49 $\pm$ 0.05 & \citet{lb10} \\
 &  &  & J & 18.69 $\pm$ 0.05 & \citet{lb10} \\
 &  &  & K & 18.24 $\pm$ 0.05 & \citet{lb10} \\
 &  &  & F160W & 18.42 $\pm$ 0.02 & \citet{fb13} \\
091109B & \ra{07}{30}{56.55} & \dec{-54}{05}{23.22} & F110W & 27.81 $\pm$ 0.24 & This work \\
100117 & \ra{00}{45}{04.661} & \dec{-01}{35}{42.02} & g & 26.27 $\pm$ 0.30 & \citet{fbc+11} \\
 &  &  & r & 24.40 $\pm$ 0.10 & \citet{fbc+11} \\
 &  &  & i & 22.90 $\pm$ 0.10 & \citet{fbc+11} \\
 &  &  & z & 22.37 $\pm$ 0.10 & \citet{fbc+11} \\
 &  &  & J & 21.89 $\pm$ 0.25 & \citet{fbc+11} \\
 &  &  & H & 21.27 $\pm$ 0.21 & \citet{fbc+11} \\
 &  &  & K & 21.25 $\pm$ 0.20 & \citet{fbc+11} \\
 &  &  & F160W & 21.38 $\pm$ 0.04 & \citet{fb13} \\
100206A & \ra{03}{08}{39.142} & \dec{+13}{09}{29.34} & g & 23.90 $\pm$ 0.17 & \citet{pmm+12} \\
 &  &  & R & 21.53 $\pm$ 0.09 & \citet{pmm+12} \\
 &  &  & i & 20.86 $\pm$ 0.08 & \citet{pmm+12} \\
 &  &  & z & 20.20 $\pm$ 0.05 & \citet{pmm+12} \\
 &  &  & J & 19.41 $\pm$ 0.12 & \citet{pmm+12} \\
 &  &  & H & 18.63 $\pm$ 0.09 & \citet{pmm+12} \\
 &  &  & K & 18.17 $\pm$ 0.11 & \citet{pmm+12} \\
 &  &  & W1 & 18.51 $\pm$ 0.06 & \citet{WISE} \\
 &  &  & W2 & 18.54 $\pm$ 0.11 & \citet{WISE} \\
 &  &  & W3 & 16.38 $\pm$ 0.16 & \citet{WISE} \\
 &  &  & W4 & $>$ 15.18 & \citet{WISE} \\
100625A & \ra{01}{03}{10.918} & \dec{-39}{05}{18.44} & g & 23.91 $\pm$ 0.19 & \citet{fb13} \\
 &  &  & r & 22.66 $\pm$ 0.09 & \citet{fb13} \\
 &  &  & i & 22.16 $\pm$ 0.04 & \citet{fb13} \\
 &  &  & z & 22.09 $\pm$ 0.10 & \citet{fb13} \\
 &  &  & J & 21.49 $\pm$ 0.05 & \citet{fb13} \\
 &  &  & K & 20.76 $\pm$ 0.10 & \citet{fb13} \\
101219A & \ra{04}{58}{20.497} & \dec{-02}{32}{22.45} & g & 24.76 $\pm$ 0.08 & \citet{fb13} \\
 &  &  & r & 24.08 $\pm$ 0.05 & \citet{fb13} \\
 &  &  & i & 23.29 $\pm$ 0.08 & \citet{fb13} \\
 &  &  & z & 23.29 $\pm$ 0.16 & \citet{fb13} \\
 &  &  & J & 22.47 $\pm$ 0.13 & \citet{fb13} \\
 &  &  & K & 21.57 $\pm$ 0.21 & \citet{fb13} \\
101224A & \ra{19}{03}{41.919} & \dec{+45}{42}{48.86} & g & 22.66 $\pm$ 0.07 & This work \\
 &  &  & r & 22.07 $\pm$ 0.05 & This work \\
 &  &  & i & 21.61 $\pm$ 0.05 & This work \\
 &  &  & z & 21.72 $\pm$ 0.09 & This work \\
 &  &  & Y & 21.79 $\pm$ 0.18 & This work \\
 &  &  & J & 21.72 $\pm$ 0.01 & This work \\
 &  &  & H & $>$ 22.10 & This work \\
 &  &  & K & 21.43 $\pm$ 0.18 & This work \\
110112A & \nod & \nod & F110W & $>$ 28.00 & This work \\
111117A & \ra{00}{50}{46.268} & \dec{+23}{00}{41.41} & g & 24.08 $\pm$ 0.09 & \citet{skm+18} \\
 &  &  & r & 23.79 $\pm$ 0.11 & \citet{mbf+12} \\
 &  &  & i & 23.71 $\pm$ 0.08 & \citet{mbf+12} \\
 &  &  & z & 23.08 $\pm$ 0.18 & \citet{mbf+12} \\
 &  &  & J & 23.13 $\pm$ 0.18 & \citet{skm+18} \\
 &  &  & H & 22.94 $\pm$ 0.29 & \citet{skm+18} \\
 &  &  & K & 23.07 $\pm$ 0.32 & \citet{skm+18} \\
120305A & \ra{03}{10}{08.754} & \dec{+28}{29}{35.87} & g & 23.54 $\pm$ 0.23 & This work \\
 &  &  & r & 22.40 $\pm$ 0.05 & This work \\
 &  &  & i & 21.42 $\pm$ 0.02 & This work \\
 &  &  & z & 21.60 $\pm$ 0.05 & This work \\
 &  &  & J & 21.23 $\pm$ 0.17 & This work \\
 &  &  & H & 21.56 $\pm$ 0.13 & This work \\
120804A & \ra{15}{35}{47.51} & \dec{-28}{46}{56.11} & r & 26.41 $\pm$ 0.20 & \citet{bzl+13} \\
 &  &  & i & 25.18 $\pm$ 0.15 & \citet{bzl+13} \\
 &  &  & Y & 23.93 $\pm$ 0.30 & \citet{bzl+13} \\
 &  &  & J & 23.16 $\pm$ 0.20 & \citet{bzl+13} \\
 &  &  & K & 22.07 $\pm$ 0.10 & \citet{bzl+13} \\
121226A & \ra{11}{14}{34.121} & \dec{-30}{24}{22.84} & g & 24.82 $\pm$ 0.10 & This work \\
 &  &  & r & 24.31 $\pm$ 0.06 & This work \\
 &  &  & i & 24.00 $\pm$ 0.10 & This work \\
 &  &  & z & 23.82 $\pm$ 0.15 & This work \\
 &  &  & J & 22.82 $\pm$ 0.10 & This work \\
 &  &  & H & 22.45 $\pm$ 0.11 & This work \\
 &  &  & K & 21.37 $\pm$ 0.10 & This work \\
130515A & \ra{18}{53}{45.021} & \dec{-54}{16}{50.72} & g & 22.46 $\pm$ 0.06 & This work \\
 &  &  & r & 22.65 $\pm$ 0.04 & This work \\
 &  &  & i & 20.90 $\pm$ 0.02 & This work \\
 &  &  & z & 20.89 $\pm$ 0.10 & This work \\
 &  &  & J & 20.56 $\pm$ 0.21 & This work \\
 &  &  & H & 20.56 $\pm$ 0.09 & This work \\
 &  &  & K & 20.29 $\pm$ 0.04 & This work \\
130603B & \ra{11}{28}{48.231} & \dec{+17}{04}{18.61} & g & 22.11 $\pm$ 0.06 & \citet{dtr+14} \\
 &  &  & r & 21.06 $\pm$ 0.06 & \citet{dtr+14} \\
 &  &  & i & 20.72 $\pm$ 0.06 & \citet{dtr+14} \\
 &  &  & z & 20.36 $\pm$ 0.06 & \citet{dtr+14} \\
 &  &  & J & 20.12 $\pm$ 0.07 & \citet{dtr+14} \\
 &  &  & H & 19.82 $\pm$ 0.06 & \citet{dtr+14} \\
 &  &  & K & 19.59 $\pm$ 0.07 & \citet{dtr+14} \\
 &  &  & F160W & 19.84 $\pm$ 0.02 & \citet{fb13} \\
 &  &  & F606W & 21.14 $\pm$ 0.04 & \citet{fb13} \\
130716A & \ra{11}{58}{17.862} & \dec{+63}{03}{15.35} & g & 24.67 $\pm$ 0.14 & This work \\
 &  &  & r & 24.89 $\pm$ 0.34 & This work \\
 &  &  & i & 24.82 $\pm$ 0.26 & This work \\
 &  &  & z & $>$ 24.20 & This work \\
 &  &  & J & $>$ 22.40 & This work \\
130822A & \ra{01}{51}{42.708} & \dec{-03}{12}{25.447} & g & 18.86 $\pm$ 0.07 & This work \\
 &  &  & r & 18.25 $\pm$ 0.06 & This work \\
 &  &  & i & 17.91 $\pm$ 0.03 & This work \\
 &  &  & z & 17.55 $\pm$ 0.08 & This work \\
 &  &  & J & 17.29 $\pm$ 0.05 & This work \\
130912A & \ra{03}{10}{22.2} & \dec{+13}{59}{48.74} & F110W & 27.47 $\pm$ 0.23 & This work \\
131004A & \ra{19}{44}{27.064} & \dec{-02}{57}{30.429} & F110W & 25.46 $\pm$ 0.09 & This work \\
140129B & \ra{21}{47}{01.649} & \dec{+26}{12}{23.27} & g & 24.18 $\pm$ 0.10 & This work \\
 &  &  & r & 23.55 $\pm$ 0.07 & This work \\
 &  &  & i & 22.68 $\pm$ 0.07 & This work \\
 &  &  & z & 22.95 $\pm$ 0.14 & This work \\
 &  &  & Y & 22.68 $\pm$ 0.21 & This work \\
 &  &  & J & 22.78 $\pm$ 0.19 & This work \\
 &  &  & H & 22.21 $\pm$ 0.33 & This work \\
 &  &  & K & 21.99 $\pm$ 0.34 & This work \\
140516A & \nod & \nod & $i$ & $>$ 26.1 & This work \\
 &  &  & K & $>$ 23.6 & This work \\
140622A & \ra{21}{08}{41.744} & \dec{-14}{25}{06.166} & g & 23.35 $\pm$ 0.06 & This work \\
 &  &  & r & 22.70 $\pm$ 0.04 & This work \\
 &  &  & i & 22.09 $\pm$ 0.03 & This work \\
 &  &  & z & 21.72 $\pm$ 0.06 & This work \\
 &  &  & J & 21.54 $\pm$ 0.10 & This work \\
 &  &  & H & 21.19 $\pm$ 0.10 & This work \\
 &  &  & K & 20.41 $\pm$ 0.08 & This work \\
140903A & \ra{15}{52}{03.265} & \dec{+27}{36}{10.71} & g & 21.97 $\pm$ 0.16 & \citet{tsc+16} \\
 &  &  & r & 21.37 $\pm$ 0.19 & This work \\
 &  &  & i & 20.46 $\pm$ 0.15 & This work \\
 &  &  & z & 19.64 $\pm$ 0.13 & \citet{tsc+16} \\
 &  &  & J & 19.03 $\pm$ 0.08 & This work \\
 &  &  & H & 18.32 $\pm$ 0.06 & This work \\
 &  &  & K & 18.19 $\pm$ 0.05 & This work \\
140930B & \ra{00}{25}{23.473} & \dec{+24}{17}{37.931} & g & 24.45 $\pm$ 0.23 & This work \\
 &  &  & r & 24.21 $\pm$ 0.25 & This work \\
 &  &  & i & 24.09 $\pm$ 0.21 & This work \\
 &  &  & z & $>$ 23.20 & This work \\
 &  &  & Y & 23.59 $\pm$ 0.17 & This work \\
 &  &  & J & 23.65 $\pm$ 0.33 & This work \\
 &  &  & H & 23.16 $\pm$ 0.17 & This work \\
 &  &  & K & 22.62 $\pm$ 0.16 & This work \\
141212A & \ra{02}{36}{29.957} & \dec{+18}{08}{47.228} & g & 24.04 $\pm$ 0.10 & This work \\
 &  &  & r & 22.95 $\pm$ 0.06 & This work \\
 &  &  & i & 22.29 $\pm$ 0.05 & This work \\
 &  &  & z & 22.06 $\pm$ 0.06 & This work \\
 &  &  & Y & 21.53 $\pm$ 0.30 & This work \\
 &  &  & J & 21.95 $\pm$ 0.24 & This work \\
 &  &  & H & 21.44 $\pm$ 0.16 & This work \\
 &  &  & K & 21.65 $\pm$ 0.24 & This work \\
150101B & \ra{12}{32}{04.973} & \dec{-10}{56}{00.5} & g & 17.56 $\pm$ 0.04 & \citet{fmc+16} \\
 &  &  & r & 16.60 $\pm$ 0.04 & \citet{fmc+16} \\
 &  &  & i & 16.15 $\pm$ 0.04 & \citet{fmc+16} \\
 &  &  & z & 15.82 $\pm$ 0.05 & \citet{fmc+16} \\
 &  &  & J & 15.50 $\pm$ 0.05 & \citet{fmc+16} \\
 &  &  & K & 15.12 $\pm$ 0.05 & \citet{fmc+16} \\
 &  &  & F160W & 15.11 $\pm$ 0.01 & \citet{fmc+16} \\
 &  &  & F606W & 16.67 $\pm$ 0.01 & \citet{fmc+16} \\
150120A & \ra{00}{41}{16.563} & \dec{+33}{59}{42.598} & g & 23.35 $\pm$ 0.06 & This work \\
 &  &  & r & 22.05 $\pm$ 0.06 & This work \\
 &  &  & i & 21.44 $\pm$ 0.05 & This work \\
 &  &  & z & 21.02 $\pm$ 0.06 & This work \\
 &  &  & Y & 20.63 $\pm$ 0.10 & This work \\
 &  &  & J & 20.44 $\pm$ 0.08 & This work \\
 &  &  & H & 20.31 $\pm$ 0.05 & This work \\
 &  &  & K & 19.45 $\pm$ 0.06 & This work \\
150423A & \nod & \nod & F110W & $\gtrsim 28.1$ & This work \\
150424A & \ra{10}{09}{13.406} & \dec{-26}{37}{51.745} & F125W & 26.29 $\pm$ 0.15 & This work \\
 &  &  & F160W & 25.89 $\pm$ 0.14 & This work \\
150728A & \ra{19}{28}{54.808} & \dec{+33}{54}{58.22} & g & 22.45 $\pm$ 0.07 & This work \\
 &  &  & r & 21.42 $\pm$ 0.05 & This work \\
 &  &  & i & 20.99 $\pm$ 0.05 & This work \\
 &  &  & z & 20.97 $\pm$ 0.06 & This work \\
 &  &  & Y & 20.84 $\pm$ 0.11 & This work \\
 &  &  & J & 20.66 $\pm$ 0.07 & This work \\
 &  &  & H & 20.32 $\pm$ 0.05 & This work \\
 &  &  & K & 20.26 $\pm$ 0.08 & This work \\
150831A & \ra{14}{44}{05.939} & \dec{-25}{38}{05.78} & g & 26.00 $\pm$ 0.32 & This work \\
 &  &  & r & 24.43 $\pm$ 0.45 & This work \\
 &  &  & i & 24.67 $\pm$ 0.10 & This work \\
 &  &  & z & 23.74 $\pm$ 0.10 & This work \\
 &  &  & Y & $>$ 21.93 & This work \\
 &  &  & J & 23.59 $\pm$ 0.30 & This work \\
 &  &  & H & 22.87 $\pm$ 0.32 & This work \\
 &  &  & K & 22.63 $\pm$ 0.30 & This work \\
151229A & \ra{21}{57}{28.701} & \dec{-20}{43}{54.8} & g & $>$ 23.89 & This work \\
 &  &  & r & $>$ 24.49 & This work \\
 &  &  & i & 24.92 $\pm$ 0.13 & This work \\
 &  &  & z & 24.46 $\pm$ 0.08 & This work \\
 &  &  & Y & 24.41 $\pm$ 0.35 & This work \\
 &  &  & J & 24.16 $\pm$ 0.39 & This work \\
 &  &  & H & $>$ 21.38 & This work \\
 &  &  & K & 22.80 $\pm$ 0.26 & This work \\
160303A & \ra{11}{14}{48.119} & \dec{+22}{44}{33.42} & g & $>$ 25.75 & This work \\
 &  &  & r & 25.80 $\pm$ 0.30 & \citet{gcn19160} \\
 &  &  & i & 25.33 $\pm$ 0.24 & This work \\
 &  &  & z & 23.67 $\pm$ 0.16 & This work \\
 &  &  & F110W & 23.77 $\pm$ 0.02 & This work \\
 &  &  & J & $>$ 22.01 & This work \\
 &  &  & K & 22.83 $\pm$ 0.39 & This work \\
160408A & \ra{08}{10}{29.56} & \dec{+71}{07}{44.978} & g & 25.52 $\pm$ 0.27 & This work \\
 &  &  & r & 25.74 $\pm$ 0.16 & This work \\
 &  &  & i & $>$ 25.20 & This work \\
 &  &  & z & 25.23 $\pm$ 0.27 & This work \\
 &  &  & J & $>$ 22.80 & This work \\
160410A & \nod & \nod & r & $>$ 27.22 & \citet{atk+21} \\
 &  &  & z & $>$ 25.20 & This work \\
 &  &  & J & $>$ 22.20 & This work \\
 &  &  & K & $>$ 23.80 & This work \\
160411A & \ra{23}{17}{25.355} & \dec{-40}{14}{30.56} & g & 25.26 $\pm$ 0.35 & This work \\
 &  &  & r & 24.58 $\pm$ 0.13 & This work \\
 &  &  & i & 24.18 $\pm$ 0.12 & This work \\
 &  &  & z & 24.04 $\pm$ 0.21 & This work \\
 &  &  & J & 23.16 $\pm$ 0.17 & This work \\
 &  &  & H & $>$ 23.31 & This work \\
 &  &  & K & 23.19 $\pm$ 0.24 & This work \\
160525B & \ra{09}{57}{32.227} & \dec{+51}{12}{24.813} & i & 24.08 $\pm$ 0.30 & \citet{cmm+16} \\
160601A & \ra{15}{39}{43.949} & \dec{+64}{32}{30.604} & i & $>$ 24.10 & This work \\
 &  &  & z & 24.95 $\pm$ 0.34 & This work \\
 &  &  & J & $>$ 22.80 & This work \\
 &  &  & K & $>$ 22.80 & This work \\
160624A & \ra{22}{00}{46.145} & \dec{+29}{38}{39.336} & g & 23.09 $\pm$ 0.06 & This work \\
 &  &  & r & 21.91 $\pm$ 0.05 & This work \\
 &  &  & i & 21.59 $\pm$ 0.04 & This work \\
 &  &  & z & 21.42 $\pm$ 0.06 & This work \\
 &  &  & Y & 21.24 $\pm$ 0.12 & This work \\
 &  &  & J & 20.85 $\pm$ 0.06 & This work \\
 &  &  & H & 20.58 $\pm$ 0.08 & This work \\
 &  &  & K & 20.33 $\pm$ 0.09 & This work \\
160821B & \ra{18}{39}{53.994} & \dec{+62}{23}{34.427} & g & 20.03 $\pm$ 0.00 & This work \\
 &  &  & r & 19.55 $\pm$ 0.00 & This work \\
 &  &  & i & 19.48 $\pm$ 0.01 & This work \\
 &  &  & z & 19.34 $\pm$ 0.01 & This work \\
 &  &  & F110W & 19.30 $\pm$ 0.01 & This work \\
 &  &  & F606W & 19.67 $\pm$ 0.02 & This work \\
 &  &  & F160W & 19.19 $\pm$ 0.01 & This work \\
160927A & \nod & \nod & G & $>$ 25.70 & This work \\
 &  &  & r & $>$ 25.20 & This work \\
 &  &  & i & $>$ 24.40 & This work \\
 &  &  & J & $>$ 24.40 & This work \\
 &  &  & K & $>$ 23.80 & This work \\
161001A & \ra{04}{47}{40.53} & \dec{-57}{15}{39.184} & g & 24.11 $\pm$ 0.07 & This work \\
 &  &  & r & 22.97 $\pm$ 0.05 & This work \\
 &  &  & i & 22.19 $\pm$ 0.04 & This work \\
 &  &  & z & 21.49 $\pm$ 0.06 & This work \\
 &  &  & J & 21.84 $\pm$ 0.09 & This work \\
 &  &  & K & 21.26 $\pm$ 0.08 & This work \\
161104A & \ra{05}{11}{34.37} & \dec{-51}{27}{36.29} & g & 25.49 $\pm$ 0.25 & \citet{nfd+20} \\
 &  &  & r & 23.85 $\pm$ 0.10 & \citet{nfd+20} \\
 &  &  & i & 22.75 $\pm$ 0.06 & \citet{nfd+20} \\
 &  &  & z & 22.16 $\pm$ 0.07 & \citet{nfd+20} \\
 &  &  & J & 21.57 $\pm$ 0.04 & \citet{nfd+20} \\
170127B & \ra{01}{19}{54.415} & \dec{-30}{21}{29.615} & G & 25.26 $\pm$ 0.16 & This work \\
 &  &  & r & 25.32 $\pm$ 0.29 & This work \\
 &  &  & I & 25.81 $\pm$ 0.33 & This work \\
 &  &  & z & 24.27 $\pm$ 0.21 & This work \\
 &  &  & J & $>$ 23.40 & This work \\
 &  &  & K & $>$ 22.80 & This work \\
170428A & \ra{22}{00}{18.71} & \dec{+26}{54}{56.28} & g & 23.77 $\pm$ 0.18 & This work \\
 &  &  & r & 22.35 $\pm$ 0.10 & This work \\
 &  &  & i & 22.55 $\pm$ 0.07 & This work \\
 &  &  & z & 22.09 $\pm$ 0.11 & This work \\
 &  &  & Y & 21.08 $\pm$ 0.16 & This work \\
 &  &  & J & 21.48 $\pm$ 0.12 & This work \\
 &  &  & H & 21.02 $\pm$ 0.15 & This work \\
 &  &  & K & 20.93 $\pm$ 0.09 & This work \\
170728A & \ra{03}{55}{33.111} & \dec{+10}{12}{50.879} & g & 25.51 $\pm$ 0.19 & This work \\
 &  &  & R & 24.73 $\pm$ 0.14 & This work \\
 &  &  & i & 24.31 $\pm$ 0.28 & This work \\
 &  &  & z & 24.40 $\pm$ 0.33 & This work \\
 &  &  & J & 22.15 $\pm$ 0.09 & This work \\
 &  &  & K & 21.22 $\pm$ 0.12 & This work \\
170728B & \ra{15}{51}{55.529} & \dec{+70}{07}{22.038} & r & 23.31 $\pm$ 0.10 & This work \\
 &  &  & J & 22.24 $\pm$ 0.13 & This work \\
 &  &  & K & 21.73 $\pm$ 0.27 & This work \\
170817 & \ra{13}{09}{47.7} & \dec{-23}{23}{02} & NUV & 17.82 $\pm$ 0.09 & \citet{bbf+17} \\
 &  &  & g & 13.19 $\pm$ 0.02 & \citet{bbf+17} \\
 &  &  & r & 12.44 $\pm$ 0.01 & \citet{bbf+17} \\
 &  &  & i & 12.02 $\pm$ 0.01 & \citet{bbf+17} \\
 &  &  & z & 11.73 $\pm$ 0.01 & \citet{bbf+17} \\
 &  &  & Y & 11.49 $\pm$ 0.02 & \citet{bbf+17} \\
 &  &  & J & 11.07 $\pm$ 0.02 & \citet{bbf+17} \\
 &  &  & H & 10.88 $\pm$ 0.02 & \citet{bbf+17} \\
 &  &  & K & 11.06 $\pm$ 0.02 & \citet{bbf+17} \\
 &  &  & W1 & 11.94 $\pm$ 0.01 & \citet{bbf+17} \\
 &  &  & W2 & 12.61 $\pm$ 0.01 & \citet{bbf+17} \\
 &  &  & W3 & 13.70 $\pm$ 0.04 & \citet{bbf+17} \\
 &  &  & W4 & 13.86 $\pm$ 0.18 & \citet{bbf+17} \\
180418A & \ra{11}{20}{29.21} & \dec{+24}{55}{58.734} & g & $>$ 25.76 & \citet{rfv+21} \\
 &  &  & r & 25.73 $\pm$ 0.21 & \citet{rfv+21} \\
 &  &  & i & 24.85 $\pm$ 0.14 & \citet{rfv+21} \\
 &  &  & z & 24.62 $\pm$ 0.21 & \citet{rfv+21} \\
 &  &  & Y & $>$ 23.32 & \citet{rfv+21} \\
 &  &  & J & 23.35 $\pm$ 0.36 & \citet{rfv+21} \\
 &  &  & H & $>$ 22.81 & \citet{rfv+21} \\
 &  &  & K & $>$ 22.41 & \citet{rfv+21} \\
180618A & \ra{11}{19}{45.801} & \dec{+73}{50}{15.03} & i & 22.18 $\pm$ 0.08 & This work \\
 &  &  & J & 22.42 $\pm$ 0.19 & This work \\
 &  &  & K & 21.91 $\pm$ 0.19 & This work \\
180727A & \ra{23}{06}{40.038} & \dec{-63}{03}{07.088} & g & 26.62 $\pm$ 0.21 & This work \\
 &  &  & r & 26.49 $\pm$ 0.28 & This work \\
 &  &  & i & 26.05 $\pm$ 0.35 & This work \\
 &  &  & z & 25.86 $\pm$ 0.37 & This work \\
180805B & \ra{01}{43}{07.655} & \dec{-17}{29}{33.091} & G & 23.71 $\pm$ 0.09 & This work \\
 &  &  & r & 22.15 $\pm$ 0.06 & This work \\
 &  &  & I & 22.22 $\pm$ 0.06 & This work \\
 &  &  & RG850 & 22.21 $\pm$ 0.07 & This work \\
 &  &  & J & 22.09 $\pm$ 0.08 & This work \\
 &  &  & K & 21.49 $\pm$ 0.14 & This work \\
181123B & \ra{12}{17}{27.91} & \dec{+14}{35}{52.27} & g & 24.20 $\pm$ 0.23 & \citet{pfn+20} \\
 &  &  & r & 23.92 $\pm$ 0.19 & \citet{pfn+20} \\
 &  &  & i & 23.85 $\pm$ 0.19 & \citet{pfn+20} \\
 &  &  & z & 23.88 $\pm$ 0.22 & \citet{pfn+20} \\
 &  &  & Y & 22.81 $\pm$ 0.24 & \citet{pfn+20} \\
 &  &  & J & 22.88 $\pm$ 0.23 & \citet{pfn+20} \\
 &  &  & H & 22.63 $\pm$ 0.19 & \citet{pfn+20} \\
 &  &  & K & 22.34 $\pm$ 0.23 & \citet{pfn+20} \\
191031D & \ra{18}{53}{08.8988} & \dec{+47}{38}{36.538} & g & $>$ 24.90 & This work \\
 &  &  & r & 24.46 $\pm$ 0.26 & This work \\
 &  &  & i & 24.83 $\pm$ 0.30 & This work \\
 &  &  & z & 24.36 $\pm$ 0.38 & This work \\
 &  &  & Y & 23.78 $\pm$ 0.27 & This work \\
 &  &  & J & 22.87 $\pm$ 0.01 & This work \\
 &  &  & K & 21.85 $\pm$ 0.08 & This work \\
200219A & \ra{22}{50}{33.108} & \dec{-59}{07}{11.579} & g & 21.96 $\pm$ 0.01 & \citet{sdh+21} \\
 &  &  & r & 20.66 $\pm$ 0.05 & \citet{sdh+21} \\
 &  &  & z & 19.76 $\pm$ 0.03 & \citet{sdh+21} \\
 &  &  & W1 & 18.62 $\pm$ 0.02 & \citet{sdh+21} \\
 &  &  & W2 & 19.62 $\pm$ 0.07 & \citet{sdh+21} \\
200411A & \ra{03}{10}{39.135} & \dec{-52}{18}{59.545} & g & 22.51 $\pm$ 0.09 & \citet{sdh+21} \\
 &  &  & r & 22.56 $\pm$ 0.04 & \citet{sdh+21} \\
 &  &  & z & 21.12 $\pm$ 0.03 & \citet{sdh+21} \\
200522A & \ra{00}{22}{43.717} & \dec{-00}{16}{57.466} & u & 20.54 $\pm$ 0.31 & \citet{aaa+15} \\
 &  &  & G & 22.27 $\pm$ 0.02 & \citet{flr+21} \\
 &  &  & R & 21.20 $\pm$ 0.02 & \citet{flr+21} \\
 &  &  & I & 20.97 $\pm$ 0.01 & \citet{flr+21} \\
 &  &  & Z & 20.87 $\pm$ 0.01 & \citet{flr+21} \\
 &  &  & y & 20.90 $\pm$ 0.30 & \citet{cmm+16} \\
 &  &  & Spitzer-1 & 21.07 $\pm$ 0.10 & \citet{Papovich2016} \\
 &  &  & Spitzer-2 & 21.30 $\pm$ 0.10 & \citet{Timlin2016} \\
 &  &  & F160W & 20.66 $\pm$ 0.01 & \citet{flr+21} \\
 &  &  & F125W & 20.86 $\pm$ 0.01 & \citet{flr+21} \\
200907B & \ra{05}{56}{06.951} & \dec{+06}{54}{22.637} & i & 23.94 $\pm$ 0.11 & This work \\
 &  &  & z & 24.01 $\pm$ 0.40 & This work \\
 &  &  & J & 22.55 $\pm$ 0.13 & This work \\
 &  &  & K & 22.06 $\pm$ 0.11 & This work \\
201006A & \nod & \nod & K & $>$ 23.60 & This work \\
201221D & \ra{11}{24}{14.064} & \dec{+42}{08}{40.047} & g & 23.20 $\pm$ 0.20 & \citet{gcn29133} \\
 &  &  & r & 23.42 $\pm$ 0.08 & This work \\
 &  &  & i & 23.36 $\pm$ 0.13 & This work \\
 &  &  & Y & 23.01 $\pm$ 0.13 & This work \\
 &  &  & J & 22.33 $\pm$ 0.09 & This work \\
 &  &  & K & 22.21 $\pm$ 0.13 & This work \\
210323A & \ra{21}{11}{47.32} & \dec{+25}{22}{09.989} & B & 24.77 $\pm$ 0.13 & This work \\
 &  &  & R & 24.70 $\pm$ 0.11 & This work \\
 &  &  & g & 24.88 $\pm$ 0.11 & This work \\
 &  &  & r & 24.97 $\pm$ 0.25 & This work \\
 &  &  & i & 23.98 $\pm$ 0.14 & This work \\
210726A & \ra{12}{53}{09.638} & \dec{+19}{11}{27.319} & g & 22.69 $\pm$ 0.04 & \citet{sdh+21} \\
 &  &  & r & 22.03 $\pm$ 0.20 & This work \\
 &  &  & z & 21.96 $\pm$ 0.08 & \citet{sdh+21} \\
 &  &  & J & 21.89 $\pm$ 0.15 & This work \\
210919A & \ra{05}{21}{01.954} & \dec{+01}{18}{40.022} & u & 25.73 $\pm$ 0.98 & \citet{sdh+21} \\
 &  &  & g & 21.51 $\pm$ 0.07 & \citet{sdh+21} \\
 &  &  & r & 20.50 $\pm$ 0.05 & \citet{sdh+21} \\
 &  &  & i & 19.93 $\pm$ 0.05 & \citet{sdh+21} \\
 &  &  & z & 19.60 $\pm$ 0.14 & \citet{sdh+21} \\
 &  &  & J & 19.08 $\pm$ 0.07 & This work \\
 &  &  & H & 18.88 $\pm$ 0.06 & This work \\
211023B & \ra{11}{21}{14.311} & \dec{+39}{08}{08.36} & g & 24.32 $\pm$ 0.22 & \citet{sdh+21} \\
 &  &  & r & 24.36 $\pm$ 0.38 & \citet{sdh+21} \\
 &  &  & i & 23.35 $\pm$ 0.14 & This work \\
 &  &  & z & 23.31 $\pm$ 0.24 & \citet{sdh+21} \\
211106A & \ra{22}{54}{20.541} & \dec{-53}{13}{50.548} & F110W & 25.71 $\pm$ 0.02 & This work \\
 &  &  & F814W & 25.79 $\pm$ 0.07 & This work \\
211211A & \ra{14}{09}{10.467} & \dec{+27}{53}{21.05} & u & 20.93 $\pm$ 0.13 & \citet{SDSS-DR13} \\
 &  &  & g & 19.84 $\pm$ 0.06 & \citet{rgl+22} \\
 &  &  & r & 19.46 $\pm$ 0.04 & \citet{rgl+22} \\
 &  &  & i & 19.19 $\pm$ 0.05 & \citet{rgl+22} \\
 &  &  & z & 19.20 $\pm$ 0.08 & \citet{rgl+22} \\
 &  &  & J & 19.01 $\pm$ 0.01 & \citet{rgl+22} \\
 &  &  & K & 19.23 $\pm$ 0.07 & \citet{rgl+22} \\
 &  &  & F140W & 18.96 $\pm$ 0.01 & \citet{rgl+22} \\
 &  &  & F606W & 19.57 $\pm$ 0.01 & \citet{rgl+22} \\
 &  &  & W1 & 19.76 $\pm$ 0.05 & \citet{WISE} 
\enddata
\tablecomments{Host galaxy positions and photometry from the literature and this work that have been incorporated into the BRIGHT catalog. We emphasize that the literature data set is comprehensive for a given host in that we attempt to fill out the SED, but does not include all literature photometry that exists for every host galaxy. For bursts with no identified host, $3\sigma$ limits at the afterglow position are reported. Photometry is in AB magnitudes and is not corrected for extinction in the direction of the bursts. All positions and photometry are also on the BRIGHT website (\url{https://bright.ciera.northwestern.edu/}).
}
\end{deluxetable*}

\section{Burst classifications} \label{appendix}

\startlongtable 
\tabletypesize{\footnotesize}
\begin{deluxetable*}{lcccc}
\tablecolumns{10}
\tablewidth{0pc}
\tablecaption{Classifications of Events in our Sample
\label{tab:class}}
\tablehead{
\colhead {GRB}	&
\colhead {$T_{90}$}	&
\colhead {This paper Class$^{\star}$}	 &
\colhead {\citet{bnp+13} $f_{\rm NC}^{\dagger}$} 		&
\colhead {\citet{jss+20} Class$^{\ddagger}$} 	 \\	
\colhead {}	&
\colhead {(sec)}	 &
\colhead {} 		&
\colhead {} 		&
\colhead {}		    
}
\startdata
050509B & 0.024 & SGRB & 0.94 & S$^{a}$ \\
050724A & 98 & SGRB-EE & \nod & S \\
050813 & 0.38 & SGRB & 0.61 & S \\
051210 & 1.3 & SGRB & 0.81 & S \\
051221A & 1.4 & SGRB & 0.18 & S \\
060313 & 0.74 & SGRB & 0.91 & S \\
060614 & 108.7 & possible SGRB-EE & \nod & L \\
060801 & 0.49 & SGRB & 0.95 & S \\
061006 & 129.9 & SGRB-EE & \nod & S \\
061201 & 0.76 & SGRB & 0.91 & S \\
061210 & 85.3 & SGRB-EE & \nod & S \\
070429B & 0.47 & SGRB & 0.54 & S \\
070714B & 64 & SGRB-EE & \nod & L \\
070724 & 0.4 & SGRB & 0.34 & S \\
070729 & 0.9 & SGRB & 0.87 & S \\
070809 & 1.3 & SGRB & 0.09 & L \\
071227 & 142.5 & SGRB-EE & \nod & L\\
080123 & 115 & possible SGRB-EE & \nod & L \\
080503 & 170 & SGRB-EE & \nod & L \\
080905A & 1.0 & SGRB & 0.87 & S \\
081226A & 0.4 & SGRB & 0.57 & S \\
090305A & 0.54 & SGRB & 0.94 & S \\
090426A & 1.2 & SGRB & 0.1 & L \\
090510 & 5.66 & possible SGRB-EE & 0.97 & S \\
090515 & 0.036 & SGRB & 0.92 & S$^{a}$ \\ 
091109B & 0.3 & SGRB & 0.97 & S \\
100117A & 0.3 & SGRB & 0.97 & S \\ 
100206A & 0.12 & SGRB & 0.98 & S \\
100625A & 0.33 & SGRB & 0.97 & S \\
101219A & 0.83 & SGRB & 0.90 & S \\
101224A & 0.2 & SGRB & 0.98 & S \\
110112A & 0.5 & SGRB & 0.29 & S \\
111117A & 0.47 & SGRB & 0.95 & S \\
120305A & 0.1 & SGRB & 0.99 & S \\
120804A & 0.81 & SGRB & 0.36 & S \\
121226A & 1.0 & SGRB & 0.29 & L \\
130515A & 0.29 & SGRB & 0.97 & S \\
130603B & 0.18 & SGRB & 0.98 & S \\
130716A & 87.7 & possible SGRB-EE & \nod & L \\
130822A & 0.04 & SGRB & 0.76 & S$^{a}$ \\
130912A & 0.28 & SGRB & 0.69 & S \\
131004A & 1.54 & SGRB & 0.07 & L \\
140129B & 1.36 & SGRB & 0.08 & L \\
140516A & 0.26 & SGRB & 0.49 & S \\
140622A & 0.13 & SGRB & 0.64 & S \\
140903A & 0.3 & SGRB & 0.45 & S \\
141212A & 0.3 & SGRB & 0.68 & S \\
150101B & 0.018 & SGRB & 0.95 & S$^{a}$ \\
150120A & 1.2 & SGRB & 0.10 & L \\
150423A & 0.22 & SGRB & 0.98 & S \\
150424A & 81 & SGRB-EE & \nod & \nod \\
150728A & 0.83 & SGRB & 0.17 & S \\
150831A & 0.92 & SGRB & 0.88 & S \\
151229A & 1.44 & SGRB & 0.08 & L \\
160303A & 5.0 & SGRB & 0.22 & S \\
160408A & 0.32 & SGRB & 0.92 & S \\
160410A & 96 & possible SGRB-EE & \nod & L \\
160411A & 0.36 & SGRB & 0.63 & S \\
160525B & 0.29 & SGRB & 0.46 & S \\
160601A & 0.12 & SGRB & 0.98 & S \\
160624A & 0.2 & SGRB & 0.98 & S \\
160821B & 0.48 & SGRB & 0.31 & S \\
160927A & 0.48 & SGRB & 0.95 & S \\
161001A & 2.6 & SGRB & 0.55 & L \\
161104A & 0.1 & SGRB & 0.86 & S \\
170127B & 0.51 & SGRB & 0.95 & S \\
170428A & 0.2 & SGRB & 0.98 & S \\
170728B & 47.7 & possible SGRB-EE & \nod & L \\
180418A & 2.29 & possible SGRB-EE & 0.02 & L \\
180618A & 47.4 & SGRB-EE & \nod & L\\
180727A & 1.1 & SGRB & 0.27 & L \\
180805B & 122.5 & SGRB-EE & \nod & L \\
181123B & 0.26 & SGRB & 0.97 & S \\ 
191031D & 0.29 & SGRB & 0.97 & \nod \\
200219A & 288 & SGRB-EE & \nod & \nod \\
200411A & 0.22 & SGRB & 0.98 & \nod \\
200522A & 0.62 & SGRB & 0.46 & \nod \\
200907B & 0.83 & SGRB & 0.90 & \nod \\
201006A & 0.49 & SGRB & 0.54 & \nod \\
201221D & 0.16 & SGRB & 0.80 & \nod \\ 
210323A & 1.12 & SGRB & 0.25 & \nod \\
210726A & 0.39 & SGRB & 0.61 & \nod \\ 
210919A & 0.16 & SGRB & 0.80 & \nod \\
211023B & 1.3 & SGRB & 0.09 & \nod \\
211106A & 1.75 & SGRB & 0.99 & \nod \\
211211A & 51.37 & merger-driven LGRB & \nod & \nod
\enddata
\tablecomments{
$^\star$ Classification of canonical short GRB (SGRB), short GRB with extended emission (SGRB-EE), or possible short GRB with extended emission (possible SGRB-EE) according to \citet{lsb+16}. GRB\,211211A is classified as a merger-driven long-duration GRB. \\
$^{\dagger}$ Probability that a burst is not a collapsar following the methods of \citet{bnp+13}. \\
$^{\ddagger}$ Long GRB (``L'') and short GRB (``S'') classifications following the methods of \citet{jss+20}. \\
$^{a}$ These bursts have durations which are too short for the \citet{jss+20} criteria to be applied. For the purposes of our comparative analysis, we classify them as short. }
\end{deluxetable*}

\end{CJK*}
\bibliography{refs}

\end{document}